\documentclass[11pt]{article}
\usepackage{amsbsy}
\usepackage{amssymb}
\newcommand{\beq}{\begin{equation}}
\newcommand{\eeq}{\end{equation}}
\newcommand{\bdm}{\begin{displaymath}}
\newcommand{\edm}{\end{displaymath}}
\newcommand{\beqr}{\begin{eqnarray}}
\newcommand{\eeqr}{\end{eqnarray}}
\newcommand{\beqrn}{\begin{eqnarray*}}
\newcommand{\eeqrn}{\end{eqnarray*}}
\def\l{\lambda}
\def\R{{\Bbb R}}
\def\bchi{\boldsymbol{\chi}}
\def\ba{\boldsymbol{a}}

\def\ve{\varepsilon}
\def\a{\alpha}
\def\k{\kappa}
\def\l{\lambda}
\def\nn{\nonumber}

\begin{document}

\title{Quantum trigonometric Calogero-Sutherland model and irreducible characters for the exceptional algebra $E_8$}

\author{J. Fern\'andez N\'u\~{n}ez\footnote{{\it Departamento de F\'{\i}sica, Facultad de
Ciencias,  Universidad de Oviedo, E-33007 Oviedo, Spain.}} , W. Garc\'{\i}a
Fuertes$^*$,
A.M. Perelomov\footnote{\it  Institute for Theoretical and Experimental Physics, 117259, Moscow, Russia.}
}

\date{}

\maketitle

\begin{abstract}\noindent
We express the Hamiltonian of the quantum trigonometric Calogero-Sutherland model for the Lie algebra
$E_8$ and coupling constant $\kappa=1$ by using
the fundamental irreducible characters of the algebra as dynamical independent
variables. Then, we compute the second order characters of the algebra and some higher order characters.
\end{abstract}

\section{Introduction. Basic notations}
Integrable systems are important because they can be considered as  0-th
order perturbative aproximations to non-integrable systems. By 
integrability we mean here integrability in the sense of Liouville, that
is, the existence of a complete set of mutually commuting integrals of
motion. During the three last decades of the past century, a plethora of
highly nontrivial (classical and quantum) mechanical integrable systems
were discovered, see \cite{ca01, pe90} for comprehensive reviews. Among
these, the Calogero-Sutherland models form a distinguished class.
The first analysis of a system of this kind was performed by Calogero
\cite{ca71} who studied, from the quantum standpoint, the dynamics on the
infinite line of a set of particles interacting pairwise by rational
plus quadratic potentials, and found that the problem was exactly
solvable. Soon afterwards,
Sutherland \cite{su72} arrived to similar results for the quantum problem
on the circle, this time with trigonometric interaction; and later
Moser \cite{mo75} proved, in terms of Lax pairs, that the classical
counterparts of these models also
enjoyed integrability.

The identification of the
general scope of these discoveries came with the work
of Olshanetsky and Perelomov \cite{op83}, who realized that it
is possible to associate models of this kind to all the root systems
of the simple Lie algebras, and that all these models are integrable,
both in the classical and  the quantum framework, for
interactions of the type rational (or inverse-square), $q^{-2}$;
rational+quadratic, $q^{-2}+\omega^2q^2$; trigonometric, $\sin^{-2}q$;
hyperbolic, $\sinh^{-2}q$; and the most general, given by the Weierstrass
elliptic function $\mathcal{P}(q)$.

Nowadays, there is a widespread interest in this kind of integrable
systems, and many mathematical and physical applications for them
have been found, see for instance \cite{dv00}. In Physics, we mention, among
others, the  remarkable connection  established \cite{cas96} between
the different Calogero-Sutherland models and the properties of the equations
describing the physics of disordered wires (the DMPK equation); the results
are in good agreement with the experimental observations.

As the Hamiltonian $H$ of these systems is invariant under the Weyl group of the underlying Lie algebra, the main point of our approach is to express $H$ in a suitable
set of independent variables making clear that invariance, indeed the fundamental characters $z$ of $E_8$.
The use of such kind of variables has been quite useful to solve the
Schr\"{o}dinger equation for the models associated to some other algebras, as those of type $A$
\cite{pe98a}, $D_4$, $E_6$ and $E_7 $\cite{ffp03}.

Although in our previous papers   an important part of them was devoted to the computation of the full set of quadratic Clebsch-Gordan series following a strategy explained and applied in these papers, in this ocassion we will make use extensively of the computer algebra system LiE ``specialised in computations involving Lie groups and their representations'' \cite{lie}, by means of which it is  easy to do these and other  computations needed to reach the main goal: to find the Hamiltonian $H$ for $\k=1$ in terms of the $z$-variables in $E_8$.
\begin{center}
\begin{picture}(70,48)(2,-8)
\put(0,0){\circle{8}}
\put(4,0){\line(1,0){30}}
\put(38,0){\circle{8}}
\put(42,0){\line(1,0){30}}
\put(76,0){\circle{8}}
\put(0,4){\line(0,1){30}}
\put(0,38){\circle{8}}
\put(-4,0){\line(-1,0){30}}
\put(-38,0){\circle{8}}
\put(-42,0){\line(-1,0){30}}
\put(-76,0){\circle{8}}
\put(80,0){\line(1,0){30}}
\put(114,0){\circle{8}}
\put(118,0){\line(1,0){30}}
\put(152,0){\circle{8}}
\put(-90,-10){$\a_1$}
\put(-52,-10){$\a_3$}
\put(-14,-10){$\a_4$}
\put(24,-10){$\a_5$}
\put(61,-10){$\a_6$}
\put(-18,36){$\a_2$}
\put(99,-10){$\a_7$}
\put(137,-10){$\a_8$}
\end{picture}
\vspace{3mm}
\footnotesize

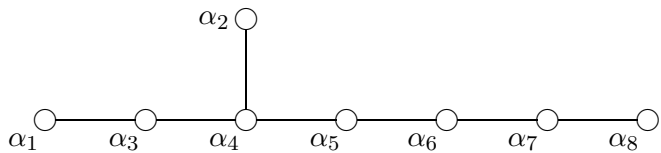
Figure 1. The Dynkin diagram for the Lie algebra $E_8$.
\normalsize
\end{center}

In order to fix the notation, we mention here only a few very basic facts about the exceptional Lie algebra $E_8$. More details can be found in the monographies, see for instance \cite{ov90,otros}.
The Dynkin diagram of $E_8$, see Figure 1,
encodes the euclidean relations $A_{ij}=(\a_i,\a_j)$ among the simple roots, which are
\beqr
\label{raices}
(\a_i,\a_i)&=&2,\hspace{1.5cm} i=1,\dots,8\nonumber\\
(\a_i,\a_{i+2})&=&-1,\hspace{1.2cm} i=1,2\\
(\a_i,\a_{i+1}) &=& -1,\hspace{1.2cm} i=3,\dots,7\nonumber\\
(\a_i,\a_j)&=&0, \hspace{1.2 cm}{\rm in\ all\ other\ cases}.\nonumber
\eeqr

The matrix $A=(A_{ij})$ is the Cartan matrix of the algebra and its inverse $A^{-1}$ gives the eight fundamental weights  by $\l_i=\sum_{j=1}^8A_{ji}^{-1}\a_j$, $i=1,\dots,8$. The characters of the basic fundamental representations $R_{\l_i}$ will be denoted by $z_i$. They are the independent variables in terms of which we want to express the Hamiltonian because of its invariance  under the Weyl group $W$ of the algebra $E_8$.

The organization of the paper is as follows. Section 2 is a brief reminder of the  main properties of the Calogero-Sutherland models associated to root systems and it is  explained how to find the Hamiltonian in the variables mentioned above. Section 3 gives the coefficients of the Hamiltonian in these variables and  finally in the appendices we include some explicit results for the characters  of $E_8$.

\section{Review of the theory}
In this section, we review briefly the general theory of the quantum
trigonometric Calogero-Sutherland model related to a root system $\cal R$
associated to a simple Lie algebra $L$ of rank $r$, and later will study
explicity the $E_8$ case. For Calogero-Sutherland systems other than
trigonometric see \cite{op83}; see also \cite{sa00}. For the other Lie algebras of the E-serie, see \cite{ffp03}.

The  trigonometric Calogero-Sutherland model related to the root system
$\cal R$ of rank $r$ is the quantum system in an Euclidean space $\R^r$
defined by the standard Hamiltonian operator
\beq
\label{ham}
H=\frac{1}{2}\sum_{j=1}^rp_j ^2+\sum_{\a\in{\cal
R}^+}\kappa_\a(\kappa_\a-1)\sin^{-2}(\a,q),
\eeq
where $q=({q_j})$ is  the Cartesian coordinate system provided by the canonical basis  of $\R^r$ and $p_j=-{\rm
i}\,\partial_{q_j}$; ${\cal R}^+$  is the set of the positive roots of $L$,
and the coupling constants $\k_\a$ are such that $\kappa_\a=\kappa_\beta$ if
$|\a|=|\beta|$. We will restrict ourselves to the case of simply-laced root
systems (as the $E$-series is),  for which the Calogero-Sutherland model
depends only on one coupling constant $\k$.

To find the stationary states, it is necessary to solve the
Schr\"odinger eigenvalue problem $H\Psi=E\Psi$. The following important
facts about this family of  quantum mechanical systems were  established
in \cite{op83}.

(a)The ground state energy and (non-normalized) wave function of these integrable systems are
\begin{eqnarray}
E_0(\k)&=&2 \rho^2\k^2\nn\\
\Psi_0^\k(q)&=&{\prod_{\a\in {\cal R}^+}\sin^\k(\a, q)},
\end{eqnarray}
 while the excited states are indexed by the highest weights $\mu=\sum
m_i\l_i\in P^+$ (where $P^+$ is the cone of dominant weights) of the irreducible
representations of $L$, that is, by the $r$-tuple of non-negative integers
${\bf m}=(m_1,\dots,m_r)$ (the quantum numbers), and the wave functions $\Psi_{\bf m}^\k$ and the energy levels $E_{\bf m}(\k)$ satisfy
\begin{eqnarray}
H\Psi^\k_{\bf m}&=&E_{\bf m}(\k)\Psi_{\bf m}^\k\nn\\
E_{\bf m}(\k)&=&2 (\mu+\k\rho,\mu+\k\rho),\label{105}
\end{eqnarray}
with $\rho$ being the Weyl vector $\rho=(1/2)\sum_{\a\in\cal R^+}\a$ of the algebra.

(b)It is natural to look for the solutions $\Psi_{\bf m}^\k$ in the form
\beq
\Psi_{\bf m}^\k(q)=\Psi_0^\k(q)\Phi_{\bf m}^\k(q),
\eeq
and consequently we are led to the eigenvalue problem
\beq
\Delta^\k\Phi_{\bf m}^\k=\ve_{\bf m}(\k)\Phi_{\bf m}^\k\,,
\label{sch}
\eeq
where $\Delta ^\k$ is the linear differential operator
\beq
\Delta^\k=-\frac{1}{2}\sum_{j=1}^r\partial_{q_j}^{\,2}-\k\sum_{\a\in {\cal
R}^+}  {\cot}(\a, q)(\a,\partial_q)
\label{4b},
\eeq
and the eigenvalues $\ve_{\bf m}(\k)$ are the energies over the ground
level, i.e.,
\beq
\label{energ}
\ve_{\bf m}(\k)=E_{\bf m}(\k)-E_0(\k)= 2(\mu, \mu+2\k\rho).
\eeq
Taking into account that $(\l_j,\l_k)=A_{jk}^{-1}$, it is possible to give a
more explicit expression for the eigenvalues $\ve_{\bf m}(\kappa)$:
\beq
\ve_{\bf m}(\k)=2\sum_{j,k=1}^r A_{jk}^{-1} m_j m_k+4\k\sum_{j,k=1}^r
A_{jk}^{-1}m_j.
\label{eigenvalues}
\eeq
We will write $\ve_j(\k)$ for the fundamental weigth $\l_j$, i.e., $\ve_j(\k)=2(A_{jj}^{-1}+2\k\sum_kA_{jk}^{-1})$ for the
quantum numbers   $(0,\dots,\buildrel(j)\over1,\dots,0)$.

(c)In the case $\k=0$ the wave functions (\ref{sch}) are (proportional to)
the monomial symmetric functions
\beq
M_\l(q)=\sum_{w\in W}e^{2i(w\cdot \l,q)},\ \l\in P^+\,,
\eeq
$W$ being the Weyl group of $L$. And the wave functions in the case $\k=1$
are (proportional to) the characters of the irreducible representations
\beq
\label{bchi}
\bchi_\l(q)=\frac{\sum_{w\in W}(\det w)e^{2i(w\cdot(\l+\rho),q)}}{\sum_{w\in
W}(\det w)e^{2i(w\cdot \rho,q)}},\ \l\in P^+\,.
\eeq
Both $M_\l$ and $\bchi_\l$ are sums over the orbit $\{w\cdot\l\}$ of $\l$ under $W$, and
consequently, $W$-invariant; as wave functions, they represent
superpositions of plane waves. 

\medskip
Due to the Weyl symmetry of the Hamiltonian, the wave functions
$\Phi_{\bf m}^\k(q)$ are   $W$-invariant, and the best way to solve the
eigenvalue problem (\ref{sch}) is to use the set of independent
$W$-invariant variables $z_k=\bchi_{\l_k}(q)$, in terms of which the wave
functions $\Phi_{\bf m}^\k$ are polynomials.
Unfortunately, the expression of these characters $z_k$ in terms of the
$q$-variables   is complicated and makes the direct change of variables
$z=z(q)$ very cumbersome. We  are thus forced to follow a much more
convenient, indirect route, which has proven to be very useful for other root
systems, \cite{ffp03}.

First of all, we need the expression of the operator $\Delta^\k$ in terms of $z$-ariables.  To this goal, the starting point is to write  $\Delta^\k$ in its general form
\beq
\Delta^\kappa=\sum_{j, k} \ba_{jk}(z)\partial_{z_j}\partial_{z_k}+\sum_{j}
\left[b_j(z)+(\kappa-1) b_j^{1}(z)\right]\partial_{z_j},
\label{deltaz}
\eeq
with $\ba_{jk}=\ba_{kj}$. Now,  if we take into account the  fact that, as
pointed above,
$b_j(z)=\Delta^1z_j=\ve_{j}(1)z_j $, the full expression for
the coefficients $b_j(z)$ appearing in $\Delta^1$ is
completely determined by the Cartan matrix of the algebra;   explicitly
\beq
\label{bes}
b_j(z)=2(A_{jj}^{-1}+2\sum_{k\ne j}A_{kj}^{-1})z_j,\ j=1,\dots,r.
\eeq

On the other hand, in order to find the coefficients $\ba_{jk}$ we   note that $\Delta^1(z_jz_k)=2\ba_{jk}(z)+b_j(z)z_k+b_k(z)z_j$; as the product character $z_jz_k$ is the character of the tensor product $R_{\l_j}\otimes R_{\l_k}$ of representations, knowing the full set of  quadratic Clebsch-Gordan series we will be able to determine the $\ba$-coefficients.
The Clebsch-Gordan series yield the formulas
\beq
\label{cgsz}
z_jz_k=\sum_{\mu\in Q_{jk}}N_{{\mu};jk}\,\bchi_{\mu}(z)
\eeq
for the products of fundamental characters $z_jz_k$, with $Q_{jk}\subset P^+$ being the  set of dominant weights  in the irreducible representation of highest weight $\l_j+\l_k$, and $N_{\l;jk}$ is the multiplicity of the irreducible representation $R_\l$ in that series; in particular,  $N_{\l_j+\l_k;jk}=1$. Consequently we
obtain the coefficients $\ba$ by applying  $\Delta^1$ to the
two members of (\ref{cgsz}). The required
Clebsch-Gordan series  will be obtained using the system LiE \cite{lie}.

\section{The Calogero-Sutherland Hamiltonian   in $E_8$ for $\k=1$}

The  coefficients $b_j(z)$ in the expression of $\Delta^1$ are easily
obtained from (\ref{bes}) (for the inverse Cartan matrix $A^{-1}$ see, for instance,  \cite{ov90,otros} or \cite{lie}):
\\

$b_1(z)=192\,z_1;\  b_2(z)=288\,z_2;\  b_3(z)=392\,z_3;\ 
b_4(z)=600\,z_4;$
\\

$b_5(z)=480\,z_5;\  b_6(z)=360\,z_6;\  b_7(z)=240\,z_7;\ 
b_8(z)=120\,z_8.$
\\

We can now follow the lines indicated above to obtain the $\ba$-coefficients. Recall that we need only to look for the operator $\Delta^1$, where only the coefficients $b_j$ appear. The best way to obtain the $\ba$'s is to start from the outer regions of the Dynkin diagram in such a way that the computation of each $\ba$ only requires the knowledge of the ones already obtained as explained in \cite{ffp03}. The ordered list of the 28 independent  coefficients $\ba_{jk}(z)$ in $\Delta^1$ (\ref{deltaz})  is
\\

$ {\ba}_{88} (z) =-4\,(31 + 7\,z_1 + z_7 + 16\,z_8 - z_8^2)$
\\

$\ba_{18} (z) =8\,(-16\,z_1 - 4\,z_2 - 10\,z_7 - 25\,z_8 + z_1\,z_8)$
\\

$\ba_{11} (z) =4\,(-31 + 3\,z_1 + 2\,z_1^2 - 4\,z_2 - z_3 - 5\,z_6 + 9\,z_7 -
6\,z_8 - 10\,z_1\,z_8 -
      19\,z_8^2)$
\\ 

$\ba_{78} (z) =
   -4\,(-31 + z_1 + 8\,z_2 + 3\,z_6 - 11\,z_7 + 19\,z_8 + 13\,z_1\,z_8 -
3\,z_7\,z_8 + 31\,z_8^2)$
\\

$
\ba_{28} (z)=-4\,(32\,z_1 + 8\,z_2 + 7\,z_3 + 15\,z_6 + 20\,z_7 - 25\,z_8 +
25\,z_1\,z_8 - 3\,z_2\,z_8)$
\\

$
\ba_{17} (z) =4\,(31 + 31\,z_1 + 24\,z_2 - 7\,z_3 + 4\,z_6 - 9\,z_7 +
4\,z_1\,z_7 + 6\,z_8 - 38\,z_1\,z_8 -
      7\,z_2\,z_8 - 19\,z_7\,z_8 - 31\,z_8^2)$
\\

$
\ba_{68} (z) =-4\,(31 - z_1 + 20\,z_2 + 7\,z_3 + 4\,z_5 - 4\,z_6 + 51\,z_7 +
12\,z_1\,z_7 + 6\,z_8 -
      6\,z_1\,z_8 + 7\,z_2\,z_8 - 4\,z_6\,z_8 + 19\,z_7\,z_8 - 31\,z_8^2)$
\\

$
\ba_{77} (z) =-4\,(31 + 6\,z_1 + 7\,z_1^2 + 6\,z_2 - 14\,z_3 + 2\,z_5 -
3\,z_6 - 22\,z_7 - 3\,z_7^2 -
      34\,z_8 - 12\,z_1\,z_8 + 8\,z_2\,z_8+ z_6\,z_8 - 10\,z_7\,z_8 + 4\,z_8^2 + 6\,z_1\,z_8^2 + 15\,z_8^3)$
\\

$
\ba_{12} (z) =-4\,(-31 - 6\,z_1 + 25\,z_1^2 - 24\,z_2 - 5\,z_1\,z_2 -
18\,z_3 + 5\,z_5 - 29\,z_6 -
      31\,z_7 + 13\,z_1\,z_7 + 19\,z_8 + 13\,z_1\,z_8 + 7\,z_2\,z_8 + 19\,z_7\,z_8 + 6\,z_8^2)$
\\

$
\ba_{38} (z) =-4\,(-7\,z_1 + 25\,z_1^2 + 28\,z_2 + 7\,z_1\,z_2 - 18\,z_3 +
5\,z_5 + 10\,z_6 - 20\,z_7 +
      13\,z_1\,z_7 + 7\,z_2\,z_8- 4\,z_3\,z_8 - 20\,z_7\,z_8 - 25\,z_8^2)$
\\

$    
\ba_{27} (z)  -4\,(-31\,z_1 + z_1^2 - 10\,z_2 + 7\,z_1\,z_2 + 13\,z_3 -
9\,z_5 - 15\,z_6 + 20\,z_7 +
      z_1\,z_7 - 6\,z_2\,z_7 + 25\,z_8 + 7\,z_1\,z_8 - 10\,z_2\,z_8 + 6\,z_3\,z_8 + 14\,z_6\,z_8 +
20\,z_7\,z_8 -
      25\,z_8^2 + 24\,z_1\,z_8^2)$
\\

$
\ba_{16} (z) =
   -4\,(33\,z_1 + z_1^2 - 18\,z_2 + 7\,z_1\,z_2 - 15\,z_3 - 9\,z_5 -
15\,z_6 - 6\,z_1\,z_6 + 18\,z_7 +
      19\,z_1\,z_7 + 6\,z_2\,z_7+ 18\,z_7^2 - 7\,z_8 - 21\,z_1\,z_8 - 10\,z_2\,z_8 +
6\,z_3\,z_8 - 4\,z_6\,z_8 +
      56\,z_7\,z_8 - 7\,z_8^2 + 6\,z_1\,z_8^2 - 18\,z_8^3)$
\\

$
\ba_{22} (z) =-4\,(31 - 4\,z_1 + 13\,z_1^2 + 3\,z_2 + 8\,z_1\,z_2 -
4\,z_2^2 - 2\,z_3 + z_4 - 9\,z_5 +
      19\,z_6 + 4\,z_1\,z_6 + 16\,z_1\,z_7+ 3\,z_2\,z_7 + 9\,z_7^2 - 30\,z_8 - 15\,z_1\,z_8 +9\,z_1^2\,z_8 -16\,z_2\,z_8 - 6\,z_3\,z_8 - 11\,z_6\,z_8 - 30\,z_7\,z_8 + 10\,z_8^2
-6\,z_1\,z_8^2 + 10\,z_8^3)$
\\

$
\ba_{13} (z) =-4\,(31 + 36\,z_1 - 27\,z_1^2 - 18\,z_2 + 16\,z_1\,z_2 +
22\,z_3 - 7\,z_1\,z_3 + 3\,z_4 -
      18\,z_5 - 9\,z_6 + 9\,z_1\,z_6+ 22\,z_7 - 36\,z_1\,z_7 + 6\,z_2\,z_7 - 9\,z_7^2 - 12\,z_8 +
      13\,z_1\,z_8 + 19\,z_1^2\,z_8 + 9\,z_2\,z_8 - 13\,z_3\,z_8 +
4\,z_6\,z_8 - 20\,z_7\,z_8 - 18\,z_8^2 + 14\,z_1\,z_8^2 - 10\,z_8^3)$
\\

$\ba_{58} (z) =
   -4\,(-31 - 12\,z_1 - 13\,z_1^2 + 6\,z_2 + 7\,z_1\,z_2 + 20\,z_3 +
5\,z_4 - 9\,z_5 + 65\,z_6 +
      11\,z_1\,z_6 - 42\,z_7 - 12\,z_1\,z_7 + 6\,z_2\,z_7 - 11\,z_7^2 - 22\,z_8 +
24\,z_2\,z_8 + 6\,z_3\,z_8 -
      5\,z_5\,z_8 + 25\,z_6\,z_8 - 42\,z_7\,z_8 + 50\,z_8^2 - 6\,z_1\,z_8^2 + 11\,z_8^3)$
\\

$\
\ba_{67} (z) =-4\,(-8\,z_1 - 8\,z_1^2 - 8\,z_2 + z_1\,z_2 + 2\,z_3 +
5\,z_4 - 9\,z_5 + 30\,z_6 - 49\,z_7 +
      7\,z_1\,z_7 + 8\,z_2\,z_7- 8\,z_6\,z_7 - 9\,z_7^2 + 22\,z_8 + 33\,z_1\,z_8 +
13\,z_1^2\,z_8 +
      24\,z_2\,z_8 - 19\,z_3\,z_8 + 3\,z_5\,z_8 - 9\,z_6\,z_8 - 40\,z_7\,z_8
+11\,z_1\,z_7\,z_8 + 7\,z_8^2 - 31\,z_1\,z_8^2 + 7\,z_2\,z_8^2
+ 19\,z_7\,z_8^2 - 2\,z_8^3)$
\\

$
  \ba_{37} (z) =-4\,(-31 - 5\,z_1 - 6\,z_1^2 + 2\,z_2 - 24\,z_1\,z_2 +
8\,z_2^2 - 19\,z_3 + 7\,z_1\,z_3 -
      16\,z_4 + 26\,z_5 - 30\,z_6 - 7\,z_1\,z_6 - 11\,z_7 + 14\,z_1\,z_7 - 14\,z_2\,z_7 -
8\,z_3\,z_7 +
      20\,z_7^2 + 31\,z_8 - 21\,z_1\,z_8 + 18\,z_1^2\,z_8 + 7\,z_2\,z_8+ 6\,z_1\,z_2\,z_8 + 2\,z_3\,z_8 +
      5\,z_5\,z_8 - 11\,z_6\,z_8 + 66\,z_7\,z_8 + 13\,z_1\,z_7\,z_8 +
36\,z_8^2 + z_1\,z_8^2 + 7\,z_2\,z_8^2- 19\,z_7\,z_8^2 - 46\,z_8^3)$
\\

$
\ba_{26} (z) =
   4\,(31 - z_1 - 31\,z_2 - 8\,z_2^2 - 20\,z_3 - 7\,z_1\,z_3 + 16\,z_4 -
31\,z_5 + 25\,z_6 + 18\,z_1\,z_6 +
      9\,z_2\,z_6 - 11\,z_7 - 34\,z_1\,z_7 - 19\,z_2\,z_7 - 5\,z_3\,z_7 -
13\,z_6\,z_7 - 22\,z_7^2 - 57\,z_8 +
      7\,z_1\,z_8 + 6\,z_1^2\,z_8 + 22\,z_2\,z_8 - 6\,z_1\,z_2\,z_8 - 6\,z_3\,z_8 + 8\,z_5\,z_8 + 24\,z_6\,z_8 -
      52\,z_7\,z_8 - 24\,z_1\,z_7\,z_8 - 12\,z_8^2 + 13\,z_1\,z_8^2 + 17\,z_2\,z_8^2 + 22\,z_8^3)$
\\

$
\ba_{15} (z) =-4\,(-19\,z_1 + 13\,z_1^2 + z_2 + 6\,z_1\,z_2 + 8\,z_2^2 -
15\,z_3 + 7\,z_1\,z_3 - 16\,z_4 -
      21\,z_5 - 8\,z_1\,z_5 - 43\,z_6 + 11\,z_1\,z_6 + 5\,z_2\,z_6 + 20\,z_7 - 19\,z_1\,z_7 -
19\,z_2\,z_7 +
      5\,z_3\,z_7 + 13\,z_6\,z_7 + 20\,z_7^2 + 32\,z_8 + z_1\,z_8 - 19\,z_1^2\,z_8 - 5\,z_2\,z_8 +
      6\,z_1\,z_2\,z_8 + 13\,z_3\,z_8 - 8\,z_5\,z_8 + 44\,z_6\,z_8 +
32\,z_7\,z_8 - 5\,z_1\,z_7\,z_8 - 25\,z_8^2 + 12\,z_1\,z_8^2 + 12\,z_2\,z_8^2 - 19\,z_7\,z_8^2 +
9\,z_8^3)$
\\

$
\ba_{23} (z) =-4\,(-36\,z_1 - 12\,z_1^2 + 24\,z_1^3 - 13\,z_2 -
3\,z_1\,z_2 - 5\,z_3 - 34\,z_1\,z_3 -
      10\,z_2\,z_3 + 19\,z_4 - 6\,z_5+ 4\,z_1\,z_5 + 51\,z_6 - 17\,z_1\,z_6 + 5\,z_2\,z_6 -
29\,z_7 -
      37\,z_1\,z_7 + 12\,z_1^2\,z_7 + 11\,z_2\,z_7 - 7\,z_3\,z_7 + z_6\,z_7
- 29\,z_7^2 - 20\,z_8 +
      43\,z_1\,z_8 - 11\,z_1^2\,z_8 + 5\,z_2\,z_8 + 14\,z_1\,z_2\,z_8 -
z_3\,z_8 - 16\,z_5\,z_8 + 12\,z_6\,z_8 - 60\,z_7\,z_8 - 4\,z_1\,z_7\,z_8 - 7\,z_8^2 - 13\,z_1\,z_8^2 -
9\,z_2\,z_8^2 + 26\,z_8^3)$
\\

$
\ba_{48} (z) =4\,(-31 + 7\,z_1 + 12\,z_1^2 + 6\,z_1^3 + 25\,z_2 -
6\,z_1\,z_2 - 8\,z_2^2 - 24\,z_3 -
      19\,z_1\,z_3 - 6\,z_2\,z_3 + 22\,z_4-37\,z_5 - 4\,z_1\,z_5 + 7\,z_6 + 14\,z_1\,z_6 - 5\,z_2\,z_6 -
      13\,z_7 + 6\,z_1^2\,z_7 - 15\,z_2\,z_7 - 11\,z_3\,z_7 + 9\,z_6\,z_7 -
2\,z_7^2 + 15\,z_8 - 17\,z_1\,z_8 +
      26\,z_1^2\,z_8 + 18\,z_2\,z_8 - 6\,z_1\,z_2\,z_8 - 44\,z_3\,z_8 +
6\,z_4\,z_8 - 20\,z_5\,z_8 + 16\,z_6\,z_8 - 32\,z_7\,z_8 + 10\,z_1\,z_7\,z_8 + 40\,z_8^2 -
21\,z_1\,z_8^2 - 5\,z_2\,z_8^2 - 20\,z_7\,z_8^2 -       12\,z_8^3) $
\\

$
\ba_{57} (z) =4\,(-31 - 11\,z_1 - 6\,z_1^2 + 6\,z_1^3 + 13\,z_1\,z_2 +
8\,z_2^2 + 18\,z_3 - 12\,z_1\,z_3 -
      6\,z_2\,z_3 - 5\,z_4 - 31\,z_5 + 6\,z_1\,z_5 + 21\,z_6 - 10\,z_1\,z_6 - 8\,z_2\,z_6 - 2\,z_7 -
      31\,z_1\,z_7 - 6\,z_1^2\,z_7 - 20\,z_2\,z_7 + 18\,z_3\,z_7 +
10\,z_5\,z_7 + 6\,z_6\,z_7 + 29\,z_7^2 +
      41\,z_8 + 18\,z_1\,z_8 + z_1^2\,z_8 - 8\,z_1\,z_2\,z_8 - z_3\,z_8 -
4\,z_4\,z_8 + 17\,z_5\,z_8 -
      23\,z_6\,z_8- 10\,z_1\,z_6\,z_8 + 51\,z_7\,z_8 + 25\,z_1\,z_7\,z_8 -
6\,z_2\,z_7\,z_8 + 10\,z_7^2\,z_8 +
      32\,z_8^2 - 7\,z_1\,z_8^2 - 17\,z_2\,z_8^2 - 6\,z_3\,z_8^2 - 24\,z_6\,z_8^2 + 21\,z_7\,z_8^2 - 36\,z_8^3 +
6\,z_1\,z_8^3 - 10\,z_8^4)$
\\

$
\ba_{66} (z) =-4\,(31 - 8\,z_1 - 3\,z_1^2 + 4\,z_1^3 - z_2 - 3\,z_1\,z_2 +
7\,z_3 - 15\,z_1\,z_3 +
      3\,z_2\,z_3 + 18\,z_4 - 17\,z_5 - 3\,z_1\,z_5 +41\,z_6 + 3\,z_1\,z_6 - 6\,z_6^2 + 19\,z_7 - 6\,z_1\,z_7 -
      3\,z_1^2\,z_7 + 17\,z_2\,z_7 + 10\,z_3\,z_7 + z_5\,z_7 - z_6\,z_7 +
18\,z_7^2 + 5\,z_1\,z_7^2 - 45\,z_8 -
      5\,z_1^2\,z_8 + 9\,z_2\,z_8 + z_1\,z_2\,z_8 + 5\,z_3\,z_8 +
2\,z_4\,z_8 - 9\,z_5\,z_8 + 35\,z_6\,z_8 -
      62\,z_7\,z_8 + 3\,z_1\,z_7\,z_8 + 7\,z_2\,z_7\,z_8 + 5\,z_7^2\,z_8 -
12\,z_8^2 + 14\,z_1\,z_8^2 +
      6\,z_1^2\,z_8^2 - 2\,z_2\,z_8^2 - 12\,z_3\,z_8^2 - 2\,z_6\,z_8^2 - 59\,z_7\,z_8^2 + 21\,z_8^3 - 12\,z_1\,z_8^3 +
14\,z_8^4)$
\\

$
\ba_{36} (z) =
   4\,(31 - 27\,z_1 - 51\,z_1^2 + 7\,z_1^3 - 42\,z_2 + 43\,z_1\,z_2 -
7\,z_1^2\,z_2 - 16\,z_2^2 + 42\,z_3 -
      8\,z_1\,z_3 - z_2\,z_3 + 32\,z_4 - 43\,z_5 + 17\,z_1\,z_5 + 79\,z_6 + 25\,z_1\,z_6 -
9\,z_2\,z_6 +
      12\,z_3\,z_6 + 2\,z_7 - 40\,z_1\,z_7 - 18\,z_1^2\,z_7 - 18\,z_2\,z_7 - 5\,z_1\,z_2\,z_7 + 29\,z_3\,z_7 -
      5\,z_5\,z_7 + 11\,z_7^2 - 13\,z_1\,z_7^2 - 57\,z_8 + 64\,z_1\,z_8 -
7\,z_1^2\,z_8 - 22\,z_2\,z_8 +
      24\,z_1\,z_2\,z_8 - 7\,z_2^2\,z_8 + 4\,z_3\,z_8 - 6\,z_1\,z_3\,z_8 +
14\,z_4\,z_8 - 48\,z_5\,z_8 +
      42\,z_6\,z_8 + 6\,z_1\,z_6\,z_8 - 55\,z_7\,z_8 - 19\,z_1\,z_7\,z_8 +
6\,z_2\,z_7\,z_8 - z_7^2\,z_8 -
      14\,z_8^2 + 20\,z_1\,z_8^2 + 6\,z_1^2\,z_8^2 + 18\,z_2\,z_8^2-18\,z_3\,z_8^2 + 20\,z_6\,z_8^2 - 58\,z_7\,z_8^2 +
20\,z_8^4)$
\\

$
\ba_{25} (z) =   4\,(31 + 10\,z_1 + 18\,z_1^2 + 7\,z_1^3 + 2\,z_2 - 5\,z_1\,z_2 -
7\,z_1^2\,z_2 + 12\,z_2^2 - 31\,z_3 -
      z_1\,z_3 - z_2\,z_3 - 30\,z_4 - 23\,z_5 + 17\,z_1\,z_5 + 12\,z_2\,z_5 - 54\,z_6 -
73\,z_1\,z_6 -
      9\,z_2\,z_6 - 4\,z_3\,z_6 - 12\,z_6^2 + 42\,z_7 + 29\,z_1\,z_7 + 18\,z_1^2\,z_7 + 2\,z_2\,z_7 -
      5\,z_1\,z_2\,z_7 - 31\,z_3\,z_7 + 7\,z_5\,z_7 - 24\,z_6\,z_7 +
11\,z_7^2 - z_1\,z_7^2 + 54\,z_8+ 20\,z_1\,z_8 - 20\,z_1^2\,z_8 - 7\,z_2\,z_8 -
4\,z_1\,z_2\,z_8 - 7\,z_2^2\,z_8 + 13\,z_3\,z_8 -
      6\,z_1\,z_3\,z_8 + 14\,z_4\,z_8 - 7\,z_5\,z_8 + 69\,z_6\,z_8 - 6\,z_1\,z_6\,z_8 + 33\,z_7\,z_8 +
      27\,z_1\,z_7\,z_8 + 6\,z_2\,z_7\,z_8 - z_7^2\,z_8 - 77\,z_8^2 -
18\,z_1\,z_8^2 + 6\,z_1^2\,z_8^2 + 4\,z_2\,z_8^2 + 6\,z_3\,z_8^2 + 8\,z_6\,z_8^2 -
72\,z_7\,z_8^2 + z_8^3 - 12\,z_1\,z_8^3 +       20\,z_8^4)$
\\

$
\ba_{14} (z) =-4\,(7\,z_1 - 46\,z_1^2 - 21\,z_1^3 - 9\,z_2 - 13\,z_1\,z_2 +
7\,z_1^2\,z_2 - 12\,z_2^2 +
      55\,z_3 + 36\,z_1\,z_3 + z_2\,z_3 - z_4- 10\,z_1\,z_4 - 26\,z_5 + 11\,z_1\,z_5 + 4\,z_2\,z_5 - 7\,z_6+ 6\,z_1\,z_6 - 20\,z_2\,z_6 + 4\,z_3\,z_6 - 4\,z_6^2 + 22\,z_7 -
8\,z_1\,z_7 - 18\,z_1^2\,z_7 -
      33\,z_2\,z_7 + 5\,z_1\,z_2\,z_7 + 9\,z_5\,z_7 + 19\,z_6\,z_7 +
22\,z_7^2 - 9\,z_1\,z_7^2 + 28\,z_8 -
      32\,z_1\,z_8 + 20\,z_1^2\,z_8 - 44\,z_2\,z_8 + 16\,z_1\,z_2\,z_8 +
7\,z_2^2\,z_8 + 41\,z_3\,z_8 +
      6\,z_1\,z_3\,z_8 - 14\,z_4\,z_8 + 4\,z_5\,z_8 + 2\,z_6\,z_8 - 12\,z_1\,z_6\,z_8 + 54\,z_7\,z_8 +
      19\,z_1\,z_7\,z_8 + 14\,z_2\,z_7\,z_8 + 9\,z_7^2\,z_8 + 5\,z_8^2 +
39\,z_1\,z_8^2 - 18\,z_1^2\,z_8^2 +
      z_2\,z_8^2 + 24\,z_3\,z_8^2 - 8\,z_6\,z_8^2 + 16\,z_7\,z_8^2 - 26\,z_8^3 +
2\,z_1\,z_8^3 + 10\,z_8^4)$
\\

$
\ba_{33} (z) =-4\,(31 - z_1 - 17\,z_1^2 - 17\,z_1^3 - 28\,z_2 - 6\,z_1\,z_2
+ 12\,z_1^2\,z_2 + 2\,z_2^2 +
      42\,z_3 + 14\,z_1\,z_3 - 8\,z_2\,z_3- 7\,z_3^2 + 23\,z_4 + z_1\,z_4 - 40\,z_5 - 14\,z_1\,z_5 +
      2\,z_2\,z_5 + 41\,z_6 - 7\,z_1\,z_6 + 4\,z_1^2\,z_6 + 5\,z_2\,z_6 -
2\,z_3\,z_6 + 3\,z_6^2 + 11\,z_7 -
      5\,z_1\,z_7 - 30\,z_1^2\,z_7 - 22\,z_2\,z_7 + 6\,z_1\,z_2\,z_7 +
21\,z_3\,z_7 - 8\,z_5\,z_7 - 7\,z_6\,z_7 +
      20\,z_7^2 - 7\,z_1\,z_7^2 - 61\,z_8 + 26\,z_1\,z_8 + 14\,z_1^2\,z_8 +
9\,z_1^3\,z_8 - 11\,z_2\,z_8 -
      3\,z_1\,z_2\,z_8 - 12\,z_1\,z_3\,z_8 + 7\,z_4\,z_8 -13\,z_5\,z_8 + 39\,z_6\,z_8 + 3\,z_1\,z_6\,z_8 -
      28\,z_7\,z_8 + 18\,z_1\,z_7\,z_8 - 6\,z_2\,z_7\,z_8 - z_7^2\,z_8 -
4\,z_8^2 + 9\,z_1\,z_8^2 +  5\,z_1^2\,z_8^2 + 3\,z_2\,z_8^2 - 8\,z_3\,z_8^2 +
9\,z_6\,z_8^2 - 50\,z_7\,z_8^2 + 6\,z_8^3 -
      16\,z_1\,z_8^3 + 20\,z_8^4)$
\\

$
\ba_{47} (z) =
   -4\,(15\,z_1 + 36\,z_1^2 + 21\,z_1^3 + 25\,z_2 - 11\,z_1\,z_2 -
7\,z_1^2\,z_2 - 16\,z_2^2 +
      7\,z_1\,z_2^2 - 62\,z_3 - 49\,z_1\,z_3+ 7\,z_3^2 + 31\,z_4 - 14\,z_1\,z_4 + 3\,z_5 + 22\,z_1\,z_5 +
      z_2\,z_5 + 17\,z_6 - z_1\,z_6 - 3\,z_1^2\,z_6 + 3\,z_2\,z_6 -
8\,z_3\,z_6 + 4\,z_6^2 - 11\,z_7 -
      9\,z_1\,z_7 + 8\,z_1^2\,z_7 + 29\,z_2\,z_7 - 6\,z_1\,z_2\,z_7 -
12\,z_4\,z_7 - 7\,z_5\,z_7 - 13\,z_6\,z_7 -
      51\,z_7^2 + 11\,z_1\,z_7^2 + 25\,z_8 + 2\,z_1\,z_8 + 2\,z_1^2\,z_8 -
19\,z_1^3\,z_8 + 43\,z_2\,z_8 +
      z_1\,z_2\,z_8 + z_2^2\,z_8 - 33\,z_3\,z_8+ 45\,z_1\,z_3\,z_8 + 5\,z_2\,z_3\,z_8 - 25\,z_4\,z_8 +
28\,z_5\,z_8 +
      4\,z_1\,z_5\,z_8 - 35\,z_6\,z_8 - 22\,z_1\,z_6\,z_8 + 5\,z_2\,z_6\,z_8
-108\,z_7\,z_8 + 51\,z_1\,z_7\,z_8 -
      5\,z_1^2\,z_7\,z_8 + 16\,z_2\,z_7\,z_8 + 10\,z_3\,z_7\,z_8 -
9\,z_6\,z_7\,z_8 + 5\,z_7^2\,z_8 - 5\,z_8^2 - 39\,z_1\,z_8^2 - 8\,z_1^2\,z_8^2 - 32\,z_2\,z_8^2 +
6\,z_1\,z_2\,z_8^2 + 33\,z_3\,z_8^2 +                19\,z_5\,z_8^2 -
18\,z_6\,z_8^2 + 71\,z_7\,z_8^2 - 11\,z_1\,z_7\,z_8^2 + 2\,z_8^3 + 9\,z_1\,z_8^3 +5\,z_2\,z_8^3
+ 19\,z_7\,z_8^3 - 8\,z_8^4)$
\\

$
\ba_{56} (z) =
   4\,(14\,z_1 + 7\,z_1^2 - 7\,z_1^3 + 6\,z_2 - 15\,z_1\,z_2 +
6\,z_1^2\,z_2 - 7\,z_1\,z_2^2 - 8\,z_3 +
      z_1\,z_3 - 5\,z_2\,z_3 - 7\,z_3^2 + 8\,z_4 + 14\,z_1\,z_4 + 21\,z_5 - 6\,z_1\,z_5 + 7\,z_2\,z_5 -
33\,z_6 +
      50\,z_1\,z_6 + 14\,z_1^2\,z_6 - 25\,z_2\,z_6 - 21\,z_3\,z_6 + 15\,z_5\,z_6 - 3\,z_6^2 + 29\,z_7 -
      7\,z_1\,z_7 + 7\,z_1^2\,z_7 - z_1\,z_2\,z_7 - 8\,z_3\,z_7 -
3\,z_4\,z_7 + 11\,z_5\,z_7 - 64\,z_6\,z_7 - 9\,z_1\,z_6\,z_7 + 38\,z_7^2 - z_1\,z_7^2 - 6\,z_2\,z_7^2 +
9\,z_7^3 - 29\,z_8 - 59\,z_1\,z_8 +
      14\,z_1^2\,z_8 + 6\,z_1^3\,z_8 + 2\,z_2\,z_8+ 19\,z_1\,z_2\,z_8 + 8\,z_2^2\,z_8 + 5\,z_3\,z_8 +
      z_1\,z_3\,z_8 - 5\,z_2\,z_3\,z_8 - 31\,z_4\,z_8 - 26\,z_5\,z_8 +
5\,z_1\,z_5\,z_8 - 74\,z_6\,z_8 - 16\,z_1\,z_6\,z_8 - 7\,z_2\,z_6\,z_8 +
70\,z_7\,z_8 - 72\,z_1\,z_7\,z_8 -                   6\,z_1^2\,z_7\,z_8 -
30\,z_2\,z_7\,z_8 + 11\,z_3\,z_7\,z_8 - 17\,z_6\,z_7\,z_8 + 79\,z_7^2\,z_8 +
61\,z_8^2 + 18\,z_1\,z_8^2 -
                  18\,z_1^2\,z_8^2 -24\,z_2\,z_8^2 - z_1\,z_2\,z_8^2 + 6\,z_3\,z_8^2 + 8\,z_5\,z_8^2 + 21\,z_6\,z_8^2 +
49\,z_7\,z_8^2 +
      19\,z_1\,z_7\,z_8^2 - 17\,z_8^3 + 19\,z_1\,z_8^3 + 7\,z_2\,z_8^3- 28\,z_7\,z_8^3 - 18\,z_8^4)$
\\

$
  \ba_{35} (z) =4\,(-25\,z_1 - 6\,z_1^2 - 13\,z_1^3 - z_2 + 13\,z_1\,z_2 +
8\,z_1^2\,z_2 - 6\,z_2^2 -
      8\,z_1\,z_2^2 + 21\,z_3 + 15\,z_1\,z_3 - 7\,z_1^2\,z_3 - 21\,z_2\,z_3 + 7\,z_3^2 + 26\,z_4 +
7\,z_1\,z_4 +
      7\,z_5 + 13\,z_1\,z_5 + 10\,z_2\,z_5 + 16\,z_3\,z_5 + 46\,z_6 + 41\,z_1\,z_6 - 7\,z_1^2\,z_6 -
      26\,z_2\,z_6 - 4\,z_1\,z_2\,z_6 + 7\,z_3\,z_6 - 5\,z_5\,z_6 -
25\,z_6^2 - 18\,z_7 - 36\,z_1\,z_7 -
      6\,z_1^2\,z_7+ 11\,z_2\,z_7 + 19\,z_1\,z_2\,z_7 - 6\,z_2^2\,z_7 +
21\,z_3\,z_7 - 5\,z_1\,z_3\,z_7 +
      12\,z_4\,z_7 - 2\,z_5\,z_7 + 52\,z_6\,z_7 - 8\,z_1\,z_6\,z_7- 36\,z_7^2 - 11\,z_1\,z_7^2 + 12\,z_2\,z_7^2 -
      18\,z_7^3 - 27\,z_8 - 9\,z_1\,z_8 + 5\,z_1^2\,z_8 + 13\,z_1^3\,z_8 -
2\,z_2\,z_8 + 9\,z_1\,z_2\,z_8 -
      6\,z_1^2\,z_2\,z_8 - 5\,z_2^2\,z_8 - 20\,z_3\,z_8 - z_1\,z_3\,z_8 -
z_2\,z_3\,z_8 + 2\,z_4\,z_8 - 21\,z_5\,z_8+ 15\,z_1\,z_5\,z_8 + 26\,z_6\,z_8 - 13\,z_1\,z_6\,z_8 -
15\,z_2\,z_6\,z_8 - 91\,z_7\,z_8 -                3\,z_1\,z_7\,z_8 +
11\,z_1^2\,z_7\,z_8 + 24\,z_2\,z_7\,z_8 - 6\,z_3\,z_7\,z_8 + 45\,z_6\,z_7\,z_8 -
64\,z_7^2\,z_8 - 16\,z_8^2 +
      45\,z_1\,z_8^2 - 19\,z_1^2\,z_8^2 - 17\,z_2\,z_8^2 + z_1\,z_2\,z_8^2 + 7\,z_3\,z_8^2 - 22\,z_5\,z_8^2 +
      17\,z_6\,z_8^2 - 10\,z_7\,z_8^2 - 5\,z_1\,z_7\,z_8^2 + 14\,z_8^3 -
5\,z_1\,z_8^3 - z_2\,z_8^3 + 18\,z_7\,z_8^3 + 14\,z_8^4)$
\\

$
\ba_{24} (z)=
   -4\,(32\,z_1 - 6\,z_1^2 - 6\,z_1^3 + 25\,z_2 - 46\,z_1\,z_2 +
12\,z_1^2\,z_2 - 2\,z_2^2 + 8\,z_1\,z_2^2 -
      17\,z_3 + 3\,z_1\,z_3 + 7\,z_1^2\,z_3 - 14\,z_2\,z_3 - 7\,z_3^2 + 4\,z_4 -
7\,z_1\,z_4 - 15\,z_2\,z_4 +
      24\,z_5 + 5\,z_1\,z_5 - 2\,z_2\,z_5 + 3\,z_3\,z_5 - 3\,z_6 + 13\,z_1\,z_6 - 17\,z_1^2\,z_6 + 31\,z_2\,z_6 +
      4\,z_1\,z_2\,z_6 + 14\,z_3\,z_6 + 5\,z_5\,z_6 + 7\,z_6^2 - 13\,z_7 -
7\,z_1\,z_7 - 14\,z_2\,z_7 + 9\,z_1\,z_2\,z_7 + 6\,z_2^2\,z_7 - 2\,z_3\,z_7 +
5\,z_1\,z_3\,z_7 - 12\,z_4\,z_7 + 25\,z_5\,z_7 -           21\,z_6\,z_7 -
13\,z_1\,z_6\,z_7 + 7\,z_7^2 - 6\,z_1\,z_7^2 + 9\,z_2\,z_7^2 - 39\,z_8 - 26\,z_1\,z_8 +
23\,z_1^2\,z_8 -
      26\,z_1^3\,z_8 - 17\,z_2\,z_8 - z_1\,z_2\,z_8 + 6\,z_1^2\,z_2\,z_8- 21\,z_2^2\,z_8 + 7\,z_3\,z_8 +
      26\,z_1\,z_3\,z_8 + z_2\,z_3\,z_8 + 28\,z_4\,z_8 - 6\,z_5\,z_8 +
6\,z_1\,z_5\,z_8 + 7\,z_6\,z_8 + 49\,z_1\,z_6\,z_8 - 14\,z_2\,z_6\,z_8 + 29\,z_7\,z_8 -
8\,z_1\,z_7\,z_8 - 22\,z_1^2\,z_7\,z_8 -
      31\,z_2\,z_7\,z_8 + 12\,z_3\,z_7\,z_8+10\,z_6\,z_7\,z_8 + 23\,z_7^2\,z_8 + 6\,z_8^2 -
52\,z_1\,z_8^2 +
      32\,z_1^2\,z_8^2 + 2\,z_2\,z_8^2 + 4\,z_1\,z_2\,z_8^2 - 10\,z_3\,z_8^2
- 20\,z_5\,z_8^2 - 4\,z_6\,z_8^2 -
      52\,z_7\,z_8^2 + 10\,z_1\,z_7\,z_8^2 + 3\,z_8^3 + 15\,z_1\,z_8^3 +
7\,z_2\,z_8^3 + 10\,z_8^4)$
\\

$
\ba_{46} (z) =-4\,(-31\,z_1 + 7\,z_1^2 - 2\,z_1^3 - 8\,z_1^4 - 35\,z_2 +
11\,z_1\,z_2 -
      6\,z_1^2\,z_2 +
      20\,z_2^2 - z_1\,z_2^2 + 8\,z_2^3+ 8\,z_3 + 19\,z_1\,z_3 + 24\,z_1^2\,z_3 + 14\,z_2\,z_3 +
      8\,z_1\,z_2\,z_3 - 8\,z_3^2 - 39\,z_4 - 22\,z_1\,z_4 - 24\,z_2\,z_4 +
12\,z_5 + 8\,z_1\,z_5 -
      8\,z_1^2\,z_5 + 37\,z_2\,z_5 + 15\,z_3\,z_5 - 17\,z_6 - 70\,z_1\,z_6 -
29\,z_2\,z_6 -       15\,z_1\,z_2\,z_6 -  18\,z_3\,z_6 - 18\,z_4\,z_6 + 3\,z_5\,z_6 + 15\,z_6^2 +
66\,z_7 - 5\,z_1\,z_7 -             10\,z_1^2\,z_7 - 34\,z_2\,z_7 +
16\,z_1\,z_2\,z_7 + 8\,z_2^2\,z_7 + 31\,z_3\,z_7 - 5\,z_1\,z_3\,z_7 +
4\,z_2\,z_3\,z_7 -
      z_5\,z_7 + 4\,z_1\,z_5\,z_7 + 41\,z_6\,z_7 - 2\,z_1\,z_6\,z_7 +
5\,z_2\,z_6\,z_7 + 6\,z_7^2 - 4\,z_1^2\,z_7^2 +  17\,z_2\,z_7^2 +
9\,z_3\,z_7^2 - 9\,z_6\,z_7^2 +
      47\,z_8 +  81\,z_1\,z_8 - 14\,z_1^2\,z_8 + 28\,z_1^3\,z_8 + 43\,z_2\,z_8 - 6\,z_1\,z_2\,z_8 -
21\,z_1^2\,z_2\,z_8 -             23\,z_2^2\,z_8 + 6\,z_1\,z_2^2\,z_8 -
20\,z_3\,z_8 - 42\,z_1\,z_3\,z_8 +
      13\,z_2\,z_3\,z_8 +6\,z_3^2\,z_8 + 25\,z_4\,z_8 -       12\,z_1\,z_4\,z_8
           + 24\,z_5\,z_8 + 29\,z_1\,z_5\,z_8 + z_2\,z_5\,z_8 +
16\,z_6\,z_8 - 26\,z_1\,z_6\,z_8 -         2\,z_1^2\,z_6\,z_8 - 8\,z_3\,z_6\,z_8 + 4\,z_6^2\,z_8
           - 60\,z_7\,z_8 + 102\,z_1\,z_7\,z_8 - 14\,z_1^2\,z_7\,z_8 +
77\,z_2\,z_7\,z_8 +
      z_1\,z_2\,z_7\,z_8 + 44\,z_3\,z_7\,z_8 + 10\,z_5\,z_7\,z_8 - 41\,z_6\,z_7\,z_8 -
      71\,z_7^2\,z_8 -
      2\,z_1\,z_7^2\,z_8 - 74\,z_8^2 + 63\,z_1\,z_8^2 + 8\,z_1^2\,z_8^2 -
12\,z_1^3\,z_8^2 +             55\,z_2\,z_8^2+3\,z_1\,z_2\,z_8^2 - 7\,z_2^2\,z_8^2 - 51\,z_3\,z_8^2 +
24\,z_1\,z_3\,z_8^2 - 2\,z_4\,z_8^2 +
      17\,z_5\,z_8^2 - 40\,z_6\,z_8^2 - 8\,z_1\,z_6\,z_8^2 - 114\,z_7\,z_8^2
+ 59\,z_1\,z_7\,z_8^2 +
      6\,z_2\,z_7\,z_8^2 + 20\,z_7^2\,z_8^2 + 21\,z_8^3 - 86\,z_1\,z_8^3 +
18\,z_1^2\,z_8^3 -             24\,z_2\,z_8^3 -
      12\,z_3\,z_8^3 - 2\,z_6\,z_8^3 +71\,z_7\,z_8^3 + 36\,z_8^4 - 12\,z_1\,z_8^4 - 20\,z_8^5)$
\\

$
  \ba_{55} (z) =-4\,(31 + 5\,z_1 + 13\,z_1^2 - 6\,z_1^3 + 3\,z_1^4 -
2\,z_2 - 9\,z_1\,z_2 + 2\,z_2^2 +
      4\,z_2^3 - 28\,z_3 + 12\,z_1\,z_3-16\,z_1^2\,z_3 + 2\,z_2\,z_3 + 4\,z_1\,z_2\,z_3 +
10\,z_3^2 -
      19\,z_4 + 16\,z_1\,z_4 - 12\,z_2\,z_4 - 7\,z_5 + 2\,z_1\,z_5 -
4\,z_1^2\,z_5 - 20\,z_2\,z_5 + 4\,z_3\,z_5 -
      10\,z_5^2 - 66\,z_6 - 7\,z_1\,z_6 - 9\,z_1^2\,z_6 - 6\,z_2\,z_6 -
z_1\,z_2\,z_6 + 40\,z_3\,z_6 +       z_4\,z_6 - z_5\,z_6 + 51\,z_6^2 + 4\,z_1\,z_6^2 + 53\,z_7 +
10\,z_1\,z_7 + 6\,z_1^2\,z_7 -             6\,z_1^3\,z_7 + 16\,z_2\,z_7 -
9\,z_1\,z_2\,z_7 - 8\,z_2^2\,z_7 - 36\,z_3\,z_7 + 12\,z_1\,z_3\,z_7 +
2\,z_2\,z_3\,z_7 +
      z_4\,z_7 + 13\,z_5\,z_7 - 2\,z_1\,z_5\,z_7 - 53\,z_6\,z_7 -
3\,z_1\,z_6\,z_7 + 6\,z_2\,z_6\,z_7 + 13\,z_7^2 +
      5\,z_1\,z_7^2 + 3\,z_1^2\,z_7^2 + 18\,z_2\,z_7^2 - 8\,z_3\,z_7^2 -
7\,z_6\,z_7^2 - 9\,z_7^3 +          41\,z_8 - 39\,z_1\,z_8 + 5\,z_1^2\,z_8 + 7\,z_1^3\,z_8 + 12\,z_2\,z_8 +
11\,z_1\,z_2\,z_8 -                   10\,z_1^2\,z_2\,z_8 +
3\,z_1\,z_2^2\,z_8 - 9\,z_3\,z_8 + 3\,z_1\,z_3\,z_8+9\,z_2\,z_3\,z_8 + 3\,z_3^2\,z_8 - 14\,z_4\,z_8 -
6\,z_1\,z_4\,z_8 - 24\,z_5\,z_8 +         10\,z_1\,z_5\,z_8 -
3\,z_2\,z_5\,z_8 - 31\,z_6\,z_8 - 25\,z_1\,z_6\,z_8 - 6\,z_1^2\,z_6\,z_8 + 29\,z_2\,z_6\,z_8
+ 12\,z_3\,z_6\,z_8 +            12\,z_6^2\,z_8 + 62\,z_7\,z_8 -
20\,z_1\,z_7\,z_8 + 18\,z_1^2\,z_7\,z_8 + 12\,z_2\,z_7\,z_8 -
2\,z_1\,z_2\,z_7\,z_8 - 29\,z_3\,z_7\,z_8 -       5\,z_5\,z_7\,z_8 -
57\,z_6\,z_7\,z_8 + 21\,z_7^2\,z_8 -z_1\,z_7^2\,z_8 - 18\,z_8^2 + 29\,z_1\,z_8^2 -
26\,z_1^2\,z_8^2 -
      8\,z_2\,z_8^2 + z_1\,z_2\,z_8^2 + 32\,z_3\,z_8^2 - 6\,z_1\,z_3\,z_8^2
+ 12\,z_4\,z_8^2 + 10\,z_5\,z_8^2 +
      94\,z_6\,z_8^2 - 24\,z_7\,z_8^2 + 23\,z_1\,z_7\,z_8^2 -
10\,z_2\,z_7\,z_8^2 + 4\,z_7^2\,z_8^2 -
      55\,z_8^3 + 7\,z_1\,z_8^3 + 3\,z_1^2\,z_8^3 + 9\,z_2\,z_8^3 + 6\,z_3\,z_8^3 + 13\,z_6\,z_8^3 -
      64\,z_7\,z_8^3 + 20\,z_8^4 - 12\,z_1\,z_8^4 + 15\,z_8^5)$
\\

$
  \ba_{34} (z) =4\,(-16\,z_1 + 40\,z_1^2 + 63\,z_1^3 + 7\,z_1^4 - 11\,z_2 +
38\,z_1\,z_2 -                   7\,z_1^3\,z_2 + 23\,z_2^2 +
12\,z_1\,z_2^2 - 27\,z_3 - 79\,z_1\,z_3 + 7\,z_1^2\,z_3 - z_2\,z_3 +
6\,z_1\,z_2\,z_3 -
      15\,z_3^2 + z_4 - 46\,z_1\,z_4 - 8\,z_2\,z_4 + 20\,z_3\,z_4 + 28\,z_5
+ 68\,z_1\,z_5 - 18\,z_1^2\,z_5 -
      5\,z_2\,z_5 - 3\,z_1\,z_2\,z_5 + 29\,z_3\,z_5 - 5\,z_5^2 - 4\,z_6 -
66\,z_1\,z_6 -             29\,z_1^2\,z_6- 48\,z_2\,z_6 + 23\,z_1\,z_2\,z_6 - 5\,z_2^2\,z_6 +
9\,z_3\,z_6 - 4\,z_1\,z_3\,z_6 +         10\,z_4\,z_6 -
      30\,z_5\,z_6 + 15\,z_6^2 + 4\,z_1\,z_6^2+42\,z_7 - 21\,z_1\,z_7 + 26\,z_1^2\,z_7 + 24\,z_1^3\,z_7 -
      10\,z_2\,z_7 + 58\,z_1\,z_2\,z_7 - 5\,z_1^2\,z_2\,z_7 - 17\,z_2^2\,z_7
+ 15\,z_3\,z_7- 30\,z_1\,z_3\,z_7 -
      z_2\,z_3\,z_7 + 34\,z_4\,z_7 - 16\,z_5\,z_7 + 67\,z_6\,z_7 -
12\,z_1\,z_6\,z_7 +             4\,z_2\,z_6\,z_7 -16\,z_7^2 
             - 45\,z_1\,z_7^2 + 5\,z_1^2\,z_7^2 + 25\,z_2\,z_7^2 +
9\,z_3\,z_7^2 - 9\,z_6\,z_7^2 -         18\,z_7^3 + 23\,z_8 + 61\,z_1\,z_8 -
24\,z_1^2\,z_8 - 26\,z_1^3\,z_8 + 62\,z_2\,z_8 - 4\,z_1\,z_2\,z_8 -
4\,z_1^2\,z_2\,z_8
      -15\,z_2^2\,z_8- 7\,z_1\,z_2^2\,z_8 - 23\,z_3\,z_8 - 17\,z_1\,z_3\,z_8
- 6\,z_1^2\,z_3\,z_8 + z_2\,z_3\,z_8 + 6\,z_3^2\,z_8 +
25\,z_4\,z_8 +
      6\,z_1\,z_4\,z_8 + 12\,z_5\,z_8 + 43\,z_1\,z_5\,z_8 + 2\,z_2\,z_5\,z_8 - z_6\,z_8 - 13\,z_1\,z_6\,z_8 +
20\,z_1^2\,z_6\,z_8 -                   9\,z_2\,z_6\,z_8 - 24\,z_3\,z_6\,z_8
+ 4\,z_6^2\,z_8 - 109\,z_7\,z_8 +54\,z_1\,z_7\,z_8 - 41\,z_1^2\,z_7\,z_8 +
      76\,z_2\,z_7\,z_8 - 2\,z_1\,z_2\,z_7\,z_8 + z_3\,z_7\,z_8 +
15\,z_5\,z_7\,z_8 -                   16\,z_6\,z_7\,z_8 - 129\,z_7^2\,z_8 - 5\,z_1\,z_7^2\,z_8 - 30\,z_8^2 - 41\,z_1\,z_8^2 -
6\,z_1^2\,z_8^2 +                   6\,z_1^3\,z_8^2 +
      37\,z_2\,z_8^2 - z_1\,z_2\,z_8^2 + 9\,z_2^2\,z_8^2 - 11\,z_3\,z_8^2 +
6\,z_1\,z_3\,z_8^2 - 26\,z_4\,z_8^2 +
      15\,z_5\,z_8^2 - 38\,z_6\,z_8^2 - 32\,z_1\,z_6\,z_8^2 - 78\,z_7\,z_8^2
+ 49\,z_1\,z_7\,z_8^2 -
      3\,z_2\,z_7\,z_8^2 + 20\,z_7^2\,z_8^2 + 17\,z_8^3 - 4\,z_1\,z_8^3 - 6\,z_1^2\,z_8^3 - 40\,z_2\,z_8^3 +
      20\,z_3\,z_8^3 - 2\,z_6\,z_8^3 + 129\,z_7\,z_8^3 + 24\,z_8^4 -
12\,z_1\,z_8^4 - 20\,z_8^5)$
\\

$
 \ba_{45} (z) =-4\,(25\,z_1^2 + 25\,z_1^3 + 13\,z_1\,z_2 - 28\,z_1^2\,z_2 -
8\,z_1^3\,z_2 -
      22\,z_2^2 - 2\,z_1\,z_2^2 - 24\,z_3 - 52\,z_1\,z_3+ 20\,z_2\,z_3 + 15\,z_1\,z_2\,z_3 + 9\,z_2^2\,z_3 - 2\,z_3^2
+
      49\,z_4 + 4\,z_1\,z_4 + 9\,z_1^2\,z_4 - 6\,z_2\,z_4 - 18\,z_3\,z_4 -
45\,z_5 + 31\,z_1\,z_5 -
      5\,z_1^2\,z_5 + 2\,z_2\,z_5 - 11\,z_1\,z_2\,z_5 + 4\,z_3\,z_5 -
24\,z_4\,z_5 + 9\,z_5^2 +             38\,z_6 - z_1\,z_6 + 12\,z_1^2\,z_6 -3\,z_1^3\,z_6 + 12\,z_1\,z_2\,z_6 + 9\,z_2^2\,z_6 +
z_3\,z_6 + 13\,z_1\,z_3\,z_6 +
      3\,z_2\,z_3\,z_6 - 35\,z_4\,z_6 + 59\,z_5\,z_6 + 4\,z_1\,z_5\,z_6- 3\,z_6^2 - 14\,z_1\,z_6^2 +
      5\,z_2\,z_6^2 - 11\,z_7 - 26\,z_1\,z_7 + 29\,z_1^2\,z_7 +
19\,z_1^3\,z_7 + 28\,z_2\,z_7 +             3\,z_1\,z_2\,z_7-13\,z_1^2\,z_2\,z_7 - 22\,z_2^2\,z_7 + 5\,z_1\,z_2^2\,z_7 -
20\,z_3\,z_7 -             33\,z_1\,z_3\,z_7 +
      5\,z_2\,z_3\,z_7 + 5\,z_3^2\,z_7 + 38\,z_4\,z_7 - 10\,z_1\,z_4\,z_7 - 24\,z_5\,z_7 +
15\,z_1\,z_5\,z_7 +
      z_2\,z_5\,z_7 - 3\,z_6\,z_7 - 4\,z_1^2\,z_6\,z_7 + 31\,z_2\,z_6\,z_7
- 5\,z_6^2\,z_7 - 42\,z_7^2 +
      3\,z_1\,z_7^2 + 4\,z_1^2\,z_7^2 + 28\,z_2\,z_7^2 - 4\,z_1\,z_2\,z_7^2
+ 4\,z_3\,z_7^2 -                9\,z_5\,z_7^2 - 10\,z_6\,z_7^2 - 31\,z_7^3 + 9\,z_1\,z_7^3 - 5\,z_8 - 11\,z_1\,z_8 -
4\,z_1^2\,z_8 - 12\,z_1^3\,z_8 +
      6\,z_1^4\,z_8 + 14\,z_2\,z_8 - 4\,z_1\,z_2\,z_8 + 3\,z_1^2\,z_2\,z_8 - 2\,z_2^2\,z_8 - z_1\,z_2^2\,z_8 +
      7\,z_2^3\,z_8 - 39\,z_3\,z_8 + 40\,z_1\,z_3\,z_8 - 31\,z_1^2\,z_3\,z_8
+ 26\,z_2\,z_3\,z_8 +7\,z_1\,z_2\,z_3\,z_8 + 19\,z_3^2\,z_8 - 71\,z_4\,z_8 +
15\,z_1\,z_4\,z_8 -                 21\,z_2\,z_4\,z_8 - 42\,z_5\,z_8 -
13\,z_1\,z_5\,z_8 - 7\,z_1^2\,z_5\,z_8 + 5\,z_2\,z_5\,z_8 + 13\,z_3\,z_5\,z_8 - 73\,z_6\,z_8 -
      50\,z_1\,z_6\,z_8 -
      16\,z_1^2\,z_6\,z_8 - 42\,z_2\,z_6\,z_8 - 7\,z_1\,z_2\,z_6\,z_8 +
50\,z_3\,z_6\,z_8+ 19\,z_5\,z_6\,z_8 +
      7\,z_6^2\,z_8 - 22\,z_7\,z_8 + 53\,z_1\,z_7\,z_8 -
30\,z_1^2\,z_7\,z_8 - 6\,z_1^3\,z_7\,z_8 +
      44\,z_2\,z_7\,z_8 - 5\,z_1\,z_2\,z_7\,z_8- 7\,z_2^2\,z_7\,z_8 - 53\,z_3\,z_7\,z_8 +
z_1\,z_3\,z_7\,z_8 +
      19\,z_4\,z_7\,z_8 - 52\,z_5\,z_7\,z_8 + 43\,z_6\,z_7\,z_8 -
10\,z_1\,z_6\,z_7\,z_8 -             37\,z_7^2\,z_8 + 17\,z_1\,z_7^2\,z_8
             + 2\,z_2\,z_7^2\,z_8 + 129\,z_8^2 - 42\,z_1\,z_8^2 -
21\,z_1^2\,z_8^2 +
      7\,z_1^3\,z_8^2 - 7\,z_2\,z_8^2 + 32\,z_1\,z_2\,z_8^2 -
13\,z_1^2\,z_2\,z_8^2+ 2\,z_2^2\,z_8^2 +
      11\,z_3\,z_8^2 + 6\,z_1\,z_3\,z_8^2 + 12\,z_2\,z_3\,z_8^2 -
25\,z_4\,z_8^2 + 28\,z_5\,z_8^2 +
      7\,z_1\,z_5\,z_8^2 - 28\,z_6\,z_8^2 + 16\,z_1\,z_6\,z_8^2 + 3\,z_2\,z_6\,z_8^2 +
      103\,z_7\,z_8^2 -
      30\,z_1\,z_7\,z_8^2 + 24\,z_1^2\,z_7\,z_8^2 + 53\,z_2\,z_7\,z_8^2 -
14\,z_3\,z_7\,z_8^2 -             14\,z_7^2\,z_8^2 - 37\,z_8^3 + 49\,z_1\,z_8^3 - 23\,z_1^2\,z_8^3 -
8\,z_2\,z_8^3 + 7\,z_1\,z_2\,z_8^3 +             38\,z_3\,z_8^3 +
      28\,z_5\,z_8^3 + 12\,z_6\,z_8^3 + 31\,z_7\,z_8^3 - 21\,z_1\,z_7\,z_8^3 - 52\,z_8^4 + 3\,z_1\,z_8^4 -
9\,z_2\,z_8^4 + 12\,z_8^5)$
\\

$
\ba_{44} (z) =
   -4\,(31 + 16\,z_1 + 18\,z_1^2 - z_1^3 + 5\,z_1^4 + 7\,z_1^5 + 5\,z_2 +
26\,z_1\,z_2 -
      5\,z_1^2\,z_2 +
      3\,z_2^2 - 30\,z_1\,z_2^2 - 5\,z_2^3 - 11\,z_3 - 21\,z_1\,z_3 - 45\,z_1^2\,z_3 -
35\,z_1^3\,z_3 -
      24\,z_2\,z_3 - 10\,z_1\,z_2\,z_3 + 35\,z_3^2 + 35\,z_1\,z_3^2 +
23\,z_4 + 35\,z_1\,z_4 + 30\,z_1^2\,z_4 +
      15\,z_2\,z_4 + 5\,z_1\,z_2\,z_4 - 30\,z_3\,z_4 - 15\,z_4^2 + 36\,z_5 +
13\,z_1\,z_5 +             3\,z_1^2\,z_5 +4\,z_1^3\,z_5 - 5\,z_2\,z_5 - 19\,z_1\,z_2\,z_5 +
3\,z_2^2\,z_5 - 16\,z_3\,z_5 -             6\,z_1\,z_3\,z_5 +
      z_2\,z_3\,z_5 - 12\,z_4\,z_5 + 31\,z_5^2+ 2\,z_1\,z_5^2 - 11\,z_6 + 20\,z_1\,z_6 + 51\,z_1^2\,z_6 +
      16\,z_1^3\,z_6 - 3\,z_2\,z_6 + 34\,z_1\,z_2\,z_6 - 7\,z_1^2\,z_2\,z_6
+ 19\,z_2^2\,z_6 + 2\,z_1\,z_2^2\,z_6 -
      12\,z_3\,z_6 - 9\,z_1\,z_3\,z_6 + 5\,z_2\,z_3\,z_6 + 2\,z_3^2\,z_6 -
24\,z_4\,z_6 -             4\,z_1\,z_4\,z_6 - 13\,z_5\,z_6 + 2\,z_1\,z_5\,z_6 + 3\,z_2\,z_5\,z_6 -
2\,z_6^2 - 22\,z_1\,z_6^2 +             2\,z_2\,z_6^2 - 4\,z_3\,z_6^2 +
2\,z_6^3 + 11\,z_7 - 35\,z_1\,z_7-4\,z_1^2\,z_7 + 3\,z_1^3\,z_7 - 3\,z_1^4\,z_7 +
      15\,z_2\,z_7 + 16\,z_1\,z_2\,z_7 - 11\,z_1^2\,z_2\,z_7 -
15\,z_2^2\,z_7 + 4\,z_1\,z_2^2\,z_7+3\,z_2^3\,z_7 +
      7\,z_3\,z_7 - 21\,z_1\,z_3\,z_7 + 9\,z_1^2\,z_3\,z_7 -
11\,z_2\,z_3\,z_7 +
      3\,z_1\,z_2\,z_3\,z_7 - 3\,z_3^2\,z_7 + 31\,z_4\,z_7 - 17\,z_1\,z_4\,z_7 - 9\,z_2\,z_4\,z_7 + 13\,z_5\,z_7 +
34\,z_1\,z_5\,z_7 -                   4\,z_1^2\,z_5\,z_7 + 22\,z_2\,z_5\,z_7
+ 9\,z_3\,z_5\,z_7 - 7\,z_6\,z_7 - 26\,z_1\,z_6\,z_7 + z_1^2\,z_6\,z_7 -
17\,z_2\,z_6\,z_7 -
      7\,z_1\,z_2\,z_6\,z_7 - 9\,z_5\,z_6\,z_7 + 4\,z_6^2\,z_7 - 18\,z_7^2 -
28\,z_1\,z_7^2 - 5\,z_1^2\,z_7^2 +
      2\,z_1^3\,z_7^2 + 12\,z_2\,z_7^2 - 2\,z_1\,z_2\,z_7^2 +
35\,z_3\,z_7^2 - 9\,z_1\,z_3\,z_7^2 +          9\,z_4\,z_7^2 -
18\,z_5\,z_7^2 + 9\,z_6\,z_7^2 + 9\,z_1\,z_6\,z_7^2 - 28\,z_7^3 - 15\,z_8 - 36\,z_1\,z_8 -
33\,z_1^2\,z_8 +
      30\,z_1^3\,z_8 + 16\,z_2\,z_8 - 23\,z_1\,z_2\,z_8 +
37\,z_1^2\,z_2\,z_8 - 3\,z_1^3\,z_2\,z_8 - 22\,z_2^2\,z_8 +
      18\,z_1\,z_2^2\,z_8 - 5\,z_2^3\,z_8 - 15\,z_3\,z_8 -
37\,z_1\,z_3\,z_8 - 45\,z_2\,z_3\,z_8 +
      5\,z_1\,z_2\,z_3\,z_8 + 4\,z_2^2\,z_3\,z_8 +38\,z_3^2\,z_8 + 73\,z_4\,z_8 - 41\,z_1\,z_4\,z_8 +
      4\,z_1^2\,z_4\,z_8 + 9\,z_2\,z_4\,z_8 - 8\,z_3\,z_4\,z_8 -
32\,z_5\,z_8 - 28\,z_1^2\,z_5\,z_8 -
      7\,z_2\,z_5\,z_8 - 2\,z_1\,z_2\,z_5\,z_8 + 18\,z_3\,z_5\,z_8 + 12\,z_5^2\,z_8 +
4\,z_6\,z_8 + 13\,z_1\,z_6\,z_8 -
      28\,z_1^2\,z_6\,z_8 - 4\,z_1^3\,z_6\,z_8 + 18\,z_2\,z_6\,z_8 +19\,z_1\,z_2\,z_6\,z_8 - 3\,z_2^2\,z_6\,z_8 +
      8\,z_3\,z_6\,z_8 + 8\,z_1\,z_3\,z_6\,z_8 - 6\,z_4\,z_6\,z_8 -
4\,z_5\,z_6\,z_8 + 4\,z_6^2\,z_8 -       4\,z_1\,z_6^2\,z_8 - 33\,z_7\,z_8 - 6\,z_1\,z_7\,z_8 - 45\,z_1^2\,z_7\,z_8 +
3\,z_1^3\,z_7\,z_8 +       50\,z_2\,z_7\,z_8 - 41\,z_1\,z_2\,z_7\,z_8 -
9\,z_1^2\,z_2\,z_7\,z_8 - 13\,z_2^2\,z_7\,z_8 + 46\,z_3\,z_7\,z_8 +
      9\,z_1\,z_3\,z_7\,z_8 +
      8\,z_2\,z_3\,z_7\,z_8 + 4\,z_4\,z_7\,z_8 + 11\,z_5\,z_7\,z_8 -
6\,z_1\,z_5\,z_7\,z_8 - 49\,z_6\,z_7\,z_8 +
      5\,z_1\,z_6\,z_7\,z_8 + 6\,z_2\,z_6\,z_7\,z_8 - 85\,z_7^2\,z_8 +
96\,z_1\,z_7^2\,z_8 + 3\,z_1^2\,z_7^2\,z_8 +
      26\,z_2\,z_7^2\,z_8 - 9\,z_6\,z_7^2\,z_8 - 9\,z_7^3\,z_8 - 42\,z_8^2 +
25\,z_1\,z_8^2 + 11\,z_1^2\,z_8^2 -
      44\,z_1^3\,z_8^2 + 6\,z_1^4\,z_8^2 + 37\,z_2\,z_8^2 -
51\,z_1\,z_2\,z_8^2 + 4\,z_1^2\,z_2\,z_8^2 +
      11\,z_2^2\,z_8^2 - 38\,z_3\,z_8^2 + 89\,z_1\,z_3\,z_8^2 -
24\,z_1^2\,z_3\,z_8^2 + 12\,z_3^2\,z_8^2 - 48\,z_4\,z_8^2 +
16\,z_1\,z_4\,z_8^2 - 63\,z_5\,z_8^2 + 2\,z_1\,z_5\,z_8^2 - 11\,z_2\,z_5\,z_8^2 -
      22\,z_6\,z_8^2 - 5\,z_1\,z_6\,z_8^2 + 14\,z_1^2\,z_6\,z_8^2 +
z_2\,z_6\,z_8^2 + 4\,z_3\,z_6\,z_8^2 + 2\,z_6^2\,z_8^2 - 41\,z_7\,z_8^2 + 74\,z_1\,z_7\,z_8^2 -
18\,z_1^2\,z_7\,z_8^2 - 54\,z_2\,z_7\,z_8^2 -
      z_1\,z_2\,z_7\,z_8^2 - 13\,z_3\,z_7\,z_8^2 + 21\,z_5\,z_7\,z_8^2- 4\,z_6\,z_7\,z_8^2 + 94\,z_7^2\,z_8^2 -
      17\,z_1\,z_7^2\,z_8^2 + 75\,z_8^3 - 64\,z_1\,z_8^3 +
32\,z_1^2\,z_8^3 - 44\,z_2\,z_8^3 +
      26\,z_1\,z_2\,z_8^3 + 4\,z_2^2\,z_8^3 + 6\,z_3\,z_8^3 - 8\,z_4\,z_8^3 + 14\,z_5\,z_8^3 +
34\,z_6\,z_8^3 +
      4\,z_1\,z_6\,z_8^3 + 95\,z_7\,z_8^3 - 64\,z_1\,z_7\,z_8^3 -
z_2\,z_7\,z_8^3 + 9\,z_7^2\,z_8^3 + 7\,z_8^4+ 8\,z_1\,z_8^4 - 8\,z_1^2\,z_8^4 + 12\,z_2\,z_8^4 +
16\,z_3\,z_8^4 - 4\,z_6\,z_8^4 - 42\,z_7\,z_8^4 - 32\,z_8^5 + 16\,z_1\,z_8^5
+ 10\,z_8^6)$

\medskip
\section*{Appendix A: List of the characters of $E_8$ of second order.}

Once the operator $\Delta^1$ is known, we can apply the methods described in \cite{ffp03} to solve in a recursive way the eigenvalue problem $\Delta^1\bchi_{\bf m}=\ve_{\bf m}(1)\bchi_{\bf m}$ to find the characters $\bchi_{\bf m}(z)$ (\ref{bchi}). In \cite{ffp03} we describe also how the Hamiltonian can be used to compute the Clebsch-Gordan series for the $E$-algebras. In this  appendix we give the complete list of the second order characters of $E_8$, and subsequently (App. B) some  higher order characters.
\\

$\bchi_{0, 0, 0, 0, 0, 0, 0, 2}  = -1 - z_1 - z_7 - z_8 + z_8^2\\ $

 $ \bchi_{1, 0, 0, 0, 0, 0, 0, 1} = -z_1 - z_2 - z_7 - z_8 + z_1\,z_8 $\\
 
 $ \bchi_{2, 0, 0, 0, 0, 0, 0, 0} = z_1 + z_1^2 - z_3 - z_6 + z_7 + z_8 -
z_1\,z_8 - z_8^2\\ $

 $ \bchi_{0, 0, 0, 0, 0, 0, 1, 1} = 1 + z_1 - z_6 + z_7 + z_8 - z_1\,z_8 +
z_7\,z_8 - z_8^2\\ $

 $ \bchi_{0, 1, 0, 0, 0, 0, 0, 1} = -z_3 - z_6 + z_8 - z_1\,z_8 + z_2\,z_8 $\\

 $ \bchi_{1, 0, 0, 0, 0, 0, 1, 0} = z_1 + z_2 + z_6 + z_1\,z_7 - z_2\,z_8 -
z_7\,z_8\\ $

 $ \bchi_{0, 0, 0, 0, 0, 1, 0, 1} =
   -1 - z_1 - z_2 - z_5 - z_7 - z_1\,z_7 - z_8 + z_1\,z_8 + z_6\,z_8 +
z_8^2\\ $

$\bchi_{0, 0, 0, 0, 0, 0, 2, 0} =
   z_3 + z_7 + z_1\,z_7 + z_7^2 + z_8 + 2\,z_1\,z_8 - z_6\,z_8 + 2\,z_7\,z_8
+ z_8^2 - z_1\,z_8^2 - z_8^3\\ $

$\bchi_{1, 1, 0, 0, 0, 0, 0, 0} =
   -z_1 - z_1^2 + z_1\,z_2 + z_3 - z_5 + z_6 - z_1\,z_7 - z_8 + z_1\,z_8 +
z_8^2\\ $

 $ \bchi_{0, 0, 1, 0, 0, 0, 0, 1} = -z_2 - z_1\,z_2 - z_6 + z_3\,z_8 +
z_7\,z_8\\ $

 $ \bchi_{0, 1, 0, 0, 0, 0, 1, 0} =
   z_1 + z_1^2 + z_2 - z_3 + z_5 + z_1\,z_7 + z_2\,z_7 + z_2\,z_8 -
z_3\,z_8 - z_6\,z_8 - z_1\,z_8^2\\ $

$ \bchi_{1, 0, 0, 0, 0, 1, 0, 0} =
   -z_1 + z_1\,z_6 - z_7 - z_1\,z_7 - z_2\,z_7 - z_7^2 - z_8 - z_1\,z_8 +
z_6\,z_8 - 2\,z_7\,z_8 - z_8^2 + z_1\,z_8^2 + z_8^3\\ $

$\bchi_{0, 2, 0, 0, 0, 0, 0, 0} =
   -z_2 + z_2^2 - z_4 - z_6 - z_1\,z_6 + z_8 - z_1^2\,z_8 + z_3\,z_8 +
z_6\,z_8 + z_7\,z_8 - z_8^2 + z_1\,z_8^2 \\ $

$\bchi_{1, 0, 1, 0, 0, 0, 0, 0} =
   z_1 + z_1^2 + z_1\,z_3 - z_4 - z_1\,z_6 + z_7 + 2\,z_1\,z_7 + z_7^2 +
z_8 - z_1^2\,z_8 - z_2\,z_8 +
    z_3\,z_8 + z_7\,z_8 - z_8^2\\ $

$\bchi_{0, 0, 0, 0, 1, 0, 0, 1} =
   1 + z_1 - z_4 - z_6 - z_1\,z_6 + 2\,z_7 + z_1\,z_7 + z_7^2 + 2\,z_8 -
z_2\,z_8 + z_5\,z_8 - z_6\,z_8 + 2\,z_7\,z_8 - z_8^2 - z_8^3\\ $

 $ \bchi_{0, 0, 0, 0, 0, 1, 1, 0} =
   -z_1 - z_1^2 + z_3 + z_6 + z_1\,z_6 - z_1\,z_7 + z_6\,z_7 - z_8 -
2\,z_1\,z_8 - z_2\,z_8 + z_3\,z_8 -   z_5\,z_8 + z_6\,z_8 - z_7\,z_8 -
z_1\,z_7\,z_8 - z_8^2 + 2\,z_1\,z_8^2 + z_8^3\\ $

$  \bchi_{0, 0, 1, 0, 0, 0, 1, 0} =
   z_2 + z_1\,z_2 + z_4 + z_6 + z_1\,z_6 - z_7 - z_1\,z_7 + z_2\,z_7 +
z_3\,z_7 - z_7^2 - z_8 + z_1^2\,z_8 + z_2\,z_8 - z_1\,z_2\,z_8 - z_3\,z_8 -
z_6\,z_8 - 2\,z_7\,z_8 + z_8^2 - z_1\,z_8^2 + z_7\,z_8^2\\ $

 $ \bchi_{0, 1, 0, 0, 0, 1, 0, 0} =
   -z_2 - z_1\,z_2 - z_3 - z_6 + z_2\,z_6 - z_2\,z_7 - z_3\,z_7 - z_6\,z_7 +
z_8 + z_1^2\,z_8 - z_3\,z_8 +
    z_5\,z_8 - z_6\,z_8 + z_7\,z_8 + z_8^2 - z_1\,z_8^2 + z_2\,z_8^2 -
z_8^3\\ $

$\bchi_{1, 0, 0, 0, 1, 0, 0, 0} =
   z_1 + z_2 + z_5 + z_1\,z_5 - z_1\,z_6 - z_2\,z_6 + z_1\,z_7 + z_2\,z_7 +
z_8 + z_1\,z_8 + z_2\,z_8 - z_6\,z_8 +
    z_7\,z_8 + z_1\,z_7\,z_8 + z_8^2 - z_1\,z_8^2 - z_2\,z_8^2 - z_8^3\\ $

$\bchi_{0, 1, 1, 0, 0, 0, 0, 0} =
   -z_1^2 - z_1^3 - z_1\,z_2 + 2\,z_1\,z_3 + z_2\,z_3 - z_4 - z_1\,z_5 - z_6
+ z_1\,z_6 + z_7 - z_1\,z_7 -
    z_1^2\,z_7 - z_2\,z_7 + z_3\,z_7 + z_6\,z_7 + z_8 - 2\,z_1\,z_8 +
z_1^2\,z_8 - z_2\,z_8 + z_3\,z_8 + z_6\,z_8 +
    z_1\,z_7\,z_8 - z_8^2 + 2\,z_1\,z_8^2\\ $

$\bchi_{0, 0, 0, 1, 0, 0, 0, 1} =
   -1 - z_1 + z_1^2 + z_1^3 + z_1\,z_2 - 2\,z_1\,z_3 - z_2\,z_3 + z_4 - z_5
+ z_6 - z_7 + z_1^2\,z_7 -
    z_3\,z_7 - 2\,z_8 + z_1\,z_8 - z_1^2\,z_8 + z_2\,z_8 - z_3\,z_8 +
z_4\,z_8 - z_5\,z_8 + z_6\,z_8 - z_7\,z_8 -
    z_1\,z_7\,z_8 + z_8^2 - z_1\,z_8^2 - z_7\,z_8^2 + z_8^3\\ $

$\bchi_{0, 0, 0, 0, 1, 0, 1, 0} =
   z_1\,z_2 + z_3 + z_1\,z_5 + z_6 + z_3\,z_7 + z_5\,z_7 + 2\,z_1\,z_8 -
z_1^2\,z_8 + z_3\,z_8 - z_4\,z_8 +
    z_5\,z_8 - z_1\,z_6\,z_8 + z_7\,z_8 + z_1\,z_7\,z_8 + z_7^2\,z_8 +
2\,z_8^2 - z_2\,z_8^2 - z_6\,z_8^2 +
    2\,z_7\,z_8^2 - z_8^4 $\\

$\bchi_{0, 0, 0, 0, 0, 2, 0, 0} =
   z_1 + z_1^2 - z_3 - z_4 + z_5 - 2\,z_6 - z_1\,z_6 + z_6^2 - z_2\,z_7 -
z_3\,z_7 - z_5\,z_7 - z_6\,z_7 -
    z_1\,z_7^2 + 2\,z_8 - z_1\,z_8 - z_1^2\,z_8 - z_2\,z_8 - 2\,z_6\,z_8 +
z_1\,z_6\,z_8 + 2\,z_7\,z_8 -
    z_1\,z_7\,z_8 - z_1\,z_8^2 + z_3\,z_8^2 + z_6\,z_8^2 + z_7\,z_8^2 -
z_8^3 + z_1\,z_8^3 $\\

$\bchi_{0, 0, 1, 0, 0, 1, 0, 0} =
   -z_1 - z_1^2 - z_2^2 + z_3 + z_4 - z_5 + 2\,z_6 - z_2\,z_6 + z_3\,z_6 -
z_7 - z_1\,z_2\,z_7 - z_6\,z_7 -
    3\,z_8 + z_1\,z_8 - z_3\,z_8 + z_4\,z_8 - z_5\,z_8 + z_1\,z_6\,z_8 -
2\,z_7\,z_8 - z_1\,z_7\,z_8 + z_2\,z_7\,z_8 +
    z_1^2\,z_8^2 + z_2\,z_8^2 - z_3\,z_8^2 - z_7\,z_8^2 + 2\,z_8^3 -
z_1\,z_8^3\\ $

$\bchi_{0, 1, 0, 0, 1, 0, 0, 0} =
   z_1 + z_1^2 + z_2^2 - z_3 - z_4 + z_2\,z_5 - z_6 - z_1\,z_6 - z_3\,z_6 -
z_6^2 + z_7 + 2\,z_1\,z_7 +
    z_1^2\,z_7 + z_2\,z_7 - z_3\,z_7 + z_5\,z_7 + z_7^2 + z_1\,z_7^2 + z_8 +
z_1\,z_8 + z_2\,z_8 - z_1\,z_2\,z_8 +
    z_5\,z_8 + z_6\,z_8 - z_1\,z_6\,z_8 + 2\,z_7\,z_8 + z_1\,z_7\,z_8 -
z_8^2 - z_6\,z_8^2 - z_7\,z_8^2 - z_1\,z_8^3\\ $

$\bchi_{1, 0, 0, 1, 0, 0, 0, 0} =
   z_1^2 - z_3 + z_1\,z_4 + z_5 - z_1\,z_5 - z_2\,z_5 - z_6 + z_1\,z_6 +
z_2\,z_6 + z_6^2 - z_7 - z_5\,z_7 -
    z_7^2 - 2\,z_1\,z_8 - z_3\,z_8 - 2\,z_6\,z_8 + z_1\,z_6\,z_8 -
z_7\,z_8 - 2\,z_1\,z_7\,z_8 - z_2\,z_7\,z_8 -
    z_7^2\,z_8 - z_1\,z_8^2 + z_2\,z_8^2 + z_6\,z_8^2 + z_7\,z_8^2 +
z_1\,z_8^3\\  $

$\bchi_{0, 0, 2, 0, 0, 0, 0, 0} =
   z_1 + 2\,z_1^2 + z_1^3 + z_2 + z_1\,z_2 - z_1\,z_3 + z_3^2 - z_4 -
z_1\,z_4 + z_5 + z_1\,z_5 - z_1\,z_6 -
    z_1^2\,z_6 - z_2\,z_6 + z_3\,z_6 + z_7 + 3\,z_1\,z_7 + 2\,z_1^2\,z_7 +
z_2\,z_7 - z_3\,z_7 + z_5\,z_7 +
    z_7^2 + z_1\,z_7^2 + 2\,z_8 + 2\,z_1\,z_8 - z_1^2\,z_8 - z_1^3\,z_8 -
z_1\,z_2\,z_8 - z_3\,z_8 +
    2\,z_1\,z_3\,z_8 - z_4\,z_8 + z_5\,z_8 - 2\,z_6\,z_8 + 3\,z_7\,z_8 +
z_8^2 - 3\,z_1\,z_8^2 - z_2\,z_8^2 +
    z_3\,z_8^2 - 2\,z_8^3 + z_1\,z_8^3\\ $

$\bchi_{0, 0, 0, 1, 0, 0, 1, 0} =
   -z_1 - z_1^2 - z_2 + z_3 + z_4 + z_1\,z_4 - z_5 + z_6 + z_3\,z_6 -
2\,z_1\,z_7 - z_1^2\,z_7 - z_2\,z_7 +
    z_3\,z_7 + z_4\,z_7 - z_1\,z_7^2 - 2\,z_8 - 2\,z_1\,z_8 + z_1^3\,z_8 -
z_2\,z_8 + 2\,z_1\,z_2\,z_8 + z_3\,z_8 -
    2\,z_1\,z_3\,z_8 - z_2\,z_3\,z_8 + 2\,z_4\,z_8 - 2\,z_5\,z_8 +
3\,z_6\,z_8 + z_1\,z_6\,z_8 - 3\,z_7\,z_8 -
    z_1\,z_7\,z_8 + z_1^2\,z_7\,z_8 - z_3\,z_7\,z_8 - 4\,z_8^2 +
2\,z_1\,z_8^2 - z_1^2\,z_8^2 + z_2\,z_8^2 -
    z_3\,z_8^2 - z_5\,z_8^2 + 2\,z_6\,z_8^2 - 3\,z_7\,z_8^2 -
z_1\,z_7\,z_8^2 + z_8^3 - z_7\,z_8^3 + 2\,z_8^4\\ $

$\bchi_{0, 0, 0, 0, 1, 1, 0, 0} =
   -z_1\,z_2 - z_1\,z_5 - z_6 - z_3\,z_6 + z_5\,z_6 + z_7 + z_1\,z_7 +
z_3\,z_7 - z_4\,z_7 - z_5\,z_7 - z_6\,z_7 -
    z_1\,z_6\,z_7 + 2\,z_7^2 + z_1\,z_7^2 + z_7^3 + z_8 + z_1^2\,z_8 +
z_1\,z_2\,z_8 + z_3\,z_8 - z_4\,z_8 -
    z_5\,z_8 + z_1\,z_5\,z_8 - z_1\,z_6\,z_8 + 3\,z_7\,z_8 + z_1\,z_7\,z_8 -
z_2\,z_7\,z_8 + z_3\,z_7\,z_8 -
    z_6\,z_7\,z_8 + 3\,z_7^2\,z_8 + z_1\,z_8^2 - z_1^2\,z_8^2 - z_2\,z_8^2 +
z_5\,z_8^2 - z_8^3 - z_7\,z_8^3\\ $

$\bchi_{0, 0, 1, 0, 1, 0, 0, 0} =
   z_2 + z_1\,z_2 + z_5 + z_1\,z_5 + z_3\,z_5 + z_6 - z_2\,z_6 -
z_1\,z_2\,z_6 - z_6^2 - z_7 - z_1\,z_7 +
    2\,z_2\,z_7 + z_1\,z_2\,z_7 - z_3\,z_7 + z_4\,z_7 + z_5\,z_7 + z_6\,z_7
+ z_1\,z_6\,z_7 - 2\,z_7^2 - z_1\,z_7^2 +
    z_2\,z_7^2 - z_7^3 + 2\,z_2\,z_8 + z_1\,z_2\,z_8 + z_5\,z_8 + z_6\,z_8 -
z_2\,z_6\,z_8 - 3\,z_7\,z_8 +
    z_1^2\,z_7\,z_8 + 4\,z_2\,z_7\,z_8 - z_3\,z_7\,z_8 + z_6\,z_7\,z_8 -
2\,z_7^2\,z_8 - z_1\,z_2\,z_8^2 - z_5\,z_8^2 -
    z_6\,z_8^2 + z_7\,z_8^2 - z_1\,z_7\,z_8^2 - z_2\,z_8^3 + z_7\,z_8^3\\ $

$\bchi_{0, 1, 0, 1, 0, 0, 0, 0} =
   z_1\,z_3 - z_4 + z_2\,z_4 - z_3\,z_5 - z_6 + z_1^2\,z_6 - z_3\,z_6 + z_7
+ z_1\,z_7 - z_2\,z_7 - z_1\,z_2\,z_7 -
    z_6\,z_7 + z_1\,z_6\,z_7 + z_7^2 - z_2\,z_7^2 + 2\,z_8 - 2\,z_2\,z_8 -
z_1\,z_2\,z_8 + z_2^2\,z_8 - z_4\,z_8 +
    z_5\,z_8 - z_1\,z_5\,z_8 - 2\,z_6\,z_8 - z_1\,z_6\,z_8 + z_2\,z_6\,z_8 +
3\,z_7\,z_8 - z_2\,z_7\,z_8 - z_6\,z_7\,z_8 +
    z_7^2\,z_8 + 2\,z_8^2 + z_5\,z_8^2 - z_6\,z_8^2 + 2\,z_7\,z_8^2 - z_8^3
+ z_2\,z_8^3 - z_8^4\\ $

$\bchi_{0, 0, 0, 1, 0, 1, 0, 0} =
   z_1^2 + z_1^3 + z_1\,z_2 - 2\,z_1\,z_3 - z_2\,z_3 - z_1\,z_4 + z_1\,z_5 -
z_3\,z_5 + z_4\,z_6 + 2\,z_1^2\,z_7 +
    z_1^3\,z_7 + z_1\,z_2\,z_7 - 2\,z_1\,z_3\,z_7 - z_2\,z_3\,z_7 - z_4\,z_7
+ z_7^2 + z_1\,z_7^2 + z_1^2\,z_7^2 -
    z_3\,z_7^2 + z_7^3 + z_1\,z_8 - z_1^2\,z_8 - z_1\,z_2\,z_8 - z_4\,z_8 +
z_1\,z_4\,z_8 - z_1\,z_5\,z_8 -
    z_6\,z_8 - z_1\,z_6\,z_8 + z_3\,z_6\,z_8 + 3\,z_7\,z_8 + z_1\,z_7\,z_8 -
2\,z_1^2\,z_7\,z_8 - z_2\,z_7\,z_8 -
    z_5\,z_7\,z_8 + 3\,z_7^2\,z_8 - 2\,z_1\,z_7^2\,z_8 + 2\,z_8^2 -
2\,z_1\,z_8^2 - z_1^2\,z_8^2 - z_2\,z_8^2 +
    z_1\,z_2\,z_8^2 + z_3\,z_8^2 + z_4\,z_8^2 - z_5\,z_8^2 + z_6\,z_8^2 +
z_1\,z_6\,z_8^2 + z_7\,z_8^2 -
    2\,z_1\,z_7\,z_8^2 - z_7^2\,z_8^2 - 2\,z_8^3 + z_6\,z_8^3 -
3\,z_7\,z_8^3 - z_8^4 + z_1\,z_8^4 + z_8^5\\ $

$\bchi_{0, 0, 0, 0, 2, 0, 0, 0} =
   z_2^2 + 2\,z_2\,z_5 + z_5^2 + z_1\,z_6 + z_1^2\,z_6 - 2\,z_3\,z_6 -
z_4\,z_6 - 2\,z_6^2 - z_1\,z_6^2 -
    z_1\,z_7 - z_1^2\,z_7 + z_1\,z_2\,z_7 + 2\,z_3\,z_7 + z_4\,z_7 +
z_1\,z_5\,z_7 + 4\,z_6\,z_7 + 2\,z_1\,z_6\,z_7 -
    2\,z_7^2 - z_1\,z_7^2 + z_3\,z_7^2 + z_6\,z_7^2 - z_7^3 - z_1\,z_8 -
z_1^2\,z_8 + 2\,z_2\,z_8 -
    z_1\,z_2\,z_8 + 2\,z_3\,z_8 + z_4\,z_8 + 2\,z_5\,z_8 - z_1\,z_5\,z_8 +
4\,z_6\,z_8 + z_1\,z_6\,z_8 - z_2\,z_6\,z_8 -
    z_3\,z_6\,z_8 - z_6^2\,z_8 - 4\,z_7\,z_8 - z_1^2\,z_7\,z_8 +
z_2\,z_7\,z_8 + 3\,z_3\,z_7\,z_8 + 4\,z_6\,z_7\,z_8 -
    4\,z_7^2\,z_8 - z_8^2 + z_1^2\,z_8^2 + z_2\,z_8^2 - z_4\,z_8^2 -
2\,z_6\,z_8^2 - z_1\,z_6\,z_8^2 +
    z_1\,z_7\,z_8^2 + z_7^2\,z_8^2 + 3\,z_8^3 - z_2\,z_8^3 - z_3\,z_8^3 -
2\,z_6\,z_8^3 + 4\,z_7\,z_8^3 - z_8^5\\ $

$\bchi_{0, 0, 1, 1, 0, 0, 0, 0} =
   -z_2^2 + z_2\,z_3 + z_4 + z_3\,z_4 - 2\,z_2\,z_5 - z_1\,z_2\,z_5 + z_6 +
z_1\,z_6 + z_2\,z_6 + z_1\,z_2\,z_6 +
    z_3\,z_6 + z_4\,z_6 - z_5\,z_6 + z_6^2 + z_1\,z_6^2 - z_7 + z_2\,z_7 -
z_2^2\,z_7 + z_4\,z_7 - z_5\,z_7 +
    z_6\,z_7 - z_1\,z_6\,z_7 - z_7^2 + z_2\,z_7^2 - z_6\,z_7^2 - 2\,z_8 -
z_2^2\,z_8 - 2\,z_3\,z_8 + z_1\,z_5\,z_8 -
    z_6\,z_8 - z_1\,z_6\,z_8 + z_1^2\,z_6\,z_8 + z_2\,z_6\,z_8 -
z_3\,z_6\,z_8 - 4\,z_7\,z_8 - z_1\,z_2\,z_7\,z_8 -
    2\,z_3\,z_7\,z_8 + z_5\,z_7\,z_8 - 3\,z_6\,z_7\,z_8 - 2\,z_7^2\,z_8 +
z_2^2\,z_8^2 - z_4\,z_8^2 + z_5\,z_8^2 -
    z_6\,z_8^2 - z_1\,z_6\,z_8^2 + z_7\,z_8^2 + z_7^2\,z_8^2 + 3\,z_8^3 +
z_3\,z_8^3 + 3\,z_7\,z_8^3 - z_8^5\\ $

$\bchi_{0, 0, 0, 1, 1, 0, 0, 0} =
  z_1\,z_2 + z_1^2\,z_2 - z_2\,z_3 + z_3^2 + z_2\,z_4 + z_1\,z_5 +
z_1^2\,z_5 - z_3\,z_5 + z_4\,z_5 - z_1\,z_6 +
   z_1^3\,z_6 - z_2\,z_6 + z_1\,z_2\,z_6 + z_3\,z_6 - 2\,z_1\,z_3\,z_6 -
z_2\,z_3\,z_6 + 2\,z_4\,z_6 - 2\,z_5\,z_6 +
   z_6^2 - z_1\,z_7 - z_1^2\,z_7 - z_2\,z_7 + z_3\,z_7 + z_1\,z_4\,z_7 +
z_1\,z_5\,z_7 - z_6\,z_7 - z_1\,z_6\,z_7 +
   z_1^2\,z_6\,z_7 - 2\,z_1\,z_7^2 - z_1^2\,z_7^2 - z_2\,z_7^2 +
z_3\,z_7^2 - z_6\,z_7^2 - z_1\,z_7^3 -
   2\,z_2\,z_8 + z_2^2\,z_8 - z_2\,z_3\,z_8 + z_4\,z_8 - z_1\,z_4\,z_8 -
z_5\,z_8 + z_1\,z_5\,z_8 + z_2\,z_5\,z_8 -
   z_3\,z_5\,z_8 - 2\,z_6\,z_8 + z_1\,z_6\,z_8 - z_1^2\,z_6\,z_8 +
2\,z_2\,z_6\,z_8 - z_3\,z_6\,z_8 - z_5\,z_6\,z_8 +
   z_6^2\,z_8 - z_7\,z_8 - z_1\,z_7\,z_8 - 2\,z_2\,z_7\,z_8 +
z_1\,z_2\,z_7\,z_8 + z_3\,z_7\,z_8 - z_4\,z_7\,z_8 +
   z_5\,z_7\,z_8 - 2\,z_6\,z_7\,z_8 + z_7^3\,z_8 - 2\,z_8^2 +
2\,z_1\,z_8^2 - z_1\,z_2\,z_8^2 - z_1\,z_5\,z_8^2 +
   2\,z_6\,z_8^2 - 2\,z_1\,z_6\,z_8^2 + 3\,z_1\,z_7\,z_8^2 -
z_2\,z_7\,z_8^2 - z_6\,z_7\,z_8^2 + 2\,z_7^2\,z_8^2 +
   z_2\,z_8^3 + z_6\,z_8^3 + z_1\,z_7\,z_8^3 + z_8^4 - z_1\,z_8^4 -
z_7\,z_8^4\\ $

$\bchi_{0, 0, 0, 2, 0, 0, 0, 0} =
  -z_4 - z_1\,z_4 + z_4^2 + z_1^2\,z_5 + z_1^3\,z_5 - z_2\,z_5 + z_3\,z_5 -
2\,z_1\,z_3\,z_5 - z_2\,z_3\,z_5 +
   z_4\,z_5 - z_5^2 - z_6 - 2\,z_1\,z_6 - z_1^2\,z_6 - z_2^2\,z_6 + z_3\,z_6
+ 3\,z_4\,z_6 + z_1\,z_4\,z_6 -
   z_5\,z_6 + 3\,z_6^2 + 2\,z_1\,z_6^2 + z_3\,z_6^2 + z_7 + z_1\,z_7 -
3\,z_4\,z_7 - z_1\,z_4\,z_7 +
   z_1^2\,z_5\,z_7 - z_2\,z_5\,z_7 - 2\,z_3\,z_5\,z_7 - 3\,z_6\,z_7 -
3\,z_1\,z_6\,z_7 - z_1^2\,z_6\,z_7 + z_3\,z_6\,z_7 +
   z_6^2\,z_7 + 3\,z_7^2 + 2\,z_1\,z_7^2 - 2\,z_4\,z_7^2 - 2\,z_6\,z_7^2 -
z_1\,z_6\,z_7^2 + 3\,z_7^3 +
   z_1\,z_7^3 + z_7^4 + 2\,z_8 + 2\,z_1\,z_8 - 4\,z_4\,z_8 + z_2\,z_4\,z_8 +
2\,z_1\,z_5\,z_8 - z_2\,z_5\,z_8 -
   2\,z_3\,z_5\,z_8 - z_5^2\,z_8 - 7\,z_6\,z_8 - 3\,z_1\,z_6\,z_8 +
z_1\,z_2\,z_6\,z_8 + 2\,z_4\,z_6\,z_8 -
   2\,z_5\,z_6\,z_8 + 3\,z_6^2\,z_8 + z_1\,z_6^2\,z_8 + 8\,z_7\,z_8 +
4\,z_1\,z_7\,z_8 - z_2\,z_7\,z_8 - 4\,z_4\,z_7\,z_8 +
   z_5\,z_7\,z_8 - z_1\,z_5\,z_7\,z_8 - 9\,z_6\,z_7\,z_8 -
3\,z_1\,z_6\,z_7\,z_8 + 10\,z_7^2\,z_8 + 2\,z_1\,z_7^2\,z_8 -
   z_2\,z_7^2\,z_8 - 2\,z_6\,z_7^2\,z_8 + 4\,z_7^3\,z_8 + 4\,z_8^2 -
2\,z_2\,z_8^2 + z_4\,z_8^2 + z_5\,z_8^2 -
   2\,z_1\,z_5\,z_8^2 + z_2\,z_5\,z_8^2 - 3\,z_6\,z_8^2 + z_1\,z_6\,z_8^2 +
z_2\,z_6\,z_8^2 + z_6^2\,z_8^2 +
   7\,z_7\,z_8^2 - 2\,z_2\,z_7\,z_8^2 - 4\,z_6\,z_7\,z_8^2 +
3\,z_7^2\,z_8^2 - 3\,z_8^3 - z_1\,z_8^3 +
   2\,z_4\,z_8^3 - z_5\,z_8^3 + 4\,z_6\,z_8^3 + z_1\,z_6\,z_8^3 -
5\,z_7\,z_8^3 - z_1\,z_7\,z_8^3 -    2\,z_7^2\,z_8^3 - 4\,z_8^4 + z_2\,z_8^4
+ 2\,z_6\,z_8^4 - 4\,z_7\,z_8^4 + z_8^5 + z_8^6 $

\medskip
\section*{Appendix B: List of other characters.}
\

 $ \bchi_{0, 0, 0, 0, 0, 0, 0, 3} = z_1 + z_2 + z_6 - z_8 - z_1\,z_8 -
2\,z_7\,z_8 - z_8^2 + z_8^3\\ $

 $ \bchi_{1, 0, 0, 0, 0, 0, 0, 2} = -z_1 - z_1^2 + z_3 + z_6 - z_1\,z_7 -
z_2\,z_8 - z_7\,z_8 + z_1\,z_8^2\\ $

  $\bchi_{2, 0, 0, 0, 0, 0, 0, 1} =
   -z_1\,z_2 - z_3 + z_5 - z_6 + z_8 + z_1^2\,z_8 + z_2\,z_8 - z_3\,z_8 -
z_6\,z_8 + 2\,z_7\,z_8 + z_8^2 -
    z_1\,z_8^2 - z_8^3\\ $

$ \bchi_{0, 0, 0, 0, 0, 0, 1, 2} =
   z_1^2 - z_3 + z_5 - z_6 - z_7 - z_7^2 + z_8 + z_1\,z_8 + z_2\,z_8 -
z_6\,z_8 + z_7\,z_8 + z_8^2 -
    z_1\,z_8^2 + z_7\,z_8^2 - z_8^3\\ $

$\bchi_{0, 1, 0, 0, 0, 0, 0, 2} =
   z_1 + z_1^2 + z_7 + z_1\,z_7 - z_2\,z_7 - z_2\,z_8 - z_3\,z_8 -
z_6\,z_8 - z_1\,z_8^2 + z_2\,z_8^2\\ $

$  \bchi_{1, 0, 0, 0, 0, 0, 1, 1} =
   z_1\,z_2 - z_5 - z_1\,z_6 - z_8 - z_1^2\,z_8 + z_2\,z_8 + z_3\,z_8 +
2\,z_6\,z_8 - z_7\,z_8 + z_1\,z_7\,z_8 -
    z_8^2 + z_1\,z_8^2 - z_2\,z_8^2 - z_7\,z_8^2 + z_8^3\\ $

 $ \bchi_{1, 1, 0, 0, 0, 0, 0, 1} =
   -z_1 - z_1^2 - z_2^2 + z_3 - z_1\,z_3 + z_4 - z_5 + 2\,z_6 + z_1\,z_6 -
z_7 - z_1\,z_7 - 2\,z_8 -
    z_2\,z_8 + z_1\,z_2\,z_8 - z_5\,z_8 + z_6\,z_8 - 2\,z_7\,z_8 -
z_1\,z_7\,z_8 + z_1\,z_8^2 + z_8^3\\ $

$  \bchi_{2, 0, 0, 0, 0, 0, 1, 0} =
   z_1 + z_1^2 + z_1\,z_2 + z_2^2 - z_4 + z_5 - z_6 + z_7 + z_1\,z_7 +
z_1^2\,z_7 - z_3\,z_7 - z_6\,z_7 +
    z_7^2 + 2\,z_8 - z_1\,z_2\,z_8 - z_3\,z_8 + z_5\,z_8 - 2\,z_6\,z_8 +
2\,z_7\,z_8 - z_1\,z_7\,z_8 - z_1\,z_8^2 +
    z_2\,z_8^2 - z_8^3\\ $

$ \bchi_{3, 0, 0, 0, 0, 0, 0, 0} =
   z_1^2 + z_1^3 + z_1\,z_2 - 2\,z_1\,z_3 + z_4 - z_5 - z_1\,z_6 + z_1\,z_7
+ z_2\,z_7 - z_8 + 2\,z_1\,z_8 -
    z_1^2\,z_8 + z_2\,z_8 - z_3\,z_8 + z_8^2 - 2\,z_1\,z_8^2\\ $

$  \bchi_{0, 0, 0, 0, 0, 1, 0, 2} =
   -z_1^2 + z_2 + z_4 + z_6 - z_7 + z_2\,z_7 - z_6\,z_7 - 2\,z_8 - z_1\,z_8
+ z_1^2\,z_8 - z_3\,z_8 -
    z_5\,z_8 - z_6\,z_8 - 2\,z_7\,z_8 - z_1\,z_7\,z_8 + z_6\,z_8^2 + z_8^3\\ $

 $ \bchi_{0, 0, 1, 0, 0, 0, 0, 2} =
   z_2^2 - z_3 - z_4 + z_5 - z_6 - z_3\,z_7 + 2\,z_8 - z_1\,z_2\,z_8 -
z_6\,z_8 + 2\,z_7\,z_8 + z_3\,z_8^2 +
    z_7\,z_8^2 - z_8^3\\ $

$ \bchi_{0, 1, 0, 0, 0, 0, 1, 1} =
   -z_1\,z_2 + z_1\,z_3 - z_4 + z_5 - z_6 - z_1\,z_6 - z_2\,z_6 + 2\,z_8 +
z_1\,z_8 + z_1^2\,z_8 + z_2\,z_8 +
    z_5\,z_8 + 2\,z_7\,z_8 + 2\,z_1\,z_7\,z_8 + z_2\,z_7\,z_8 - z_3\,z_8^2 -
z_6\,z_8^2 - z_8^3 - z_1\,z_8^3\\ $

$  \bchi_{1, 0, 0, 0, 0, 1, 0, 1} =
   -z_2 - z_1\,z_2 - z_5 - z_1\,z_5 - z_6 - z_1\,z_6 + z_7 - z_1^2\,z_7 +
z_3\,z_7 + z_6\,z_7 + z_8 -
    z_1^2\,z_8 + z_1\,z_2\,z_8 + z_3\,z_8 - z_5\,z_8 + z_6\,z_8 +
z_1\,z_6\,z_8 - z_2\,z_7\,z_8 - z_7^2\,z_8 -
    z_8^2 - z_1\,z_8^2 + z_6\,z_8^2 - 2\,z_7\,z_8^2 - z_8^3 + z_1\,z_8^3 +
z_8^4\\ $

 $  \bchi_{0, 0, 0, 0, 0, 0, 0, 4} =
   -z_2 - z_5 + z_7 + z_1\,z_7 + z_7^2 + 2\,z_1\,z_8 + z_2\,z_8 +
2\,z_6\,z_8 + z_7\,z_8 - z_8^2 - z_1\,z_8^2 -
    3\,z_7\,z_8^2 - z_8^3 + z_8^4\\ $

$ \bchi_{0, 0, 0, 0, 0, 0, 2, 1} =
   -z_1 - z_1^2 - z_2 - z_1\,z_2 - z_1\,z_7 - z_2\,z_7 - z_6\,z_7 - z_1\,z_8
+ z_1^2\,z_8 - z_2\,z_8 + z_5\,z_8 -
    z_6\,z_8 + z_1\,z_7\,z_8 + z_7^2\,z_8 + z_8^2 + 2\,z_1\,z_8^2 +
z_2\,z_8^2 - z_6\,z_8^2 + 2\,z_7\,z_8^2 +
    z_8^3 - z_1\,z_8^3 - z_8^4\\ $

$ \bchi_{1, 0, 0, 0, 0, 0, 0, 3} =
   z_2 + z_1\,z_2 + z_3 + z_6 + z_1\,z_6 + z_7 + z_1\,z_7 + z_2\,z_7 +
z_7^2 - z_1\,z_8 - z_1^2\,z_8 +
    z_3\,z_8 + z_6\,z_8 + z_7\,z_8 - 2\,z_1\,z_7\,z_8 - z_2\,z_8^2 -
z_7\,z_8^2 + z_1\,z_8^3\\ $

$  \bchi_{1, 0, 0, 0, 0, 0, 2, 0} =
   z_1 + z_1^2 - z_3 + z_1\,z_6 + z_2\,z_6 + 2\,z_1\,z_7 + z_1^2\,z_7 -
z_3\,z_7 + z_6\,z_7 + z_1\,z_7^2 -
    z_2\,z_8 + z_1\,z_2\,z_8 - z_5\,z_8 + z_6\,z_8 - z_1\,z_6\,z_8 -
z_1\,z_7\,z_8 - z_2\,z_7\,z_8 - z_7^2\,z_8 -
    z_8^2 - z_1\,z_8^2 - z_1^2\,z_8^2 + z_3\,z_8^2 + z_6\,z_8^2 -
2\,z_7\,z_8^2 - z_8^3 + z_1\,z_8^3 + z_8^4\\ $

$  \bchi_{2, 0, 0, 0, 0, 0, 0, 2} =
   -z_1^2 - z_1^3 - z_1\,z_2 + 2\,z_1\,z_3 + z_5 + z_1\,z_6 - z_7 -
z_1\,z_7 - z_1^2\,z_7 + z_3\,z_7 +
    z_6\,z_7 - z_7^2 - z_1\,z_8 + z_1^2\,z_8 - z_1\,z_2\,z_8 + z_5\,z_8 -
z_6\,z_8 - z_7\,z_8 + z_1\,z_7\,z_8 +
    z_8^2 + z_1\,z_8^2 + z_1^2\,z_8^2 + z_2\,z_8^2 - z_3\,z_8^2 - z_6\,z_8^2
+ 3\,z_7\,z_8^2 + z_8^3 -
    z_1\,z_8^3 - z_8^4\\ $

$  \bchi_{0, 2, 0, 0, 0, 0, 0, 1} =
   z_1 + 2\,z_1^2 + z_1^3 + z_2 + z_1\,z_2 - z_1\,z_3 - z_2\,z_3 + z_5 +
z_1\,z_5 - z_1\,z_6 - z_2\,z_6 +
    2\,z_1\,z_7 + z_1^2\,z_7 + z_2\,z_7 + z_1\,z_8 - z_1^2\,z_8 -
z_1\,z_2\,z_8 + z_2^2\,z_8 - z_4\,z_8 -
    2\,z_6\,z_8 - z_1\,z_6\,z_8 + z_8^2 - 2\,z_1\,z_8^2 - z_1^2\,z_8^2 -
z_2\,z_8^2 + z_3\,z_8^2 + z_6\,z_8^2 +
    z_7\,z_8^2 - z_8^3 + z_1\,z_8^3\\ $

$ \bchi_{1, 0, 1, 0, 0, 0, 0, 1} =
   z_1^2 + z_1^3 - z_2 - z_1\,z_2 - z_1^2\,z_2 - z_3 - 2\,z_1\,z_3 +
z_1\,z_5 - z_6 - 2\,z_1\,z_6 + z_2\,z_6 +
    z_1^2\,z_7 - z_2\,z_7 - z_3\,z_7 - z_6\,z_7 + 2\,z_1\,z_8 +
z_1\,z_2\,z_8 - z_3\,z_8 + z_1\,z_3\,z_8 - z_4\,z_8 +
    z_5\,z_8 - z_1\,z_6\,z_8 + z_7\,z_8 + 2\,z_1\,z_7\,z_8 + z_7^2\,z_8 +
z_8^2 - z_1\,z_8^2 - z_1^2\,z_8^2 +
    z_3\,z_8^2 - z_8^3\\ $

$ \bchi_{1, 1, 0, 0, 0, 0, 1, 0} =
   -z_1 - z_1^2 + z_1\,z_2 + z_3 + z_2\,z_3 + z_4 - z_5 + z_6 + z_1\,z_6 +
z_2\,z_6 - 2\,z_1\,z_7 -
    z_1^2\,z_7 + z_1\,z_2\,z_7 + z_3\,z_7 - z_5\,z_7 - z_1\,z_7^2 - 2\,z_8 -
z_1\,z_8 - z_1^2\,z_8 + z_1\,z_2\,z_8 -
    z_2^2\,z_8 + z_3\,z_8 - z_1\,z_3\,z_8 + z_4\,z_8 - z_5\,z_8 +
2\,z_6\,z_8 + z_1\,z_6\,z_8 - 2\,z_7\,z_8 -
    2\,z_1\,z_7\,z_8 - z_8^2 + z_1\,z_8^2 + z_1^2\,z_8^2 - z_3\,z_8^2 -
z_7\,z_8^2 + 2\,z_8^3\\ $

$  \bchi_{2, 1, 0, 0, 0, 0, 0, 0} =
   -z_1 - 2\,z_1^2 - z_1^3 + z_1^2\,z_2 + z_1\,z_3 - z_2\,z_3 - z_1\,z_5 +
z_6 + 2\,z_1\,z_6 - z_7 -
    2\,z_1\,z_7 - z_1^2\,z_7 + z_6\,z_7 - z_7^2 - z_8 - 2\,z_1\,z_8 +
z_1^2\,z_8 + z_3\,z_8 - z_5\,z_8 +
    2\,z_6\,z_8 - 2\,z_7\,z_8 - z_8^2 + 3\,z_1\,z_8^2 - z_2\,z_8^2 + z_8^3\\ $

 $ \bchi_{2, 0, 0, 0, 0, 1, 0, 0} =
   -z_1\,z_2 + z_1\,z_3 - z_4 + z_5 - z_6 + z_1^2\,z_6 - z_2\,z_6 -
z_3\,z_6 - z_6^2 + z_7 + z_1\,z_7 -
    z_1\,z_2\,z_7 + z_5\,z_7 + z_7^2 + 2\,z_8 - z_1\,z_2\,z_8 + z_2^2\,z_8 +
z_3\,z_8 - z_4\,z_8 + 2\,z_5\,z_8 -
    z_6\,z_8 - z_1\,z_6\,z_8 + 4\,z_7\,z_8 + z_2\,z_7\,z_8 + z_7^2\,z_8 +
2\,z_8^2 - 2\,z_6\,z_8^2 + 2\,z_7\,z_8^2 -
    z_8^3 - z_8^4\\ $

 $\bchi_{3, 0, 0, 0, 0, 0, 0, 1} =
   -z_1 - z_1^2 - z_1\,z_2 - z_1^2\,z_2 - z_2^2 + z_3 + z_2\,z_3 + z_4 - z_5
+ z_1\,z_5 + z_6 - z_7 -
    2\,z_1\,z_7 - z_2\,z_7 + z_3\,z_7 - z_7^2 - 2\,z_8 + z_1^2\,z_8 +
z_1^3\,z_8 - z_2\,z_8 + 2\,z_1\,z_2\,z_8 +
    z_3\,z_8 - 2\,z_1\,z_3\,z_8 + z_4\,z_8 - z_5\,z_8 + z_6\,z_8 -
z_1\,z_6\,z_8 - 3\,z_7\,z_8 + 2\,z_1\,z_7\,z_8 +
    z_2\,z_7\,z_8 - z_8^2 + 3\,z_1\,z_8^2 - z_1^2\,z_8^2 + z_2\,z_8^2 -
z_3\,z_8^2 + 2\,z_8^3 - 2\,z_1\,z_8^3\\ $

$  \bchi_{0, 0, 1, 0, 0, 0, 1, 1} =
   -z_1^2 - z_1^3 - z_1\,z_2 + z_1^2\,z_2 + z_1\,z_3 - z_2\,z_3 - z_4 +
z_5 - z_1\,z_5 - z_6 - z_2\,z_6 -
    z_3\,z_6 - z_1^2\,z_7 + z_8 - 2\,z_1\,z_8 + z_1^2\,z_8 + z_2^2\,z_8 +
z_5\,z_8 - z_6\,z_8 + z_1\,z_6\,z_8 -
    z_1\,z_7\,z_8 + z_2\,z_7\,z_8 + z_3\,z_7\,z_8 - z_7^2\,z_8 +
2\,z_1\,z_8^2 + z_1^2\,z_8^2 + z_2\,z_8^2 -
    z_1\,z_2\,z_8^2 - z_3\,z_8^2 - z_6\,z_8^2 + z_7\,z_8^2 + z_8^3 -
z_1\,z_8^3 + z_7\,z_8^3 - z_8^4\\ $

 $ \bchi_{0, 1, 0, 0, 0, 1, 0, 1} =
   -z_1^2 - z_1^3 + z_1\,z_3 - z_1\,z_5 - z_2\,z_5 + z_6 + 2\,z_1\,z_6 +
z_2\,z_6 + z_6\,z_7 + z_1\,z_7^2 -
    z_8 - z_1\,z_8 + 2\,z_1^2\,z_8 - 2\,z_2\,z_8 - z_1\,z_2\,z_8 - z_3\,z_8
+ z_1\,z_3\,z_8 - z_4\,z_8 - z_6\,z_8 -
    z_1\,z_6\,z_8 + z_2\,z_6\,z_8 - z_7\,z_8 + z_1\,z_7\,z_8 -
2\,z_2\,z_7\,z_8 - z_3\,z_7\,z_8 - z_6\,z_7\,z_8 + z_8^2 +
    2\,z_1\,z_8^2 + z_5\,z_8^2 - z_6\,z_8^2 + 2\,z_7\,z_8^2 + z_8^3 -
z_1\,z_8^3 + z_2\,z_8^3 - z_8^4\\ $

$  \bchi_{0, 1, 1, 0, 0, 0, 0, 1} =
   1 - 2\,z_1^2 - z_1^3 + z_1\,z_2 + z_1^2\,z_2 - z_1\,z_2^2 + z_3 +
z_1\,z_3 - z_2\,z_3 - z_3^2 + z_1\,z_4 +
    z_2\,z_5 + 2\,z_1\,z_6 + 2\,z_1^2\,z_6 - z_2\,z_6 - 2\,z_3\,z_6 - z_6^2
+ z_7 - 2\,z_1\,z_7 - 2\,z_1^2\,z_7 +
    z_1\,z_2\,z_7 + z_3\,z_7 + z_6\,z_7 - z_1\,z_7^2 + z_8 - 2\,z_1\,z_8 +
z_1^3\,z_8 + z_2\,z_8 + 2\,z_3\,z_8 -
    2\,z_1\,z_3\,z_8 + z_2\,z_3\,z_8 - z_1\,z_5\,z_8 + z_6\,z_8 -
z_1\,z_6\,z_8 + z_7\,z_8 - z_1\,z_7\,z_8 -
    z_1^2\,z_7\,z_8 + z_3\,z_7\,z_8 + z_6\,z_7\,z_8 - z_8^2 +
2\,z_1\,z_8^2 - z_2\,z_8^2 - z_3\,z_8^2 -
    z_6\,z_8^2 + z_1\,z_7\,z_8^2\\  $

$\bchi_{1, 0, 0, 0, 1, 0, 0, 1} =
   -z_1\,z_4 - z_1^2\,z_6 + z_3\,z_6 + z_6^2 - z_7 - z_1\,z_7 +
z_1\,z_2\,z_7 - z_5\,z_7 + z_6\,z_7 - z_7^2 -
    z_8 + z_1^2\,z_8 - z_5\,z_8 + z_1\,z_5\,z_8 - z_1\,z_6\,z_8 -
z_2\,z_6\,z_8 - 2\,z_7\,z_8 + 2\,z_1\,z_7\,z_8 +
    z_2\,z_7\,z_8 - z_7^2\,z_8 - z_8^2 + 2\,z_1\,z_8^2 - z_1^2\,z_8^2 +
z_2\,z_8^2 + z_3\,z_8^2 + z_6\,z_8^2 +
    z_7\,z_8^2 + z_1\,z_7\,z_8^2 + z_8^3 - z_1\,z_8^3 - z_2\,z_8^3\\ $

$  \bchi_{1, 1, 0, 0, 0, 0, 0, 2} =
   z_1 + z_1^2 + z_2 + z_1\,z_2 + z_1\,z_3 + z_2\,z_3 + z_7 + 3\,z_1\,z_7 +
z_1^2\,z_7 + z_2\,z_7 -
    z_1\,z_2\,z_7 + z_5\,z_7 - z_6\,z_7 + z_7^2 + z_1\,z_7^2 - z_1\,z_8 -
z_1^2\,z_8 - z_1\,z_2\,z_8 - z_2^2\,z_8 +
    z_3\,z_8 - z_1\,z_3\,z_8 + z_4\,z_8 + z_6\,z_8 + z_1\,z_6\,z_8 -
2\,z_1\,z_7\,z_8 - 2\,z_8^2 - z_1\,z_8^2 -
    z_2\,z_8^2 + z_1\,z_2\,z_8^2 - z_5\,z_8^2 + z_6\,z_8^2 - 3\,z_7\,z_8^2 -
z_1\,z_7\,z_8^2 + z_1\,z_8^3 + z_8^4\\ $

$  \bchi_{1, 2, 0, 0, 0, 0, 0, 0} =
   -z_2 - 2\,z_1\,z_2 - z_1^2\,z_2 + z_1\,z_2^2 + z_2\,z_3 - z_1\,z_4 -
z_2\,z_5 - z_6 - 2\,z_1\,z_6 -
    z_1^2\,z_6 + z_2\,z_6 + z_3\,z_6 + z_6^2 - z_2\,z_7 - z_1\,z_2\,z_7 -
z_6\,z_7 + z_8 + 2\,z_1\,z_8 -
    z_1^3\,z_8 - z_2\,z_8 - z_3\,z_8 + 2\,z_1\,z_3\,z_8 - z_4\,z_8 +
z_5\,z_8 - 3\,z_6\,z_8 + 2\,z_1\,z_6\,z_8 +
    2\,z_7\,z_8 + z_1\,z_7\,z_8 - z_2\,z_7\,z_8 + 2\,z_8^2 - 3\,z_1\,z_8^2 +
z_1^2\,z_8^2 + z_2\,z_8^2 +
    z_3\,z_8^2 + z_6\,z_8^2 + z_7\,z_8^2 - 2\,z_8^3 + z_1\,z_8^3\\ $

$  \bchi_{0, 0, 0, 0, 0, 0, 1, 3} =
   -z_1\,z_2 - z_3 - z_4 + z_5 - z_6 - z_7 - z_1\,z_7 + 2\,z_6\,z_7 - z_7^2
+ z_8 - 2\,z_1\,z_8 + z_1^2\,z_8 -
    z_2\,z_8 - z_3\,z_8 + z_5\,z_8 - z_6\,z_8 - 2\,z_7\,z_8 +
z_1\,z_7\,z_8 - 2\,z_7^2\,z_8 + 2\,z_1\,z_8^2 +
    z_2\,z_8^2 - z_6\,z_8^2 + 2\,z_7\,z_8^2 + z_8^3 - z_1\,z_8^3 +
z_7\,z_8^3 - z_8^4\\ $

 $ \bchi_{0, 0, 0, 0, 0, 1, 1, 1} =
   z_1 + z_1^2 + z_2 + z_1\,z_2 - z_1\,z_3 + z_4 + z_6 - z_2\,z_6 - z_6^2 +
z_1\,z_7 + z_2\,z_7 - z_8 +
    z_1\,z_8 - 2\,z_1^2\,z_8 + 2\,z_2\,z_8 - z_1\,z_2\,z_8 + z_4\,z_8 +
2\,z_6\,z_8 + z_1\,z_6\,z_8 - z_7\,z_8 -
    z_1\,z_7\,z_8 + z_2\,z_7\,z_8 + z_6\,z_7\,z_8 - z_8^2 - 3\,z_1\,z_8^2 +
z_1^2\,z_8^2 - z_2\,z_8^2 -
    z_5\,z_8^2 - 2\,z_7\,z_8^2 - z_1\,z_7\,z_8^2 + z_1\,z_8^3 + z_8^4 \\$

$ \bchi_{0, 1, 0, 0, 0, 0, 0, 3} =
   -z_1 - z_1^2 - z_2 + z_3 + z_4 - z_5 + z_6 + z_1\,z_6 + z_2\,z_6 - z_7 -
2\,z_1\,z_7 - z_2\,z_7 +
    z_3\,z_7 + z_6\,z_7 - z_7^2 - 2\,z_8 + z_1\,z_8 + z_1^2\,z_8 + z_3\,z_8
+ 2\,z_6\,z_8 - 2\,z_7\,z_8 +
    2\,z_1\,z_7\,z_8 - 2\,z_2\,z_7\,z_8 + z_1\,z_8^2 - z_2\,z_8^2 -
z_3\,z_8^2 - z_6\,z_8^2 + z_8^3 - z_1\,z_8^3 +
    z_2\,z_8^3 \\$

$ \bchi_{0, 1, 0, 0, 0, 0, 2, 0} =
   -z_1\,z_2 - z_1^2\,z_2 - z_3 + z_2\,z_3 - z_4 + z_5 + z_1\,z_5 - 2\,z_6 -
z_1\,z_6 + z_3\,z_6 + z_6^2 +
    z_2\,z_7 - z_3\,z_7 + z_5\,z_7 + z_2\,z_7^2 + 2\,z_8 + z_2\,z_8 -
z_3\,z_8 + z_1\,z_3\,z_8 - z_4\,z_8 +
    2\,z_5\,z_8 - 3\,z_6\,z_8 - z_2\,z_6\,z_8 + 2\,z_7\,z_8 + z_1\,z_7\,z_8
+ 2\,z_2\,z_7\,z_8 - z_3\,z_7\,z_8 -
    z_6\,z_7\,z_8 + 2\,z_8^2 + z_2\,z_8^2 + z_3\,z_8^2 + 3\,z_7\,z_8^2 -
z_8^3 - z_2\,z_8^3 - z_8^4\\ $

 $ \bchi_{1, 0, 0, 0, 0, 0, 1, 2} =
   z_1^2 + z_1^3 - z_2 - z_2^2 - 2\,z_1\,z_3 + z_4 - z_5 + z_1\,z_5 -
z_1\,z_6 - z_1\,z_7 - z_2\,z_7 -
    z_6\,z_7 - z_1\,z_7^2 - z_8 + 2\,z_1\,z_8 - z_1^2\,z_8 +
2\,z_1\,z_2\,z_8 - z_3\,z_8 - z_5\,z_8 - z_1\,z_6\,z_8 +
    z_2\,z_7\,z_8 + z_7^2\,z_8 - 2\,z_1\,z_8^2 - z_1^2\,z_8^2 + z_2\,z_8^2 +
z_3\,z_8^2 + 2\,z_6\,z_8^2 -
    z_7\,z_8^2 + z_1\,z_7\,z_8^2 - z_8^3 + z_1\,z_8^3 - z_2\,z_8^3 -
z_7\,z_8^3 + z_8^4\\ $

$  \bchi_{1, 0, 0, 0, 0, 1, 1, 0} =
   z_1^2 + z_1^3 + z_2 + z_1\,z_2 + z_2^2 - z_3 - z_1\,z_3 - z_2\,z_3 + z_5
+ 2\,z_1\,z_5 + z_2\,z_5 +
    z_1^2\,z_6 - z_3\,z_6 - z_6^2 - z_7 - z_1\,z_7 + z_1^2\,z_7 - z_3\,z_7 +
z_5\,z_7 + z_1\,z_6\,z_7 -
    2\,z_7^2 - z_1\,z_7^2 - z_2\,z_7^2 - z_7^3 + z_2\,z_8 -
2\,z_1\,z_2\,z_8 - z_3\,z_8 + z_5\,z_8 -
    z_1\,z_5\,z_8 - 2\,z_1\,z_6\,z_8 - 2\,z_7\,z_8 - z_1^2\,z_7\,z_8 -
z_2\,z_7\,z_8 + z_3\,z_7\,z_8 + 2\,z_6\,z_7\,z_8 -
    2\,z_7^2\,z_8 - z_1^2\,z_8^2 - z_2\,z_8^2 + z_1\,z_2\,z_8^2 +
z_3\,z_8^2 - z_5\,z_8^2 + 2\,z_1\,z_7\,z_8^2 +
    z_7\,z_8^3\\ $

$\bchi_{2, 0, 0, 0, 0, 0, 1, 1} =
   z_1^2\,z_2 - z_2\,z_3 - z_4 - z_1\,z_5 - z_1\,z_6 - z_1^2\,z_6 - z_2\,z_6
+ z_3\,z_6 + z_6^2 + z_7 +
    z_1\,z_7 - z_3\,z_7 - z_6\,z_7 + z_7^2 + z_8 + z_1\,z_8 + z_1^2\,z_8 -
z_1^3\,z_8 - z_2\,z_8 - z_1\,z_2\,z_8 +
    z_2^2\,z_8 - z_3\,z_8 + 2\,z_1\,z_3\,z_8 - z_4\,z_8 + z_5\,z_8 -
3\,z_6\,z_8 + 2\,z_1\,z_6\,z_8 + 3\,z_7\,z_8 +
    z_1^2\,z_7\,z_8 - z_3\,z_7\,z_8 - z_6\,z_7\,z_8 + z_7^2\,z_8 +
2\,z_8^2 - z_1\,z_8^2 + z_1^2\,z_8^2 -
    z_2\,z_8^2 - z_1\,z_2\,z_8^2 + z_5\,z_8^2 - z_6\,z_8^2 + 2\,z_7\,z_8^2 -
z_1\,z_7\,z_8^2 - z_8^3 +
    z_2\,z_8^3 - z_8^4\\ $

  $\bchi_{0, 0, 0, 0, 0, 0, 3, 0} =
   -z_1^2 - z_1^3 + z_3 + 2\,z_1\,z_3 - z_5 - z_1\,z_5 + z_1\,z_6 + z_2\,z_6
+ z_6^2 + z_7 + z_1\,z_7 -
    z_1^2\,z_7 + 2\,z_3\,z_7 - z_5\,z_7 + 2\,z_7^2 + z_1\,z_7^2 + z_7^3 -
z_2\,z_8 - z_1\,z_2\,z_8 + z_3\,z_8 -
    z_5\,z_8 + z_1\,z_6\,z_8 + 2\,z_7\,z_8 + z_1\,z_7\,z_8 -
2\,z_2\,z_7\,z_8 - 2\,z_6\,z_7\,z_8 + 2\,z_7^2\,z_8 +
    z_1^2\,z_8^2 - z_2\,z_8^2 - z_3\,z_8^2 + z_5\,z_8^2 - z_1\,z_7\,z_8^2 +
z_2\,z_8^3 - z_7\,z_8^3\\ $

 $ \bchi_{0, 2, 0, 0, 0, 0, 1, 0} =
   z_1^2 + z_1^3 - z_3 - 2\,z_1\,z_3 - z_2\,z_3 + z_3^2 - z_1\,z_4 +
z_1\,z_5 + z_2\,z_5 - 2\,z_1\,z_6 -
    z_1^2\,z_6 - z_2\,z_6 + z_3\,z_6 - z_7 + z_1^2\,z_7 - z_2\,z_7 +
z_2^2\,z_7 - z_3\,z_7 - z_4\,z_7 - z_6\,z_7 -
    z_1\,z_6\,z_7 - z_7^2 + 2\,z_1\,z_8 + z_2\,z_8 + z_1\,z_2\,z_8 +
z_2^2\,z_8 - z_3\,z_8 + z_1\,z_3\,z_8 -
    z_2\,z_3\,z_8 + z_1\,z_5\,z_8 - z_1\,z_6\,z_8 - z_2\,z_6\,z_8 +
2\,z_1\,z_7\,z_8 + 2\,z_2\,z_7\,z_8 + z_3\,z_7\,z_8 +
    z_6\,z_7\,z_8 + z_7^2\,z_8 - z_1^2\,z_8^2 + z_2\,z_8^2 - z_1\,z_2\,z_8^2
+ z_3\,z_8^2 + z_1\,z_7\,z_8^2 -
    z_1\,z_8^3 - z_2\,z_8^3\\ $

$ \bchi_{1, 0, 1, 0, 0, 0, 1, 0} =
   -z_1^2 - z_1^3 + z_1\,z_2^2 + z_3 + 2\,z_1\,z_3 + 2\,z_2\,z_3 + z_4 -
z_5 - z_1\,z_5 - z_2\,z_5 + z_6 +
    2\,z_1\,z_6 + 2\,z_2\,z_6 - z_1^2\,z_7 + 2\,z_3\,z_7 + z_1\,z_3\,z_7 -
z_4\,z_7 - z_5\,z_7 + z_6\,z_7 -
    z_1\,z_6\,z_7 + z_7^2 + z_1\,z_7^2 + z_7^3 - z_8 - z_1\,z_8 + z_1^2\,z_8
+ z_1^3\,z_8 - 2\,z_2\,z_8 +
    z_1\,z_2\,z_8 - z_1^2\,z_2\,z_8 - z_2^2\,z_8 + 2\,z_3\,z_8 -
2\,z_1\,z_3\,z_8 - 2\,z_5\,z_8 + z_1\,z_5\,z_8 +
    2\,z_6\,z_8 - 2\,z_1\,z_6\,z_8 + z_2\,z_6\,z_8 + z_1\,z_7\,z_8 -
3\,z_2\,z_7\,z_8 - z_6\,z_7\,z_8 + 2\,z_7^2\,z_8 -
    z_8^2 + 3\,z_1\,z_8^2 - z_1^2\,z_8^2 - z_2\,z_8^2 + z_1\,z_2\,z_8^2 -
z_3\,z_8^2 + z_5\,z_8^2 + z_8^3 -
    z_1\,z_8^3 + z_2\,z_8^3 - z_7\,z_8^3\\  $

$\bchi_{1, 1, 0, 0, 0, 1, 0, 0} =
   -z_2 - 2\,z_1\,z_2 - z_1^2\,z_2 - z_2^2 - z_1\,z_3 + z_2\,z_3 - z_5 -
z_1\,z_5 - z_2\,z_5 - z_1\,z_6 -
    z_1^2\,z_6 + z_2\,z_6 + z_1\,z_2\,z_6 + z_3\,z_6 - z_5\,z_6 + z_6^2 -
z_2\,z_7 - z_1\,z_2\,z_7 - z_2^2\,z_7 -
    z_1\,z_3\,z_7 + z_4\,z_7 - z_5\,z_7 + z_6\,z_7 - 2\,z_2\,z_8 +
z_1\,z_2\,z_8 - z_2^2\,z_8 + z_2\,z_3\,z_8 +
    z_4\,z_8 - z_5\,z_8 + z_6\,z_8 + 2\,z_1\,z_6\,z_8 + z_2\,z_6\,z_8 -
2\,z_7\,z_8 + z_1\,z_7\,z_8 + z_1^2\,z_7\,z_8 -
    z_2\,z_7\,z_8 - z_3\,z_7\,z_8 - z_6\,z_7\,z_8 - z_7^2\,z_8 - 2\,z_8^2 +
z_1\,z_2\,z_8^2 + z_6\,z_8^2 -
    z_7\,z_8^2 - 2\,z_1\,z_7\,z_8^2 + z_2\,z_8^3 + z_8^4\\ $

 $ \bchi_{2, 0, 1, 0, 0, 0, 0, 0} =
   z_1 + 2\,z_1^2 + z_1^3 + 2\,z_1\,z_2 + z_1^2\,z_2 + z_2^2 + z_1^2\,z_3 -
z_2\,z_3 - z_3^2 - z_1\,z_4 -
    z_5 + z_2\,z_5 - z_1\,z_6 - z_1^2\,z_6 - z_3\,z_6 + z_7 + 3\,z_1\,z_7 +
2\,z_1^2\,z_7 + z_2\,z_7 +
    z_1\,z_2\,z_7 + z_7^2 + z_1\,z_7^2 + 2\,z_1\,z_8 - z_1^2\,z_8 -
z_1^3\,z_8 + z_2\,z_8 - 2\,z_1\,z_2\,z_8 +
    z_6\,z_8 - z_2\,z_7\,z_8 - z_7^2\,z_8 - z_8^2 - 3\,z_1\,z_8^2 -
z_2\,z_8^2 + z_6\,z_8^2 - 3\,z_7\,z_8^2 -
    z_8^3 + z_1\,z_8^3 + z_8^4\\  $

  $  \bchi_{2, 1, 0, 0, 0, 0, 0, 1} =
   z_1 + z_1^2 + z_2 + z_1\,z_2 - z_1\,z_2^2 - z_3 - z_1\,z_3 - z_1^2\,z_3 -
z_2\,z_3 + z_3^2 + z_1\,z_4 +
    z_5 + z_1\,z_5 + z_2\,z_5 + z_1^2\,z_6 - 2\,z_2\,z_6 + z_3\,z_6 - z_6^2
+ z_1\,z_7 + z_2\,z_7 - z_3\,z_7 +
    z_5\,z_7 + z_8 - z_1\,z_8 - 2\,z_1^2\,z_8 + 2\,z_2\,z_8 - z_1\,z_2\,z_8
+ z_1^2\,z_2\,z_8 - 2\,z_3\,z_8 +
    z_1\,z_3\,z_8 - z_2\,z_3\,z_8 + z_4\,z_8 - z_1\,z_5\,z_8 +
z_1\,z_6\,z_8 - 3\,z_1\,z_7\,z_8 - z_1^2\,z_7\,z_8 +
    z_2\,z_7\,z_8 + z_6\,z_7\,z_8 - z_7^2\,z_8 - z_8^2 - 3\,z_1\,z_8^2 +
z_1^2\,z_8^2 + z_3\,z_8^2 - z_5\,z_8^2 +
    z_6\,z_8^2 - 2\,z_7\,z_8^2 - z_8^3 + 2\,z_1\,z_8^3 - z_2\,z_8^3 +
z_8^4\\ $

$  \bchi_{0, 0, 0, 0, 0, 1, 0, 3} =
   z_1 + z_1^2 + z_2 + z_1\,z_2 - z_2\,z_3 + z_5 + z_1\,z_5 + z_6^2 + z_7 +
3\,z_1\,z_7 + z_1^2\,z_7 +
    z_2\,z_7 - z_3\,z_7 + z_5\,z_7 - z_6\,z_7 + z_7^2 + z_1\,z_7^2 + z_8 +
2\,z_1\,z_8 - z_1^2\,z_8 +
    2\,z_2\,z_8 - z_1\,z_2\,z_8 - z_3\,z_8 + z_4\,z_8 + z_5\,z_8 - z_6\,z_8
+ z_7\,z_8 - z_1\,z_7\,z_8 + z_2\,z_7\,z_8 -
    2\,z_6\,z_7\,z_8 - z_8^2 - 3\,z_1\,z_8^2 + z_1^2\,z_8^2 - z_3\,z_8^2 -
z_5\,z_8^2 - z_6\,z_8^2 -
    3\,z_7\,z_8^2 - z_1\,z_7\,z_8^2 - z_8^3 + z_6\,z_8^3 + z_8^4\\ $

$  \bchi_{0, 0, 0, 0, 0, 2, 0, 1} =
   z_1\,z_3 + z_2\,z_3 - z_4 - z_6 - z_1\,z_6 - z_1^2\,z_6 + z_3\,z_6 -
z_5\,z_6 + z_6^2 + z_2\,z_7 +
    z_1\,z_2\,z_7 + z_4\,z_7 + z_6\,z_7 + z_2\,z_7^2 + z_8 + z_1\,z_8 +
2\,z_1^2\,z_8 + z_2\,z_8 + 2\,z_1\,z_2\,z_8 -
    z_3\,z_8 - z_1\,z_3\,z_8 + 2\,z_5\,z_8 - 2\,z_6\,z_8 - z_2\,z_6\,z_8 +
z_6^2\,z_8 + z_1\,z_7\,z_8 +
    z_1^2\,z_7\,z_8 + 2\,z_2\,z_7\,z_8 - 2\,z_3\,z_7\,z_8 - z_5\,z_7\,z_8 -
2\,z_6\,z_7\,z_8 - z_7^2\,z_8 -
    z_1\,z_7^2\,z_8 + z_8^2 - z_1^2\,z_8^2 + z_2\,z_8^2 - z_1\,z_2\,z_8^2 -
z_3\,z_8^2 - 2\,z_6\,z_8^2 +
    z_1\,z_6\,z_8^2 + 2\,z_7\,z_8^2 - 2\,z_1\,z_7\,z_8^2 - 2\,z_1\,z_8^3 -
z_2\,z_8^3 + z_3\,z_8^3 + z_6\,z_8^3 +
    z_7\,z_8^3 - z_8^4 + z_1\,z_8^4\\  $

   $ \bchi_{0, 0, 0, 0, 1, 0, 1, 1} =
   -z_1 + z_1^3 - z_2 - z_1\,z_2 - z_2^2 - z_1\,z_3 + z_4 - z_5 - z_1\,z_5 -
z_2\,z_5 - z_1\,z_6 - z_5\,z_6 -
    z_2\,z_7 - z_1\,z_2\,z_7 - z_6\,z_7 - z_8 - z_1^3\,z_8 - z_2\,z_8 +
z_1\,z_2\,z_8 + z_1\,z_3\,z_8 + z_2\,z_3\,z_8 +
    3\,z_1\,z_6\,z_8 + z_2\,z_6\,z_8 - z_7\,z_8 - z_1\,z_7\,z_8 -
z_2\,z_7\,z_8 + z_3\,z_7\,z_8 + z_5\,z_7\,z_8 -
    2\,z_1\,z_8^2 + z_3\,z_8^2 - z_4\,z_8^2 + z_5\,z_8^2 + z_6\,z_8^2 -
z_1\,z_6\,z_8^2 - z_7\,z_8^2 +
    z_7^2\,z_8^2 + z_8^3 + 2\,z_1\,z_8^3 - z_6\,z_8^3 + 2\,z_7\,z_8^3 +
z_8^4 - z_8^5\\ $

$  \bchi_{0, 0, 0, 1, 0, 0, 0, 2} =
   z_1 + z_1^2 + z_1^3 - 2\,z_1\,z_2 - z_1^2\,z_2 + z_1\,z_2^2 - z_3 -
z_1\,z_3 + z_2\,z_3 + z_3^2 - z_4 -
    2\,z_1\,z_4 + z_5 + 2\,z_1\,z_5 - z_6 - 3\,z_1\,z_6 - z_1^2\,z_6 +
z_3\,z_6 + z_1\,z_7 + z_1^2\,z_7 -
    z_1\,z_2\,z_7 - z_3\,z_7 - z_4\,z_7 + z_5\,z_7 - z_6\,z_7 + z_8 +
2\,z_1\,z_8 - z_1^3\,z_8 + z_1\,z_2\,z_8 -
    z_3\,z_8 + 2\,z_1\,z_3\,z_8 - z_2\,z_3\,z_8 - z_4\,z_8 + z_5\,z_8 -
z_6\,z_8 + z_1\,z_6\,z_8 + 2\,z_7\,z_8 +
    2\,z_1\,z_7\,z_8 + z_1^2\,z_7\,z_8 - z_3\,z_7\,z_8 + z_7^2\,z_8 -
4\,z_1\,z_8^2 + z_3\,z_8^2 + z_4\,z_8^2 -
    z_5\,z_8^2 + z_6\,z_8^2 - z_1\,z_7\,z_8^2 - 2\,z_8^3 + z_1\,z_8^3 -
z_7\,z_8^3 + z_8^4\\ $

$  \bchi_{0, 0, 1, 0, 0, 0, 0, 3} =
   z_2 + z_1\,z_2 + z_5 + z_1\,z_5 + z_6 - z_2\,z_6 + z_3\,z_6 + z_2\,z_7 +
z_1\,z_2\,z_7 + z_6\,z_7 + z_2^2\,z_8 -
    2\,z_3\,z_8 - z_4\,z_8 + z_5\,z_8 - 2\,z_6\,z_8 - z_7\,z_8 -
2\,z_3\,z_7\,z_8 - z_7^2\,z_8 + 2\,z_8^2 -
    z_1\,z_2\,z_8^2 - z_6\,z_8^2 + 2\,z_7\,z_8^2 + z_3\,z_8^3 + z_7\,z_8^3 -
z_8^4\\ $

$  \bchi_{0, 0, 1, 0, 0, 0, 2, 0} =
   2\,z_1 + z_1^2 - z_1^3 + z_2 + z_1\,z_2 - z_1\,z_2^2 - z_3 + z_1\,z_3 -
z_1^2\,z_3 + z_2\,z_3 + z_3^2 -
    z_4 + 2\,z_1\,z_4 + z_5 - z_1\,z_5 - z_6 + 3\,z_1\,z_6 + z_1^2\,z_6 +
z_2\,z_6 + z_1\,z_2\,z_6 + z_3\,z_6 +
    z_6^2 + z_7 + 2\,z_2\,z_7 + z_1\,z_2\,z_7 - z_2^2\,z_7 + 2\,z_4\,z_7 +
2\,z_6\,z_7 + z_1\,z_6\,z_7 - z_7^2 -
    z_1\,z_7^2 + z_2\,z_7^2 + z_3\,z_7^2 - z_7^3 + 2\,z_8 - 4\,z_1\,z_8 +
z_2\,z_8 - z_1\,z_2\,z_8 +
    z_1^2\,z_2\,z_8 - z_2^2\,z_8 - z_3\,z_8 + z_1\,z_3\,z_8 - z_2\,z_3\,z_8
+ 2\,z_5\,z_8 - z_1\,z_5\,z_8 -
    2\,z_6\,z_8 + z_1\,z_6\,z_8 - z_2\,z_6\,z_8 - z_3\,z_6\,z_8 -
2\,z_7\,z_8 - 4\,z_1\,z_7\,z_8 + z_2\,z_7\,z_8 -
    z_1\,z_2\,z_7\,z_8 - z_3\,z_7\,z_8 - 2\,z_6\,z_7\,z_8 - 4\,z_7^2\,z_8 -
2\,z_1\,z_8^2 + z_1^2\,z_8^2 -
    z_1\,z_2\,z_8^2 + z_2^2\,z_8^2 - z_4\,z_8^2 + z_5\,z_8^2 - 3\,z_6\,z_8^2
+ z_7\,z_8^2 - z_1\,z_7\,z_8^2 +
    z_7^2\,z_8^2 + z_8^3 + 2\,z_1\,z_8^3 + 4\,z_7\,z_8^3 - z_8^5\\ $

 $ \bchi_{0, 0, 1, 0, 0, 1, 0, 1} =
   z_1 + 2\,z_1^2 + z_1^3 + z_1\,z_2 + z_2^2 - z_3 - z_1\,z_3 - z_2\,z_3 -
z_4 - z_1\,z_4 + z_1\,z_5 +
    z_2\,z_5 - z_3\,z_5 - z_6 - 2\,z_1\,z_6 - z_1^2\,z_6 + z_3\,z_6 + z_6^2
+ z_7 + 3\,z_1\,z_7 + 2\,z_1^2\,z_7 +
    z_2^2\,z_7 - 2\,z_3\,z_7 - z_4\,z_7 + z_5\,z_7 - 2\,z_6\,z_7 + z_7^2 +
z_1\,z_7^2 + 2\,z_8 + 2\,z_1\,z_8 -
    4\,z_1^2\,z_8 - 2\,z_1^3\,z_8 + z_2\,z_8 - 2\,z_1\,z_2\,z_8 +
z_1^2\,z_2\,z_8 + 3\,z_1\,z_3\,z_8 - z_2\,z_3\,z_8 +
    z_5\,z_8 - z_1\,z_5\,z_8 + 2\,z_1\,z_6\,z_8 - 2\,z_2\,z_6\,z_8 +
z_3\,z_6\,z_8 + 3\,z_7\,z_8 - z_1\,z_7\,z_8 -
    z_1^2\,z_7\,z_8 + z_2\,z_7\,z_8 - z_1\,z_2\,z_7\,z_8 - z_6\,z_7\,z_8 +
z_7^2\,z_8 - z_8^2 - 5\,z_1\,z_8^2 +
    2\,z_1^2\,z_8^2 - z_1\,z_2\,z_8^2 + z_3\,z_8^2 + z_4\,z_8^2 - z_5\,z_8^2
+ z_6\,z_8^2 + z_1\,z_6\,z_8^2 -
    3\,z_7\,z_8^2 - z_1\,z_7\,z_8^2 + z_2\,z_7\,z_8^2 - 3\,z_8^3 +
3\,z_1\,z_8^3 + z_1^2\,z_8^3 - z_3\,z_8^3 -
    z_7\,z_8^3 + 2\,z_8^4 - z_1\,z_8^4\\  $

  $  \bchi_{0, 0, 2, 0, 0, 0, 0, 1} =
   -1 - z_1 + z_1^3 + z_1^4 - z_2 - 2\,z_1\,z_2 - z_3 - 2\,z_1\,z_3 -
3\,z_1^2\,z_3 - z_2\,z_3 -
    z_1\,z_2\,z_3 + z_3^2 + 2\,z_1\,z_4 + z_2\,z_4 - z_1\,z_5 + z_1^2\,z_5 -
z_2\,z_5 - z_3\,z_5 - z_6 - z_1\,z_6 -
    z_1^2\,z_6 + z_1\,z_2\,z_6 + z_3\,z_6 - z_5\,z_6 + z_6^2 - z_7 -
2\,z_1\,z_7 + z_1^3\,z_7 - 2\,z_2\,z_7 -
    z_1\,z_2\,z_7 - z_3\,z_7 - 2\,z_1\,z_3\,z_7 + z_4\,z_7 - z_5\,z_7 -
z_6\,z_7 - z_1\,z_6\,z_7 - z_1\,z_7^2 -
    z_2\,z_7^2 - 2\,z_1\,z_8 + 2\,z_1^2\,z_8 - z_2\,z_8 + 2\,z_1\,z_2\,z_8 +
z_1^2\,z_2\,z_8 + z_2^2\,z_8 -
    z_3\,z_8 - z_1\,z_3\,z_8 - z_2\,z_3\,z_8 + z_3^2\,z_8 + z_4\,z_8 -
z_1\,z_4\,z_8 - z_6\,z_8 + 2\,z_1\,z_6\,z_8 -
    z_1^2\,z_6\,z_8 + z_3\,z_6\,z_8 + z_1^2\,z_7\,z_8 - z_3\,z_7\,z_8 +
z_5\,z_7\,z_8 + z_1\,z_7^2\,z_8 + z_8^2 +
    2\,z_1\,z_8^2 - 2\,z_1^2\,z_8^2 - z_1^3\,z_8^2 + 2\,z_2\,z_8^2 -
z_1\,z_2\,z_8^2 + z_3\,z_8^2 +
    2\,z_1\,z_3\,z_8^2 - z_4\,z_8^2 + z_6\,z_8^2 + 2\,z_7\,z_8^2 -
z_1\,z_7\,z_8^2 + z_8^3 - z_1\,z_8^3 -
    z_2\,z_8^3 + z_3\,z_8^3 - z_8^4 + z_1\,z_8^4\\ $

$  \bchi_{0, 1, 0, 0, 0, 0, 1, 2} =
   -z_1 - z_1^2 - z_2 + z_2^2 - z_1\,z_3 - z_2\,z_3 - z_5 + z_2\,z_5 - z_6 -
z_1\,z_6 - 2\,z_1\,z_7 -
    z_1^2\,z_7 - z_2\,z_7 - z_5\,z_7 - z_1\,z_7^2 - z_2\,z_7^2 - z_1\,z_8 -
z_1^2\,z_8 - z_2\,z_8 - z_1\,z_2\,z_8 +
    z_3\,z_8 + z_1\,z_3\,z_8 - z_5\,z_8 - z_2\,z_6\,z_8 - 2\,z_1\,z_7\,z_8 -
2\,z_2\,z_7\,z_8 + z_3\,z_7\,z_8 +
    z_6\,z_7\,z_8 - z_7^2\,z_8 + 2\,z_1\,z_8^2 + z_1^2\,z_8^2 + z_2\,z_8^2 +
z_3\,z_8^2 + z_5\,z_8^2 +
    2\,z_6\,z_8^2 + 3\,z_1\,z_7\,z_8^2 + z_2\,z_7\,z_8^2 + z_1\,z_8^3 -
z_3\,z_8^3 - z_6\,z_8^3 - z_1\,z_8^4\\ $

 $ \bchi_{0, 1, 0, 0, 0, 1, 1, 0} =
   -z_1^2 - z_1^3 + z_1^2\,z_2 + z_3 + z_1\,z_3 - z_3^2 + z_4 + z_1\,z_4 -
z_1\,z_5 + z_3\,z_5 + z_6 +
    2\,z_1\,z_6 + z_1^2\,z_6 - 2\,z_3\,z_6 + z_5\,z_6 - z_6^2 - z_1^2\,z_7 -
z_2\,z_7 - z_1\,z_2\,z_7 - z_3\,z_7 -
    2\,z_6\,z_7 - z_1\,z_6\,z_7 + z_2\,z_6\,z_7 - z_2\,z_7^2 - z_3\,z_7^2 -
z_6\,z_7^2 - z_8 - z_1\,z_8 +
    z_1^2\,z_8 + z_1^3\,z_8 - z_2\,z_8 - z_1\,z_2\,z_8 - z_1^2\,z_2\,z_8 -
3\,z_1\,z_3\,z_8 + z_2\,z_3\,z_8 -
    z_5\,z_8 + z_1\,z_5\,z_8 - z_2\,z_5\,z_8 - z_1\,z_6\,z_8 +
2\,z_2\,z_6\,z_8 + z_7\,z_8 - 2\,z_1\,z_7\,z_8 +
    z_1^2\,z_7\,z_8 - z_2\,z_7\,z_8 - z_3\,z_7\,z_8 + z_5\,z_7\,z_8 +
z_7^2\,z_8 + z_1\,z_7^2\,z_8 + z_8^2 +
    z_1\,z_8^2 + z_1^2\,z_8^2 - z_2\,z_8^2 + z_1\,z_2\,z_8^2 - 2\,z_3\,z_8^2
+ z_1\,z_3\,z_8^2 - z_4\,z_8^2 +
    z_5\,z_8^2 - z_6\,z_8^2 - z_1\,z_6\,z_8^2 + z_1\,z_7\,z_8^2 + z_8^3 -
z_1^2\,z_8^3 + z_2\,z_8^3 +
    z_3\,z_8^3 - z_8^4\\  $

  $  \bchi_{0, 1, 0, 0, 1, 0, 0, 1} =
   z_2 + z_1\,z_2 + z_1^2\,z_2 + z_1\,z_3 - z_2\,z_4 + z_5 + 2\,z_1\,z_6 +
z_1^2\,z_6 + z_2\,z_7 + z_1\,z_3\,z_7 -
    z_4\,z_7 + z_5\,z_7 + z_8 - z_1\,z_8 - z_1^2\,z_8 - z_1^3\,z_8 +
z_2\,z_8 - 2\,z_1\,z_2\,z_8 - z_3\,z_8 +
    z_1\,z_3\,z_8 - z_4\,z_8 + z_5\,z_8 - z_1\,z_5\,z_8 + z_2\,z_5\,z_8 -
z_6\,z_8 + z_1\,z_6\,z_8 - z_3\,z_6\,z_8 -
    z_6^2\,z_8 + 2\,z_7\,z_8 - 3\,z_1\,z_7\,z_8 - z_1^2\,z_7\,z_8 +
z_5\,z_7\,z_8 + z_6\,z_7\,z_8 + z_1\,z_7^2\,z_8 +
    2\,z_8^2 - 2\,z_1\,z_8^2 + 2\,z_1^2\,z_8^2 - z_2\,z_8^2 + z_6\,z_8^2 -
z_1\,z_6\,z_8^2 + 2\,z_1\,z_7\,z_8^2 -
    2\,z_8^3 + 3\,z_1\,z_8^3 - z_6\,z_8^3 - z_1\,z_8^4\\ $

 $ \bchi_{0, 1, 1, 0, 0, 0, 1, 0} =
   -z_1^2 - 2\,z_1^3 - z_1^4 - z_1\,z_2 - z_1^2\,z_2 + 2\,z_1\,z_3 +
2\,z_1^2\,z_3 + z_2\,z_3 + z_1\,z_2\,z_3 -
    z_1\,z_4 - z_1\,z_5 - z_1^2\,z_5 + z_3\,z_5 + z_3\,z_6 - z_1\,z_7 -
3\,z_1^2\,z_7 - 2\,z_1^3\,z_7 - z_2\,z_7 -
    z_1\,z_2\,z_7 + 2\,z_3\,z_7 + 3\,z_1\,z_3\,z_7 + z_2\,z_3\,z_7 -
z_4\,z_7 - z_5\,z_7 - z_1\,z_5\,z_7 + z_6\,z_7 +
    z_1\,z_6\,z_7 - z_1\,z_7^2 - z_1^2\,z_7^2 - z_2\,z_7^2 + z_3\,z_7^2 +
z_6\,z_7^2 - 2\,z_1^2\,z_8 -
    z_1^3\,z_8 - z_2\,z_8 - z_1\,z_2\,z_8 + z_1^2\,z_2\,z_8 -
z_1\,z_2^2\,z_8 + z_3\,z_8 + 4\,z_1\,z_3\,z_8 -
    z_3^2\,z_8 - z_4\,z_8 + z_1\,z_4\,z_8 - z_5\,z_8 + z_2\,z_5\,z_8 +
3\,z_1\,z_6\,z_8 + 2\,z_1^2\,z_6\,z_8 -
    z_2\,z_6\,z_8 - 2\,z_3\,z_6\,z_8 - z_6^2\,z_8 - 2\,z_1\,z_7\,z_8 -
z_1^2\,z_7\,z_8 - 3\,z_2\,z_7\,z_8 +
    z_1\,z_2\,z_7\,z_8 + 3\,z_3\,z_7\,z_8 + 4\,z_6\,z_7\,z_8 - z_7^2\,z_8 +
z_8^2 - 2\,z_1\,z_8^2 + 3\,z_1^2\,z_8^2 +
    2\,z_1^3\,z_8^2 + z_1\,z_2\,z_8^2 + 3\,z_3\,z_8^2 - 4\,z_1\,z_3\,z_8^2 +
z_4\,z_8^2 + 4\,z_6\,z_8^2 -
    2\,z_1\,z_6\,z_8^2 - z_7\,z_8^2 + 4\,z_1\,z_7\,z_8^2 + z_2\,z_7\,z_8^2 -
2\,z_8^3 + 5\,z_1\,z_8^3 -
    z_1^2\,z_8^3 - 2\,z_3\,z_8^3 - 2\,z_6\,z_8^3 + z_8^4 - 2\,z_1\,z_8^4\\ $

  $\bchi_{0, 2, 0, 0, 0, 0, 0, 2} =
   -z_1 - 2\,z_1^2 - z_1^3 - z_1\,z_2 - z_2^2 + z_1\,z_3 + z_2\,z_3 + z_4 +
z_1\,z_4 - z_1\,z_5 - z_2\,z_5 +
    z_6 + 2\,z_1\,z_6 + z_2\,z_6 + z_3\,z_6 + z_6^2 - z_7 - 3\,z_1\,z_7 -
2\,z_1^2\,z_7 - z_2^2\,z_7 + z_3\,z_7 +
    z_4\,z_7 - z_5\,z_7 + z_6\,z_7 + z_1\,z_6\,z_7 - z_7^2 - z_1\,z_7^2 -
z_8 - z_1\,z_8 + 3\,z_1^2\,z_8 +
    z_1^3\,z_8 + z_2\,z_8 + 3\,z_1\,z_2\,z_8 - z_2^2\,z_8 - z_3\,z_8 -
z_1\,z_3\,z_8 - z_2\,z_3\,z_8 + z_4\,z_8 +
    z_1\,z_5\,z_8 + z_1\,z_6\,z_8 - z_2\,z_6\,z_8 - 3\,z_7\,z_8 +
z_1\,z_7\,z_8 + 2\,z_1^2\,z_7\,z_8 + z_2\,z_7\,z_8 -
    z_3\,z_7\,z_8 - z_6\,z_7\,z_8 - z_7^2\,z_8 + 2\,z_1\,z_8^2 +
z_2\,z_8^2 - z_1\,z_2\,z_8^2 + z_2^2\,z_8^2 -
    z_3\,z_8^2 - z_4\,z_8^2 - 2\,z_6\,z_8^2 - z_1\,z_6\,z_8^2 + z_7\,z_8^2 -
z_1\,z_7\,z_8^2 + 2\,z_8^3 -
    2\,z_1\,z_8^3 - z_1^2\,z_8^3 - z_2\,z_8^3 + z_3\,z_8^3 + z_6\,z_8^3 +
z_7\,z_8^3 - z_8^4 + z_1\,z_8^4\\ $

  $\bchi_{0, 3, 0, 0, 0, 0, 0, 0} =
   z_1^2 + z_1^3 + z_2 + 2\,z_1\,z_2 + z_1^2\,z_2 - z_2^2 + z_2^3 + z_3 +
z_1^2\,z_3 - z_2\,z_3 - z_3^2 +
    z_4 - z_1\,z_4 - 2\,z_2\,z_4 + z_1\,z_5 + z_2\,z_5 + z_3\,z_5 + 2\,z_6 -
3\,z_2\,z_6 - 2\,z_1\,z_2\,z_6 -
    2\,z_3\,z_6 + z_5\,z_6 - 2\,z_6^2 + 2\,z_1\,z_7 + 2\,z_1^2\,z_7 +
2\,z_2\,z_7 + z_1\,z_2\,z_7 + z_1\,z_3\,z_7 -
    z_4\,z_7 + z_5\,z_7 - z_6\,z_7 + z_7^2 + z_1\,z_7^2 - 2\,z_8 +
2\,z_1\,z_8 + 3\,z_2\,z_8 - z_1\,z_2\,z_8 -
    2\,z_1^2\,z_2\,z_8 + z_3\,z_8 - z_1\,z_3\,z_8 + 2\,z_2\,z_3\,z_8 -
z_5\,z_8 + z_1\,z_5\,z_8 + 2\,z_6\,z_8 -
    3\,z_1\,z_6\,z_8 + z_2\,z_6\,z_8 + 2\,z_2\,z_7\,z_8 + z_3\,z_7\,z_8 +
z_6\,z_7\,z_8 + z_7^2\,z_8 - z_1^2\,z_8^2 -
    3\,z_2\,z_8^2 + z_1\,z_2\,z_8^2 - z_3\,z_8^2 - z_6\,z_8^2 -
2\,z_7\,z_8^2 + z_1\,z_7\,z_8^2 + z_8^3 -
    z_1\,z_8^3\\  $

   $ \bchi_{1, 0, 0, 0, 0, 0, 2, 1} =
   -z_1 - z_1^2 - 2\,z_1\,z_2 - z_1^2\,z_2 - z_2^2 + z_2\,z_3 + z_4 -
z_1\,z_5 - z_2\,z_5 + z_6 - z_6^2 -
    2\,z_7 - z_1\,z_7 - z_2\,z_7 - z_1\,z_2\,z_7 - z_5\,z_7 -
z_1\,z_6\,z_7 - z_7^2 - 2\,z_8 + 3\,z_1^2\,z_8 +
    z_1^3\,z_8 - z_2\,z_8 + z_1\,z_2\,z_8 - z_2^2\,z_8 - z_3\,z_8 -
2\,z_1\,z_3\,z_8 + z_4\,z_8 - 2\,z_5\,z_8 +
    z_1\,z_5\,z_8 + z_6\,z_8 + 2\,z_2\,z_6\,z_8 - 3\,z_7\,z_8 +
3\,z_1\,z_7\,z_8 + z_1^2\,z_7\,z_8 - z_3\,z_7\,z_8 +
    2\,z_6\,z_7\,z_8 + z_1\,z_7^2\,z_8 - z_8^2 + 3\,z_1\,z_8^2 -
z_1^2\,z_8^2 + 2\,z_1\,z_2\,z_8^2 - z_3\,z_8^2 -
    z_5\,z_8^2 + z_6\,z_8^2 - z_1\,z_6\,z_8^2 - z_1\,z_7\,z_8^2 -
z_2\,z_7\,z_8^2 - z_7^2\,z_8^2 + z_8^3 -
    3\,z_1\,z_8^3 - z_1^2\,z_8^3 + z_3\,z_8^3 + z_6\,z_8^3 - 2\,z_7\,z_8^3 -
z_8^4 + z_1\,z_8^4 + z_8^5\\ $

 $ \bchi_{1, 0, 0, 0, 0, 1, 0, 2} =
   -z_1 - z_1^2 - z_1^2\,z_2 + z_3 + z_2\,z_3 + z_4 + z_1\,z_4 + z_1\,z_5 +
z_6 + z_1\,z_6 + z_2\,z_6 -
    z_1\,z_7 + z_2\,z_7 + z_1\,z_2\,z_7 + 2\,z_3\,z_7 + z_6\,z_7 -
z_1\,z_6\,z_7 + z_7^2 + z_1\,z_7^2 + z_2\,z_7^2 +
    z_7^3 - z_8 + 2\,z_1^2\,z_8 + z_1^3\,z_8 + z_1\,z_2\,z_8 - z_2^2\,z_8 +
z_3\,z_8 - 2\,z_1\,z_3\,z_8 +
    z_4\,z_8 - z_5\,z_8 - z_1\,z_5\,z_8 + z_6\,z_8 - 2\,z_1\,z_6\,z_8 +
z_7\,z_8 + 3\,z_1\,z_7\,z_8 - z_1^2\,z_7\,z_8 +
    z_3\,z_7\,z_8 + 2\,z_7^2\,z_8 + 3\,z_1\,z_8^2 - 2\,z_1^2\,z_8^2 +
z_1\,z_2\,z_8^2 - z_3\,z_8^2 - z_5\,z_8^2 +
    z_1\,z_6\,z_8^2 - z_1\,z_7\,z_8^2 - z_2\,z_7\,z_8^2 - z_7^2\,z_8^2 -
3\,z_1\,z_8^3 + z_6\,z_8^3 -
    3\,z_7\,z_8^3 - z_8^4 + z_1\,z_8^4 + z_8^5\\ $

 $ \bchi_{1, 0, 0, 0, 1, 0, 1, 0} =
   -z_1^2 - z_1^3 + z_1\,z_3 + z_1\,z_4 + z_2\,z_4 + z_1\,z_5 + z_1^2\,z_5 -
z_3\,z_5 - z_2\,z_6 + z_3\,z_6 -
    z_5\,z_6 - z_1^2\,z_7 + z_4\,z_7 + z_1\,z_5\,z_7 + z_6\,z_7 -
z_1\,z_6\,z_7 - z_2\,z_6\,z_7 + z_1^2\,z_8 +
    z_1^3\,z_8 + z_1\,z_2\,z_8 + z_2^2\,z_8 - z_3\,z_8 - z_1\,z_3\,z_8 -
z_2\,z_3\,z_8 + 2\,z_4\,z_8 - z_1\,z_4\,z_8 -
    z_1^2\,z_6\,z_8 + z_3\,z_6\,z_8 + z_6^2\,z_8 - z_7\,z_8 +
z_1^2\,z_7\,z_8 + z_1\,z_2\,z_7\,z_8 - z_3\,z_7\,z_8 -
    z_5\,z_7\,z_8 - 2\,z_7^2\,z_8 + z_1\,z_7^2\,z_8 - z_8^2 + z_1\,z_8^2 +
z_1^2\,z_8^2 + z_2\,z_8^2 -
    z_1\,z_2\,z_8^2 - z_3\,z_8^2 - 2\,z_5\,z_8^2 - 3\,z_7\,z_8^2 +
z_1\,z_7\,z_8^2 - z_2\,z_7\,z_8^2 -
    z_7^2\,z_8^2 - z_8^3 - z_1\,z_8^3 - z_1^2\,z_8^3 + z_3\,z_8^3 +
2\,z_6\,z_8^3 - z_7\,z_8^3 + z_8^5\\ $

  $\bchi_{1, 0, 0, 1, 0, 0, 0, 1} =
   z_1^3 + z_1^4 - z_1\,z_2 + z_1^2\,z_2 - 2\,z_1^2\,z_3 - z_1\,z_2\,z_3 +
z_1\,z_4 - 2\,z_1\,z_5 + z_3\,z_5 -
    z_1^2\,z_6 + z_1\,z_2\,z_6 + z_1\,z_7 + z_1^2\,z_7 + z_1^3\,z_7 -
z_1\,z_3\,z_7 - z_5\,z_7 - 2\,z_1\,z_6\,z_7 +
    z_1\,z_7^2 - z_1\,z_8 + 2\,z_1^2\,z_8 - z_1^3\,z_8 - z_2\,z_8 -
z_1\,z_3\,z_8 + z_1\,z_4\,z_8 - z_2\,z_5\,z_8 -
    z_6\,z_8 + z_1\,z_6\,z_8 + z_2\,z_6\,z_8 + z_6^2\,z_8 - 2\,z_7\,z_8 -
z_1\,z_7\,z_8 - 2\,z_1^2\,z_7\,z_8 -
    z_2\,z_7\,z_8 + z_3\,z_7\,z_8 - z_5\,z_7\,z_8 + z_6\,z_7\,z_8 -
2\,z_7^2\,z_8 - z_1\,z_8^2 - 2\,z_1^2\,z_8^2 -
    z_2\,z_8^2 + z_1\,z_2\,z_8^2 - z_6\,z_8^2 + z_1\,z_6\,z_8^2 - 2\,z_7\,z_
8^2 - 2\,z_1\,z_7\,z_8^2 -
    z_2\,z_7\,z_8^2 - z_7^2\,z_8^2 + z_2\,z_8^3 + z_6\,z_8^3 + 2\,z_7\,z_8^3
+ z_1\,z_8^4\\ $

 $ \bchi_{1, 0, 1, 0, 0, 0, 0, 2} =
   -z_1 - z_1^2 - z_2 - z_1\,z_2 - z_1^2\,z_2 + z_1\,z_2^2 - z_1\,z_3 +
z_2\,z_3 + z_4 - z_5 - z_2\,z_5 +
    2\,z_2\,z_6 - z_7 - 3\,z_1\,z_7 - z_1^2\,z_7 - z_2\,z_7 -
z_1\,z_2\,z_7 - z_1\,z_3\,z_7 + z_4\,z_7 +
    z_1\,z_6\,z_7 - 2\,z_7^2 - 2\,z_1\,z_7^2 - z_7^3 - z_8 + z_1\,z_8 +
z_1^2\,z_8 + z_1^3\,z_8 - z_2\,z_8 +
    z_1\,z_2\,z_8 - z_1^2\,z_2\,z_8 - z_3\,z_8 - 2\,z_1\,z_3\,z_8 +
z_1\,z_5\,z_8 - 2\,z_1\,z_6\,z_8 + z_2\,z_6\,z_8 -
    2\,z_7\,z_8 + z_1\,z_7\,z_8 + 2\,z_1^2\,z_7\,z_8 - z_2\,z_7\,z_8 -
2\,z_3\,z_7\,z_8 - z_6\,z_7\,z_8 - z_7^2\,z_8 +
    z_8^2 + 2\,z_1\,z_8^2 + z_2\,z_8^2 + z_1\,z_2\,z_8^2 - z_3\,z_8^2 +
z_1\,z_3\,z_8^2 - z_4\,z_8^2 +
    z_5\,z_8^2 - z_1\,z_6\,z_8^2 + 2\,z_7\,z_8^2 + 2\,z_1\,z_7\,z_8^2 +
z_7^2\,z_8^2 + z_8^3 - z_1\,z_8^3 -
    z_1^2\,z_8^3 + z_3\,z_8^3 - z_8^4\\  $

$    \bchi_{1, 1, 0, 0, 0, 0, 1, 1} =
   z_1 + 2\,z_1^2 + z_1^3 + z_1^2\,z_3 - z_2\,z_3 - z_3^2 - z_4 - z_1\,z_4 -
z_6 - 2\,z_1\,z_6 - z_2\,z_6 -
    z_1\,z_2\,z_6 - 3\,z_3\,z_6 + z_5\,z_6 - z_6^2 + z_7 + 3\,z_1\,z_7 +
2\,z_1^2\,z_7 - z_6\,z_7 + z_1\,z_6\,z_7 +
    z_7^2 + z_1\,z_7^2 + 2\,z_8 + 3\,z_1\,z_8 - 2\,z_1^2\,z_8 - z_1^3\,z_8 +
z_2\,z_8 + z_2^2\,z_8 +
    2\,z_3\,z_8 + z_1\,z_3\,z_8 + 2\,z_2\,z_3\,z_8 - z_4\,z_8 -
z_1\,z_6\,z_8 + z_2\,z_6\,z_8 + 4\,z_7\,z_8 +
    2\,z_1\,z_7\,z_8 - z_1^2\,z_7\,z_8 + z_2\,z_7\,z_8 + z_1\,z_2\,z_7\,z_8
+ 2\,z_3\,z_7\,z_8 - z_5\,z_7\,z_8 +
    2\,z_7^2\,z_8 - z_1\,z_7^2\,z_8 + z_8^2 - 4\,z_1\,z_8^2 - z_1^2\,z_8^2 -
z_2\,z_8^2 - z_2^2\,z_8^2 +
    z_3\,z_8^2 - z_1\,z_3\,z_8^2 + z_4\,z_8^2 + z_1\,z_6\,z_8^2 -
2\,z_1\,z_7\,z_8^2 - 3\,z_8^3 + z_1\,z_8^3 +
    z_1^2\,z_8^3 - z_3\,z_8^3 - z_7\,z_8^3 + z_8^4\\ $

$ \bchi_{1, 1, 1, 0, 0, 0, 0, 0} =
   -z_1^2 - z_1^3 - z_1^4 - z_1\,z_2 - z_1^2\,z_2 + 2\,z_1^2\,z_3 +
z_1\,z_2\,z_3 - 2\,z_1\,z_4 - z_2\,z_4 +
    z_1\,z_5 - z_1^2\,z_5 + z_2\,z_5 - z_1\,z_6 + z_1^2\,z_6 + z_5\,z_6 -
z_1^2\,z_7 - z_1^3\,z_7 - z_1\,z_2\,z_7 +
    z_1\,z_3\,z_7 - z_4\,z_7 + z_5\,z_7 - z_6\,z_7 + z_1\,z_6\,z_7 +
3\,z_1\,z_8 + z_1^3\,z_8 + z_2\,z_8 -
    z_1\,z_2\,z_8 + z_1\,z_3\,z_8 - z_4\,z_8 + z_5\,z_8 + z_1\,z_5\,z_8 -
z_6\,z_8 - z_1\,z_6\,z_8 - z_2\,z_6\,z_8 +
    2\,z_7\,z_8 + 3\,z_1\,z_7\,z_8 + z_1^2\,z_7\,z_8 + z_2\,z_7\,z_8 -
z_6\,z_7\,z_8 + 2\,z_7^2\,z_8 + 3\,z_8^2 +
    z_1\,z_8^2 + 2\,z_1^2\,z_8^2 - z_3\,z_8^2 + z_5\,z_8^2 - 3\,z_6\,z_8^2 +
3\,z_7\,z_8^2 + z_1\,z_7\,z_8^2 +
    z_8^3 - 3\,z_1\,z_8^3 - 2\,z_8^4\\  $

  $  \bchi_{1, 2, 0, 0, 0, 0, 0, 1} =
   -1 - z_1 + z_1^3 + z_1^4 + 2\,z_1\,z_2 + z_1^2\,z_2 + z_2^2 - z_2^3 +
z_3 - 2\,z_1^2\,z_3 + z_2\,z_3 -
    z_1\,z_2\,z_3 + z_4 + 3\,z_1\,z_4 + 2\,z_2\,z_4 - z_5 - z_1\,z_5 +
z_1^2\,z_5 - 2\,z_2\,z_5 + 2\,z_6 +
    3\,z_1\,z_6 - z_1^2\,z_6 + 3\,z_2\,z_6 + z_1\,z_2\,z_6 + z_3\,z_6 -
z_5\,z_6 + z_6^2 - z_7 - z_1\,z_7 +
    z_1^2\,z_7 + z_1^3\,z_7 + z_2\,z_7 + 2\,z_1\,z_2\,z_7 - z_1\,z_3\,z_7 +
2\,z_4\,z_7 - z_5\,z_7 + 2\,z_6\,z_7 -
    z_1\,z_6\,z_7 + z_2\,z_7^2 - 3\,z_8 - 3\,z_1\,z_8 + 2\,z_1^2\,z_8 -
3\,z_2\,z_8 + z_1\,z_2\,z_8 - z_2^2\,z_8 +
    z_1\,z_2^2\,z_8 - 3\,z_1\,z_3\,z_8 - z_2\,z_3\,z_8 + 2\,z_4\,z_8 -
z_1\,z_4\,z_8 - z_5\,z_8 - z_2\,z_5\,z_8 +
    2\,z_6\,z_8 - 2\,z_1\,z_6\,z_8 - z_1^2\,z_6\,z_8 + z_3\,z_6\,z_8 +
z_6^2\,z_8 - 5\,z_7\,z_8 - 2\,z_1\,z_7\,z_8 -
    2\,z_2\,z_7\,z_8 - z_1\,z_2\,z_7\,z_8 - 2\,z_3\,z_7\,z_8 -
2\,z_6\,z_7\,z_8 - 2\,z_7^2\,z_8 - z_8^2 +
    5\,z_1\,z_8^2 - 3\,z_1^2\,z_8^2 - z_1^3\,z_8^2 + z_2\,z_8^2 -
z_1\,z_2\,z_8^2 - 2\,z_3\,z_8^2 +
    2\,z_1\,z_3\,z_8^2 - z_4\,z_8^2 - 2\,z_6\,z_8^2 + 2\,z_1\,z_6\,z_8^2 -
z_1\,z_7\,z_8^2 - z_2\,z_7\,z_8^2 +
    4\,z_8^3 - 3\,z_1\,z_8^3 + z_1^2\,z_8^3 + z_2\,z_8^3 + z_3\,z_8^3 +
z_6\,z_8^3 + z_7\,z_8^3 - z_8^4 +
    z_1\,z_8^4\\  $

 $   \bchi_{2, 0, 0, 0, 0, 0, 0, 3} =
   z_2 + 2\,z_1\,z_2 + z_1^2\,z_2 + z_2^2 - z_2\,z_3 - z_1\,z_5 + z_6 +
2\,z_1\,z_6 + z_1^2\,z_6 - z_3\,z_6 -
    z_6^2 - z_7 + z_2\,z_7 + z_1\,z_2\,z_7 - z_5\,z_7 + 2\,z_6\,z_7 -
z_7^2 - z_8 - z_1\,z_8 - z_1^2\,z_8 -
    z_1^3\,z_8 + z_2\,z_8 - 2\,z_1\,z_2\,z_8 + 2\,z_1\,z_3\,z_8 +
2\,z_6\,z_8 - 4\,z_7\,z_8 - 2\,z_1\,z_7\,z_8 -
    2\,z_1^2\,z_7\,z_8 - z_2\,z_7\,z_8 + 2\,z_3\,z_7\,z_8 +
2\,z_6\,z_7\,z_8 - 3\,z_7^2\,z_8 - z_8^2 + z_1^2\,z_8^2 -
    z_2\,z_8^2 - z_1\,z_2\,z_8^2 + z_5\,z_8^2 - 2\,z_6\,z_8^2 -
2\,z_7\,z_8^2 + 2\,z_1\,z_7\,z_8^2 + 2\,z_8^3 +
    z_1\,z_8^3 + z_1^2\,z_8^3 + z_2\,z_8^3 - z_3\,z_8^3 - z_6\,z_8^3 +
4\,z_7\,z_8^3 + z_8^4 - z_1\,z_8^4 -
    z_8^5\\  $

   $ \bchi_{2, 0, 0, 0, 0, 1, 0, 1} =
   -z_1^2 - z_1^3 - z_1\,z_2^2 + z_1\,z_3 + z_1\,z_4 - 2\,z_1\,z_5 -
z_1^2\,z_5 + z_2\,z_5 + z_3\,z_5 +
    z_1\,z_6 + z_5\,z_6 - z_1^2\,z_7 - z_1^3\,z_7 + 2\,z_1\,z_3\,z_7 +
z_5\,z_7 + z_1\,z_6\,z_7 - 2\,z_1\,z_8 +
    z_2\,z_8 + z_1^2\,z_2\,z_8 + z_1\,z_3\,z_8 - z_2\,z_3\,z_8 - z_4\,z_8 +
z_5\,z_8 + 2\,z_1\,z_6\,z_8 +
    z_1^2\,z_6\,z_8 - 2\,z_2\,z_6\,z_8 - z_3\,z_6\,z_8 - z_6^2\,z_8 +
z_7\,z_8 - 2\,z_1\,z_7\,z_8 + z_1^2\,z_7\,z_8 +
    z_2\,z_7\,z_8 - z_1\,z_2\,z_7\,z_8 + z_5\,z_7\,z_8 + z_7^2\,z_8 +
2\,z_8^2 + z_1^2\,z_8^2 + z_2\,z_8^2 -
    2\,z_1\,z_2\,z_8^2 + z_2^2\,z_8^2 - z_4\,z_8^2 + 2\,z_5\,z_8^2 -
z_6\,z_8^2 - z_1\,z_6\,z_8^2 +
    3\,z_7\,z_8^2 + z_2\,z_7\,z_8^2 + z_7^2\,z_8^2 + 2\,z_8^3 + z_1\,z_8^3 -
z_2\,z_8^3 - 2\,z_6\,z_8^3 +
    2\,z_7\,z_8^3 - z_8^4 - z_8^5\\  $

  $ \bchi_{2, 0, 0, 0, 1, 0, 0, 0} =
   z_1\,z_2 - z_1\,z_3 - z_2\,z_3 + 2\,z_1\,z_5 + z_1^2\,z_5 - z_3\,z_5 -
z_1\,z_2\,z_6 - z_3\,z_6 - 2\,z_1\,z_7 -
    z_2\,z_7 + z_1\,z_2\,z_7 + z_2^2\,z_7 - z_4\,z_7 + z_5\,z_7 -
2\,z_6\,z_7 - z_1\,z_7^2 + z_1\,z_8 + z_1^2\,z_8 +
    z_1\,z_2\,z_8 + z_2^2\,z_8 - z_3\,z_8 + z_1\,z_3\,z_8 - z_4\,z_8 +
2\,z_5\,z_8 - z_1\,z_5\,z_8 - 2\,z_6\,z_8 -
    z_1\,z_6\,z_8 + z_7\,z_8 + z_1\,z_7\,z_8 + z_2\,z_7\,z_8 + z_3\,z_7\,z_8
+ 2\,z_7^2\,z_8 + 2\,z_8^2 + z_2\,z_8^2 -
    2\,z_1\,z_2\,z_8^2 - z_6\,z_8^2 + 2\,z_7\,z_8^2 + z_1\,z_7\,z_8^2 -
z_1\,z_8^3 - z_8^4\\ $

 $ \bchi_{3, 0, 0, 0, 0, 0, 1, 0} =
   z_1\,z_2^2 + z_2\,z_3 - z_1\,z_4 - z_2\,z_5 + z_2\,z_6 + z_1\,z_7 +
z_1^3\,z_7 + z_2\,z_7 - 2\,z_1\,z_3\,z_7 +
    z_4\,z_7 - z_5\,z_7 + z_6\,z_7 - z_1\,z_6\,z_7 + z_1\,z_7^2 + z_2\,z_7^2
+ z_1\,z_8 - z_1^2\,z_8 -
    z_1^2\,z_2\,z_8 - z_2^2\,z_8 + z_3\,z_8 + z_2\,z_3\,z_8 - z_5\,z_8 +
z_1\,z_5\,z_8 + z_6\,z_8 - z_1\,z_6\,z_8 +
    2\,z_1\,z_7\,z_8 - z_1^2\,z_7\,z_8 - z_2\,z_8^2 + z_1\,z_2\,z_8^2 +
z_3\,z_8^2 + z_7\,z_8^2 - z_1\,z_7\,z_8^2\\ $

$ \bchi_{0, 0, 0, 0, 0, 0, 0, 5} =
   -z_1 - z_1^2 - z_2 + z_3 + z_4 - z_5 - z_1\,z_6 - z_1\,z_7 - z_2\,z_7 -
2\,z_6\,z_7 - z_8 + z_1\,z_8 -
    2\,z_2\,z_8 - 2\,z_5\,z_8 - z_6\,z_8 + 2\,z_7\,z_8 + 2\,z_1\,z_7\,z_8 +
3\,z_7^2\,z_8 + z_8^2 + 2\,z_1\,z_8^2 +
    z_2\,z_8^2 + 3\,z_6\,z_8^2 + 2\,z_7\,z_8^2 - z_8^3 - z_1\,z_8^3 -
4\,z_7\,z_8^3 - z_8^4 + z_8^5\\ $

$  \bchi_{0, 0, 0, 0, 0, 0, 2, 2} =
   z_1 + z_1^2 + z_2 + 2\,z_1\,z_2 + z_2^2 - z_4 + z_7 + z_2\,z_7 - z_3\,z_7
+ z_5\,z_7 - z_7^2 -
    z_1\,z_7^2 - z_7^3 + z_8 - z_1\,z_8 - 2\,z_1^2\,z_8 - 2\,z_1\,z_2\,z_8 -
z_4\,z_8 + z_5\,z_8 -
    4\,z_1\,z_7\,z_8 - z_2\,z_7\,z_8 - 2\,z_7^2\,z_8 - 2\,z_1\,z_8^2 +
z_1^2\,z_8^2 - 2\,z_2\,z_8^2 + z_5\,z_8^2 -
    z_6\,z_8^2 - z_7\,z_8^2 + 2\,z_1\,z_7\,z_8^2 + z_7^2\,z_8^2 +
3\,z_1\,z_8^3 + z_2\,z_8^3 - z_6\,z_8^3 +
    3\,z_7\,z_8^3 + z_8^4 - z_1\,z_8^4 - z_8^5\\ $

$  \bchi_{0, 0, 0, 0, 0, 1, 2, 0} =
   -z_1 + z_1^3 - z_2 + z_1\,z_2 + z_1^2\,z_2 + z_3 - z_1\,z_3 - z_2\,z_3 -
z_1\,z_4 - z_5 + z_1\,z_5 +
    z_2\,z_5 + z_6 + z_5\,z_6 - 2\,z_1\,z_7 - z_2\,z_7 + z_3\,z_7 - z_4\,z_7
+ z_6\,z_7 + 2\,z_1\,z_6\,z_7 -
    z_1\,z_7^2 - z_2\,z_7^2 + z_6\,z_7^2 - z_8 + z_1\,z_8 - z_1^2\,z_8 -
z_1^3\,z_8 + z_3\,z_8 + z_1\,z_3\,z_8 -
    z_5\,z_8 + 2\,z_6\,z_8 - z_2\,z_6\,z_8 - z_6^2\,z_8 - z_7\,z_8 -
z_1\,z_7\,z_8 - z_1^2\,z_7\,z_8 - z_2\,z_7\,z_8 +
    2\,z_3\,z_7\,z_8 - z_5\,z_7\,z_8 + 2\,z_6\,z_7\,z_8 - z_1\,z_7^2\,z_8 -
z_8^2 + z_1\,z_8^2 + z_2\,z_8^2 -
    z_1\,z_2\,z_8^2 + z_4\,z_8^2 + z_6\,z_8^2 - 2\,z_7\,z_8^2 +
3\,z_1\,z_7\,z_8^2 + z_2\,z_7\,z_8^2 +
    z_1^2\,z_8^3 - z_3\,z_8^3 - z_6\,z_8^3 + z_8^4 - z_1\,z_8^4\\ $

$  \bchi_{1, 0, 0, 0, 0, 0, 0, 4} =
   z_1 + z_1^2 - z_1\,z_2 - z_3 - z_4 - z_1\,z_5 - 2\,z_6 - 2\,z_1\,z_6 -
z_2\,z_6 + z_7 + 2\,z_1\,z_7 +
    z_1^2\,z_7 - z_3\,z_7 - 2\,z_6\,z_7 + z_7^2 + z_1\,z_7^2 + 2\,z_8 +
z_2\,z_8 + z_1\,z_2\,z_8 + z_3\,z_8 -
    z_6\,z_8 + 2\,z_1\,z_6\,z_8 + 4\,z_7\,z_8 + z_1\,z_7\,z_8 +
2\,z_2\,z_7\,z_8 + 2\,z_7^2\,z_8 - 2\,z_1\,z_8^2 -
    z_1^2\,z_8^2 + z_3\,z_8^2 + z_6\,z_8^2 + z_7\,z_8^2 -
3\,z_1\,z_7\,z_8^2 - z_8^3 - z_2\,z_8^3 -
    z_7\,z_8^3 + z_1\,z_8^4\\  $

   $ \bchi_{1, 0, 0, 0, 0, 2, 0, 0} =
   -z_2^2 + z_3^2 - z_1\,z_4 - z_2\,z_5 - 2\,z_1\,z_6 - z_1^2\,z_6 -
z_2\,z_6 - z_1\,z_2\,z_6 + z_3\,z_6 +
    z_1\,z_6^2 - z_1^2\,z_7 + z_2\,z_7 + z_3\,z_7 - z_5\,z_7 -
z_1\,z_5\,z_7 - 2\,z_1\,z_6\,z_7 - z_2\,z_6\,z_7 -
    z_1^2\,z_7^2 + z_2\,z_7^2 + z_3\,z_7^2 + 2\,z_1\,z_8 + z_2\,z_8 +
z_1\,z_2\,z_8 + z_2^2\,z_8 - z_3\,z_8 +
    z_1\,z_3\,z_8 - z_2\,z_3\,z_8 - z_4\,z_8 + z_1\,z_5\,z_8 +
z_2\,z_5\,z_8 - 2\,z_6\,z_8 - 2\,z_1\,z_6\,z_8 +
    z_1^2\,z_6\,z_8 - z_3\,z_6\,z_8 + 3\,z_1\,z_7\,z_8 + 2\,z_2\,z_7\,z_8 +
z_1\,z_2\,z_7\,z_8 + z_1\,z_7^2\,z_8 +
    2\,z_8^2 + z_2\,z_8^2 - z_1\,z_2\,z_8^2 + z_5\,z_8^2 + 2\,z_7\,z_8^2 +
z_1\,z_7\,z_8^2 - z_2\,z_7\,z_8^2 -
    z_1\,z_8^3 - z_2\,z_8^3 - z_8^4\\  $

  $  \bchi_{2, 0, 0, 0, 0, 0, 2, 0} =
   z_1^2 + z_1^3 + 2\,z_1\,z_2 + z_2^2 - 2\,z_1\,z_3 - z_2\,z_3 - z_1\,z_4 +
2\,z_1\,z_5 + z_2\,z_5 - z_1\,z_6 +
    z_1\,z_2\,z_6 + z_3\,z_6 - z_5\,z_6 + z_6^2 + z_1\,z_7 + 3\,z_1^2\,z_7 +
z_1^3\,z_7 + z_1\,z_2\,z_7 +
    z_2^2\,z_7 - z_3\,z_7 - 2\,z_1\,z_3\,z_7 - z_4\,z_7 + z_5\,z_7 -
2\,z_6\,z_7 - z_1\,z_6\,z_7 + z_7^2 +
    2\,z_1\,z_7^2 + z_1^2\,z_7^2 - z_3\,z_7^2 - z_6\,z_7^2 + z_7^3 +
3\,z_1\,z_8 - z_1^2\,z_8 + 2\,z_2\,z_8 -
    z_1\,z_2\,z_8 + z_1^2\,z_2\,z_8 + z_2^2\,z_8 - 2\,z_3\,z_8 -
z_2\,z_3\,z_8 - z_4\,z_8 + 2\,z_5\,z_8 -
    z_1\,z_5\,z_8 - 2\,z_6\,z_8 - z_1\,z_6\,z_8 - z_1^2\,z_6\,z_8 -
2\,z_2\,z_6\,z_8 + z_3\,z_6\,z_8 + z_6^2\,z_8 +
    2\,z_7\,z_8 + 3\,z_1\,z_7\,z_8 - z_1^2\,z_7\,z_8 + z_2\,z_7\,z_8 -
z_1\,z_2\,z_7\,z_8 - 3\,z_3\,z_7\,z_8 +
    z_5\,z_7\,z_8 - 4\,z_6\,z_7\,z_8 + 3\,z_7^2\,z_8 - z_1\,z_7^2\,z_8 +
2\,z_8^2 - 2\,z_1\,z_8^2 - z_1^3\,z_8^2 +
    z_2\,z_8^2 - 2\,z_1\,z_2\,z_8^2 - z_3\,z_8^2 + 2\,z_1\,z_3\,z_8^2 -
2\,z_6\,z_8^2 + 2\,z_1\,z_6\,z_8^2 +
    2\,z_7\,z_8^2 - 3\,z_1\,z_7\,z_8^2 + z_2\,z_7\,z_8^2 - 2\,z_1\,z_8^3 +
z_1^2\,z_8^3 - z_2\,z_8^3 +
    z_3\,z_8^3 + z_6\,z_8^3 - z_7\,z_8^3 - z_8^4 + z_1\,z_8^4\\ $

$  \bchi_{3, 0, 0, 0, 0, 0, 0, 2} =
   -z_1^3 - z_1^4 - z_1^2\,z_2 + z_1\,z_3 + 3\,z_1^2\,z_3 - z_3^2 - z_1\,z_4
+ z_1\,z_5 + z_1\,z_6 +
    z_1^2\,z_6 + z_2\,z_6 - 2\,z_3\,z_6 - z_1\,z_7 - z_1^2\,z_7 -
z_1^3\,z_7 - z_2\,z_7 - z_1\,z_2\,z_7 + z_3\,z_7 +
    2\,z_1\,z_3\,z_7 - z_4\,z_7 + z_1\,z_6\,z_7 - z_1\,z_7^2 - z_2\,z_7^2 -
z_1\,z_8 - 2\,z_1^2\,z_8 + z_1^3\,z_8 -
    z_1\,z_2\,z_8 - z_1^2\,z_2\,z_8 - z_2^2\,z_8 + 2\,z_3\,z_8 +
z_2\,z_3\,z_8 - z_5\,z_8 + z_1\,z_5\,z_8 +
    2\,z_6\,z_8 - z_1\,z_6\,z_8 - z_7\,z_8 - 3\,z_1\,z_7\,z_8 +
z_1^2\,z_7\,z_8 - 3\,z_2\,z_7\,z_8 + 2\,z_3\,z_7\,z_8 -
    z_7^2\,z_8 - 2\,z_8^2 + 2\,z_1^2\,z_8^2 + z_1^3\,z_8^2 - z_2\,z_8^2 +
2\,z_1\,z_2\,z_8^2 -
    2\,z_1\,z_3\,z_8^2 + z_4\,z_8^2 - z_5\,z_8^2 + z_6\,z_8^2 -
z_1\,z_6\,z_8^2 - 4\,z_7\,z_8^2 +
    4\,z_1\,z_7\,z_8^2 + z_2\,z_7\,z_8^2 - z_8^3 + 3\,z_1\,z_8^3 -
z_1^2\,z_8^3 + z_2\,z_8^3 - z_3\,z_8^3 +
    2\,z_8^4 - 2\,z_1\,z_8^4\\  $

  $  \bchi_{4, 0, 0, 0, 0, 0, 0, 0} =
   -z_1^2 + z_1^4 + z_1^2\,z_2 - 3\,z_1^2\,z_3 + z_2\,z_3 + z_3^2 +
2\,z_1\,z_4 - 3\,z_1\,z_5 - z_2\,z_5 +
    z_1\,z_6 - z_1^2\,z_6 + z_2\,z_6 + 2\,z_3\,z_6 + z_6^2 - z_1\,z_7 -
z_1^2\,z_7 + z_1\,z_2\,z_7 - z_5\,z_7 -
    z_1\,z_7^2 - 4\,z_1\,z_8 + z_1^2\,z_8 - z_1^3\,z_8 - z_2\,z_8 +
2\,z_1\,z_2\,z_8 - z_2^2\,z_8 + z_4\,z_8 -
    2\,z_5\,z_8 + 3\,z_1\,z_6\,z_8 - z_7\,z_8 - 4\,z_1\,z_7\,z_8 -
z_2\,z_7\,z_8 - z_7^2\,z_8 - 2\,z_8^2 +
    z_1\,z_8^2 - z_1^2\,z_8^2 + z_3\,z_8^2 + 2\,z_6\,z_8^2 - 2\,z_7\,z_8^2 +
2\,z_1\,z_8^3 + z_8^4\\ $

  $\bchi_{0, 2, 0, 0, 0, 1, 0, 0} =
   z_1 + z_1^2 + z_2 - z_3 - z_1\,z_3 - z_1^2\,z_3 - z_2\,z_3 + z_3^2 - z_4
+ z_5 - z_3\,z_5 - z_6 -
    z_2\,z_6 + z_1\,z_2\,z_6 + z_2^2\,z_6 + z_3\,z_6 - z_4\,z_6 - z_5\,z_6 -
z_1\,z_6^2 + z_1\,z_7 + z_1^2\,z_7 +
    z_1^3\,z_7 + 2\,z_2\,z_7 + z_1\,z_2\,z_7 - z_2^2\,z_7 -
2\,z_1\,z_3\,z_7 - z_2\,z_3\,z_7 + z_4\,z_7 +
    z_1\,z_5\,z_7 + 3\,z_6\,z_7 - z_2\,z_6\,z_7 - z_7^2 + z_1\,z_7^2 +
z_1^2\,z_7^2 + z_2\,z_7^2 + z_8 -
    z_1\,z_8 + z_1^2\,z_8 + z_1^3\,z_8 + 2\,z_2\,z_8 + z_1^2\,z_2\,z_8 -
z_2^2\,z_8 - z_3\,z_8 - z_1\,z_3\,z_8 -
    2\,z_2\,z_3\,z_8 + z_3^2\,z_8 + z_4\,z_8 - z_1\,z_4\,z_8 + z_5\,z_8 +
z_2\,z_5\,z_8 + z_6\,z_8 + z_1\,z_6\,z_8 -
    2\,z_1^2\,z_6\,z_8 - 2\,z_2\,z_6\,z_8 + 2\,z_3\,z_6\,z_8 + z_6^2\,z_8 -
3\,z_7\,z_8 + 2\,z_1\,z_7\,z_8 +
    z_1^2\,z_7\,z_8 + z_2\,z_7\,z_8 - z_1\,z_2\,z_7\,z_8 -
2\,z_3\,z_7\,z_8 - z_6\,z_7\,z_8 - 3\,z_7^2\,z_8 - 2\,z_8^2 -
    z_1^3\,z_8^2 - z_1\,z_2\,z_8^2 + z_2^2\,z_8^2 - z_3\,z_8^2 +
2\,z_1\,z_3\,z_8^2 - z_5\,z_8^2 -
    2\,z_6\,z_8^2 + z_1\,z_6\,z_8^2 - 3\,z_1\,z_7\,z_8^2 + z_8^3 -
2\,z_1\,z_8^3 - z_2\,z_8^3 + z_3\,z_8^3 +
    z_6\,z_8^3 + z_7\,z_8^3 + z_1\,z_8^4\\  $

  $  \bchi_{1, 0, 1, 0, 0, 1, 0, 0} =
   z_1^2 + z_1^3 + z_2 + 2\,z_1\,z_2 + z_1^2\,z_2 - z_1\,z_2^2 +
z_1^2\,z_3 - z_2\,z_3 - z_3^2 - z_2\,z_4 +
    z_5 + z_1\,z_5 + z_2\,z_5 + z_3\,z_5 + 2\,z_1\,z_6 + 2\,z_1^2\,z_6 -
2\,z_2\,z_6 - 2\,z_1\,z_2\,z_6 -
    3\,z_3\,z_6 + z_1\,z_3\,z_6 - z_4\,z_6 + z_5\,z_6 - 2\,z_6^2 -
z_1\,z_6^2 + z_1\,z_7 + 3\,z_1^2\,z_7 +
    z_1^3\,z_7 + z_2\,z_7 + z_1\,z_2\,z_7 - z_1^2\,z_2\,z_7 - z_3\,z_7 -
z_1\,z_3\,z_7 + z_5\,z_7 + z_1\,z_5\,z_7 +
    z_1\,z_6\,z_7 + z_2\,z_6\,z_7 + z_1\,z_7^2 + z_1^2\,z_7^2 - z_3\,z_7^2 -
z_1\,z_8 - z_1^2\,z_8 + 2\,z_2\,z_8 -
    2\,z_1\,z_2\,z_8 - 2\,z_1^2\,z_2\,z_8 - z_2^2\,z_8 + z_1\,z_2^2\,z_8 +
z_3\,z_8 - 2\,z_1\,z_3\,z_8 +
    3\,z_2\,z_3\,z_8 + z_4\,z_8 - z_5\,z_8 - z_2\,z_5\,z_8 + 3\,z_6\,z_8 -
2\,z_1\,z_6\,z_8 - z_1^2\,z_6\,z_8 +
    2\,z_2\,z_6\,z_8 + z_3\,z_6\,z_8 - 2\,z_1\,z_7\,z_8 - 2\,z_1^2\,z_7\,z_8
+ z_2\,z_7\,z_8 + z_1\,z_2\,z_7\,z_8 +
    z_3\,z_7\,z_8 + 2\,z_6\,z_7\,z_8 - z_1\,z_7^2\,z_8 - 2\,z_8^2 -
z_1^2\,z_8^2 - 3\,z_2\,z_8^2 +
    3\,z_1\,z_2\,z_8^2 - z_2^2\,z_8^2 - z_5\,z_8^2 - 3\,z_7\,z_8^2 +
z_1\,z_8^3 + z_8^4\\ $

 $ \bchi_{1, 1, 0, 0, 1, 0, 0, 0} =
   z_1^2 + z_1^3 - z_2 - z_1^2\,z_2 + z_2^2 + z_1\,z_2^2 - z_1\,z_3 +
z_2\,z_3 - z_1\,z_4 - z_5 -
    z_1^2\,z_5 + z_1\,z_2\,z_5 + z_3\,z_5 - z_5^2 - z_6 - 4\,z_1\,z_6 -
2\,z_1^2\,z_6 + 2\,z_2\,z_6 - z_2^2\,z_6 +
    z_3\,z_6 - z_1\,z_3\,z_6 + z_4\,z_6 + z_5\,z_6 + 2\,z_6^2 + z_1\,z_6^2 +
z_7 + z_1\,z_7 + z_1^2\,z_7 -
    z_2\,z_7 + z_2^2\,z_7 - z_3\,z_7 + z_2\,z_3\,z_7 - z_1\,z_5\,z_7 -
4\,z_6\,z_7 - z_1\,z_6\,z_7 + z_2\,z_6\,z_7 +
    z_7^2 - z_6\,z_7^2 + 3\,z_1\,z_8 - z_1^2\,z_8 - z_2\,z_8 +
z_1\,z_2\,z_8 - z_1^2\,z_2\,z_8 + z_2\,z_3\,z_8 -
    z_4\,z_8 - 2\,z_5\,z_8 + 2\,z_1\,z_5\,z_8 - z_2\,z_5\,z_8 -
3\,z_6\,z_8 - z_1\,z_6\,z_8 + z_1^2\,z_6\,z_8 +
    z_2\,z_6\,z_8 - z_3\,z_6\,z_8 + 4\,z_7\,z_8 + z_1\,z_7\,z_8 +
z_1^2\,z_7\,z_8 - z_2\,z_7\,z_8 - z_3\,z_7\,z_8 -
    4\,z_6\,z_7\,z_8 + 3\,z_7^2\,z_8 - z_1\,z_7^2\,z_8 + z_8^2 -
z_1\,z_8^2 - z_1^2\,z_8^2 + z_1\,z_2\,z_8^2 -
    z_2^2\,z_8^2 + z_3\,z_8^2 + 2\,z_5\,z_8^2 + 3\,z_6\,z_8^2 + z_7\,z_8^2 +
z_2\,z_7\,z_8^2 + z_7^2\,z_8^2 -
    z_8^3 - 2\,z_7\,z_8^3\\  $

  $  \bchi_{2, 0, 1, 0, 0, 0, 0, 1} =
   -z_1 - z_1^2 + z_1^3 + z_1^4 - z_2 - 2\,z_1\,z_2 - z_1^2\,z_2 -
z_1^3\,z_2 - z_2^2 - z_1\,z_2^2 + z_3 -
    2\,z_1^2\,z_3 + z_2\,z_3 + z_1\,z_2\,z_3 + z_1\,z_4 - z_5 - z_1\,z_5 +
z_1^2\,z_5 - 2\,z_1\,z_6 -
    2\,z_1^2\,z_6 - z_2\,z_6 + z_1\,z_2\,z_6 + 2\,z_3\,z_6 - z_5\,z_6 +
z_6^2 - z_1\,z_7 + z_1^3\,z_7 - z_2\,z_7 -
    z_1\,z_2\,z_7 - z_2^2\,z_7 - z_1\,z_3\,z_7 + z_4\,z_7 - z_5\,z_7 -
z_1\,z_6\,z_7 - z_8 + 3\,z_1^2\,z_8 -
    z_2\,z_8 + 2\,z_1\,z_2\,z_8 + 2\,z_1^2\,z_2\,z_8 - z_1\,z_3\,z_8 +
z_1^2\,z_3\,z_8 - z_3^2\,z_8 + 2\,z_4\,z_8 -
    z_1\,z_4\,z_8 - 2\,z_5\,z_8 + z_2\,z_5\,z_8 + 2\,z_1\,z_6\,z_8 -
z_1^2\,z_6\,z_8 - z_3\,z_6\,z_8 - z_7\,z_8 +
    3\,z_1\,z_7\,z_8 + 2\,z_1^2\,z_7\,z_8 + 2\,z_2\,z_7\,z_8 +
z_1\,z_2\,z_7\,z_8 - z_3\,z_7\,z_8 + z_1\,z_7^2\,z_8 -
    2\,z_8^2 + 3\,z_1\,z_8^2 - 2\,z_1^2\,z_8^2 - z_1^3\,z_8^2 +
2\,z_2\,z_8^2 - z_5\,z_8^2 + 3\,z_6\,z_8^2 -
    z_7\,z_8^2 - 2\,z_1\,z_7\,z_8^2 - z_2\,z_7\,z_8^2 - z_7^2\,z_8^2 -
3\,z_1\,z_8^3 - z_2\,z_8^3 + z_6\,z_8^3 -
    3\,z_7\,z_8^3 + z_1\,z_8^4 + z_8^5\\ $

    $ \bchi_{2, 1, 0, 0, 0, 0, 1, 0} =
   z_1^2 + z_1^3 + z_1^2\,z_2 + z_1\,z_2^2 + z_2^3 - z_3 - z_1\,z_3 -
z_2\,z_3 + z_1\,z_2\,z_3 + z_3^2 -
    z_4 - 2\,z_1\,z_4 - 2\,z_2\,z_4 + z_5 + z_1\,z_5 + z_2\,z_5 - z_3\,z_5 -
z_6 - 2\,z_1\,z_6 - z_1^2\,z_6 -
    2\,z_2\,z_6 - z_1\,z_2\,z_6 + z_3\,z_6 + z_1\,z_7 - z_1^3\,z_7 +
z_2\,z_7 + z_1^2\,z_2\,z_7 - z_3\,z_7 +
    z_1\,z_3\,z_7 - z_2\,z_3\,z_7 - 2\,z_4\,z_7 + z_5\,z_7 - z_1\,z_5\,z_7 +
z_1\,z_6\,z_7 - z_1^2\,z_7^2 +
    z_6\,z_7^2 + z_8 + 2\,z_1\,z_8 - z_1^2\,z_8 - 2\,z_1^3\,z_8 +
3\,z_2\,z_8 - z_1\,z_2\,z_8 - z_1^2\,z_2\,z_8 -
    z_1\,z_2^2\,z_8 - z_3\,z_8 + 2\,z_1\,z_3\,z_8 - z_1^2\,z_3\,z_8 +
z_3^2\,z_8 - z_4\,z_8 + z_1\,z_4\,z_8 +
    z_5\,z_8 + z_2\,z_5\,z_8 + z_1\,z_6\,z_8 + z_1^2\,z_6\,z_8 -
z_2\,z_6\,z_8 + z_3\,z_6\,z_8 - z_6^2\,z_8 +
    2\,z_7\,z_8 - z_1^2\,z_7\,z_8 + 2\,z_2\,z_7\,z_8 + z_3\,z_7\,z_8 +
3\,z_6\,z_7\,z_8 - 2\,z_1\,z_8^2 +
    z_1^3\,z_8^2 - z_2\,z_8^2 + z_1\,z_2\,z_8^2 + z_3\,z_8^2 + z_4\,z_8^2 -
z_5\,z_8^2 + 2\,z_6\,z_8^2 -
    z_1\,z_6\,z_8^2 - 2\,z_7\,z_8^2 + 3\,z_1\,z_7\,z_8^2 - 2\,z_8^3 +
2\,z_1\,z_8^3 - z_2\,z_8^3 - z_6\,z_8^3 +
    z_8^4 - z_1\,z_8^4\\  $

   $ \bchi_{3, 1, 0, 0, 0, 0, 0, 0} =
   -z_1 - 2\,z_1^2 - 2\,z_1^3 - z_1^4 - z_1\,z_2 + z_1^3\,z_2 + z_1\,z_3 +
2\,z_1^2\,z_3 - z_2\,z_3 -
    2\,z_1\,z_2\,z_3 - z_3^2 + z_1\,z_4 + z_2\,z_4 + z_5 - z_1^2\,z_5 +
z_3\,z_5 + 2\,z_1\,z_6 + 3\,z_1^2\,z_6 -
    z_2\,z_6 - 2\,z_3\,z_6 - z_6^2 - z_7 - 4\,z_1\,z_7 - 3\,z_1^2\,z_7 -
z_1^3\,z_7 - z_2\,z_7 - z_1\,z_2\,z_7 +
    z_1\,z_3\,z_7 + z_4\,z_7 + z_6\,z_7 + 2\,z_1\,z_6\,z_7 - 2\,z_7^2 -
3\,z_1\,z_7^2 - z_2\,z_7^2 - z_7^3 -
    2\,z_1\,z_8 + 2\,z_1^3\,z_8 + z_1\,z_2\,z_8 + 2\,z_2^2\,z_8 -
z_1\,z_3\,z_8 - z_2\,z_3\,z_8 + z_5\,z_8 -
    z_1\,z_5\,z_8 - 2\,z_7\,z_8 - 2\,z_1\,z_7\,z_8 + z_1^2\,z_7\,z_8 +
z_2\,z_7\,z_8 + z_6\,z_7\,z_8 - 2\,z_7^2\,z_8 +
    z_8^2 + 2\,z_1\,z_8^2 + 2\,z_1^2\,z_8^2 + 2\,z_2\,z_8^2 -
2\,z_1\,z_2\,z_8^2 - 2\,z_3\,z_8^2 -
    2\,z_6\,z_8^2 + 2\,z_7\,z_8^2 + z_1\,z_7\,z_8^2 + z_8^3 - z_1\,z_8^3 +
z_7\,z_8^3 - z_8^4\\ $

$\bchi_{0, 0, 0, 0, 0, 0, 1, 4} =
  -z_1^2 - z_1^3 + z_2 + z_3 + 2\,z_1\,z_3 + z_2\,z_3 - z_1\,z_5 + 2\,z_6 +
2\,z_1\,z_6 - z_6^2 -
   2\,z_1^2\,z_7 + 2\,z_3\,z_7 - 2\,z_5\,z_7 + 3\,z_6\,z_7 + z_7^2 + z_7^3 -
z_8 - z_1\,z_8 - z_1\,z_2\,z_8 +
   z_3\,z_8 - z_4\,z_8 + 2\,z_6\,z_8 - z_1\,z_6\,z_8 - 3\,z_7\,z_8 -
z_1\,z_7\,z_8 - z_2\,z_7\,z_8 + 4\,z_6\,z_7\,z_8 -
   2\,z_7^2\,z_8 + z_1^2\,z_8^2 - 2\,z_2\,z_8^2 - z_3\,z_8^2 + z_5\,z_8^2 -
2\,z_6\,z_8^2 - 3\,z_7\,z_8^2 +
   2\,z_1\,z_7\,z_8^2 - 3\,z_7^2\,z_8^2 + z_8^3 + 2\,z_1\,z_8^3 +
z_2\,z_8^3 - z_6\,z_8^3 + 3\,z_7\,z_8^3 +
   z_8^4 - z_1\,z_8^4 + z_7\,z_8^4 - z_8^5\\ $

$\bchi_{0, 0, 0, 0, 0, 0, 3, 1} =
  z_1^2\,z_2 - z_2\,z_3 + z_1\,z_4 + z_5 - z_6 - z_1\,z_6 - z_2\,z_6 -
z_3\,z_6 - z_5\,z_6 + z_1\,z_2\,z_7 +
   z_4\,z_7 - z_6\,z_7 - z_1\,z_6\,z_7 - z_6\,z_7^2 + z_8 - z_1^3\,z_8 +
z_2\,z_8 + 2\,z_1\,z_2\,z_8 + z_2^2\,z_8 +
   z_3\,z_8 + 2\,z_1\,z_3\,z_8 - 2\,z_1\,z_5\,z_8 - 2\,z_6\,z_8 +
z_2\,z_6\,z_8 + 2\,z_6^2\,z_8 + 2\,z_7\,z_8 +
   2\,z_1\,z_7\,z_8 - z_1^2\,z_7\,z_8 + 2\,z_2\,z_7\,z_8 +
2\,z_3\,z_7\,z_8 - 2\,z_6\,z_7\,z_8 + 2\,z_7^2\,z_8 +
   z_1\,z_7^2\,z_8 + z_7^3\,z_8 + z_8^2 - z_1^2\,z_8^2 - 2\,z_1\,z_2\,z_8^2
+ z_3\,z_8^2 - z_4\,z_8^2 -
   z_5\,z_8^2 - z_6\,z_8^2 + 2\,z_1\,z_6\,z_8^2 + 4\,z_7\,z_8^2 -
z_1\,z_7\,z_8^2 - 2\,z_2\,z_7\,z_8^2 -
   2\,z_6\,z_7\,z_8^2 + 2\,z_7^2\,z_8^2 - z_1\,z_8^3 + z_1^2\,z_8^3 -
2\,z_2\,z_8^3 - z_3\,z_8^3 + z_5\,z_8^3 +
   z_6\,z_8^3 - z_1\,z_7\,z_8^3 - z_8^4 + z_1\,z_8^4 + z_2\,z_8^4 -
z_7\,z_8^4\\ $

$\bchi_{0, 0, 0, 0, 0, 1, 1, 2} =
  -z_2 - z_1\,z_2 + z_1^2\,z_2 - z_3 + z_1\,z_3 - z_4 - z_1\,z_5 - z_6 +
z_5\,z_6 + z_1\,z_7 + z_1^2\,z_7 -
   z_2\,z_7 - z_3\,z_7 - z_6\,z_7 - z_1\,z_6\,z_7 + z_1\,z_7^2 - z_6\,z_7^2
+ z_8 + 3\,z_1^2\,z_8 + 3\,z_1\,z_2\,z_8 +
   z_2^2\,z_8 - z_1\,z_3\,z_8 - z_2\,z_3\,z_8 + z_1\,z_5\,z_8 -
z_1\,z_6\,z_8 - z_2\,z_6\,z_8 - z_6^2\,z_8 + 2\,z_7\,z_8 +
   4\,z_1\,z_7\,z_8 + z_1^2\,z_7\,z_8 + 2\,z_2\,z_7\,z_8 - 2\,z_3\,z_7\,z_8
+ z_5\,z_7\,z_8 - 2\,z_6\,z_7\,z_8 +
   z_7^2\,z_8 + z_1\,z_7^2\,z_8 + 3\,z_1\,z_8^2 - 3\,z_1^2\,z_8^2 +
3\,z_2\,z_8^2 - 2\,z_1\,z_2\,z_8^2 +
   z_4\,z_8^2 + z_5\,z_8^2 + z_6\,z_8^2 + z_1\,z_6\,z_8^2 + 2\,z_7\,z_8^2 -
3\,z_1\,z_7\,z_8^2 + z_2\,z_7\,z_8^2 +
   z_6\,z_7\,z_8^2 - z_8^3 - 4\,z_1\,z_8^3 + z_1^2\,z_8^3 - z_2\,z_8^3 -
z_5\,z_8^3 - 3\,z_7\,z_8^3 -
   z_1\,z_7\,z_8^3 - z_8^4 + z_1\,z_8^4 + z_8^5\\ $

$\bchi_{0, 0, 0, 0, 0, 2, 1, 0} =
  z_1^2 + z_1^3 + z_1\,z_2 + z_2^2 - z_1\,z_3 + z_1^2\,z_3 - z_2\,z_3 -
z_3^2 - z_1\,z_4 + 2\,z_1\,z_5 +
   z_1^2\,z_5 + 2\,z_2\,z_5 - z_3\,z_5 + z_5^2 - 2\,z_1\,z_6 - z_2\,z_6 -
z_3\,z_6 - z_4\,z_6 - z_5\,z_6 - z_6^2 +
   z_1\,z_7 + 2\,z_1^2\,z_7 + z_1^3\,z_7 - z_2\,z_7 - z_3\,z_7 -
2\,z_1\,z_3\,z_7 + z_5\,z_7 + z_1\,z_5\,z_7 -
   z_6\,z_7 - 2\,z_1\,z_6\,z_7 - z_2\,z_6\,z_7 + z_6^2\,z_7 - z_2\,z_7^2 -
z_3\,z_7^2 - z_5\,z_7^2 - z_6\,z_7^2 -
   z_1\,z_7^3 + 2\,z_1\,z_8 - 2\,z_1^2\,z_8 - 2\,z_1\,z_2\,z_8 +
z_1^2\,z_2\,z_8 + z_3\,z_8 - z_1\,z_4\,z_8 + z_5\,z_8 -
   z_1\,z_5\,z_8 + z_2\,z_5\,z_8 + z_6\,z_8 - 3\,z_1\,z_6\,z_8 -
2\,z_1^2\,z_6\,z_8 - 2\,z_2\,z_6\,z_8 + 2\,z_3\,z_6\,z_8 -
   z_5\,z_6\,z_8 + 2\,z_6^2\,z_8 + z_7\,z_8 + 2\,z_1\,z_7\,z_8 -
2\,z_1^2\,z_7\,z_8 + z_1\,z_2\,z_7\,z_8 - z_3\,z_7\,z_8 +
   z_4\,z_7\,z_8 - z_6\,z_7\,z_8 + z_1\,z_6\,z_7\,z_8 + z_7^2\,z_8 -
z_1\,z_7^2\,z_8 + z_2\,z_7^2\,z_8 - z_1^3\,z_8^2 +
   z_1\,z_2\,z_8^2 + z_1\,z_3\,z_8^2 - z_5\,z_8^2 - 2\,z_6\,z_8^2 +
3\,z_1\,z_6\,z_8^2 - z_2\,z_6\,z_8^2 -
   z_1\,z_7\,z_8^2 + z_1^2\,z_7\,z_8^2 + 3\,z_2\,z_7\,z_8^2 - z_1\,z_8^3 +
z_1^2\,z_8^3 + z_2\,z_8^3 -
   z_1\,z_2\,z_8^3 + z_6\,z_8^3 - z_2\,z_8^4\\ $

$\bchi_{0, 0, 0, 0, 1, 0, 0, 3} =
  z_1^2\,z_2 - z_1\,z_2^2 - z_3 - z_2\,z_3 - z_3^2 + 2\,z_1\,z_4 - z_1\,z_5
+ z_2\,z_5 - z_6 + z_1\,z_6 +
   2\,z_1^2\,z_6 - z_2\,z_6 - 2\,z_3\,z_6 + z_5\,z_6 - z_6^2 - z_7 -
2\,z_1\,z_7 - z_1^2\,z_7 - z_3\,z_7 +
   z_4\,z_7 + z_1\,z_6\,z_7 - 2\,z_7^2 - 2\,z_1\,z_7^2 - z_7^3 - 4\,z_1\,z_8
+ z_2\,z_8 + z_2\,z_3\,z_8 -
   z_1\,z_5\,z_8 + z_6\,z_8 + 2\,z_1\,z_6\,z_8 + z_2\,z_6\,z_8 -
2\,z_7\,z_8 - 3\,z_1\,z_7\,z_8 + z_2\,z_7\,z_8 -
   2\,z_5\,z_7\,z_8 + z_6\,z_7\,z_8 - 2\,z_7^2\,z_8 - z_2\,z_8^2 -
z_4\,z_8^2 - z_6\,z_8^2 - z_1\,z_6\,z_8^2 +
   z_7\,z_8^2 + z_7^2\,z_8^2 + 2\,z_8^3 + 2\,z_1\,z_8^3 + z_5\,z_8^3 -
z_6\,z_8^3 + 3\,z_7\,z_8^3 - z_8^5\\ $

$\bchi_{0, 0, 0, 0, 1, 0, 2, 0} =
  z_1^3 + z_1^4 - z_1\,z_3 - 2\,z_1^2\,z_3 - z_2\,z_3 - z_1\,z_2\,z_3 +
z_3^2 + z_1\,z_4 + z_2\,z_4 + z_1\,z_5 +
   z_1^2\,z_5 - 2\,z_1\,z_6 - 2\,z_1^2\,z_6 + z_3\,z_6 + z_4\,z_6 + z_6^2 +
z_1\,z_6^2 + z_1\,z_7 + z_1^2\,z_7 +
   z_1^3\,z_7 + z_2\,z_7 + z_1\,z_2\,z_7 + z_3\,z_7 - z_1\,z_3\,z_7 -
z_2\,z_3\,z_7 + z_4\,z_7 + z_5\,z_7 +
   2\,z_1\,z_5\,z_7 - 3\,z_1\,z_6\,z_7 - z_2\,z_6\,z_7 + z_1\,z_7^2 +
z_2\,z_7^2 + z_3\,z_7^2 + z_5\,z_7^2 -
   z_6\,z_7^2 + z_1\,z_8 + z_1^2\,z_8 - z_2\,z_8 + 2\,z_1\,z_2\,z_8 +
z_1^2\,z_2\,z_8 + z_3\,z_8 - z_1\,z_3\,z_8 -
   2\,z_2\,z_3\,z_8 + z_4\,z_8 - z_5\,z_8 - z_2\,z_5\,z_8 -
2\,z_1\,z_6\,z_8 - z_5\,z_6\,z_8 + z_6^2\,z_8 + z_7\,z_8 +
   5\,z_1\,z_7\,z_8 + z_2\,z_7\,z_8 - z_1\,z_2\,z_7\,z_8 - z_4\,z_7\,z_8 +
z_5\,z_7\,z_8 - 4\,z_6\,z_7\,z_8 -
   z_1\,z_6\,z_7\,z_8 + 2\,z_7^2\,z_8 + 2\,z_1\,z_7^2\,z_8 + z_7^3\,z_8 -
z_8^2 + 2\,z_1\,z_8^2 - 3\,z_1^2\,z_8^2 -
   z_1^3\,z_8^2 - z_2\,z_8^2 - z_1\,z_2\,z_8^2 + z_1\,z_3\,z_8^2 +
z_2\,z_3\,z_8^2 - z_5\,z_8^2 - z_1\,z_5\,z_8^2 +
   3\,z_1\,z_6\,z_8^2 + z_2\,z_6\,z_8^2 + 3\,z_7\,z_8^2 -
4\,z_1\,z_7\,z_8^2 - 3\,z_2\,z_7\,z_8^2 - z_6\,z_7\,z_8^2 +
   2\,z_7^2\,z_8^2 - 4\,z_1\,z_8^3 + z_1^2\,z_8^3 - z_2\,z_8^3 +
2\,z_6\,z_8^3 - 3\,z_7\,z_8^3 - z_1\,z_7\,z_8^3 -
   z_8^4 + 2\,z_1\,z_8^4 + z_2\,z_8^4 - z_7\,z_8^4 + z_8^5\\ $

$\bchi_{0, 0, 0, 0, 1, 1, 0, 1} =
  -z_1^2 - z_1^3 - z_1\,z_2 - z_1^2\,z_2 + z_1\,z_3 + z_2\,z_3 + z_1\,z_4 -
z_1\,z_5 - z_1^2\,z_5 - z_2\,z_5 +
   z_3\,z_5 - z_5^2 + 2\,z_1\,z_6 + z_1^2\,z_6 + 2\,z_2\,z_6 + z_1\,z_2\,z_6
+ z_5\,z_6 + z_6^2 - z_1\,z_7 -
   z_1^2\,z_7 - z_1^3\,z_7 - z_2\,z_7 - z_1\,z_2\,z_7 + 2\,z_1\,z_3\,z_7 +
z_2\,z_3\,z_7 - z_4\,z_7 - z_1\,z_5\,z_7 -
   z_6\,z_7 + 2\,z_1\,z_6\,z_7 + z_2\,z_6\,z_7 - z_1\,z_7^2 - z_2\,z_7^2 -
3\,z_1\,z_8 + z_1^3\,z_8 - 2\,z_2\,z_8 -
   z_1\,z_2\,z_8 - z_2^2\,z_8 + z_3\,z_8 + z_2\,z_3\,z_8 - 2\,z_5\,z_8 -
z_1\,z_5\,z_8 - z_2\,z_5\,z_8 - 2\,z_6\,z_8 +
   z_2\,z_6\,z_8 - z_3\,z_6\,z_8 + z_5\,z_6\,z_8 + 2\,z_7\,z_8 -
2\,z_1\,z_7\,z_8 - 3\,z_2\,z_7\,z_8 - z_1\,z_2\,z_7\,z_8 +
   2\,z_3\,z_7\,z_8 - z_4\,z_7\,z_8 - z_5\,z_7\,z_8 - 2\,z_6\,z_7\,z_8 -
z_1\,z_6\,z_7\,z_8 + 3\,z_7^2\,z_8 + z_7^3\,z_8 +
   z_8^2 + 2\,z_1\,z_8^2 + 2\,z_1\,z_2\,z_8^2 + z_3\,z_8^2 - z_1\,z_3\,z_8^2
+ z_1\,z_5\,z_8^2 + 4\,z_7\,z_8^2 +
   z_1\,z_7\,z_8^2 + z_3\,z_7\,z_8^2 - z_6\,z_7\,z_8^2 + 3\,z_7^2\,z_8^2 +
z_8^3 + z_2\,z_8^3 - z_3\,z_8^3 +
   z_5\,z_8^3 - z_7\,z_8^3 - z_8^4 - z_7\,z_8^4\\ $

$\bchi_{0, 0, 0, 1, 0, 0, 1, 1} =
  -(z_1^2\,z_2) - z_1\,z_3 - z_1\,z_4 - z_2\,z_4 + z_1\,z_5 - 2\,z_1\,z_6 -
z_1^2\,z_6 - z_1\,z_2\,z_6 - z_4\,z_6 +
   z_2\,z_7 + z_1\,z_2\,z_7 - z_1\,z_3\,z_7 + z_4\,z_7 + z_6\,z_7 +
z_1\,z_6\,z_7 + z_2\,z_7^2 + 4\,z_1\,z_8 +
   2\,z_1^2\,z_8 + 2\,z_1^3\,z_8 + z_2\,z_8 - z_1^2\,z_2\,z_8 - z_2^2\,z_8 +
z_1\,z_2^2\,z_8 - 2\,z_1\,z_3\,z_8 +
   z_2\,z_3\,z_8 + z_3^2\,z_8 + z_4\,z_8 - z_1\,z_4\,z_8 + 2\,z_1\,z_5\,z_8
+ z_6\,z_8 - 5\,z_1\,z_6\,z_8 -
   z_1^2\,z_6\,z_8 - z_2\,z_6\,z_8 + 2\,z_3\,z_6\,z_8 + z_7\,z_8 +
6\,z_1\,z_7\,z_8 + 2\,z_1^2\,z_7\,z_8 +
   2\,z_2\,z_7\,z_8 - z_1\,z_2\,z_7\,z_8 - z_3\,z_7\,z_8 + z_4\,z_7\,z_8 +
z_5\,z_7\,z_8 - 2\,z_6\,z_7\,z_8 + z_7^2\,z_8 -
   z_1\,z_7^2\,z_8 + 5\,z_1\,z_8^2 - 2\,z_1^2\,z_8^2 - z_1^3\,z_8^2 +
z_1\,z_2\,z_8^2 - z_3\,z_8^2 +
   2\,z_1\,z_3\,z_8^2 - z_2\,z_3\,z_8^2 - z_6\,z_8^2 + 2\,z_1\,z_6\,z_8^2 +
3\,z_7\,z_8^2 + z_1^2\,z_7\,z_8^2 -
   z_3\,z_7\,z_8^2 + z_7^2\,z_8^2 - 8\,z_1\,z_8^3 - z_2\,z_8^3 +
z_3\,z_8^3 - z_5\,z_8^3 + 2\,z_6\,z_8^3 -
   3\,z_7\,z_8^3 - z_1\,z_7\,z_8^3 - 3\,z_8^4 + 2\,z_1\,z_8^4 - z_7\,z_8^4 +
2\,z_8^5\\ $

$\bchi_{0, 0, 1, 0, 0, 0, 1, 2} =
  z_1 + z_1^2 + z_2 + 2\,z_1\,z_2 + z_1^2\,z_2 - z_1\,z_2^2 + z_3 +
z_1\,z_3 - z_2\,z_3 + z_2\,z_5 + z_3\,z_5 +
   z_6 + z_1\,z_6 - z_2\,z_6 + z_7 + 3\,z_1\,z_7 + z_1^2\,z_7 - z_4\,z_7 -
z_1\,z_6\,z_7 + 2\,z_7^2 +
   2\,z_1\,z_7^2 - z_2\,z_7^2 - z_3\,z_7^2 + z_7^3 + z_1\,z_8 -
2\,z_1^2\,z_8 - z_1^3\,z_8 + z_2\,z_8 -
   3\,z_1\,z_2\,z_8 + z_1^2\,z_2\,z_8 + z_3\,z_8 + z_1\,z_3\,z_8 -
z_2\,z_3\,z_8 - z_4\,z_8 + z_5\,z_8 - z_1\,z_6\,z_8 -
   2\,z_2\,z_6\,z_8 - z_3\,z_6\,z_8 + 2\,z_7\,z_8 - 2\,z_1^2\,z_7\,z_8 -
z_2\,z_7\,z_8 + z_1\,z_2\,z_7\,z_8 + z_3\,z_7\,z_8 +
   z_6\,z_7\,z_8 + 2\,z_7^2\,z_8 + z_8^2 - 3\,z_1\,z_8^2 + z_1^2\,z_8^2 -
2\,z_2\,z_8^2 + z_2^2\,z_8^2 +
   z_5\,z_8^2 - 2\,z_6\,z_8^2 + z_1\,z_6\,z_8^2 - 2\,z_7\,z_8^2 +
z_2\,z_7\,z_8^2 + z_3\,z_7\,z_8^2 -
   2\,z_7^2\,z_8^2 + 2\,z_1\,z_8^3 + z_1^2\,z_8^3 + z_2\,z_8^3 -
z_1\,z_2\,z_8^3 - z_3\,z_8^3 - z_6\,z_8^3 +
   z_7\,z_8^3 + z_8^4 - z_1\,z_8^4 + z_7\,z_8^4 - z_8^5\\ $

$\bchi_{0, 0, 1, 0, 0, 1, 1, 0} =
  z_1^2 + 2\,z_1^3 + z_1^4 + z_1^2\,z_2 + z_2^2 + z_1\,z_2^2 - z_3 -
z_1\,z_3 - 2\,z_1^2\,z_3 + z_2\,z_3 +
   z_3^2 - z_4 + z_1^2\,z_5 + z_2\,z_5 + z_1\,z_2\,z_5 - z_3\,z_5 - z_6 -
2\,z_1\,z_6 - 2\,z_1^2\,z_6 -
   z_2^2\,z_6 + 2\,z_3\,z_6 + z_4\,z_6 - z_5\,z_6 + z_6^2 + z_1\,z_7 +
2\,z_1^2\,z_7 + z_1^3\,z_7 - z_3\,z_7 -
   z_1\,z_3\,z_7 - z_6\,z_7 - 2\,z_1\,z_6\,z_7 - z_2\,z_6\,z_7 +
z_3\,z_6\,z_7 + z_1\,z_7^2 - z_1\,z_2\,z_7^2 -
   z_6\,z_7^2 + z_8 + 2\,z_1\,z_8 + 3\,z_1^2\,z_8 - 2\,z_1^3\,z_8 + z_2\,z_8
+ 2\,z_1\,z_2\,z_8 + z_2^2\,z_8 -
   z_1\,z_2^2\,z_8 - 3\,z_3\,z_8 - z_1^2\,z_3\,z_8 - z_2\,z_3\,z_8 +
z_3^2\,z_8 - 2\,z_4\,z_8 + z_1\,z_4\,z_8 +
   2\,z_5\,z_8 - z_1\,z_5\,z_8 + z_2\,z_5\,z_8 - z_3\,z_5\,z_8 - 4\,z_6\,z_8
+ z_1\,z_6\,z_8 - 2\,z_2\,z_6\,z_8 +
   2\,z_3\,z_6\,z_8 + z_6^2\,z_8 + 2\,z_7\,z_8 + 4\,z_1\,z_7\,z_8 +
2\,z_2\,z_7\,z_8 + z_2^2\,z_7\,z_8 - 4\,z_3\,z_7\,z_8 -
   4\,z_6\,z_7\,z_8 + z_1\,z_6\,z_7\,z_8 + z_7^2\,z_8 + z_2\,z_7^2\,z_8 +
3\,z_8^2 - 2\,z_1\,z_8^2 - 4\,z_1^2\,z_8^2 -
   z_1^3\,z_8^2 + 2\,z_2\,z_8^2 - 2\,z_1\,z_2\,z_8^2 + z_1^2\,z_2\,z_8^2 -
z_3\,z_8^2 + 3\,z_1\,z_3\,z_8^2 -
   z_2\,z_3\,z_8^2 + z_5\,z_8^2 - z_1\,z_5\,z_8^2 - 2\,z_6\,z_8^2 +
3\,z_1\,z_6\,z_8^2 - z_2\,z_6\,z_8^2 +
   5\,z_7\,z_8^2 - 6\,z_1\,z_7\,z_8^2 + 3\,z_2\,z_7\,z_8^2 -
z_3\,z_7\,z_8^2 - 4\,z_1\,z_8^3 + 2\,z_1^2\,z_8^3 -
   z_1\,z_2\,z_8^3 + 2\,z_3\,z_8^3 + z_6\,z_8^3 - z_1\,z_7\,z_8^3 - 2\,z_8^4
+ 3\,z_1\,z_8^4 - z_2\,z_8^4\\ $

$\bchi_{0, 0, 1, 0, 1, 0, 0, 1} =
  -z_1^2 - z_1^3 - z_2 - 2\,z_1\,z_2 - z_2^2 + z_1\,z_3 + z_1\,z_4 -
z_3\,z_4 - z_5 - 2\,z_1\,z_5 - z_2\,z_5 +
   z_2\,z_6 + z_1\,z_2\,z_6 + z_2^2\,z_6 - z_4\,z_6 - 2\,z_1^2\,z_7 -
z_1^3\,z_7 - z_2\,z_7 - 2\,z_1\,z_2\,z_7 +
   z_1^2\,z_2\,z_7 - z_2^2\,z_7 + z_1\,z_3\,z_7 - z_2\,z_3\,z_7 -
z_1\,z_5\,z_7 - z_2\,z_6\,z_7 - z_1^2\,z_7^2 - z_8 -
   z_1\,z_8 + 3\,z_1^2\,z_8 + z_1^3\,z_8 - z_2\,z_8 + 2\,z_1\,z_2\,z_8 +
z_1^2\,z_2\,z_8 - z_3\,z_8 - z_1\,z_3\,z_8 -
   2\,z_2\,z_3\,z_8 - z_1\,z_4\,z_8 + z_5\,z_8 + z_1\,z_5\,z_8 +
z_2\,z_5\,z_8 + z_3\,z_5\,z_8 + z_6\,z_8 - z_1\,z_6\,z_8 -
   z_1^2\,z_6\,z_8 - z_2\,z_6\,z_8 - z_1\,z_2\,z_6\,z_8 + z_3\,z_6\,z_8 -
z_7\,z_8 + z_1\,z_7\,z_8 + 3\,z_1^2\,z_7\,z_8 -
   3\,z_3\,z_7\,z_8 + z_4\,z_7\,z_8 + z_5\,z_7\,z_8 - z_6\,z_7\,z_8 +
z_1\,z_6\,z_7\,z_8 - z_7^2\,z_8 + z_2\,z_7^2\,z_8 -
   z_7^3\,z_8 + 3\,z_1\,z_8^2 - z_1^2\,z_8^2 - z_1^3\,z_8^2 +
3\,z_2\,z_8^2 - z_1\,z_2\,z_8^2 + z_2^2\,z_8^2 -
   3\,z_3\,z_8^2 + 2\,z_1\,z_3\,z_8^2 + 2\,z_5\,z_8^2 - 2\,z_6\,z_8^2 +
2\,z_1\,z_6\,z_8^2 - z_2\,z_6\,z_8^2 -
   z_7\,z_8^2 - z_1\,z_7\,z_8^2 + z_1^2\,z_7\,z_8^2 + 3\,z_2\,z_7\,z_8^2 -
z_3\,z_7\,z_8^2 + z_6\,z_7\,z_8^2 -
   2\,z_7^2\,z_8^2 + 2\,z_8^3 - 4\,z_1\,z_8^3 + z_1^2\,z_8^3 + z_2\,z_8^3 -
z_1\,z_2\,z_8^3 + z_3\,z_8^3 -
   z_5\,z_8^3 - z_1\,z_7\,z_8^3 - z_8^4 + z_1\,z_8^4 - z_2\,z_8^4 +
z_7\,z_8^4\\ $

$\bchi_{0, 0, 2, 0, 0, 0, 1, 0} =
  z_1^2 + z_1^3 + z_1\,z_2 - 2\,z_1^2\,z_2 - z_1^3\,z_2 + z_2^2 + z_1\,z_2^2
+ z_1^2\,z_2^2 + z_3 - z_1\,z_3 +
   2\,z_2\,z_3 + 2\,z_1\,z_2\,z_3 - z_2^2\,z_3 + z_3^2 - 2\,z_1\,z_4 -
z_1^2\,z_4 + 2\,z_3\,z_4 + 2\,z_1\,z_5 -
   z_1\,z_2\,z_5 - 2\,z_1\,z_6 - 3\,z_1^2\,z_6 - z_1^3\,z_6 + 2\,z_2\,z_6 +
z_1\,z_2\,z_6 - z_2^2\,z_6 + 4\,z_3\,z_6 +
   2\,z_1\,z_3\,z_6 + z_4\,z_6 - z_5\,z_6 + 3\,z_6^2 + z_1\,z_6^2 + z_7 +
2\,z_1\,z_7 + 3\,z_1^2\,z_7 + z_1^3\,z_7 +
   z_1\,z_2\,z_7 - z_1^2\,z_2\,z_7 + z_2^2\,z_7 - z_1\,z_3\,z_7 +
z_2\,z_3\,z_7 + z_3^2\,z_7 - z_4\,z_7 - z_1\,z_4\,z_7 +
   z_5\,z_7 + z_1\,z_5\,z_7 - z_6\,z_7 - z_1\,z_6\,z_7 - z_1^2\,z_6\,z_7 +
z_3\,z_6\,z_7 + 2\,z_7^2 + 3\,z_1\,z_7^2 +
   2\,z_1^2\,z_7^2 - z_3\,z_7^2 + z_5\,z_7^2 + z_7^3 + z_1\,z_7^3 +
4\,z_1\,z_8 + z_1^2\,z_8 - z_2\,z_8 -
   2\,z_1\,z_2\,z_8 - z_1\,z_2^2\,z_8 - 4\,z_3\,z_8 - z_1\,z_2\,z_3\,z_8 -
z_4\,z_8 + 2\,z_1\,z_4\,z_8 + z_2\,z_4\,z_8 +
   2\,z_5\,z_8 + z_1^2\,z_5\,z_8 - 2\,z_2\,z_5\,z_8 - z_3\,z_5\,z_8 -
5\,z_6\,z_8 + z_1^2\,z_6\,z_8 - z_2\,z_6\,z_8 +
   z_1\,z_2\,z_6\,z_8 - z_5\,z_6\,z_8 + z_6^2\,z_8 + 3\,z_7\,z_8 +
4\,z_1\,z_7\,z_8 + z_1^2\,z_7\,z_8 - 3\,z_2\,z_7\,z_8 -
   2\,z_1\,z_2\,z_7\,z_8 - 4\,z_3\,z_7\,z_8 + z_5\,z_7\,z_8 -
5\,z_6\,z_7\,z_8 - z_1\,z_6\,z_7\,z_8 + 3\,z_7^2\,z_8 -
   z_2\,z_7^2\,z_8 + 2\,z_8^2 - 5\,z_1\,z_8^2 - z_2\,z_8^2 + z_1\,z_2\,z_8^2
+ z_1^2\,z_2\,z_8^2 - z_3\,z_8^2 -
   z_2\,z_3\,z_8^2 + z_4\,z_8^2 - z_1\,z_5\,z_8^2 - z_6\,z_8^2 +
2\,z_1\,z_6\,z_8^2 + z_2\,z_6\,z_8^2 + z_7\,z_8^2 -
   5\,z_1\,z_7\,z_8^2 - z_1^2\,z_7\,z_8^2 - z_2\,z_7\,z_8^2 +
z_3\,z_7\,z_8^2 - z_7^2\,z_8^2 - 2\,z_8^3 -
   2\,z_1\,z_8^3 - z_1^2\,z_8^3 + 2\,z_2\,z_8^3 + 2\,z_3\,z_8^3 - z_5\,z_8^3
+ 3\,z_6\,z_8^3 - 3\,z_7\,z_8^3 -
   z_8^4 + 2\,z_1\,z_8^4 + z_8^5\\ $

$\bchi_{0, 1, 0, 0, 0, 0, 0, 4} =
  -z_1 - z_1^2 - z_2 - z_1\,z_2 - z_2^2 - z_1\,z_3 + z_4 - z_5 - z_1\,z_5 -
z_2\,z_5 + z_6 - z_3\,z_6 -
   z_6^2 - z_7 - 2\,z_1\,z_7 - 2\,z_1^2\,z_7 + z_2\,z_7 + z_3\,z_7 +
z_6\,z_7 - z_7^2 - z_1\,z_7^2 + z_2\,z_7^2 -
   2\,z_8 - z_1\,z_8 - z_1^2\,z_8 - z_2\,z_8 + z_1\,z_2\,z_8 + z_3\,z_8 +
z_4\,z_8 - 2\,z_5\,z_8 + 3\,z_6\,z_8 +
   2\,z_2\,z_6\,z_8 - 3\,z_7\,z_8 - 2\,z_1\,z_7\,z_8 + 2\,z_3\,z_7\,z_8 +
2\,z_6\,z_7\,z_8 - z_7^2\,z_8 - 2\,z_8^2 +
   2\,z_1\,z_8^2 + z_1^2\,z_8^2 + z_3\,z_8^2 + 2\,z_6\,z_8^2 - 2\,z_7\,z_8^2
+ 3\,z_1\,z_7\,z_8^2 -
   3\,z_2\,z_7\,z_8^2 + z_8^3 + z_1\,z_8^3 - z_2\,z_8^3 - z_3\,z_8^3 -
z_6\,z_8^3 + z_8^4 - z_1\,z_8^4 +
   z_2\,z_8^4\\ $

$\bchi_{0, 1, 0, 0, 0, 0, 2, 1} =
  z_1 + z_1^2 + z_2 + z_1\,z_2 + z_1\,z_2^2 + z_2\,z_3 - z_1\,z_4 -
z_2\,z_5 - z_3\,z_5 + z_6 - z_1\,z_6 -
   2\,z_1^2\,z_6 + 2\,z_3\,z_6 - 2\,z_5\,z_6 + z_6^2 + z_7 + 3\,z_1\,z_7 +
z_1^2\,z_7 + 2\,z_2\,z_7 + z_1\,z_2\,z_7 -
   z_4\,z_7 - 2\,z_1\,z_6\,z_7 - z_2\,z_6\,z_7 + 2\,z_7^2 + 2\,z_1\,z_7^2 +
z_2\,z_7^2 + z_7^3 + 2\,z_1\,z_8 -
   z_1^2\,z_2\,z_8 - 2\,z_3\,z_8 - z_1\,z_3\,z_8 - z_4\,z_8 + z_5\,z_8 +
z_2\,z_5\,z_8 - 3\,z_6\,z_8 - 2\,z_1\,z_6\,z_8 -
   z_2\,z_6\,z_8 + 2\,z_3\,z_6\,z_8 + 2\,z_6^2\,z_8 + 2\,z_7\,z_8 +
2\,z_1\,z_7\,z_8 + z_2\,z_7\,z_8 - 3\,z_3\,z_7\,z_8 +
   z_5\,z_7\,z_8 - 2\,z_6\,z_7\,z_8 + 2\,z_7^2\,z_8 + z_2\,z_7^2\,z_8 +
2\,z_8^2 - 2\,z_1\,z_8^2 - z_1^2\,z_8^2 -
   z_3\,z_8^2 + z_1\,z_3\,z_8^2 - 3\,z_6\,z_8^2 + 2\,z_1\,z_6\,z_8^2 -
z_2\,z_6\,z_8^2 + z_7\,z_8^2 -
   2\,z_1\,z_7\,z_8^2 + z_2\,z_7\,z_8^2 - z_3\,z_7\,z_8^2 -
z_6\,z_7\,z_8^2 - z_7^2\,z_8^2 - z_1\,z_8^3 +
   z_2\,z_8^3 + 2\,z_3\,z_8^3 + 2\,z_6\,z_8^3 - z_8^4 + z_1\,z_8^4 -
z_2\,z_8^4\\ $

$\bchi_{0, 1, 0, 0, 0, 1, 0, 2} =
  z_2 + 2\,z_1\,z_2 + z_3 - z_1\,z_3 - z_1^2\,z_3 + z_3^2 + z_4 + z_1\,z_4 +
z_2\,z_4 + z_1\,z_5 + 2\,z_6 +
   z_1\,z_6 + z_1^2\,z_6 - z_2\,z_6 + z_3\,z_6 - z_6^2 - z_7 - 2\,z_1\,z_7 -
z_1^2\,z_7 + z_2\,z_7 + z_1\,z_2\,z_7 +
   2\,z_3\,z_7 + z_4\,z_7 + 3\,z_6\,z_7 + z_1\,z_6\,z_7 - z_2\,z_6\,z_7 -
2\,z_7^2 - 2\,z_1\,z_7^2 + z_3\,z_7^2 +
   z_6\,z_7^2 - z_7^3 - 2\,z_8 - z_1\,z_8 - 3\,z_1^2\,z_8 + z_2^2\,z_8 -
z_2\,z_3\,z_8 + z_4\,z_8 - z_5\,z_8 -
   z_1\,z_5\,z_8 - z_2\,z_5\,z_8 + 3\,z_6\,z_8 + z_1\,z_6\,z_8 +
z_2\,z_6\,z_8 - 6\,z_7\,z_8 - 3\,z_1\,z_7\,z_8 -
   z_1^2\,z_7\,z_8 + z_3\,z_7\,z_8 - z_5\,z_7\,z_8 + 4\,z_6\,z_7\,z_8 -
4\,z_7^2\,z_8 + z_1\,z_7^2\,z_8 - 2\,z_8^2 +
   2\,z_1^2\,z_8^2 - 2\,z_2\,z_8^2 - z_1\,z_2\,z_8^2 - z_3\,z_8^2 +
z_1\,z_3\,z_8^2 - z_4\,z_8^2 - z_5\,z_8^2 -
   z_6\,z_8^2 - z_1\,z_6\,z_8^2 + z_2\,z_6\,z_8^2 - 2\,z_7\,z_8^2 +
3\,z_1\,z_7\,z_8^2 - 3\,z_2\,z_7\,z_8^2 -
   z_3\,z_7\,z_8^2 - z_6\,z_7\,z_8^2 + 3\,z_8^3 + 2\,z_1\,z_8^3 +
z_5\,z_8^3 - z_6\,z_8^3 + 4\,z_7\,z_8^3 +
   z_8^4 - z_1\,z_8^4 + z_2\,z_8^4 - z_8^5\\ $

$\bchi_{0, 1, 0, 0, 0, 2, 0, 0} =
  -z_1^3 - z_1^4 - z_1^2\,z_2 + z_1\,z_3 + 2\,z_1^2\,z_3 + z_1\,z_2\,z_3 -
z_1\,z_4 - z_2\,z_4 - z_1\,z_5 -
   2\,z_1^2\,z_5 - z_2\,z_5 + z_3\,z_5 - z_5^2 - z_1\,z_6 - z_1\,z_3\,z_6 +
z_4\,z_6 + z_6^2 + z_1\,z_6^2 +
   z_2\,z_6^2 + z_1\,z_7 - 2\,z_1^3\,z_7 + z_2\,z_7 + z_1^2\,z_2\,z_7 +
3\,z_1\,z_3\,z_7 - z_2\,z_3\,z_7 - z_4\,z_7 +
   z_5\,z_7 - 3\,z_1\,z_5\,z_7 - z_2\,z_5\,z_7 + 2\,z_1\,z_6\,z_7 -
z_3\,z_6\,z_7 - z_6^2\,z_7 + 2\,z_1\,z_7^2 +
   z_2\,z_7^2 + z_6\,z_7^2 + z_1\,z_7^3 + z_1\,z_8 - 2\,z_1^2\,z_8 -
z_1^3\,z_8 + z_2\,z_8 - z_1\,z_2\,z_8 +
   z_1^2\,z_2\,z_8 - z_2^2\,z_8 + z_3\,z_8 + 4\,z_1\,z_3\,z_8 - z_3^2\,z_8 +
z_1\,z_4\,z_8 - z_1\,z_5\,z_8 -
   z_2\,z_5\,z_8 + z_3\,z_5\,z_8 + 4\,z_1\,z_6\,z_8 + 3\,z_1^2\,z_6\,z_8 -
4\,z_3\,z_6\,z_8 + 2\,z_5\,z_6\,z_8 -
   2\,z_6^2\,z_8 - z_1\,z_7\,z_8 - z_1\,z_2\,z_7\,z_8 + 2\,z_3\,z_7\,z_8 +
z_1\,z_3\,z_7\,z_8 - z_4\,z_7\,z_8 -
   z_5\,z_7\,z_8 - 2\,z_1\,z_6\,z_7\,z_8 - z_2\,z_7^2\,z_8 - z_8^2 -
3\,z_1\,z_8^2 + 3\,z_1^2\,z_8^2 +
   2\,z_1^3\,z_8^2 - z_2\,z_8^2 - z_1^2\,z_2\,z_8^2 + 2\,z_3\,z_8^2 -
4\,z_1\,z_3\,z_8^2 + z_2\,z_3\,z_8^2 -
   z_5\,z_8^2 + 2\,z_1\,z_5\,z_8^2 + 2\,z_6\,z_8^2 - 4\,z_1\,z_6\,z_8^2 +
2\,z_2\,z_6\,z_8^2 - z_7\,z_8^2 +
   z_1\,z_7\,z_8^2 - z_1^2\,z_7\,z_8^2 - 2\,z_2\,z_7\,z_8^2 +
z_3\,z_7\,z_8^2 + z_7^2\,z_8^2 + 4\,z_1\,z_8^3 -
   z_1^2\,z_8^3 - z_2\,z_8^3 + z_1\,z_2\,z_8^3 - 2\,z_3\,z_8^3 +
z_5\,z_8^3 - 2\,z_6\,z_8^3 + z_7\,z_8^3 +
   z_1\,z_7\,z_8^3 + 2\,z_8^4 - 2\,z_1\,z_8^4 + z_2\,z_8^4 - z_8^5\\ $

$\bchi_{0, 1, 0, 0, 1, 0, 1, 0} =
  z_1^2 + z_1^3 - z_1\,z_2 - z_1^2\,z_2 - z_2^2 + z_1^2\,z_3 + z_2\,z_3 -
z_3^2 - z_1\,z_4 + z_3\,z_4 -
   z_1\,z_5 - z_1^2\,z_5 - z_2\,z_5 + z_3\,z_5 - 3\,z_1\,z_6 -
2\,z_1^2\,z_6 - z_2\,z_6 - z_1\,z_2\,z_6 + z_4\,z_6 +
   5\,z_1\,z_7 + 4\,z_1^2\,z_7 + z_2\,z_7 - z_1\,z_2\,z_7 - z_1^2\,z_2\,z_7
+ z_2^2\,z_7 - 2\,z_3\,z_7 + z_1\,z_3\,z_7 +
   z_2\,z_3\,z_7 - 2\,z_4\,z_7 + z_2\,z_5\,z_7 - 4\,z_6\,z_7 -
3\,z_1\,z_6\,z_7 - z_3\,z_6\,z_7 - z_6^2\,z_7 + 4\,z_7^2 +
   5\,z_1\,z_7^2 + z_1^2\,z_7^2 + z_2\,z_7^2 - z_3\,z_7^2 + z_5\,z_7^2 +
2\,z_7^3 + z_1\,z_7^3 + 4\,z_1\,z_8 -
   2\,z_1^3\,z_8 + z_2\,z_8 + z_1\,z_2\,z_8 + z_2^2\,z_8 - z_3\,z_8 +
3\,z_1\,z_3\,z_8 + 2\,z_2\,z_3\,z_8 - z_3^2\,z_8 -
   2\,z_4\,z_8 + 2\,z_1\,z_4\,z_8 - z_2\,z_4\,z_8 + z_5\,z_8 + z_1\,z_5\,z_8
+ z_2\,z_5\,z_8 - 3\,z_6\,z_8 +
   2\,z_1\,z_6\,z_8 + 3\,z_1^2\,z_6\,z_8 + z_2\,z_6\,z_8 -
2\,z_3\,z_6\,z_8 - z_6^2\,z_8 + 6\,z_7\,z_8 + 5\,z_1\,z_7\,z_8 -
   z_1^2\,z_7\,z_8 + 3\,z_2\,z_7\,z_8 + z_1\,z_3\,z_7\,z_8 - z_4\,z_7\,z_8 +
3\,z_5\,z_7\,z_8 + z_6\,z_7\,z_8 -
   z_1\,z_6\,z_7\,z_8 + 6\,z_7^2\,z_8 + 2\,z_1\,z_7^2\,z_8 + 3\,z_8^2 -
5\,z_1\,z_8^2 - 2\,z_1^2\,z_8^2 +
   z_1^3\,z_8^2 + z_2\,z_8^2 - z_1\,z_2\,z_8^2 - z_2^2\,z_8^2 +
2\,z_3\,z_8^2 - 3\,z_1\,z_3\,z_8^2 + z_4\,z_8^2 +
   2\,z_5\,z_8^2 - z_1\,z_5\,z_8^2 + 2\,z_6\,z_8^2 + z_7\,z_8^2 -
6\,z_1\,z_7\,z_8^2 - 2\,z_1^2\,z_7\,z_8^2 -
   z_2\,z_7\,z_8^2 + z_3\,z_7\,z_8^2 - 3\,z_7^2\,z_8^2 - 2\,z_8^3 +
z_1^2\,z_8^3 - z_2\,z_8^3 + z_1\,z_2\,z_8^3 -
   z_3\,z_8^3 - z_5\,z_8^3 - z_6\,z_8^3 - 4\,z_7\,z_8^3 + z_1\,z_8^4 +
z_7\,z_8^4\\ $

$\bchi_{0, 1, 0, 1, 0, 0, 0, 1} =
  -z_1^2 - z_1^3 + z_1^2\,z_2 + z_1^3\,z_2 - z_2\,z_3 - 2\,z_1\,z_2\,z_3 -
z_2^2\,z_3 + z_1\,z_4 + z_2\,z_4 +
   z_1\,z_5 + z_1^2\,z_5 + z_2\,z_5 + z_1\,z_2\,z_5 + 2\,z_1\,z_6 +
z_1^2\,z_6 - z_2\,z_6 - 2\,z_1\,z_2\,z_6 -
   z_3\,z_6 + z_1\,z_3\,z_6 - z_4\,z_6 + z_5\,z_6 - z_6^2 - z_1\,z_6^2 -
2\,z_1\,z_7 - 2\,z_1^2\,z_7 - z_1^3\,z_7 -
   2\,z_2\,z_7 + z_1^2\,z_2\,z_7 + z_2^2\,z_7 - z_3\,z_7 + z_1\,z_3\,z_7 -
z_2\,z_3\,z_7 + z_5\,z_7 + 2\,z_1\,z_6\,z_7 -
   z_7^2 - 2\,z_1\,z_7^2 - z_1^2\,z_7^2 - z_2\,z_7^2 + z_6\,z_7^2 - z_7^3 -
2\,z_1\,z_8 + 3\,z_1\,z_2\,z_8 +
   2\,z_2^2\,z_8 + z_1\,z_3\,z_8 - z_2\,z_3\,z_8 + z_2\,z_4\,z_8 -
z_1\,z_5\,z_8 - z_2\,z_5\,z_8 - z_3\,z_5\,z_8 +
   z_6\,z_8 + 2\,z_1\,z_6\,z_8 - 2\,z_1\,z_7\,z_8 + z_1^2\,z_7\,z_8 +
z_2\,z_7\,z_8 + z_3\,z_7\,z_8 - z_5\,z_7\,z_8 +
   2\,z_6\,z_7\,z_8 + z_1\,z_6\,z_7\,z_8 + z_1\,z_7^2\,z_8 - z_2\,z_7^2\,z_8
+ z_1^2\,z_8^2 - 4\,z_1\,z_2\,z_8^2 -
   z_4\,z_8^2 - z_1\,z_5\,z_8^2 - 2\,z_6\,z_8^2 - z_1\,z_6\,z_8^2 +
z_2\,z_6\,z_8^2 + z_7\,z_8^2 + z_1\,z_7\,z_8^2 -
   2\,z_2\,z_7\,z_8^2 - z_6\,z_7\,z_8^2 + z_7^2\,z_8^2 + 2\,z_8^3 +
z_1\,z_8^3 - z_2\,z_8^3 + z_5\,z_8^3 -
   z_6\,z_8^3 + 2\,z_7\,z_8^3 + z_2\,z_8^4 - z_8^5\\ $

$\bchi_{0, 1, 1, 0, 0, 0, 0, 2} =
  2\,z_1^2 - z_1^4 + 2\,z_1\,z_2 + z_2^3 + 3\,z_1^2\,z_3 - z_2\,z_3 +
z_1\,z_2\,z_3 - z_3^2 - 2\,z_1\,z_4 -
   2\,z_2\,z_4 + 2\,z_1\,z_5 + 2\,z_2\,z_5 - z_1\,z_6 - 2\,z_2\,z_6 -
2\,z_1\,z_2\,z_6 - 2\,z_3\,z_6 + z_5\,z_6 -
   z_6^2 + 3\,z_1\,z_7 + 3\,z_1^2\,z_7 + 2\,z_2\,z_7 + 2\,z_1\,z_2\,z_7 -
z_2\,z_3\,z_7 - z_4\,z_7 + z_5\,z_7 +
   z_1\,z_5\,z_7 - z_6\,z_7 - z_1\,z_6\,z_7 + z_7^2 + 3\,z_1\,z_7^2 +
z_1^2\,z_7^2 + z_2\,z_7^2 - z_3\,z_7^2 -
   z_6\,z_7^2 + z_7^3 + 4\,z_1\,z_8 - 2\,z_1^2\,z_8 + 3\,z_2\,z_8 -
2\,z_1\,z_2\,z_8 - z_1^2\,z_2\,z_8 -
   z_1\,z_2^2\,z_8 + z_3\,z_8 + z_1\,z_3\,z_8 + z_2\,z_3\,z_8 - z_3^2\,z_8 -
z_4\,z_8 + z_1\,z_4\,z_8 + z_1\,z_5\,z_8 +
   z_2\,z_5\,z_8 - 2\,z_1\,z_6\,z_8 + 2\,z_1^2\,z_6\,z_8 -
2\,z_3\,z_6\,z_8 - z_6^2\,z_8 + 3\,z_7\,z_8 + 2\,z_1\,z_7\,z_8 -
   2\,z_1^2\,z_7\,z_8 + 2\,z_2\,z_7\,z_8 + z_1\,z_2\,z_7\,z_8 +
z_3\,z_7\,z_8 - z_6\,z_7\,z_8 + 3\,z_7^2\,z_8 -
   2\,z_1\,z_7^2\,z_8 + 2\,z_8^2 - z_1\,z_8^2 + z_1^3\,z_8^2 - z_2\,z_8^2 +
z_1\,z_2\,z_8^2 + z_3\,z_8^2 -
   2\,z_1\,z_3\,z_8^2 + z_2\,z_3\,z_8^2 + z_5\,z_8^2 - z_1\,z_5\,z_8^2 -
z_1\,z_6\,z_8^2 + 2\,z_7\,z_8^2 -
   z_1\,z_7\,z_8^2 - z_1^2\,z_7\,z_8^2 + z_3\,z_7\,z_8^2 + z_6\,z_7\,z_8^2 -
z_1\,z_8^3 - z_2\,z_8^3 - z_3\,z_8^3 -
   z_6\,z_8^3 - z_7\,z_8^3 + z_1\,z_7\,z_8^3 - z_8^4\\ $

$\bchi_{0, 1, 1, 0, 0, 1, 0, 0} =
  -z_1 - z_1^2 + z_1^3 + z_1^4 - z_2 + z_1^2\,z_2 + z_3 - z_1\,z_3 -
2\,z_1^2\,z_3 - z_2\,z_3 - z_1\,z_2\,z_3 +
   z_4 + z_1\,z_4 - z_3\,z_4 - z_5 + z_1^2\,z_5 + z_6 + z_1\,z_6 -
z_1^2\,z_6 - z_1^3\,z_6 + z_2\,z_6 -
   z_1\,z_2\,z_6 - z_3\,z_6 + 2\,z_1\,z_3\,z_6 + z_2\,z_3\,z_6 - 2\,z_4\,z_6
+ z_5\,z_6 - z_1\,z_5\,z_6 - z_6^2 +
   z_1\,z_6^2 - 2\,z_1\,z_7 - z_1^2\,z_7 + z_1^3\,z_7 + 3\,z_1\,z_2\,z_7 +
z_1^2\,z_2\,z_7 - z_1\,z_2^2\,z_7 +
   2\,z_3\,z_7 - 2\,z_1\,z_3\,z_7 - 2\,z_2\,z_3\,z_7 - z_3^2\,z_7 + z_4\,z_7
+ z_1\,z_4\,z_7 - z_5\,z_7 + z_1\,z_5\,z_7 +
   z_2\,z_5\,z_7 + 2\,z_6\,z_7 + z_1^2\,z_6\,z_7 - 2\,z_2\,z_6\,z_7 -
z_3\,z_6\,z_7 - 2\,z_1\,z_7^2 - z_1^2\,z_7^2 +
   z_2\,z_7^2 + z_1\,z_2\,z_7^2 + z_3\,z_7^2 + z_6\,z_7^2 - z_1\,z_7^3 - z_8
+ 2\,z_1^2\,z_8 - z_1^4\,z_8 +
   3\,z_1\,z_2\,z_8 - z_1^2\,z_2\,z_8 + 2\,z_3\,z_8 - 4\,z_1\,z_3\,z_8 +
2\,z_1^2\,z_3\,z_8 - 2\,z_2\,z_3\,z_8 +
   z_1\,z_2\,z_3\,z_8 + 2\,z_4\,z_8 - 2\,z_1\,z_4\,z_8 - z_5\,z_8 +
2\,z_1\,z_5\,z_8 - z_1^2\,z_5\,z_8 + z_3\,z_5\,z_8 +
   4\,z_6\,z_8 - 5\,z_1\,z_6\,z_8 + z_1^2\,z_6\,z_8 - 2\,z_2\,z_6\,z_8 +
z_3\,z_6\,z_8 - 3\,z_7\,z_8 - z_1\,z_7\,z_8 -
   z_1^2\,z_7\,z_8 + z_1^3\,z_7\,z_8 + 2\,z_2\,z_7\,z_8 + z_1\,z_2\,z_7\,z_8
+ 3\,z_3\,z_7\,z_8 - 3\,z_1\,z_3\,z_7\,z_8 +
   z_4\,z_7\,z_8 - z_5\,z_7\,z_8 + 4\,z_6\,z_7\,z_8 - z_1\,z_6\,z_7\,z_8 -
2\,z_7^2\,z_8 - z_1\,z_7^2\,z_8 +
   z_2\,z_7^2\,z_8 - 2\,z_8^2 + 6\,z_1\,z_8^2 - 4\,z_1^2\,z_8^2 +
2\,z_2\,z_8^2 - 2\,z_1\,z_2\,z_8^2 -
   2\,z_3\,z_8^2 + 3\,z_1\,z_3\,z_8^2 + z_2\,z_3\,z_8^2 - z_4\,z_8^2 -
2\,z_6\,z_8^2 + 2\,z_1\,z_6\,z_8^2 -
   3\,z_7\,z_8^2 + 5\,z_1\,z_7\,z_8^2 - z_1^2\,z_7\,z_8^2 -
2\,z_2\,z_7\,z_8^2 - z_3\,z_7\,z_8^2 - z_7^2\,z_8^2 +
   2\,z_8^3 - 4\,z_1\,z_8^3 + 2\,z_1^2\,z_8^3 - z_2\,z_8^3 + 2\,z_7\,z_8^3\\ $

$\bchi_{0, 2, 0, 0, 0, 0, 1, 1} =
  -z_2 - z_1\,z_2 - z_1^2\,z_2 + z_2^2 - z_3 - z_1\,z_3 - z_1^2\,z_3 +
z_3^2 - z_4 - z_1\,z_4 + z_1\,z_5 +
   z_2\,z_5 - z_3\,z_5 - 2\,z_6 - 2\,z_1\,z_6 + z_2\,z_6 - z_2^2\,z_6 +
z_3\,z_6 + z_4\,z_6 + z_6^2 + z_1\,z_6^2 -
   z_2\,z_7 - z_1\,z_2\,z_7 + z_2^2\,z_7 - z_3\,z_7 - z_1\,z_3\,z_7 -
z_4\,z_7 + z_5\,z_7 - 2\,z_6\,z_7 + z_8 -
   z_1^2\,z_8 - z_1\,z_2\,z_8 + z_1^2\,z_2\,z_8 - z_3\,z_8 - z_2\,z_3\,z_8 +
z_3^2\,z_8 + z_5\,z_8 - z_1\,z_5\,z_8 -
   z_6\,z_8 - z_2\,z_6\,z_8 + z_3\,z_6\,z_8 + z_7\,z_8 - z_1\,z_7\,z_8 -
z_1^2\,z_7\,z_8 - z_2\,z_7\,z_8 +
   z_2^2\,z_7\,z_8 - z_4\,z_7\,z_8 - z_5\,z_7\,z_8 - 2\,z_6\,z_7\,z_8 -
z_1\,z_6\,z_7\,z_8 - z_1\,z_7^2\,z_8 - z_8^2 +
   z_1^2\,z_8^2 + 2\,z_2\,z_8^2 + 2\,z_1\,z_2\,z_8^2 + z_1\,z_3\,z_8^2 -
z_2\,z_3\,z_8^2 + z_4\,z_8^2 - z_5\,z_8^2 +
   z_1\,z_5\,z_8^2 + 2\,z_6\,z_8^2 - z_2\,z_6\,z_8^2 - z_7\,z_8^2 +
z_1\,z_7\,z_8^2 + 2\,z_2\,z_7\,z_8^2 +
   z_3\,z_7\,z_8^2 + z_6\,z_7\,z_8^2 + z_7^2\,z_8^2 - z_8^3 + z_1\,z_8^3 +
z_2\,z_8^3 - z_1\,z_2\,z_8^3 -
   z_7\,z_8^3 + z_1\,z_7\,z_8^3 + z_8^4 - z_1\,z_8^4 - z_2\,z_8^4\\ $

$\bchi_{0, 2, 1, 0, 0, 0, 0, 0} =
  -z_1^2 - z_1^3 - z_1\,z_2 - z_1^2\,z_2 - z_1^3\,z_2 - z_1\,z_2^2 +
z_1\,z_3 + 2\,z_1\,z_2\,z_3 + z_2^2\,z_3 -
   z_2\,z_4 - z_3\,z_4 + z_1^2\,z_5 + z_2\,z_5 - z_1\,z_2\,z_5 - z_3\,z_5 +
z_5^2 - z_2\,z_6 - z_3\,z_6 - z_4\,z_6 -
   z_6^2 - z_1^2\,z_7 - z_1\,z_2\,z_7 - z_1^2\,z_2\,z_7 + z_3\,z_7 +
z_2\,z_3\,z_7 - z_4\,z_7 + z_1\,z_5\,z_7 -
   z_1\,z_6\,z_7 + z_7^2 + z_1\,z_7^2 + z_7^3 + 2\,z_2\,z_8 - z_1\,z_2\,z_8
+ z_1^2\,z_2\,z_8 - z_2^2\,z_8 +
   2\,z_3\,z_8 + z_1\,z_3\,z_8 + z_2\,z_3\,z_8 + z_4\,z_8 - z_1\,z_4\,z_8 +
z_2\,z_5\,z_8 + 3\,z_6\,z_8 - z_2\,z_6\,z_8 -
   z_6^2\,z_8 + 2\,z_7\,z_8 + 3\,z_1\,z_7\,z_8 + z_2\,z_7\,z_8 +
z_1\,z_2\,z_7\,z_8 + 2\,z_3\,z_7\,z_8 + z_5\,z_7\,z_8 +
   z_6\,z_7\,z_8 + 3\,z_7^2\,z_8 + z_1\,z_7^2\,z_8 - 2\,z_8^2 +
2\,z_1\,z_8^2 + z_1^2\,z_8^2 + z_1\,z_2\,z_8^2 -
   z_5\,z_8^2 + z_6\,z_8^2 - z_1\,z_6\,z_8^2 - z_7\,z_8^2 + z_1\,z_7\,z_8^2
+ z_2\,z_7\,z_8^2 - z_2\,z_8^3 -
   z_3\,z_8^3 - z_6\,z_8^3 - 2\,z_7\,z_8^3 + z_8^4 - z_1\,z_8^4\\ $

$\bchi_{0, 3, 0, 0, 0, 0, 0, 1} =
  1 - 2\,z_1^2 - 2\,z_1^3 - z_1^4 - z_2 - z_1\,z_2 + z_1^3\,z_2 + z_1\,z_3 +
2\,z_1^2\,z_3 + z_2\,z_3 -
   z_1\,z_2\,z_3 - z_2^2\,z_3 - z_3^2 + z_3\,z_4 - z_1\,z_5 - 2\,z_1^2\,z_5
+ z_1\,z_2\,z_5 + 2\,z_3\,z_5 - z_5^2 -
   z_6 + z_1\,z_6 + 2\,z_1^2\,z_6 + 2\,z_2\,z_6 - z_2^2\,z_6 + 2\,z_4\,z_6 +
z_5\,z_6 + z_6^2 + z_1\,z_6^2 + z_7 -
   3\,z_1\,z_7 - 4\,z_1^2\,z_7 - 2\,z_1^3\,z_7 - 2\,z_2\,z_7 - z_1\,z_2\,z_7
+ z_1^2\,z_2\,z_7 + 2\,z_1\,z_3\,z_7 -
   2\,z_1\,z_5\,z_7 - z_6\,z_7 + 2\,z_1\,z_6\,z_7 + z_2\,z_6\,z_7 - z_7^2 -
3\,z_1\,z_7^2 - z_1^2\,z_7^2 -
   z_2\,z_7^2 - z_7^3 + z_8 - 2\,z_1\,z_8 - 2\,z_1^2\,z_8 + z_1^3\,z_8 -
z_2\,z_8 + 3\,z_1\,z_2\,z_8 +
   z_1^2\,z_2\,z_8 - z_1\,z_2^2\,z_8 + z_2^3\,z_8 + z_3\,z_8 + z_1\,z_3\,z_8
+ z_1^2\,z_3\,z_8 - 2\,z_3^2\,z_8 -
   z_4\,z_8 + 2\,z_1\,z_4\,z_8 - 2\,z_2\,z_4\,z_8 - z_5\,z_8 +
z_3\,z_5\,z_8 - z_6\,z_8 + 4\,z_1\,z_6\,z_8 +
   3\,z_1^2\,z_6\,z_8 - z_2\,z_6\,z_8 - 2\,z_1\,z_2\,z_6\,z_8 -
5\,z_3\,z_6\,z_8 + z_5\,z_6\,z_8 - 3\,z_6^2\,z_8 +
   z_7\,z_8 - 4\,z_1\,z_7\,z_8 + z_1^2\,z_7\,z_8 + 2\,z_1\,z_2\,z_7\,z_8 +
z_1\,z_3\,z_7\,z_8 - z_4\,z_7\,z_8 - z_8^2 -
   2\,z_1\,z_8^2 + 3\,z_1^2\,z_8^2 + 2\,z_1^3\,z_8^2 + 3\,z_2\,z_8^2 -
2\,z_1^2\,z_2\,z_8^2 - z_2^2\,z_8^2 +
   3\,z_3\,z_8^2 - 5\,z_1\,z_3\,z_8^2 + 2\,z_2\,z_3\,z_8^2 + z_4\,z_8^2 +
z_1\,z_5\,z_8^2 + 6\,z_6\,z_8^2 -
   5\,z_1\,z_6\,z_8^2 + z_2\,z_6\,z_8^2 + z_1\,z_7\,z_8^2 +
2\,z_2\,z_7\,z_8^2 + z_3\,z_7\,z_8^2 + z_6\,z_7\,z_8^2 +
   z_7^2\,z_8^2 - 3\,z_8^3 + 6\,z_1\,z_8^3 - 2\,z_1^2\,z_8^3 - 2\,z_2\,z_8^3
+ z_1\,z_2\,z_8^3 - 2\,z_3\,z_8^3 -
   2\,z_6\,z_8^3 - 2\,z_7\,z_8^3 + z_1\,z_7\,z_8^3 + 2\,z_8^4 -
2\,z_1\,z_8^4\\ $

$\bchi_{1, 0, 0, 0, 0, 0, 1, 3} =
  -z_1 - z_1^2 - z_2 - z_1\,z_2 - z_1^2\,z_2 + z_3 + z_1\,z_3 + z_2\,z_3 -
z_1\,z_4 + z_1\,z_5 + z_1\,z_6 +
   2\,z_2\,z_6 + z_6^2 - z_1\,z_7 - z_2\,z_7 - z_1\,z_2\,z_7 + z_3\,z_7 +
z_5\,z_7 + 2\,z_1\,z_6\,z_7 + z_1\,z_8 +
   2\,z_1^2\,z_8 + z_1^3\,z_8 - 2\,z_2\,z_8 - z_2^2\,z_8 + z_3\,z_8 -
2\,z_1\,z_3\,z_8 + z_1\,z_5\,z_8 - z_6\,z_8 -
   2\,z_1\,z_6\,z_8 - z_2\,z_6\,z_8 + 2\,z_7\,z_8 + z_1\,z_7\,z_8 +
z_1^2\,z_7\,z_8 - 2\,z_2\,z_7\,z_8 - z_3\,z_7\,z_8 -
   4\,z_6\,z_7\,z_8 + 2\,z_7^2\,z_8 - 2\,z_1\,z_7^2\,z_8 + z_8^2 +
3\,z_1\,z_8^2 - z_1^2\,z_8^2 + z_2\,z_8^2 +
   2\,z_1\,z_2\,z_8^2 - z_3\,z_8^2 - z_5\,z_8^2 - z_6\,z_8^2 -
z_1\,z_6\,z_8^2 + 4\,z_7\,z_8^2 - z_1\,z_7\,z_8^2 +
   2\,z_2\,z_7\,z_8^2 + 2\,z_7^2\,z_8^2 - 3\,z_1\,z_8^3 - z_1^2\,z_8^3 +
z_2\,z_8^3 + z_3\,z_8^3 + 2\,z_6\,z_8^3 -
   2\,z_7\,z_8^3 + z_1\,z_7\,z_8^3 - 2\,z_8^4 + z_1\,z_8^4 - z_2\,z_8^4 -
z_7\,z_8^4 + z_8^5\\ $

$\bchi_{1, 0, 0, 0, 0, 0, 3, 0} =
  -z_1^3 - z_1^4 + z_1\,z_2 - z_1^2\,z_2 + z_2^2 + z_1\,z_3 +
3\,z_1^2\,z_3 - z_3^2 - z_1\,z_4 + z_1\,z_5 -
   z_1^2\,z_5 + z_2\,z_5 + z_3\,z_5 + z_1^2\,z_6 + z_2\,z_6 - z_3\,z_6 +
2\,z_5\,z_6 + z_1\,z_6^2 + z_1\,z_7 +
   z_1^2\,z_7 - z_1^3\,z_7 + z_2^2\,z_7 - z_3\,z_7 + 2\,z_1\,z_3\,z_7 -
z_4\,z_7 + z_5\,z_7 - z_1\,z_5\,z_7 +
   2\,z_1\,z_6\,z_7 + z_2\,z_6\,z_7 + 2\,z_1\,z_7^2 + z_1^2\,z_7^2 -
z_3\,z_7^2 + z_6\,z_7^2 + z_1\,z_7^3 +
   z_1\,z_8 - 2\,z_1^2\,z_8 + z_1^3\,z_8 - 2\,z_1\,z_2\,z_8 -
z_1^2\,z_2\,z_8 + z_3\,z_8 + z_2\,z_3\,z_8 -
   z_2\,z_5\,z_8 + 2\,z_6\,z_8 - z_1\,z_6\,z_8 + z_1^2\,z_6\,z_8 -
z_3\,z_6\,z_8 - 2\,z_6^2\,z_8 - 2\,z_1\,z_7\,z_8 +
   z_1^2\,z_7\,z_8 - 2\,z_2\,z_7\,z_8 - z_1\,z_2\,z_7\,z_8 + z_3\,z_7\,z_8 -
2\,z_5\,z_7\,z_8 + 2\,z_6\,z_7\,z_8 -
   2\,z_1\,z_6\,z_7\,z_8 - z_7^2\,z_8 - z_1\,z_7^2\,z_8 - z_2\,z_7^2\,z_8 -
z_7^3\,z_8 - z_8^2 - z_1\,z_8^2 +
   2\,z_1^2\,z_8^2 + z_1^3\,z_8^2 - z_2\,z_8^2 + z_1\,z_2\,z_8^2 -
z_2^2\,z_8^2 - 2\,z_1\,z_3\,z_8^2 +
   z_4\,z_8^2 - z_5\,z_8^2 + z_1\,z_5\,z_8^2 + 2\,z_6\,z_8^2 -
2\,z_1\,z_6\,z_8^2 + z_2\,z_6\,z_8^2 - 5\,z_7\,z_8^2 +
   2\,z_1\,z_7\,z_8^2 - z_1^2\,z_7\,z_8^2 + z_3\,z_7\,z_8^2 +
2\,z_6\,z_7\,z_8^2 - 2\,z_7^2\,z_8^2 - 2\,z_8^3 +
   3\,z_1\,z_8^3 - z_1^2\,z_8^3 + z_2\,z_8^3 + z_1\,z_2\,z_8^3 -
z_3\,z_8^3 - z_6\,z_8^3 - z_7\,z_8^3 +
   z_1\,z_7\,z_8^3 + 2\,z_8^4 - 2\,z_1\,z_8^4 + z_7\,z_8^4\\ $

$\bchi_{1, 0, 0, 0, 0, 1, 1, 1} =
  -z_1^2 - z_1^3 - z_1\,z_2 + z_1\,z_2^2 - z_1\,z_4 - z_2\,z_4 - z_1\,z_5 -
z_2\,z_5 + z_1\,z_6 - z_1\,z_2\,z_6 -
   z_3\,z_6 - z_6^2 - z_1\,z_6^2 - z_1\,z_7 - 2\,z_1^2\,z_7 - z_4\,z_7 -
z_5\,z_7 + z_1\,z_6\,z_7 + z_2\,z_6\,z_7 -
   z_1\,z_7^2 + z_6\,z_7^2 - 2\,z_1\,z_8 - z_1^2\,z_8 + z_1^3\,z_8 +
z_1\,z_2\,z_8 - 2\,z_1^2\,z_2\,z_8 + z_3\,z_8 +
   z_2\,z_3\,z_8 + z_1\,z_4\,z_8 - z_5\,z_8 + 3\,z_1\,z_5\,z_8 +
z_2\,z_5\,z_8 + 3\,z_6\,z_8 + 2\,z_1\,z_6\,z_8 +
   z_1^2\,z_6\,z_8 + 2\,z_2\,z_6\,z_8 - z_3\,z_6\,z_8 - 2\,z_6^2\,z_8 -
z_7\,z_8 - 5\,z_1\,z_7\,z_8 + z_1^2\,z_7\,z_8 -
   z_2\,z_7\,z_8 + z_1\,z_2\,z_7\,z_8 + 2\,z_3\,z_7\,z_8 + z_5\,z_7\,z_8 +
4\,z_6\,z_7\,z_8 + z_1\,z_6\,z_7\,z_8 -
   2\,z_7^2\,z_8 - z_1\,z_7^2\,z_8 - z_2\,z_7^2\,z_8 - z_7^3\,z_8 - 2\,z_8^2
+ 2\,z_1^2\,z_8^2 + z_1^3\,z_8^2 +
   z_1\,z_2\,z_8^2 - z_2^2\,z_8^2 + z_3\,z_8^2 - 2\,z_1\,z_3\,z_8^2 +
z_4\,z_8^2 + z_5\,z_8^2 - z_1\,z_5\,z_8^2 +
   4\,z_6\,z_8^2 - 4\,z_1\,z_6\,z_8^2 - 5\,z_7\,z_8^2 + 3\,z_1\,z_7\,z_8^2 -
z_1^2\,z_7\,z_8^2 - z_2\,z_7\,z_8^2 +
   z_3\,z_7\,z_8^2 + 2\,z_6\,z_7\,z_8^2 - 2\,z_7^2\,z_8^2 - z_8^3 +
4\,z_1\,z_8^3 - 2\,z_1^2\,z_8^3 +
   z_1\,z_2\,z_8^3 - z_3\,z_8^3 - z_5\,z_8^3 - 2\,z_6\,z_8^3 - z_7\,z_8^3 +
2\,z_1\,z_7\,z_8^3 + 2\,z_8^4 -
   2\,z_1\,z_8^4 + z_7\,z_8^4\\ $

$\bchi_{1, 0, 0, 0, 1, 0, 0, 2} =
  -z_1^3 - z_1^4 + z_1\,z_2 + 2\,z_1^2\,z_3 + z_1\,z_2\,z_3 - z_1\,z_4 -
z_1^2\,z_5 + z_1\,z_6 + 2\,z_1^2\,z_6 +
   z_1\,z_2\,z_6 - z_1\,z_7 - z_1^2\,z_7 - z_2\,z_7 - z_1\,z_2\,z_7 -
z_2^2\,z_7 - z_1\,z_3\,z_7 + z_4\,z_7 -
   2\,z_5\,z_7 - z_1\,z_5\,z_7 + z_6\,z_7 + z_1\,z_6\,z_7 + z_2\,z_6\,z_7 -
z_1\,z_7^2 - z_2\,z_7^2 - 2\,z_1^2\,z_8 +
   2\,z_1^3\,z_8 - z_2\,z_8 - 3\,z_1\,z_2\,z_8 - z_1^2\,z_2\,z_8 -
z_2^2\,z_8 - z_3\,z_8 - z_1\,z_3\,z_8 +
   z_2\,z_3\,z_8 + z_4\,z_8 - z_1\,z_4\,z_8 - z_5\,z_8 + z_1\,z_5\,z_8 -
z_6\,z_8 - z_1\,z_6\,z_8 - z_1^2\,z_6\,z_8 +
   z_2\,z_6\,z_8 + z_3\,z_6\,z_8 + z_6^2\,z_8 - 3\,z_7\,z_8 -
3\,z_1\,z_7\,z_8 - 2\,z_2\,z_7\,z_8 + z_1\,z_2\,z_7\,z_8 -
   z_3\,z_7\,z_8 - z_5\,z_7\,z_8 + z_6\,z_7\,z_8 - 3\,z_7^2\,z_8 -
z_1\,z_7^2\,z_8 - z_8^2 + z_1\,z_8^2 +
   3\,z_1^2\,z_8^2 + 2\,z_1\,z_2\,z_8^2 - z_3\,z_8^2 + z_1\,z_5\,z_8^2 -
z_6\,z_8^2 - z_1\,z_6\,z_8^2 -
   z_2\,z_6\,z_8^2 - z_7\,z_8^2 + 3\,z_1\,z_7\,z_8^2 + 2\,z_2\,z_7\,z_8^2 -
z_7^2\,z_8^2 + z_8^3 - z_1^2\,z_8^3 +
   2\,z_2\,z_8^3 + z_3\,z_8^3 + z_6\,z_8^3 + 2\,z_7\,z_8^3 +
z_1\,z_7\,z_8^3 - z_1\,z_8^4 - z_2\,z_8^4\\ $

$\bchi_{1, 0, 0, 0, 1, 1, 0, 0} =
  -(z_1\,z_2^2) - z_1\,z_3 - z_1^2\,z_3 - z_2\,z_3 + z_2\,z_5 -
z_1\,z_2\,z_5 - z_3\,z_5 + z_5^2 + z_6 +
   2\,z_1\,z_6 + z_1^2\,z_6 - z_4\,z_6 + z_5\,z_6 + z_1\,z_5\,z_6 - z_6^2 -
z_1\,z_6^2 - z_2\,z_6^2 - z_7 -
   2\,z_1\,z_7 - z_1^2\,z_7 - z_2\,z_7 + z_1\,z_2\,z_7 - z_1\,z_3\,z_7 +
z_4\,z_7 - z_1\,z_4\,z_7 - 2\,z_5\,z_7 +
   3\,z_6\,z_7 + 2\,z_1\,z_6\,z_7 - z_1^2\,z_6\,z_7 + 2\,z_2\,z_6\,z_7 +
z_3\,z_6\,z_7 + z_6^2\,z_7 - 2\,z_7^2 -
   2\,z_1\,z_7^2 - z_2\,z_7^2 + z_1\,z_2\,z_7^2 - z_5\,z_7^2 + z_6\,z_7^2 -
z_7^3 - z_8 - 2\,z_1\,z_8 -
   z_1^2\,z_8 - z_2\,z_8 + z_1^2\,z_2\,z_8 + z_3^2\,z_8 + z_4\,z_8 -
z_1\,z_4\,z_8 + z_2\,z_4\,z_8 - z_5\,z_8 +
   z_1^2\,z_5\,z_8 - z_3\,z_5\,z_8 + 4\,z_6\,z_8 + z_1\,z_6\,z_8 -
z_1^2\,z_6\,z_8 + 2\,z_3\,z_6\,z_8 - z_5\,z_6\,z_8 -
   z_6^2\,z_8 - 6\,z_7\,z_8 - 3\,z_1\,z_7\,z_8 - z_1^2\,z_7\,z_8 -
z_2\,z_7\,z_8 + z_1\,z_2\,z_7\,z_8 + z_3\,z_7\,z_8 +
   z_4\,z_7\,z_8 - 2\,z_5\,z_7\,z_8 + 6\,z_6\,z_7\,z_8 +
z_1\,z_6\,z_7\,z_8 - 6\,z_7^2\,z_8 - z_2\,z_7^2\,z_8 -
   z_7^3\,z_8 - 3\,z_8^2 + 2\,z_1\,z_8^2 + z_1^2\,z_8^2 + z_2^2\,z_8^2 +
z_1\,z_3\,z_8^2 - z_2\,z_3\,z_8^2 -
   2\,z_1\,z_6\,z_8^2 - z_2\,z_6\,z_8^2 - 4\,z_7\,z_8^2 + 4\,z_1\,z_7\,z_8^2
+ 2\,z_6\,z_7\,z_8^2 - 2\,z_7^2\,z_8^2 +
   2\,z_8^3 + 2\,z_1\,z_8^3 + z_2\,z_8^3 - z_1\,z_2\,z_8^3 - 2\,z_6\,z_8^3 +
4\,z_7\,z_8^3 + z_1\,z_7\,z_8^3 +
   2\,z_8^4 - 2\,z_1\,z_8^4 + z_7\,z_8^4 - z_8^5\\ $

$\bchi_{1, 0, 0, 1, 0, 0, 1, 0} =
  z_1 + z_1^2 - z_1^3 - z_1^4 + z_2 + z_1\,z_2 - z_1^2\,z_2 - z_1^3\,z_2 +
z_2^2 - z_1\,z_2^2 - z_3 +
   z_1\,z_3 + 2\,z_1^2\,z_3 + 3\,z_1\,z_2\,z_3 + z_2^2\,z_3 - z_4 +
z_1^2\,z_4 - z_2\,z_4 - z_3\,z_4 + z_5 -
   z_1^2\,z_5 + z_2\,z_5 - z_1\,z_2\,z_5 - z_6 + z_1^2\,z_6 - z_2\,z_6 +
z_1\,z_2\,z_6 - z_4\,z_6 + 2\,z_1\,z_7 +
   z_1^2\,z_7 - z_1^3\,z_7 + 2\,z_2\,z_7 - z_1\,z_2\,z_7 - z_1^2\,z_2\,z_7 -
2\,z_3\,z_7 + 2\,z_1\,z_3\,z_7 +
   z_2\,z_3\,z_7 - z_4\,z_7 + z_1\,z_4\,z_7 + 2\,z_5\,z_7 - z_1\,z_5\,z_7 -
z_2\,z_5\,z_7 - z_6\,z_7 + z_1\,z_6\,z_7 +
   z_2\,z_6\,z_7 + z_6^2\,z_7 - z_7^2 + z_1\,z_7^2 - z_5\,z_7^2 +
z_6\,z_7^2 - z_7^3 + 2\,z_8 - z_1\,z_8 -
   3\,z_1^2\,z_8 + z_1^3\,z_8 + z_1^4\,z_8 + 3\,z_2\,z_8 - 5\,z_1\,z_2\,z_8
+ 2\,z_1^2\,z_2\,z_8 - z_2^2\,z_8 +
   2\,z_1\,z_3\,z_8 - 2\,z_1^2\,z_3\,z_8 + z_2\,z_3\,z_8 -
z_1\,z_2\,z_3\,z_8 + 2\,z_5\,z_8 - 2\,z_1\,z_5\,z_8 +
   z_3\,z_5\,z_8 + z_1\,z_6\,z_8 - z_1^2\,z_6\,z_8 - 2\,z_2\,z_6\,z_8 +
z_1\,z_2\,z_6\,z_8 + z_3\,z_6\,z_8 + 2\,z_7\,z_8 -
   3\,z_1\,z_7\,z_8 + z_1^3\,z_7\,z_8 + z_2\,z_7\,z_8 + z_3\,z_7\,z_8 -
z_1\,z_3\,z_7\,z_8 - z_5\,z_7\,z_8 + z_6\,z_7\,z_8 -
   z_1\,z_6\,z_7\,z_8 - z_7^2\,z_8 - z_2\,z_7^2\,z_8 - z_7^3\,z_8 -
3\,z_1\,z_8^2 + 4\,z_1^2\,z_8^2 - z_1^3\,z_8^2 -
   z_2\,z_8^2 + z_1\,z_2\,z_8^2 - z_1\,z_3\,z_8^2 - z_5\,z_8^2 +
z_1\,z_5\,z_8^2 - z_6\,z_8^2 + z_1\,z_6\,z_8^2 -
   z_7\,z_8^2 - 2\,z_1^2\,z_7\,z_8^2 + z_3\,z_7\,z_8^2 +
2\,z_6\,z_7\,z_8^2 - z_7^2\,z_8^2 - z_8^3 +
   2\,z_1\,z_8^3 - z_1^2\,z_8^3 - z_2\,z_8^3 + z_1\,z_2\,z_8^3 - z_7\,z_8^3
+ z_1\,z_7\,z_8^3 + z_7\,z_8^4\\ $

$\bchi_{1, 0, 1, 0, 0, 0, 1, 1} =
  -z_1 - z_1^2 - z_1^3 - z_1^4 - 2\,z_1\,z_2 - z_1^2\,z_2 + z_1^3\,z_2 -
z_2^2 - z_2^3 + 2\,z_1^2\,z_3 -
   2\,z_1\,z_2\,z_3 - z_3^2 + z_4 + 2\,z_2\,z_4 - z_1^2\,z_5 - 2\,z_2\,z_5 +
z_6 + z_1\,z_6 + 2\,z_1^2\,z_6 +
   2\,z_2\,z_6 - 3\,z_3\,z_6 - z_1\,z_3\,z_6 + z_4\,z_6 + z_5\,z_6 - z_6^2 +
z_1\,z_6^2 - z_7 - 2\,z_1\,z_7 -
   2\,z_1^2\,z_7 - z_1^3\,z_7 - z_2\,z_7 - z_1\,z_2\,z_7 + z_3\,z_7 +
z_1\,z_3\,z_7 - z_7^2 - z_1\,z_7^2 -
   z_6\,z_7^2 - 2\,z_8 - z_1\,z_8 + z_1^3\,z_8 - 4\,z_2\,z_8 +
z_1\,z_2\,z_8 - z_1^2\,z_2\,z_8 + 2\,z_1\,z_2^2\,z_8 +
   z_3\,z_8 + z_1\,z_3\,z_8 + 2\,z_2\,z_3\,z_8 - 2\,z_5\,z_8 -
z_1\,z_5\,z_8 - 2\,z_2\,z_5\,z_8 - z_1\,z_6\,z_8 +
   z_1^2\,z_6\,z_8 + 5\,z_2\,z_6\,z_8 - z_3\,z_6\,z_8 - 2\,z_7\,z_8 -
3\,z_2\,z_7\,z_8 - z_1\,z_2\,z_7\,z_8 +
   3\,z_3\,z_7\,z_8 + z_1\,z_3\,z_7\,z_8 - z_4\,z_7\,z_8 - z_5\,z_7\,z_8 -
z_1\,z_6\,z_7\,z_8 + z_7^2\,z_8 +
   z_1\,z_7^2\,z_8 + z_7^3\,z_8 + 2\,z_1\,z_8^2 + z_1^2\,z_8^2 +
z_1^3\,z_8^2 - z_2\,z_8^2 + z_1\,z_2\,z_8^2 -
   z_1^2\,z_2\,z_8^2 - z_2^2\,z_8^2 + z_3\,z_8^2 - 2\,z_1\,z_3\,z_8^2 -
z_5\,z_8^2 + z_1\,z_5\,z_8^2 + z_6\,z_8^2 -
   2\,z_1\,z_6\,z_8^2 + z_2\,z_6\,z_8^2 + 2\,z_7\,z_8^2 +
3\,z_1\,z_7\,z_8^2 - 4\,z_2\,z_7\,z_8^2 - z_6\,z_7\,z_8^2 +
   2\,z_7^2\,z_8^2 + z_8^3 + z_1\,z_8^3 - z_1^2\,z_8^3 + z_2\,z_8^3 +
z_1\,z_2\,z_8^3 - z_3\,z_8^3 +
   z_5\,z_8^3 - z_1\,z_8^4 + z_2\,z_8^4 - z_7\,z_8^4\\ $

$\bchi_{1, 0, 2, 0, 0, 0, 0, 0} =
  2\,z_1^2 + 2\,z_1^3 + z_1^4 + z_2 + 4\,z_1\,z_2 + 3\,z_1^2\,z_2 +
z_1^3\,z_2 + z_2^2 + z_1\,z_2^2 -
   z_1^2\,z_3 - z_2\,z_3 - 2\,z_1\,z_2\,z_3 + z_1\,z_3^2 - z_1^2\,z_4 +
z_2\,z_4 - z_3\,z_4 + z_1\,z_5 + z_1^2\,z_5 +
   z_2\,z_5 + z_1\,z_2\,z_5 + z_6 + 2\,z_1\,z_6 - z_1^2\,z_6 - z_1^3\,z_6 -
2\,z_1\,z_2\,z_6 - z_3\,z_6 + z_5\,z_6 -
   z_6^2 + 2\,z_1\,z_7 + 4\,z_1^2\,z_7 + 2\,z_1^3\,z_7 + 2\,z_2\,z_7 +
4\,z_1\,z_2\,z_7 + z_1^2\,z_2\,z_7 +
   z_2^2\,z_7 - z_1\,z_3\,z_7 - z_2\,z_3\,z_7 + z_5\,z_7 + z_1\,z_5\,z_7 +
z_6\,z_7 - z_2\,z_6\,z_7 + 2\,z_1\,z_7^2 +
   z_1^2\,z_7^2 + z_2\,z_7^2 - z_8 + z_1^2\,z_8 - z_1^3\,z_8 - z_1^4\,z_8 +
z_2\,z_8 + z_1\,z_2\,z_8 -
   2\,z_1^2\,z_2\,z_8 + z_2^2\,z_8 + z_3\,z_8 - 2\,z_1\,z_3\,z_8 +
z_1^2\,z_3\,z_8 - 2\,z_2\,z_3\,z_8 + z_3^2\,z_8 -
   z_1\,z_4\,z_8 + 2\,z_6\,z_8 - 3\,z_1\,z_6\,z_8 - 2\,z_2\,z_6\,z_8 +
z_3\,z_6\,z_8 - z_7\,z_8 - z_1^2\,z_7\,z_8 +
   z_2\,z_7\,z_8 - z_1\,z_2\,z_7\,z_8 - z_5\,z_7\,z_8 - z_1\,z_7^2\,z_8 -
z_8^2 - 4\,z_1^2\,z_8^2 - 4\,z_1\,z_2\,z_8^2 -
   z_3\,z_8^2 + 2\,z_1\,z_3\,z_8^2 + z_4\,z_8^2 - z_5\,z_8^2 - z_6\,z_8^2 +
2\,z_1\,z_6\,z_8^2 - z_7\,z_8^2 -
   2\,z_1\,z_7\,z_8^2 - z_1\,z_8^3 + 2\,z_1^2\,z_8^3 + z_8^4\\ $

$\bchi_{1, 1, 0, 0, 0, 0, 0, 3} =
  z_1 + z_1^2 - z_2 - 2\,z_1\,z_2 - z_3 - z_4 - z_1\,z_5 - 2\,z_6 -
2\,z_1\,z_6 - z_1^2\,z_6 + z_1\,z_2\,z_6 +
   z_3\,z_6 - z_5\,z_6 + z_6^2 + z_7 + 2\,z_1\,z_7 + z_1^2\,z_7 -
2\,z_2\,z_7 - z_1\,z_2\,z_7 + z_2^2\,z_7 -
   2\,z_3\,z_7 + z_1\,z_3\,z_7 - z_4\,z_7 + z_5\,z_7 - 3\,z_6\,z_7 -
2\,z_1\,z_6\,z_7 + z_7^2 + z_1\,z_7^2 + 2\,z_8 +
   z_1\,z_8 + z_1^2\,z_8 + z_1\,z_2\,z_8 + z_2^2\,z_8 - z_3\,z_8 +
z_1\,z_3\,z_8 + z_2\,z_3\,z_8 + z_5\,z_8 -
   2\,z_6\,z_8 + z_1\,z_6\,z_8 + 5\,z_7\,z_8 + 3\,z_1\,z_7\,z_8 +
z_1^2\,z_7\,z_8 + 3\,z_2\,z_7\,z_8 -
   2\,z_1\,z_2\,z_7\,z_8 + 2\,z_5\,z_7\,z_8 - 2\,z_6\,z_7\,z_8 +
3\,z_7^2\,z_8 + 2\,z_1\,z_7^2\,z_8 - 2\,z_1\,z_8^2 -
   z_1^2\,z_8^2 + 2\,z_2\,z_8^2 - z_1\,z_2\,z_8^2 - z_2^2\,z_8^2 +
z_3\,z_8^2 - z_1\,z_3\,z_8^2 + z_4\,z_8^2 +
   2\,z_6\,z_8^2 + z_1\,z_6\,z_8^2 - 3\,z_1\,z_7\,z_8^2 - 3\,z_8^3 -
z_1\,z_8^3 - z_2\,z_8^3 + z_1\,z_2\,z_8^3 -
   z_5\,z_8^3 + z_6\,z_8^3 - 4\,z_7\,z_8^3 - z_1\,z_7\,z_8^3 + z_1\,z_8^4 +
z_8^5\\ $

$\bchi_{1, 1, 0, 0, 0, 0, 2, 0} =
  -2\,z_1 - 2\,z_1^2 + z_1^3 + z_1^4 - z_2 - z_1\,z_2 - z_1^3\,z_2 - z_2^2 -
z_1\,z_2^2 + z_3 - z_1\,z_3 -
   2\,z_1^2\,z_3 + z_1\,z_2\,z_3 + z_4 + 2\,z_1\,z_4 - z_5 - z_1\,z_5 +
z_1^2\,z_5 + z_6 + 3\,z_1\,z_6 +
   z_1^2\,z_6 + z_2\,z_6 + z_1\,z_2\,z_6 + z_2^2\,z_6 - z_3\,z_6 +
z_1\,z_3\,z_6 - z_4\,z_6 + z_5\,z_6 - z_6^2 -
   z_1\,z_6^2 - z_7 - 6\,z_1\,z_7 - 3\,z_1^2\,z_7 + z_1^3\,z_7 -
2\,z_2\,z_7 - z_2^2\,z_7 + z_3\,z_7 -
   2\,z_1\,z_3\,z_7 + 2\,z_4\,z_7 - 2\,z_5\,z_7 + 2\,z_6\,z_7 +
2\,z_1\,z_6\,z_7 + z_2\,z_6\,z_7 - 2\,z_7^2 -
   5\,z_1\,z_7^2 - z_1^2\,z_7^2 - z_2\,z_7^2 + z_1\,z_2\,z_7^2 -
z_5\,z_7^2 - z_7^3 - z_1\,z_7^3 - 2\,z_8 -
   4\,z_1\,z_8 + 2\,z_1^2\,z_8 - 2\,z_2\,z_8 + 2\,z_1\,z_2\,z_8 +
z_1^2\,z_2\,z_8 - z_2^2\,z_8 + z_3\,z_8 -
   2\,z_1\,z_3\,z_8 + z_1^2\,z_3\,z_8 - z_3^2\,z_8 + z_4\,z_8 -
z_1\,z_4\,z_8 - 2\,z_5\,z_8 + 3\,z_6\,z_8 +
   z_1\,z_6\,z_8 - z_1^2\,z_6\,z_8 + z_2\,z_6\,z_8 - z_1\,z_2\,z_6\,z_8 -
2\,z_3\,z_6\,z_8 + z_5\,z_6\,z_8 - z_6^2\,z_8 -
   5\,z_7\,z_8 - 4\,z_1\,z_7\,z_8 + z_1^2\,z_7\,z_8 - z_2\,z_7\,z_8 +
2\,z_1\,z_2\,z_7\,z_8 - z_2^2\,z_7\,z_8 -
   z_1\,z_3\,z_7\,z_8 + z_4\,z_7\,z_8 - 2\,z_5\,z_7\,z_8 + 2\,z_6\,z_7\,z_8
+ 2\,z_1\,z_6\,z_7\,z_8 - 3\,z_7^2\,z_8 -
   z_1\,z_7^2\,z_8 - z_8^2 + 6\,z_1\,z_8^2 - z_1^2\,z_8^2 - z_1^3\,z_8^2 +
z_2\,z_8^2 + z_1\,z_2\,z_8^2 +
   z_2^2\,z_8^2 + z_1\,z_3\,z_8^2 + z_2\,z_3\,z_8^2 - 2\,z_4\,z_8^2 +
z_5\,z_8^2 - z_6\,z_8^2 - 2\,z_1\,z_6\,z_8^2 +
   2\,z_7\,z_8^2 + 3\,z_1\,z_7\,z_8^2 + z_1^2\,z_7\,z_8^2 + z_2\,z_7\,z_8^2
+ z_7^2\,z_8^2 + 5\,z_8^3 -
   z_1\,z_8^3 - z_1\,z_2\,z_8^3 + z_5\,z_8^3 - 2\,z_6\,z_8^3 +
4\,z_7\,z_8^3 - z_8^4 - z_8^5\\ $

$\bchi_{1, 1, 0, 0, 0, 1, 0, 1} =
  -z_1 - 2\,z_1^2 - z_1^3 + z_3 + 2\,z_1\,z_3 + z_2\,z_3 - z_1\,z_2\,z_3 -
z_3^2 + z_4 + 2\,z_1\,z_4 + z_2\,z_4 +
   z_1\,z_5 + z_1^2\,z_5 - z_1\,z_2\,z_5 + z_5^2 + z_6 + 3\,z_1\,z_6 +
2\,z_1^2\,z_6 + z_2\,z_6 + z_1\,z_2\,z_6 -
   z_3\,z_6 - z_5\,z_6 - z_7 - 4\,z_1\,z_7 - 3\,z_1^2\,z_7 + 2\,z_3\,z_7 +
z_1\,z_3\,z_7 + z_2\,z_3\,z_7 + 2\,z_4\,z_7 +
   z_1\,z_5\,z_7 + 2\,z_6\,z_7 + 2\,z_1\,z_6\,z_7 - 2\,z_7^2 - 3\,z_1\,z_7^2
+ z_3\,z_7^2 - z_7^3 - 2\,z_8 -
   3\,z_1\,z_8 + 3\,z_1^2\,z_8 + 2\,z_1^3\,z_8 - z_2\,z_8 + z_1\,z_2\,z_8 +
2\,z_3\,z_8 - 3\,z_1\,z_3\,z_8 +
   z_1^2\,z_3\,z_8 - z_3^2\,z_8 + z_4\,z_8 - z_1\,z_4\,z_8 + z_5\,z_8 -
2\,z_1\,z_5\,z_8 - z_2\,z_5\,z_8 + z_6\,z_8 -
   z_1\,z_6\,z_8 - 2\,z_1^2\,z_6\,z_8 + z_1\,z_2\,z_6\,z_8 - z_3\,z_6\,z_8 -
z_5\,z_6\,z_8 + z_6^2\,z_8 - 5\,z_7\,z_8 +
   3\,z_1^2\,z_7\,z_8 - 2\,z_1\,z_2\,z_7\,z_8 - z_2^2\,z_7\,z_8 -
z_1\,z_3\,z_7\,z_8 + z_4\,z_7\,z_8 - 3\,z_7^2\,z_8 +
   z_1\,z_7^2\,z_8 - z_8^2 + 5\,z_1\,z_8^2 - z_1^2\,z_8^2 - z_1^3\,z_8^2 -
2\,z_1\,z_2\,z_8^2 - z_3\,z_8^2 +
   z_2\,z_3\,z_8^2 - z_4\,z_8^2 - z_5\,z_8^2 - 2\,z_6\,z_8^2 +
z_1\,z_6\,z_8^2 + z_2\,z_6\,z_8^2 + z_7\,z_8^2 +
   z_1\,z_7\,z_8^2 + z_1^2\,z_7\,z_8^2 - z_2\,z_7\,z_8^2 - z_3\,z_7\,z_8^2 -
z_6\,z_7\,z_8^2 + 3\,z_8^3 -
   3\,z_1\,z_8^3 - z_2\,z_8^3 + z_1\,z_2\,z_8^3 + z_6\,z_8^3 +
2\,z_7\,z_8^3 - 2\,z_1\,z_7\,z_8^3 - z_8^4 +
   z_1\,z_8^4 + z_2\,z_8^4\\ $

$\bchi_{1, 1, 1, 0, 0, 0, 0, 1} =
  1 + 2\,z_1 - 2\,z_1^3 - z_1^4 + z_2 + z_1\,z_2 + z_1^2\,z_2 + z_1^3\,z_2 -
z_1^2\,z_2^2 - z_3 +
   z_1^2\,z_3 - z_2\,z_3 - z_1\,z_2\,z_3 - z_1\,z_3^2 - z_4 + z_1^2\,z_4 +
z_3\,z_4 + z_5 - z_1^2\,z_5 +
   2\,z_1\,z_2\,z_5 - z_5^2 - z_6 + 3\,z_1^2\,z_6 + 2\,z_1^3\,z_6 -
2\,z_2\,z_6 - z_1\,z_2\,z_6 - 2\,z_1\,z_3\,z_6 +
   z_4\,z_6 - z_5\,z_6 - z_6^2 - 2\,z_1\,z_6^2 + z_7 - 3\,z_1^2\,z_7 -
2\,z_1^3\,z_7 + z_2\,z_7 + z_1^2\,z_2\,z_7 -
   z_3\,z_7 + z_1\,z_3\,z_7 + z_5\,z_7 - z_1\,z_5\,z_7 + z_6\,z_7 +
3\,z_1\,z_6\,z_7 - z_7^2 - 2\,z_1\,z_7^2 -
   z_1^2\,z_7^2 + z_6\,z_7^2 - z_7^3 + 2\,z_8 - 2\,z_1\,z_8 - 5\,z_1^2\,z_8
+ z_1^4\,z_8 + 2\,z_2\,z_8 +
   z_2^2\,z_8 - z_3\,z_8 + 3\,z_1\,z_3\,z_8 - 2\,z_1^2\,z_3\,z_8 +
z_1\,z_2\,z_3\,z_8 - z_4\,z_8 + z_1\,z_4\,z_8 -
   z_2\,z_4\,z_8 + z_5\,z_8 - 2\,z_1\,z_5\,z_8 - z_1^2\,z_5\,z_8 +
z_2\,z_5\,z_8 + 4\,z_1\,z_6\,z_8 - 2\,z_1^2\,z_6\,z_8 +
   z_3\,z_6\,z_8 + z_5\,z_6\,z_8 + z_6^2\,z_8 - 4\,z_1\,z_7\,z_8 -
z_1^2\,z_7\,z_8 - z_1^3\,z_7\,z_8 + 2\,z_2\,z_7\,z_8 +
   z_1\,z_2\,z_7\,z_8 + z_1\,z_3\,z_7\,z_8 - z_4\,z_7\,z_8 - z_5\,z_7\,z_8 +
2\,z_6\,z_7\,z_8 + z_1\,z_6\,z_7\,z_8 -
   2\,z_7^2\,z_8 - 2\,z_8^2 - 4\,z_1\,z_8^2 + 6\,z_1^2\,z_8^2 -
2\,z_1\,z_2\,z_8^2 + z_3\,z_8^2 - z_1\,z_3\,z_8^2 -
   z_5\,z_8^2 + z_1\,z_5\,z_8^2 - 2\,z_1\,z_6\,z_8^2 - z_2\,z_6\,z_8^2 -
z_7\,z_8^2 + 2\,z_1\,z_7\,z_8^2 +
   z_1^2\,z_7\,z_8^2 - z_6\,z_7\,z_8^2 + z_7^2\,z_8^2 - z_8^3 +
6\,z_1\,z_8^3 - z_1^2\,z_8^3 - z_2\,z_8^3 +
   z_5\,z_8^3 - z_6\,z_8^3 + 2\,z_7\,z_8^3 + z_1\,z_7\,z_8^3 + 2\,z_8^4 -
2\,z_1\,z_8^4 - z_8^5\\ $

$\bchi_{1, 2, 0, 0, 0, 0, 1, 0} =
  z_1^2 + 2\,z_1^3 + z_1^4 - z_1\,z_2 - z_2^2 - z_1\,z_3 - z_1^2\,z_3 -
z_2\,z_3 - z_1\,z_2\,z_3 + z_2^2\,z_3 +
   z_1\,z_3^2 - z_1\,z_4 - z_1^2\,z_4 - z_3\,z_4 - z_1\,z_5 - z_2\,z_5 +
z_3\,z_5 - 2\,z_1\,z_6 - 2\,z_1^2\,z_6 -
   z_1^3\,z_6 - 2\,z_2\,z_6 + z_2^2\,z_6 - z_3\,z_6 - z_4\,z_6 - z_6^2 +
z_1\,z_7 + 3\,z_1^2\,z_7 + 2\,z_1^3\,z_7 -
   z_2\,z_7 - 3\,z_1\,z_2\,z_7 - z_1^2\,z_2\,z_7 - z_2^2\,z_7 +
z_1\,z_2^2\,z_7 - z_1\,z_3\,z_7 - z_1\,z_4\,z_7 -
   z_2\,z_5\,z_7 - z_6\,z_7 - 3\,z_1\,z_6\,z_7 - z_1^2\,z_6\,z_7 +
z_3\,z_6\,z_7 + z_6^2\,z_7 + z_1\,z_7^2 +
   z_1^2\,z_7^2 - z_2\,z_7^2 - z_1\,z_2\,z_7^2 - z_6\,z_7^2 + z_1\,z_8 +
2\,z_1^2\,z_8 - z_2\,z_8 -
   2\,z_1\,z_2\,z_8 - z_1^2\,z_2\,z_8 + z_2^2\,z_8 + 2\,z_1\,z_2^2\,z_8 -
z_2^3\,z_8 - z_1\,z_3\,z_8 - z_1^2\,z_3\,z_8 +
   z_2\,z_3\,z_8 - z_1\,z_2\,z_3\,z_8 + 2\,z_3^2\,z_8 + 2\,z_2\,z_4\,z_8 +
z_1^2\,z_5\,z_8 - 2\,z_2\,z_5\,z_8 - z_6\,z_8 -
   z_1\,z_6\,z_8 - 3\,z_1^2\,z_6\,z_8 + 4\,z_2\,z_6\,z_8 +
z_1\,z_2\,z_6\,z_8 + 4\,z_3\,z_6\,z_8 - z_5\,z_6\,z_8 +
   3\,z_6^2\,z_8 + 2\,z_1\,z_7\,z_8 + z_1^2\,z_7\,z_8 - z_2\,z_7\,z_8 +
z_1\,z_3\,z_7\,z_8 + z_4\,z_7\,z_8 +
   z_1\,z_6\,z_7\,z_8 + z_1\,z_7^2\,z_8 - 3\,z_1\,z_8^2 - z_1^2\,z_8^2 -
2\,z_1^3\,z_8^2 - 3\,z_2\,z_8^2 +
   2\,z_1\,z_2\,z_8^2 + z_1^2\,z_2\,z_8^2 - 2\,z_3\,z_8^2 +
3\,z_1\,z_3\,z_8^2 - 2\,z_2\,z_3\,z_8^2 + z_4\,z_8^2 -
   2\,z_6\,z_8^2 + 5\,z_1\,z_6\,z_8^2 - z_2\,z_6\,z_8^2 - 2\,z_7\,z_8^2 -
4\,z_1\,z_7\,z_8^2 + z_1^2\,z_7\,z_8^2 -
   2\,z_2\,z_7\,z_8^2 - z_3\,z_7\,z_8^2 - 2\,z_7^2\,z_8^2 - 2\,z_8^3 -
2\,z_1\,z_8^3 + 3\,z_2\,z_8^3 -
   z_1\,z_2\,z_8^3 + z_3\,z_8^3 - z_5\,z_8^3 + 3\,z_6\,z_8^3 - z_7\,z_8^3 -
z_1\,z_7\,z_8^3 + 2\,z_1\,z_8^4 + z_8^5\\ $

$\bchi_{2, 0, 0, 0, 0, 0, 1, 2} =
  -z_1^2 + z_1^4 + z_1^2\,z_2 - z_2^2 - z_1\,z_2^2 - 3\,z_1^2\,z_3 + z_3^2 +
2\,z_1\,z_4 + z_1^2\,z_5 -
   z_3\,z_5 - z_1^2\,z_6 - 2\,z_2\,z_6 + 3\,z_3\,z_6 - z_5\,z_6 + z_6^2 -
2\,z_1\,z_7 - 2\,z_1^2\,z_7 + z_2\,z_7 -
   z_2^2\,z_7 + z_4\,z_7 + z_6\,z_7 - z_7^2 - 2\,z_1\,z_7^2 - z_1^2\,z_7^2 +
z_3\,z_7^2 + z_6\,z_7^2 - z_7^3 -
   4\,z_1\,z_8 - z_1^2\,z_8 - z_1^3\,z_8 + z_2\,z_8 + 2\,z_1\,z_2\,z_8 +
2\,z_1^2\,z_2\,z_8 - 2\,z_2\,z_3\,z_8 +
   z_4\,z_8 - 3\,z_1\,z_5\,z_8 + z_6\,z_8 + 4\,z_1\,z_6\,z_8 -
z_1^2\,z_6\,z_8 - z_2\,z_6\,z_8 + z_3\,z_6\,z_8 +
   z_6^2\,z_8 - 3\,z_7\,z_8 - 4\,z_1\,z_7\,z_8 + 2\,z_2\,z_7\,z_8 +
z_1\,z_2\,z_7\,z_8 - z_3\,z_7\,z_8 - z_5\,z_7\,z_8 +
   z_6\,z_7\,z_8 - 3\,z_7^2\,z_8 + z_1\,z_7^2\,z_8 - 2\,z_8^2 +
z_1^2\,z_8^2 - z_1^3\,z_8^2 - 2\,z_1\,z_2\,z_8^2 +
   z_2^2\,z_8^2 + 2\,z_1\,z_3\,z_8^2 - z_4\,z_8^2 - z_5\,z_8^2 +
2\,z_1\,z_6\,z_8^2 - z_7\,z_8^2 +
   z_1^2\,z_7\,z_8^2 - z_2\,z_7\,z_8^2 - z_3\,z_7\,z_8^2 - z_6\,z_7\,z_8^2 +
z_7^2\,z_8^2 + 2\,z_8^3 +
   2\,z_1\,z_8^3 + z_1^2\,z_8^3 - 2\,z_2\,z_8^3 - z_1\,z_2\,z_8^3 +
z_5\,z_8^3 - z_6\,z_8^3 + 3\,z_7\,z_8^3 -
   z_1\,z_7\,z_8^3 + z_8^4 + z_2\,z_8^4 - z_8^5\\ $

$\bchi_{2, 0, 0, 0, 0, 1, 1, 0} =
  -z_2 - z_1\,z_2 + z_1\,z_2^2 + z_3 + z_1\,z_3 + z_2\,z_3 - z_5 -
z_1\,z_5 - z_2\,z_5 + z_1\,z_2\,z_5 + z_3\,z_5 -
   z_5^2 - z_1\,z_6 + z_1^3\,z_6 + z_2\,z_6 + 2\,z_1\,z_2\,z_6 + z_3\,z_6 -
2\,z_1\,z_3\,z_6 - z_5\,z_6 + z_6^2 -
   z_1\,z_6^2 + z_7 + z_1\,z_7 - z_2\,z_7 - 2\,z_1\,z_2\,z_7 + z_3\,z_7 +
z_1\,z_3\,z_7 - z_4\,z_7 + z_5\,z_7 -
   z_6\,z_7 - z_1\,z_6\,z_7 + z_1^2\,z_6\,z_7 - z_2\,z_6\,z_7 -
z_3\,z_6\,z_7 - z_6^2\,z_7 + 2\,z_7^2 + z_1\,z_7^2 -
   z_1\,z_2\,z_7^2 + z_5\,z_7^2 + z_7^3 + z_1\,z_8 - z_1^3\,z_8 -
2\,z_2\,z_8 - z_1\,z_2^2\,z_8 + z_3\,z_8 +
   z_1\,z_3\,z_8 - z_2\,z_3\,z_8 + z_1\,z_5\,z_8 - z_1^2\,z_5\,z_8 +
z_2\,z_5\,z_8 + z_3\,z_5\,z_8 - z_6\,z_8 +
   2\,z_1\,z_6\,z_8 - z_1^2\,z_6\,z_8 - z_2\,z_6\,z_8 - z_3\,z_6\,z_8 +
z_5\,z_6\,z_8 + 3\,z_7\,z_8 + 2\,z_1\,z_7\,z_8 -
   z_1^3\,z_7\,z_8 - z_2\,z_7\,z_8 - 2\,z_1\,z_2\,z_7\,z_8 + z_2^2\,z_7\,z_8
+ 2\,z_1\,z_3\,z_7\,z_8 - z_4\,z_7\,z_8 +
   3\,z_5\,z_7\,z_8 - 3\,z_6\,z_7\,z_8 + 4\,z_7^2\,z_8 + z_1\,z_7^2\,z_8 +
z_2\,z_7^2\,z_8 + z_7^3\,z_8 +
   2\,z_2\,z_8^2 + z_1^2\,z_2\,z_8^2 + z_2^2\,z_8^2 - z_3\,z_8^2 -
z_2\,z_3\,z_8^2 + z_5\,z_8^2 - z_2\,z_6\,z_8^2 +
   2\,z_7\,z_8^2 - 2\,z_1\,z_7\,z_8^2 + z_1^2\,z_7\,z_8^2 +
2\,z_2\,z_7\,z_8^2 - 2\,z_6\,z_7\,z_8^2 +
   2\,z_7^2\,z_8^2 - z_1\,z_8^3 + z_1^2\,z_8^3 + 2\,z_2\,z_8^3 -
z_1\,z_2\,z_8^3 - z_3\,z_8^3 - 2\,z_7\,z_8^3 -
   z_2\,z_8^4 - z_7\,z_8^4\\ $

$\bchi_{2, 0, 0, 0, 1, 0, 0, 1} =
  -(z_1\,z_3) - z_1^2\,z_3 + z_3^2 - z_1\,z_4 - z_1^2\,z_4 + z_3\,z_4 -
z_1^2\,z_6 - z_1^3\,z_6 + 2\,z_3\,z_6 +
   2\,z_1\,z_3\,z_6 + z_4\,z_6 + z_6^2 + z_1\,z_6^2 + z_7 + z_1\,z_7 -
z_1\,z_2\,z_7 + z_1^2\,z_2\,z_7 - z_3\,z_7 -
   z_1\,z_3\,z_7 - z_2\,z_3\,z_7 - 2\,z_4\,z_7 + z_5\,z_7 - z_1\,z_5\,z_7 -
2\,z_6\,z_7 - z_2\,z_6\,z_7 + 2\,z_7^2 +
   z_1\,z_7^2 - z_3\,z_7^2 + z_7^3 + z_8 + z_1\,z_8 - z_1^2\,z_8 -
z_1^3\,z_8 + 2\,z_1^2\,z_2\,z_8 -
   z_1\,z_2^2\,z_8 - z_3\,z_8 + z_1\,z_3\,z_8 - 2\,z_2\,z_3\,z_8 -
2\,z_4\,z_8 + 2\,z_1\,z_4\,z_8 - z_1\,z_5\,z_8 +
   z_1^2\,z_5\,z_8 + z_2\,z_5\,z_8 - z_3\,z_5\,z_8 - 2\,z_6\,z_8 +
z_1\,z_6\,z_8 + z_1^2\,z_6\,z_8 - 2\,z_2\,z_6\,z_8 -
   z_1\,z_2\,z_6\,z_8 + 6\,z_7\,z_8 - 2\,z_1\,z_7\,z_8 - 2\,z_2\,z_7\,z_8 +
z_1\,z_2\,z_7\,z_8 + z_2^2\,z_7\,z_8 -
   2\,z_3\,z_7\,z_8 - z_4\,z_7\,z_8 + z_5\,z_7\,z_8 - 4\,z_6\,z_7\,z_8 +
5\,z_7^2\,z_8 - 2\,z_1\,z_7^2\,z_8 + 2\,z_8^2 -
   4\,z_1\,z_8^2 + 2\,z_1^2\,z_8^2 - z_2\,z_8^2 + z_1\,z_2\,z_8^2 +
z_2^2\,z_8^2 + z_1\,z_3\,z_8^2 + z_4\,z_8^2 +
   z_5\,z_8^2 - z_1\,z_5\,z_8^2 - z_6\,z_8^2 + 2\,z_1\,z_6\,z_8^2 +
2\,z_7\,z_8^2 - 3\,z_1\,z_7\,z_8^2 +
   z_2\,z_7\,z_8^2 + z_3\,z_7\,z_8^2 + z_7^2\,z_8^2 - 2\,z_8^3 +
2\,z_1\,z_8^3 + z_2\,z_8^3 - 2\,z_1\,z_2\,z_8^3 -
   3\,z_7\,z_8^3 + z_1\,z_7\,z_8^3\\ $

$\bchi_{2, 0, 0, 1, 0, 0, 0, 0} =
  z_1\,z_2 + z_1^2\,z_2 + z_3 + z_1\,z_3 + z_1\,z_4 + z_1^2\,z_4 -
z_3\,z_4 - z_5 - z_1\,z_5 - z_1\,z_2\,z_5 -
   z_3\,z_5 + z_5^2 + z_6 + 2\,z_1\,z_6 + z_1^2\,z_6 + z_2\,z_6 +
z_1\,z_2\,z_6 + z_2^2\,z_6 - z_3\,z_6 -
   2\,z_4\,z_6 - z_6^2 + z_7 + z_1\,z_7 + z_1\,z_2\,z_7 + 3\,z_3\,z_7 +
z_1\,z_3\,z_7 - z_5\,z_7 + 2\,z_6\,z_7 -
   z_1\,z_6\,z_7 - z_2\,z_6\,z_7 + 2\,z_7^2 + z_1\,z_7^2 + z_3\,z_7^2 +
z_7^3 - 2\,z_1\,z_8 - 2\,z_1^2\,z_8 -
   z_2\,z_8 - z_1\,z_2\,z_8 - z_2^2\,z_8 + 3\,z_3\,z_8 - z_1\,z_3\,z_8 -
z_2\,z_3\,z_8 + z_4\,z_8 - z_1\,z_4\,z_8 +
   z_2\,z_5\,z_8 + 4\,z_6\,z_8 - 2\,z_1\,z_6\,z_8 - z_2\,z_6\,z_8 -
2\,z_1\,z_7\,z_8 - 2\,z_1^2\,z_7\,z_8 -
   3\,z_2\,z_7\,z_8 - 2\,z_1\,z_2\,z_7\,z_8 + 2\,z_3\,z_7\,z_8 +
z_5\,z_7\,z_8 - z_6\,z_7\,z_8 - 2\,z_8^2 + 2\,z_1\,z_8^2 -
   z_2\,z_8^2 + z_2^2\,z_8^2 - z_3\,z_8^2 - z_4\,z_8^2 + z_5\,z_8^2 -
3\,z_6\,z_8^2 + z_1\,z_6\,z_8^2 -
   2\,z_7\,z_8^2 + z_1\,z_7\,z_8^2 + z_2\,z_7\,z_8^2 + z_7^2\,z_8^2 +
2\,z_8^3 + z_1\,z_8^3 + z_1^2\,z_8^3 +
   z_2\,z_8^3 - z_3\,z_8^3 - z_6\,z_8^3 + 2\,z_7\,z_8^3 + z_8^4 -
z_1\,z_8^4 - z_8^5\\ $

$\bchi_{2, 1, 0, 0, 0, 0, 0, 2} =
  z_1 + 2\,z_1^2 + z_1^3 + z_2 + 2\,z_1\,z_2 + z_1^2\,z_2 + z_2^2 +
z_1\,z_2^2 + z_1\,z_2\,z_3 + z_3^2 - z_4 -
   2\,z_1\,z_4 + z_5 + z_1\,z_5 - z_3\,z_5 - z_1\,z_6 - z_1^2\,z_6 -
z_1\,z_2\,z_6 + z_3\,z_6 + z_7 + 4\,z_1\,z_7 +
   4\,z_1^2\,z_7 + z_1^3\,z_7 + z_2\,z_7 + 2\,z_1\,z_2\,z_7 -
z_1^2\,z_2\,z_7 + z_2^2\,z_7 - z_1\,z_3\,z_7 +
   z_2\,z_3\,z_7 - z_4\,z_7 + z_1\,z_5\,z_7 - z_6\,z_7 - 2\,z_1\,z_6\,z_7 +
2\,z_7^2 + 3\,z_1\,z_7^2 + z_1^2\,z_7^2 -
   z_6\,z_7^2 + z_7^3 + z_8 + 4\,z_1\,z_8 + 2\,z_1^2\,z_8 - z_1^3\,z_8 -
z_1^2\,z_2\,z_8 + z_2^2\,z_8 -
   z_1\,z_2^2\,z_8 - 2\,z_3\,z_8 + z_1\,z_3\,z_8 - z_1^2\,z_3\,z_8 -
z_2\,z_3\,z_8 + z_3^2\,z_8 - 2\,z_4\,z_8 +
   z_1\,z_4\,z_8 + 2\,z_5\,z_8 + z_1\,z_5\,z_8 + z_2\,z_5\,z_8 -
3\,z_6\,z_8 - 2\,z_1\,z_6\,z_8 + z_1^2\,z_6\,z_8 -
   2\,z_2\,z_6\,z_8 + z_3\,z_6\,z_8 - z_6^2\,z_8 + 4\,z_7\,z_8 +
6\,z_1\,z_7\,z_8 - z_1^2\,z_7\,z_8 - 2\,z_3\,z_7\,z_8 +
   2\,z_5\,z_7\,z_8 - 2\,z_6\,z_7\,z_8 + 3\,z_7^2\,z_8 + 3\,z_8^2 -
2\,z_1\,z_8^2 - 3\,z_1^2\,z_8^2 + z_2\,z_8^2 -
   2\,z_1\,z_2\,z_8^2 + z_1^2\,z_2\,z_8^2 - z_3\,z_8^2 + z_1\,z_3\,z_8^2 -
z_2\,z_3\,z_8^2 + z_4\,z_8^2 +
   z_5\,z_8^2 - z_1\,z_5\,z_8^2 - z_6\,z_8^2 + z_1\,z_6\,z_8^2 +
3\,z_7\,z_8^2 - 5\,z_1\,z_7\,z_8^2 -
   z_1^2\,z_7\,z_8^2 + 2\,z_2\,z_7\,z_8^2 + z_6\,z_7\,z_8^2 - z_7^2\,z_8^2 -
2\,z_8^3 - 4\,z_1\,z_8^3 +
   z_1^2\,z_8^3 + z_2\,z_8^3 + z_3\,z_8^3 - z_5\,z_8^3 + z_6\,z_8^3 -
3\,z_7\,z_8^3 - 2\,z_8^4 + 2\,z_1\,z_8^4 -
   z_2\,z_8^4 + z_8^5\\ $

$\bchi_{2, 2, 0, 0, 0, 0, 0, 0} =
  -z_1 + z_1^3 + z_1^4 - z_2 - 2\,z_1\,z_2 - 2\,z_1^2\,z_2 - z_1^3\,z_2 -
z_2^2 + z_1^2\,z_2^2 - z_1\,z_3 -
   2\,z_1^2\,z_3 + z_2\,z_3 - z_2^2\,z_3 + z_4 + z_1\,z_4 - z_1^2\,z_4 +
z_2\,z_4 + z_3\,z_4 - z_5 - z_1\,z_5 +
   z_1^2\,z_5 - z_2\,z_5 - z_1\,z_2\,z_5 - 3\,z_1^2\,z_6 - z_1^3\,z_6 +
2\,z_2\,z_6 + z_1\,z_2\,z_6 + z_3\,z_6 +
   2\,z_1\,z_3\,z_6 + z_4\,z_6 + z_6^2 + z_1\,z_6^2 - z_7 - z_1\,z_7 +
2\,z_1^2\,z_7 + 2\,z_1^3\,z_7 - 2\,z_2\,z_7 -
   z_1\,z_2\,z_7 - z_1^2\,z_2\,z_7 - z_3\,z_7 - 3\,z_1\,z_3\,z_7 -
z_2\,z_3\,z_7 + z_4\,z_7 - z_5\,z_7 + z_1\,z_5\,z_7 -
   z_6\,z_7 - 3\,z_1\,z_6\,z_7 - z_7^2 + z_1\,z_7^2 + z_1^2\,z_7^2 -
z_3\,z_7^2 - z_6\,z_7^2 - 2\,z_8 -
   2\,z_1\,z_8 + 3\,z_1^2\,z_8 - z_1^4\,z_8 - 3\,z_2\,z_8 + z_1\,z_2\,z_8 -
z_3\,z_8 - 3\,z_1\,z_3\,z_8 +
   3\,z_1^2\,z_3\,z_8 - z_3^2\,z_8 + z_4\,z_8 - 2\,z_5\,z_8 +
z_1\,z_5\,z_8 - z_2\,z_5\,z_8 - z_1\,z_6\,z_8 +
   2\,z_1^2\,z_6\,z_8 + 2\,z_2\,z_6\,z_8 - z_3\,z_6\,z_8 - 3\,z_7\,z_8 +
z_1\,z_7\,z_8 + z_1^2\,z_7\,z_8 - z_2\,z_7\,z_8 -
   z_1\,z_2\,z_7\,z_8 - 3\,z_3\,z_7\,z_8 - 2\,z_6\,z_7\,z_8 - 2\,z_8^2 +
2\,z_1\,z_8^2 - 4\,z_1^2\,z_8^2 +
   z_1^3\,z_8^2 + z_2\,z_8^2 + 2\,z_1\,z_2\,z_8^2 + z_3\,z_8^2 +
3\,z_6\,z_8^2 - 3\,z_1\,z_7\,z_8^2 + z_8^3 +
   z_2\,z_8^3 - z_7\,z_8^3 + z_8^4\\ $

$\bchi_{3, 0, 0, 0, 0, 0, 1, 1} =
  z_1^2\,z_2 + z_1^3\,z_2 - 2\,z_1\,z_2\,z_3 - z_3^2 + z_4 + z_1\,z_4 -
z_1\,z_5 - z_1^2\,z_5 + 2\,z_3\,z_5 +
   z_1\,z_6 - z_1^3\,z_6 - 2\,z_1\,z_2\,z_6 - 2\,z_3\,z_6 +
2\,z_1\,z_3\,z_6 - z_4\,z_6 + 2\,z_5\,z_6 - z_6^2 +
   z_1\,z_6^2 - z_7 - z_1\,z_7 + z_2\,z_7 + 2\,z_1\,z_2\,z_7 - z_1\,z_3\,z_7
+ 2\,z_4\,z_7 - z_5\,z_7 + z_6\,z_7 -
   z_2\,z_6\,z_7 - 2\,z_7^2 - z_1\,z_7^2 + z_2\,z_7^2 - z_7^3 - z_8 -
z_1\,z_8 + 2\,z_1^2\,z_8 + z_1^3\,z_8 -
   z_1^4\,z_8 + 2\,z_1\,z_2\,z_8 - 2\,z_1^2\,z_2\,z_8 + z_1\,z_2^2\,z_8 +
z_3\,z_8 - 4\,z_1\,z_3\,z_8 +
   3\,z_1^2\,z_3\,z_8 + z_2\,z_3\,z_8 - z_3^2\,z_8 + 2\,z_4\,z_8 -
2\,z_1\,z_4\,z_8 - 2\,z_5\,z_8 + 2\,z_1\,z_5\,z_8 -
   z_2\,z_5\,z_8 + 2\,z_6\,z_8 - 4\,z_1\,z_6\,z_8 + 2\,z_1^2\,z_6\,z_8 +
z_2\,z_6\,z_8 - z_3\,z_6\,z_8 - 4\,z_7\,z_8 +
   z_1\,z_7\,z_8 + z_1^3\,z_7\,z_8 + 3\,z_2\,z_7\,z_8 + z_3\,z_7\,z_8 -
2\,z_1\,z_3\,z_7\,z_8 + z_4\,z_7\,z_8 -
   2\,z_5\,z_7\,z_8 + 2\,z_6\,z_7\,z_8 - z_1\,z_6\,z_7\,z_8 - 3\,z_7^2\,z_8
+ z_1\,z_7^2\,z_8 + z_2\,z_7^2\,z_8 -
   2\,z_8^2 + 5\,z_1\,z_8^2 - 4\,z_1^2\,z_8^2 + z_1^3\,z_8^2 + z_2\,z_8^2 -
z_1\,z_2\,z_8^2 - z_1^2\,z_2\,z_8^2 -
   z_2^2\,z_8^2 + z_2\,z_3\,z_8^2 - z_4\,z_8^2 + z_1\,z_5\,z_8^2 -
z_7\,z_8^2 + 3\,z_1\,z_7\,z_8^2 -
   z_1^2\,z_7\,z_8^2 - z_2\,z_7\,z_8^2 + 2\,z_8^3 - 3\,z_1\,z_8^3 +
z_1^2\,z_8^3 - z_2\,z_8^3 + z_1\,z_2\,z_8^3 +
   2\,z_7\,z_8^3 - z_1\,z_7\,z_8^3\\ $

$\bchi_{3, 0, 0, 0, 0, 1, 0, 0} =
  -(z_1\,z_2) - z_1^2\,z_2 - z_2^3 + z_2\,z_3 + z_4 + z_1\,z_4 +
2\,z_2\,z_4 - z_5 - z_1\,z_5 - 2\,z_2\,z_5 -
   z_3\,z_5 + z_6 + z_1\,z_6 + z_1^2\,z_6 + z_1^3\,z_6 + 3\,z_2\,z_6 +
2\,z_1\,z_2\,z_6 - 2\,z_1\,z_3\,z_6 + z_4\,z_6 -
   z_5\,z_6 - z_1\,z_6^2 - z_7 - z_1\,z_7 - z_2\,z_7 - z_1\,z_2\,z_7 -
z_1^2\,z_2\,z_7 - z_2^2\,z_7 + z_3\,z_7 +
   z_2\,z_3\,z_7 + 2\,z_4\,z_7 - 2\,z_5\,z_7 + z_1\,z_5\,z_7 + 2\,z_6\,z_7 +
z_1\,z_6\,z_7 + z_2\,z_6\,z_7 - 2\,z_7^2 -
   z_1\,z_7^2 - z_2\,z_7^2 + z_3\,z_7^2 - z_7^3 - 2\,z_8 - 2\,z_1\,z_8 -
4\,z_2\,z_8 - z_2^2\,z_8 +
   z_1\,z_2^2\,z_8 + z_4\,z_8 - z_1\,z_4\,z_8 - z_5\,z_8 + z_1\,z_5\,z_8 -
z_2\,z_5\,z_8 + z_6\,z_8 + 2\,z_1\,z_6\,z_8 -
   z_1^2\,z_6\,z_8 + 3\,z_2\,z_6\,z_8 - z_3\,z_6\,z_8 - 7\,z_7\,z_8 -
z_1\,z_7\,z_8 - 5\,z_2\,z_7\,z_8 + z_6\,z_7\,z_8 -
   5\,z_7^2\,z_8 + z_1\,z_7^2\,z_8 - 2\,z_8^2 + 2\,z_1\,z_8^2 +
z_1\,z_2\,z_8^2 - z_4\,z_8^2 + z_5\,z_8^2 +
   z_6\,z_8^2 - 3\,z_1\,z_6\,z_8^2 - z_7\,z_8^2 + 2\,z_1\,z_7\,z_8^2 +
z_7^2\,z_8^2 + 3\,z_8^3 + z_1\,z_8^3 +
   2\,z_2\,z_8^3 - z_6\,z_8^3 + 4\,z_7\,z_8^3 + z_8^4 - z_1\,z_8^4 - z_8^5\\ $

$\bchi_{4, 0, 0, 0, 0, 0, 0, 1} =
  z_1 + 2\,z_1^2 + z_1^3 - z_1^2\,z_2 - z_1^3\,z_2 - z_1\,z_2^2 - z_3 -
z_1\,z_3 - z_2\,z_3 + 2\,z_1\,z_2\,z_3 +
   z_3^2 - z_4 - z_1\,z_4 - z_2\,z_4 + z_5 + 2\,z_1\,z_5 + z_1^2\,z_5 +
z_2\,z_5 - z_3\,z_5 - z_6 - 2\,z_1\,z_6 -
   z_1^2\,z_6 - z_2\,z_6 + z_3\,z_6 + z_7 + 3\,z_1\,z_7 + 2\,z_1^2\,z_7 -
z_1\,z_2\,z_7 - z_3\,z_7 + z_1\,z_3\,z_7 -
   2\,z_4\,z_7 + 2\,z_5\,z_7 - 2\,z_6\,z_7 - z_1\,z_6\,z_7 + 2\,z_7^2 +
2\,z_1\,z_7^2 + z_7^3 + 2\,z_8 +
   3\,z_1\,z_8 - z_1^2\,z_8 - z_1^3\,z_8 + z_1^4\,z_8 + z_2\,z_8 -
3\,z_1\,z_2\,z_8 + 2\,z_1^2\,z_2\,z_8 -
   2\,z_3\,z_8 + 3\,z_1\,z_3\,z_8 - 3\,z_1^2\,z_3\,z_8 + z_2\,z_3\,z_8 +
z_3^2\,z_8 - 2\,z_4\,z_8 + 2\,z_1\,z_4\,z_8 +
   3\,z_5\,z_8 - 3\,z_1\,z_5\,z_8 - z_2\,z_5\,z_8 - 3\,z_6\,z_8 -
z_1^2\,z_6\,z_8 + z_2\,z_6\,z_8 + 2\,z_3\,z_6\,z_8 +
   z_6^2\,z_8 + 6\,z_7\,z_8 + z_1\,z_7\,z_8 - z_1^2\,z_7\,z_8 +
z_1\,z_2\,z_7\,z_8 - z_5\,z_7\,z_8 - z_6\,z_7\,z_8 +
   4\,z_7^2\,z_8 - z_1\,z_7^2\,z_8 + 3\,z_8^2 - 7\,z_1\,z_8^2 +
z_1^2\,z_8^2 - z_1^3\,z_8^2 - 2\,z_2\,z_8^2 +
   3\,z_1\,z_2\,z_8^2 - z_2^2\,z_8^2 + z_3\,z_8^2 + z_4\,z_8^2 -
2\,z_5\,z_8^2 - z_6\,z_8^2 + 3\,z_1\,z_6\,z_8^2 +
   z_7\,z_8^2 - 4\,z_1\,z_7\,z_8^2 - z_2\,z_7\,z_8^2 - z_7^2\,z_8^2 -
4\,z_8^3 + z_1\,z_8^3 - z_1^2\,z_8^3 +
   z_3\,z_8^3 + 2\,z_6\,z_8^3 - 3\,z_7\,z_8^3 - z_8^4 + 2\,z_1\,z_8^4 +
z_8^5\\ $

$\bchi_{0, 0, 0, 0, 1, 1, 1, 0} =
  -z_1^2 - 2\,z_1^3 - z_1^4 - z_1\,z_2 - z_1^2\,z_2 + z_1\,z_3 + z_1^2\,z_3
+ z_1\,z_4 + z_1^2\,z_4 + z_2\,z_4 -
   z_3\,z_4 - z_1\,z_5 - z_1^2\,z_5 + z_4\,z_5 + 2\,z_1\,z_6 + 4\,z_1^2\,z_6
+ 2\,z_1^3\,z_6 + z_1\,z_2\,z_6 -
   2\,z_3\,z_6 - 4\,z_1\,z_3\,z_6 - z_2\,z_3\,z_6 + z_5\,z_6 +
2\,z_1\,z_5\,z_6 - 2\,z_6^2 - 3\,z_1\,z_6^2 -
   z_2\,z_6^2 - 2\,z_1\,z_7 - 4\,z_1^2\,z_7 - 2\,z_1^3\,z_7 - z_2\,z_7 -
2\,z_1\,z_2\,z_7 + 2\,z_3\,z_7 +
   3\,z_1\,z_3\,z_7 + z_2\,z_3\,z_7 - z_4\,z_7 - 2\,z_5\,z_7 -
3\,z_1\,z_5\,z_7 + 2\,z_1\,z_6\,z_7 + z_2\,z_6\,z_7 -
   z_3\,z_6\,z_7 + z_5\,z_6\,z_7 + z_7^2 - z_1\,z_7^2 - z_1^2\,z_7^2 -
z_2\,z_7^2 + 2\,z_3\,z_7^2 - z_4\,z_7^2 -
   2\,z_5\,z_7^2 - z_1\,z_6\,z_7^2 + 2\,z_7^3 + z_1\,z_7^3 + z_7^4 -
2\,z_1\,z_8 - z_1^2\,z_8 + 2\,z_1^3\,z_8 +
   z_1^4\,z_8 + 2\,z_2^2\,z_8 + z_3\,z_8 + 2\,z_1\,z_3\,z_8 -
z_1^2\,z_3\,z_8 - z_1\,z_2\,z_3\,z_8 - z_4\,z_8 -
   z_1\,z_4\,z_8 + z_2\,z_4\,z_8 - z_5\,z_8 - z_1\,z_5\,z_8 + z_2\,z_5\,z_8
+ z_3\,z_5\,z_8 - z_5^2\,z_8 + 2\,z_6\,z_8 +
   3\,z_1\,z_6\,z_8 - 3\,z_1^2\,z_6\,z_8 + 3\,z_2\,z_6\,z_8 +
z_1\,z_2\,z_6\,z_8 - 2\,z_3\,z_6\,z_8 + 2\,z_5\,z_6\,z_8 -
   z_6^2\,z_8 - 2\,z_1\,z_7\,z_8 + 3\,z_1^2\,z_7\,z_8 - z_1^3\,z_7\,z_8 -
3\,z_2\,z_7\,z_8 + 4\,z_3\,z_7\,z_8 +
   3\,z_1\,z_3\,z_7\,z_8 + z_2\,z_3\,z_7\,z_8 - 2\,z_4\,z_7\,z_8 -
2\,z_5\,z_7\,z_8 + 3\,z_6\,z_7\,z_8 + 2\,z_1\,z_6\,z_7\,z_8 +
   z_2\,z_6\,z_7\,z_8 + 3\,z_7^2\,z_8 + 3\,z_1\,z_7^2\,z_8 -
2\,z_2\,z_7^2\,z_8 + z_3\,z_7^2\,z_8 - z_6\,z_7^2\,z_8 +
   3\,z_7^3\,z_8 + z_1\,z_8^2 + 4\,z_1^2\,z_8^2 - z_1^3\,z_8^2 -
z_2\,z_8^2 - z_1\,z_2\,z_8^2 +
   z_1^2\,z_2\,z_8^2 + z_3\,z_8^2 - z_2\,z_3\,z_8^2 - z_4\,z_8^2 +
z_1\,z_5\,z_8^2 - z_2\,z_5\,z_8^2 + z_6\,z_8^2 -
   5\,z_1\,z_6\,z_8^2 - z_2\,z_6\,z_8^2 + 6\,z_1\,z_7\,z_8^2 -
3\,z_2\,z_7\,z_8^2 - z_1\,z_2\,z_7\,z_8^2 +
   z_3\,z_7\,z_8^2 + z_5\,z_7\,z_8^2 - z_6\,z_7\,z_8^2 + 2\,z_7^2\,z_8^2 -
z_1\,z_7^2\,z_8^2 + 3\,z_1\,z_8^3 -
   3\,z_1^2\,z_8^3 - z_1\,z_3\,z_8^3 + z_4\,z_8^3 + z_5\,z_8^3 -
2\,z_6\,z_8^3 + z_1\,z_6\,z_8^3 + z_7\,z_8^3 -
   z_1\,z_7\,z_8^3 + z_2\,z_7\,z_8^3 - z_7^2\,z_8^3 - 2\,z_1\,z_8^4 +
z_1^2\,z_8^4 + z_2\,z_8^4 - z_3\,z_8^4 -
   z_7\,z_8^4\\ $

$\bchi_{0, 0, 0, 0, 2, 0, 0, 1} =
  z_1\,z_2 + z_1^2\,z_2 - z_2\,z_4 + z_1\,z_5 + z_1^2\,z_5 - z_4\,z_5 +
z_1\,z_6 - z_1^3\,z_6 - z_2\,z_6 -
   z_1\,z_2\,z_6 + 3\,z_1\,z_3\,z_6 + z_2\,z_3\,z_6 - 2\,z_4\,z_6 -
z_5\,z_6 - z_1\,z_5\,z_6 + z_1\,z_6^2 + z_2\,z_6^2 -
   z_1\,z_7 + z_1^3\,z_7 + z_2\,z_7 - z_2^2\,z_7 - 2\,z_1\,z_3\,z_7 -
z_2\,z_3\,z_7 + 2\,z_4\,z_7 + z_5\,z_7 -
   z_2\,z_5\,z_7 + 2\,z_6\,z_7 - 2\,z_2\,z_6\,z_7 - 2\,z_7^2 + z_2\,z_7^2 -
z_1\,z_2\,z_7^2 - z_7^3 + z_2\,z_8 -
   2\,z_1\,z_2\,z_8 - z_1^2\,z_2\,z_8 - z_2^2\,z_8 - 2\,z_1\,z_3\,z_8 +
z_4\,z_8 + z_1\,z_4\,z_8 + z_5\,z_8 -
   2\,z_1\,z_5\,z_8 - z_1^2\,z_5\,z_8 + z_3\,z_5\,z_8 + z_5^2\,z_8 +
2\,z_6\,z_8 - z_1\,z_6\,z_8 + z_1^2\,z_6\,z_8 -
   2\,z_2\,z_6\,z_8 + z_1\,z_2\,z_6\,z_8 - z_4\,z_6\,z_8 + z_5\,z_6\,z_8 -
z_1\,z_6^2\,z_8 - 4\,z_7\,z_8 - z_1\,z_7\,z_8 -
   z_1^2\,z_7\,z_8 + z_2\,z_7\,z_8 - z_1\,z_2\,z_7\,z_8 - z_1\,z_3\,z_7\,z_8
+ 2\,z_4\,z_7\,z_8 - z_5\,z_7\,z_8 +
   z_1\,z_5\,z_7\,z_8 + z_6\,z_7\,z_8 + 2\,z_1\,z_6\,z_7\,z_8 -
5\,z_7^2\,z_8 - 2\,z_1\,z_7^2\,z_8 + z_3\,z_7^2\,z_8 +
   z_6\,z_7^2\,z_8 - z_7^3\,z_8 - 2\,z_8^2 - 2\,z_1\,z_8^2 - z_2\,z_8^2 +
z_3\,z_8^2 + z_1\,z_3\,z_8^2 +
   z_2\,z_3\,z_8^2 - z_5\,z_8^2 + 4\,z_1\,z_6\,z_8^2 + z_2\,z_6\,z_8^2 -
z_3\,z_6\,z_8^2 - z_6^2\,z_8^2 -
   4\,z_7\,z_8^2 - 2\,z_1\,z_7\,z_8^2 + 2\,z_3\,z_7\,z_8^2 +
3\,z_6\,z_7\,z_8^2 - 2\,z_7^2\,z_8^2 + z_1\,z_2\,z_8^3 +
   z_3\,z_8^3 - z_4\,z_8^3 + z_5\,z_8^3 - z_1\,z_6\,z_8^3 + 3\,z_7\,z_8^3 +
z_7^2\,z_8^3 + 3\,z_8^4 +
   z_1\,z_8^4 - z_3\,z_8^4 - 2\,z_6\,z_8^4 + 3\,z_7\,z_8^4 - z_8^6\\ $

$\bchi_{0, 0, 0, 1, 0, 1, 0, 1} =
  z_1\,z_2^2 + z_2\,z_3 - z_1\,z_4 - z_1^2\,z_4 + z_3\,z_4 + z_1\,z_5 +
z_1^2\,z_5 + z_1\,z_2\,z_5 - z_4\,z_5 +
   z_2\,z_6 + z_4\,z_6 - z_1\,z_7 + z_1^3\,z_7 - z_2\,z_7 -
2\,z_1\,z_2\,z_7 - z_1^2\,z_2\,z_7 + z_1\,z_2^2\,z_7 -
   z_1\,z_3\,z_7 + 2\,z_2\,z_3\,z_7 + z_3^2\,z_7 - z_4\,z_7 -
2\,z_1\,z_4\,z_7 + z_5\,z_7 + 2\,z_1\,z_5\,z_7 -
   2\,z_6\,z_7 - 3\,z_1\,z_6\,z_7 - z_1^2\,z_6\,z_7 + z_3\,z_6\,z_7 + z_7^2
+ z_1^2\,z_7^2 - z_2\,z_7^2 -
   z_1\,z_2\,z_7^2 - z_3\,z_7^2 + z_5\,z_7^2 - 2\,z_6\,z_7^2 + z_7^3 -
2\,z_2\,z_8 - 2\,z_1^2\,z_2\,z_8 - z_3\,z_8 -
   z_1\,z_3\,z_8 + z_2\,z_3\,z_8 - 2\,z_4\,z_8 - z_2\,z_4\,z_8 +
2\,z_1\,z_5\,z_8 - z_3\,z_5\,z_8 - 2\,z_6\,z_8 -
   z_1\,z_6\,z_8 - z_1^2\,z_6\,z_8 + 2\,z_2\,z_6\,z_8 - z_1\,z_2\,z_6\,z_8 +
z_4\,z_6\,z_8 + z_6^2\,z_8 + 3\,z_7\,z_8 -
   z_1\,z_7\,z_8 + z_1^2\,z_7\,z_8 - z_1^3\,z_7\,z_8 - 4\,z_2\,z_7\,z_8 +
2\,z_1\,z_2\,z_7\,z_8 - 2\,z_3\,z_7\,z_8 +
   z_1\,z_3\,z_7\,z_8 - z_2\,z_3\,z_7\,z_8 - 2\,z_4\,z_7\,z_8 +
2\,z_5\,z_7\,z_8 - 4\,z_6\,z_7\,z_8 + 2\,z_1\,z_6\,z_7\,z_8 +
   5\,z_7^2\,z_8 + z_1\,z_7^2\,z_8 + z_1^2\,z_7^2\,z_8 + z_2\,z_7^2\,z_8 -
z_3\,z_7^2\,z_8 + 2\,z_7^3\,z_8 +
   2\,z_8^2 - z_1\,z_8^2 + 2\,z_1^2\,z_8^2 + z_1^3\,z_8^2 - z_2\,z_8^2 +
3\,z_1\,z_2\,z_8^2 - z_2^2\,z_8^2 +
   z_3\,z_8^2 - z_1\,z_3\,z_8^2 + z_1\,z_4\,z_8^2 - z_1\,z_5\,z_8^2 -
2\,z_1\,z_6\,z_8^2 - z_2\,z_6\,z_8^2 +
   z_3\,z_6\,z_8^2 + 5\,z_7\,z_8^2 + 2\,z_1\,z_7\,z_8^2 + z_1^2\,z_7\,z_8^2
+ 2\,z_2\,z_7\,z_8^2 + z_3\,z_7\,z_8^2 -
   z_5\,z_7\,z_8^2 - z_6\,z_7\,z_8^2 + 5\,z_7^2\,z_8^2 -
2\,z_1\,z_7^2\,z_8^2 - z_8^3 + 4\,z_1\,z_8^3 -
   3\,z_1^2\,z_8^3 + 2\,z_2\,z_8^3 + z_3\,z_8^3 + z_4\,z_8^3 - z_5\,z_8^3 +
z_6\,z_8^3 + z_1\,z_6\,z_8^3 -
   z_1\,z_7\,z_8^3 - z_7^2\,z_8^3 - z_8^4 - 3\,z_1\,z_8^4 - z_2\,z_8^4 +
z_6\,z_8^4 - 4\,z_7\,z_8^4 - z_8^5 +
   z_1\,z_8^5 + z_8^6\\ $

$\bchi_{0, 0, 1, 0, 0, 2, 0, 0} =
  z_1^3 + z_1^4 + z_1\,z_2 + 2\,z_1^2\,z_2 - z_1\,z_2^2 + z_1\,z_3 -
2\,z_2\,z_3 - z_1\,z_2\,z_3 + z_2^2\,z_3 -
   z_3^2 + z_1\,z_3^2 - z_1^2\,z_4 - z_2\,z_4 - 2\,z_3\,z_4 + z_1\,z_5 +
2\,z_1^2\,z_5 + z_2\,z_5 - z_1\,z_2\,z_5 -
   z_4\,z_5 + z_5^2 - z_1^2\,z_6 - z_2\,z_6 - z_1\,z_2\,z_6 -
z_1^2\,z_2\,z_6 - z_2^2\,z_6 - 2\,z_3\,z_6 -
   z_1\,z_3\,z_6 + z_2\,z_3\,z_6 + z_4\,z_6 - 2\,z_5\,z_6 + z_6^2 +
z_3\,z_6^2 + z_1\,z_7 + 2\,z_1^2\,z_7 +
   2\,z_1^3\,z_7 + z_1\,z_2\,z_7 + z_2^2\,z_7 + z_1\,z_2^2\,z_7 +
z_1\,z_3\,z_7 + z_1^2\,z_3\,z_7 - 2\,z_2\,z_3\,z_7 -
   z_3^2\,z_7 - z_4\,z_7 - 3\,z_1\,z_4\,z_7 + z_5\,z_7 + 3\,z_1\,z_5\,z_7 -
z_3\,z_5\,z_7 - z_6\,z_7 - 3\,z_1\,z_6\,z_7 -
   z_1^2\,z_6\,z_7 - z_1\,z_2\,z_6\,z_7 + 2\,z_1\,z_7^2 + 2\,z_1^2\,z_7^2 +
z_2^2\,z_7^2 - z_4\,z_7^2 + z_5\,z_7^2 -
   z_6\,z_7^2 + z_1\,z_7^3 + 2\,z_1^2\,z_8 + z_2\,z_8 - 2\,z_1\,z_2\,z_8 +
z_2^2\,z_8 + 2\,z_1\,z_2^2\,z_8 +
   2\,z_3\,z_8 - 2\,z_1\,z_3\,z_8 - z_1^2\,z_3\,z_8 - z_2\,z_3\,z_8 +
2\,z_3^2\,z_8 - z_4\,z_8 - 2\,z_1\,z_4\,z_8 +
   2\,z_5\,z_8 - z_1\,z_5\,z_8 + z_1\,z_2\,z_5\,z_8 - 2\,z_6\,z_8 -
3\,z_1\,z_6\,z_8 - 4\,z_1^2\,z_6\,z_8 +
   z_1\,z_2\,z_6\,z_8 - z_2^2\,z_6\,z_8 + z_3\,z_6\,z_8 + 2\,z_4\,z_6\,z_8 -
2\,z_5\,z_6\,z_8 + 2\,z_6^2\,z_8 +
   z_1\,z_6^2\,z_8 + z_7\,z_8 + 5\,z_1\,z_7\,z_8 + z_1^2\,z_7\,z_8 -
2\,z_1^3\,z_7\,z_8 + z_2\,z_7\,z_8 -
   2\,z_1\,z_2\,z_7\,z_8 + z_1^2\,z_2\,z_7\,z_8 + 3\,z_2^2\,z_7\,z_8 +
2\,z_3\,z_7\,z_8 + 2\,z_1\,z_3\,z_7\,z_8 -
   z_2\,z_3\,z_7\,z_8 - 3\,z_4\,z_7\,z_8 + 4\,z_5\,z_7\,z_8 -
z_1\,z_5\,z_7\,z_8 - 4\,z_6\,z_7\,z_8 - 2\,z_1\,z_6\,z_7\,z_8 +
   2\,z_7^2\,z_8 + 5\,z_1\,z_7^2\,z_8 - z_1^2\,z_7^2\,z_8 + z_7^3\,z_8 +
z_8^2 + 2\,z_1\,z_8^2 - z_1^2\,z_8^2 -
   3\,z_1^3\,z_8^2 - z_2\,z_8^2 + z_2^2\,z_8^2 - z_1\,z_2^2\,z_8^2 -
2\,z_3\,z_8^2 + 3\,z_1\,z_3\,z_8^2 -
   z_1^2\,z_3\,z_8^2 + z_3^2\,z_8^2 - z_4\,z_8^2 + 2\,z_1\,z_4\,z_8^2 -
z_1\,z_5\,z_8^2 - 2\,z_6\,z_8^2 +
   3\,z_1\,z_6\,z_8^2 + 2\,z_1^2\,z_6\,z_8^2 + 3\,z_7\,z_8^2 -
2\,z_1\,z_7\,z_8^2 - z_1\,z_2\,z_7\,z_8^2 +
   z_3\,z_7\,z_8^2 - 2\,z_6\,z_7\,z_8^2 + 2\,z_7^2\,z_8^2 - 4\,z_1\,z_8^3 +
z_1^3\,z_8^3 - z_2^2\,z_8^3 +
   z_4\,z_8^3 - z_5\,z_8^3 + 3\,z_6\,z_8^3 - 2\,z_7\,z_8^3 -
2\,z_1\,z_7\,z_8^3 - 2\,z_8^4 + 2\,z_1\,z_8^4 -
   z_7\,z_8^4 + z_8^5\\ $

$\bchi_{0, 0, 1, 0, 1, 0, 1, 0} =
  -z_1^2 - z_1^3 - 2\,z_1\,z_2 - 3\,z_1^2\,z_2 - z_1^3\,z_2 - z_2^2 -
z_1\,z_2^2 - z_2^3 - z_1^2\,z_3 +
   2\,z_2\,z_3 + z_1\,z_2\,z_3 + z_3^2 + z_1\,z_4 + 3\,z_2\,z_4 +
z_1\,z_2\,z_4 + z_3\,z_4 - 2\,z_2\,z_5 -
   z_1\,z_2\,z_5 - z_2^2\,z_5 + z_4\,z_5 - z_5^2 - z_1^2\,z_6 + 2\,z_2\,z_6
+ 2\,z_1\,z_2\,z_6 + 3\,z_3\,z_6 +
   2\,z_4\,z_6 - z_5\,z_6 + 2\,z_6^2 - z_1\,z_7 - z_1^2\,z_7 - z_1^3\,z_7 +
z_2\,z_7 - z_1\,z_2\,z_7 -
   z_1^2\,z_2\,z_7 - z_1\,z_2^2\,z_7 - 2\,z_3\,z_7 + z_1\,z_3\,z_7 -
z_1^2\,z_3\,z_7 + 2\,z_2\,z_3\,z_7 + z_3^2\,z_7 -
   z_4\,z_7 + 2\,z_1\,z_4\,z_7 + 4\,z_5\,z_7 + z_1\,z_5\,z_7 +
z_3\,z_5\,z_7 - z_6\,z_7 + 3\,z_1\,z_6\,z_7 +
   z_1^2\,z_6\,z_7 - z_2\,z_6\,z_7 - z_1\,z_2\,z_6\,z_7 + z_3\,z_6\,z_7 -
z_6^2\,z_7 - z_7^2 - 2\,z_1\,z_7^2 +
   2\,z_2\,z_7^2 + z_1\,z_2\,z_7^2 - 2\,z_3\,z_7^2 + z_4\,z_7^2 +
2\,z_5\,z_7^2 + z_6\,z_7^2 + z_1\,z_6\,z_7^2 -
   2\,z_7^3 - z_1\,z_7^3 + z_2\,z_7^3 - z_7^4 - z_1\,z_8 + z_1^2\,z_8 +
z_1^3\,z_8 + z_1^4\,z_8 - 3\,z_2\,z_8 +
   3\,z_1^2\,z_2\,z_8 - z_2^2\,z_8 + z_1\,z_2^2\,z_8 - 2\,z_3\,z_8 +
2\,z_1\,z_3\,z_8 - 2\,z_1^2\,z_3\,z_8 +
   z_2\,z_3\,z_8 + z_3^2\,z_8 - z_4\,z_8 + z_1\,z_4\,z_8 - z_3\,z_4\,z_8 -
z_1\,z_5\,z_8 + z_1^2\,z_5\,z_8 -
   2\,z_2\,z_5\,z_8 - z_3\,z_5\,z_8 - 3\,z_6\,z_8 + 3\,z_1\,z_6\,z_8 -
2\,z_1^2\,z_6\,z_8 + z_2\,z_6\,z_8 +
   z_1\,z_2\,z_6\,z_8 + z_2^2\,z_6\,z_8 + 2\,z_3\,z_6\,z_8 - z_4\,z_6\,z_8 -
z_5\,z_6\,z_8 + z_6^2\,z_8 + 2\,z_7\,z_8 -
   3\,z_1\,z_7\,z_8 + 4\,z_1^2\,z_7\,z_8 + z_1^3\,z_7\,z_8 +
2\,z_2\,z_7\,z_8 + z_1^2\,z_2\,z_7\,z_8 - z_2^2\,z_7\,z_8 -
   4\,z_3\,z_7\,z_8 - z_2\,z_3\,z_7\,z_8 - z_4\,z_7\,z_8 +
3\,z_5\,z_7\,z_8 - z_1\,z_5\,z_7\,z_8 - 4\,z_6\,z_7\,z_8 -
   z_1\,z_6\,z_7\,z_8 - 2\,z_2\,z_6\,z_7\,z_8 - z_1\,z_7^2\,z_8 +
5\,z_2\,z_7^2\,z_8 - z_3\,z_7^2\,z_8 + z_6\,z_7^2\,z_8 -
   2\,z_7^3\,z_8 + z_8^2 - z_1\,z_8^2 + 4\,z_1^2\,z_8^2 - z_1^3\,z_8^2 +
z_2\,z_8^2 + 2\,z_1\,z_2\,z_8^2 +
   z_1^2\,z_2\,z_8^2 - z_1\,z_3\,z_8^2 - 3\,z_2\,z_3\,z_8^2 -
z_1\,z_4\,z_8^2 + z_5\,z_8^2 - z_1\,z_5\,z_8^2 +
   z_2\,z_5\,z_8^2 + z_6\,z_8^2 - z_1\,z_6\,z_8^2 - z_1^2\,z_6\,z_8^2 -
z_2\,z_6\,z_8^2 + z_3\,z_6\,z_8^2 +
   z_6^2\,z_8^2 + 2\,z_7\,z_8^2 + z_1\,z_7\,z_8^2 + z_1^2\,z_7\,z_8^2 +
z_2\,z_7\,z_8^2 - 2\,z_1\,z_2\,z_7\,z_8^2 -
   2\,z_3\,z_7\,z_8^2 - z_5\,z_7\,z_8^2 - 4\,z_6\,z_7\,z_8^2 +
3\,z_7^2\,z_8^2 - z_8^3 + 2\,z_1\,z_8^3 -
   3\,z_1^2\,z_8^3 - z_1^3\,z_8^3 + z_2\,z_8^3 - 2\,z_1\,z_2\,z_8^3 +
z_2^2\,z_8^3 - z_3\,z_8^3 +
   2\,z_1\,z_3\,z_8^3 - z_6\,z_8^3 + 2\,z_1\,z_6\,z_8^3 - z_7\,z_8^3 -
2\,z_1\,z_7\,z_8^3 - 2\,z_2\,z_7\,z_8^3 +
   z_7^2\,z_8^3 - 2\,z_1\,z_8^4 + z_1^2\,z_8^4 + z_3\,z_8^4 + z_6\,z_8^4 -
z_7\,z_8^4 + z_1\,z_8^5\\ $

$\bchi_{0, 0, 1, 1, 0, 0, 0, 1} =
  z_1^2 + z_1^3 - z_1\,z_2 - z_1^2\,z_2 - z_2^2 + z_2^3 + z_1^2\,z_3 +
z_1^3\,z_3 + z_2\,z_3 + z_1\,z_2\,z_3 -
   z_3^2 - 2\,z_1\,z_3^2 - z_2\,z_3^2 - z_1\,z_4 - 2\,z_2\,z_4 + z_3\,z_4 -
z_1\,z_5 - z_1^2\,z_5 + z_2^2\,z_5 +
   z_1\,z_3\,z_5 - z_4\,z_5 + z_5^2 - 2\,z_1\,z_6 - 2\,z_1^2\,z_6 -
z_1^3\,z_6 - 2\,z_2\,z_6 - 2\,z_1\,z_2\,z_6 +
   z_1^2\,z_2\,z_6 - z_2\,z_3\,z_6 - z_4\,z_6 - z_1\,z_5\,z_6 - z_6^2 -
z_2\,z_6^2 + 3\,z_1\,z_7 + 3\,z_1^2\,z_7 +
   z_1^3\,z_7 + 2\,z_2\,z_7 + z_1\,z_2\,z_7 - z_3\,z_7 + z_1^2\,z_3\,z_7 -
z_3^2\,z_7 - z_1\,z_4\,z_7 - z_5\,z_7 +
   z_2\,z_5\,z_7 - 2\,z_1\,z_6\,z_7 - 2\,z_1^2\,z_6\,z_7 + z_6^2\,z_7 +
z_7^2 + 4\,z_1\,z_7^2 + 2\,z_1^2\,z_7^2 +
   2\,z_2\,z_7^2 - z_3\,z_7^2 - z_6\,z_7^2 + z_7^3 + z_1\,z_7^3 +
3\,z_1\,z_8 - z_1^3\,z_8 + 2\,z_2\,z_8 -
   z_1^2\,z_2\,z_8 - z_2^2\,z_8 - z_1\,z_2^2\,z_8 + 2\,z_1\,z_3\,z_8 -
z_1^2\,z_3\,z_8 + 4\,z_2\,z_3\,z_8 - z_3^2\,z_8 +
   2\,z_1\,z_4\,z_8 + z_3\,z_4\,z_8 - z_1\,z_5\,z_8 - z_1\,z_2\,z_5\,z_8 +
z_6\,z_8 + z_1\,z_6\,z_8 + z_1^2\,z_6\,z_8 +
   2\,z_2\,z_6\,z_8 + z_1\,z_2\,z_6\,z_8 + 2\,z_3\,z_6\,z_8 +
z_4\,z_6\,z_8 - 2\,z_5\,z_6\,z_8 + 2\,z_6^2\,z_8 +
   z_1\,z_6^2\,z_8 + z_7\,z_8 + 4\,z_1\,z_7\,z_8 - z_1^2\,z_7\,z_8 -
z_1^3\,z_7\,z_8 + 5\,z_2\,z_7\,z_8 -
   z_1\,z_2\,z_7\,z_8 - z_2^2\,z_7\,z_8 - 2\,z_3\,z_7\,z_8 +
z_1\,z_3\,z_7\,z_8 + 2\,z_4\,z_7\,z_8 + z_5\,z_7\,z_8 +
   2\,z_6\,z_7\,z_8 + z_1\,z_6\,z_7\,z_8 - z_1\,z_7^2\,z_8 +
z_2\,z_7^2\,z_8 - z_6\,z_7^2\,z_8 - z_7^3\,z_8 -
   4\,z_1\,z_8^2 - z_1^2\,z_8^2 - z_2\,z_8^2 + 2\,z_1\,z_2\,z_8^2 +
z_1^2\,z_2\,z_8^2 - z_2^2\,z_8^2 -
   z_3\,z_8^2 - 2\,z_1\,z_3\,z_8^2 - z_2\,z_3\,z_8^2 + z_4\,z_8^2 +
z_5\,z_8^2 - z_6\,z_8^2 + 4\,z_1\,z_6\,z_8^2 +
   z_1^2\,z_6\,z_8^2 - z_2\,z_6\,z_8^2 - z_3\,z_6\,z_8^2 - 4\,z_7\,z_8^2 -
8\,z_1\,z_7\,z_8^2 + z_1^2\,z_7\,z_8^2 -
   z_1\,z_2\,z_7\,z_8^2 - 3\,z_3\,z_7\,z_8^2 + z_5\,z_7\,z_8^2 -
2\,z_6\,z_7\,z_8^2 - 6\,z_7^2\,z_8^2 - 2\,z_8^3 -
   2\,z_1\,z_8^3 + z_1^2\,z_8^3 - z_1\,z_2\,z_8^3 + z_2^2\,z_8^3 -
z_3\,z_8^3 - z_4\,z_8^3 - z_6\,z_8^3 -
   z_1\,z_6\,z_8^3 - 2\,z_7\,z_8^3 - z_1\,z_7\,z_8^3 + z_7^2\,z_8^3 +
2\,z_8^4 + 2\,z_1\,z_8^4 + z_3\,z_8^4 +
   4\,z_7\,z_8^4 + z_8^5 - z_8^6\\ $

$\bchi_{0, 0, 2, 0, 0, 1, 0, 0} =
  z_1^2 + z_1^3 + 4\,z_1\,z_2 + 5\,z_1^2\,z_2 + 2\,z_1^3\,z_2 + z_2^2 +
z_1\,z_2^2 + z_2^3 - z_1\,z_3 +
   z_1^2\,z_3 + z_1^3\,z_3 - 3\,z_2\,z_3 - z_1\,z_2\,z_3 - 2\,z_2^2\,z_3 -
2\,z_1\,z_3^2 - z_1\,z_4 - 2\,z_2\,z_4 -
   z_1\,z_2\,z_4 + 2\,z_3\,z_4 + 2\,z_1\,z_5 + z_1^2\,z_5 + 3\,z_2\,z_5 +
2\,z_1\,z_2\,z_5 + z_2^2\,z_5 - 2\,z_3\,z_5 +
   z_1\,z_3\,z_5 - z_4\,z_5 + z_5^2 + z_1\,z_6 + 4\,z_1^2\,z_6 +
2\,z_1^3\,z_6 - 2\,z_2\,z_6 - 3\,z_1\,z_2\,z_6 -
   z_1^2\,z_2\,z_6 - z_2^2\,z_6 - 3\,z_1\,z_3\,z_6 + z_3^2\,z_6 - z_4\,z_6 -
z_1\,z_4\,z_6 + z_1\,z_5\,z_6 - z_6^2 -
   3\,z_1\,z_6^2 - z_1^2\,z_6^2 - z_2\,z_6^2 + z_3\,z_6^2 - z_1\,z_7 +
z_1^2\,z_7 + 2\,z_1^3\,z_7 + z_1^4\,z_7 -
   z_2\,z_7 + 4\,z_1\,z_2\,z_7 + 3\,z_1^2\,z_2\,z_7 + z_2^2\,z_7 -
3\,z_1\,z_3\,z_7 - 2\,z_1^2\,z_3\,z_7 -
   3\,z_2\,z_3\,z_7 - z_1\,z_2\,z_3\,z_7 + z_4\,z_7 + 2\,z_1\,z_4\,z_7 +
z_2\,z_4\,z_7 - z_5\,z_7 + z_1^2\,z_5\,z_7 -
   z_3\,z_5\,z_7 + z_6\,z_7 + 4\,z_1\,z_6\,z_7 + 2\,z_1^2\,z_6\,z_7 +
z_2\,z_6\,z_7 + z_1\,z_2\,z_6\,z_7 - 2\,z_3\,z_6\,z_7 -
   2\,z_1\,z_7^2 + z_1^3\,z_7^2 - 2\,z_2\,z_7^2 - 2\,z_1\,z_3\,z_7^2 +
z_4\,z_7^2 - z_5\,z_7^2 + z_6\,z_7^2 -
   z_1\,z_7^3 - z_2\,z_7^3 - z_1\,z_8 - 5\,z_1^2\,z_8 - z_1^3\,z_8 +
z_1^4\,z_8 + 2\,z_2\,z_8 - 3\,z_1^2\,z_2\,z_8 -
   2\,z_1^3\,z_2\,z_8 + 2\,z_2^2\,z_8 - 2\,z_1\,z_2^2\,z_8 +
z_1^2\,z_2^2\,z_8 + z_2^3\,z_8 + 3\,z_1\,z_3\,z_8 -
   3\,z_1^2\,z_3\,z_8 + 3\,z_1\,z_2\,z_3\,z_8 - z_2^2\,z_3\,z_8 -
2\,z_3^2\,z_8 + z_4\,z_8 + 2\,z_1\,z_4\,z_8 -
   z_1^2\,z_4\,z_8 - z_2\,z_4\,z_8 + 2\,z_3\,z_4\,z_8 - 3\,z_1\,z_5\,z_8 +
2\,z_2\,z_5\,z_8 - z_1\,z_2\,z_5\,z_8 +
   2\,z_6\,z_8 + 4\,z_1\,z_6\,z_8 - 4\,z_1^2\,z_6\,z_8 -
2\,z_1^3\,z_6\,z_8 - z_2\,z_6\,z_8 - z_2^2\,z_6\,z_8 -
   z_3\,z_6\,z_8 + 4\,z_1\,z_3\,z_6\,z_8 + z_1\,z_6^2\,z_8 - z_7\,z_8 -
5\,z_1\,z_7\,z_8 - 4\,z_1^2\,z_7\,z_8 -
   z_2\,z_7\,z_8 + 2\,z_1\,z_2\,z_7\,z_8 + 2\,z_2^2\,z_7\,z_8 -
3\,z_1\,z_3\,z_7\,z_8 + 3\,z_4\,z_7\,z_8 - 2\,z_5\,z_7\,z_8 -
   z_1\,z_5\,z_7\,z_8 + 4\,z_6\,z_7\,z_8 + z_1\,z_6\,z_7\,z_8 +
2\,z_2\,z_6\,z_7\,z_8 - 2\,z_7^2\,z_8 - 4\,z_1\,z_7^2\,z_8 -
   z_1^2\,z_7^2\,z_8 - 3\,z_2\,z_7^2\,z_8 - z_7^3\,z_8 - z_8^2 -
3\,z_1\,z_8^2 + 4\,z_1^2\,z_8^2 - z_1^4\,z_8^2 -
   z_2\,z_8^2 - 4\,z_1\,z_2\,z_8^2 - z_1\,z_2^2\,z_8^2 + 2\,z_3\,z_8^2 -
5\,z_1\,z_3\,z_8^2 + 3\,z_1^2\,z_3\,z_8^2 +
   3\,z_2\,z_3\,z_8^2 - z_3^2\,z_8^2 + z_4\,z_8^2 - 2\,z_5\,z_8^2 +
z_1\,z_5\,z_8^2 - z_2\,z_5\,z_8^2 -
   4\,z_1\,z_6\,z_8^2 + 2\,z_1^2\,z_6\,z_8^2 - 3\,z_7\,z_8^2 +
z_1\,z_7\,z_8^2 + 2\,z_2\,z_7\,z_8^2 - z_3\,z_7\,z_8^2 -
   2\,z_7^2\,z_8^2 + 6\,z_1\,z_8^3 - 2\,z_1^2\,z_8^3 + z_1^3\,z_8^3 -
z_2\,z_8^3 + 2\,z_1\,z_2\,z_8^3 -
   z_2^2\,z_8^3 - z_4\,z_8^3 + z_5\,z_8^3 - z_6\,z_8^3 + 2\,z_7\,z_8^3 +
2\,z_1\,z_7\,z_8^3 + z_2\,z_7\,z_8^3 +
   2\,z_8^4 - 2\,z_1\,z_8^4 + z_7\,z_8^4 - z_8^5\\ $

$\bchi_{0, 1, 2, 0, 0, 0, 0, 0} =
  z_1^2 + z_1^3 - z_2 - z_1\,z_2 - z_1^2\,z_2 + z_2^2 - z_3 - 2\,z_1\,z_3 -
z_1^2\,z_3 - z_1^3\,z_3 -
   z_1\,z_2\,z_3 + 2\,z_1\,z_3^2 + z_2\,z_3^2 - z_4 - z_1\,z_4 - z_2\,z_4 -
z_1\,z_2\,z_4 - z_3\,z_4 + z_1\,z_5 +
   2\,z_2\,z_5 + z_1\,z_2\,z_5 - z_1\,z_3\,z_5 + z_4\,z_5 - 2\,z_6 -
3\,z_1\,z_6 - z_1^2\,z_6 + z_2\,z_6 - z_2^2\,z_6 +
   z_1\,z_3\,z_6 + z_5\,z_6 + z_1\,z_5\,z_6 + z_6^2 - z_2\,z_6^2 +
z_1^2\,z_7 - z_2\,z_7 - z_1\,z_2\,z_7 +
   z_2^2\,z_7 - 2\,z_3\,z_7 - z_1\,z_3\,z_7 - z_1^2\,z_3\,z_7 -
z_2\,z_3\,z_7 + z_3^2\,z_7 - 2\,z_4\,z_7 + z_5\,z_7 +
   z_2\,z_5\,z_7 - 3\,z_6\,z_7 - z_1\,z_6\,z_7 + z_2\,z_6\,z_7 +
z_3\,z_6\,z_7 - z_3\,z_7^2 + 2\,z_8 + 2\,z_1\,z_8 +
   z_2\,z_8 + z_1\,z_2\,z_8 - 2\,z_1\,z_3\,z_8 + z_1^2\,z_3\,z_8 +
z_3^2\,z_8 + z_1\,z_5\,z_8 + z_1^2\,z_5\,z_8 +
   z_2\,z_5\,z_8 - z_3\,z_5\,z_8 - z_6\,z_8 - z_1\,z_2\,z_6\,z_8 +
z_3\,z_6\,z_8 - z_5\,z_6\,z_8 - z_6^2\,z_8 +
   4\,z_7\,z_8 + z_1\,z_7\,z_8 + 2\,z_2\,z_7\,z_8 + z_1\,z_2\,z_7\,z_8 +
z_2^2\,z_7\,z_8 + z_1\,z_3\,z_7\,z_8 +
   2\,z_7^2\,z_8 + z_2\,z_7^2\,z_8 - z_1^2\,z_8^2 + 2\,z_2\,z_8^2 +
2\,z_1\,z_2\,z_8^2 - z_2^2\,z_8^2 +
   2\,z_1\,z_3\,z_8^2 + z_4\,z_8^2 + z_5\,z_8^2 - z_1\,z_5\,z_8^2 +
3\,z_6\,z_8^2 - 2\,z_2\,z_6\,z_8^2 - z_7\,z_8^2 +
   z_1\,z_7\,z_8^2 + 2\,z_2\,z_7\,z_8^2 + z_3\,z_7\,z_8^2 +
z_6\,z_7\,z_8^2 - 3\,z_8^3 - z_1\,z_8^3 -
   z_1\,z_2\,z_8^3 - z_5\,z_8^3 - 3\,z_7\,z_8^3 - z_2\,z_8^4 + z_8^5\\ $

$\bchi_{0, 2, 0, 0, 0, 1, 0, 1} =
  z_1 - z_1^3 - z_1\,z_2 - z_1^2\,z_2 - z_1^3\,z_2 + z_2^2 - z_2^3 - z_3 +
z_1\,z_3 - z_1^2\,z_3 +
   2\,z_2\,z_3 + 2\,z_1\,z_2\,z_3 + z_3^2 - z_1\,z_3^2 - z_4 + 2\,z_1\,z_4 +
z_1^2\,z_4 + 2\,z_2\,z_4 + z_3\,z_4 +
   z_1^2\,z_5 - z_1\,z_2\,z_5 - z_2^2\,z_5 - 2\,z_3\,z_5 + z_4\,z_5 - z_6 +
z_1^3\,z_6 + 3\,z_2\,z_6 +
   3\,z_1\,z_2\,z_6 + 3\,z_3\,z_6 + z_4\,z_6 - z_5\,z_6 + z_1\,z_5\,z_6 +
2\,z_6^2 + z_7 - z_1\,z_7 - 3\,z_1^2\,z_7 -
   z_1^3\,z_7 - z_2\,z_7 - z_1\,z_2\,z_7 - z_1^2\,z_2\,z_7 - z_2^2\,z_7 +
2\,z_2\,z_3\,z_7 + 2\,z_4\,z_7 +
   z_1\,z_4\,z_7 - z_5\,z_7 - z_2\,z_5\,z_7 + z_6\,z_7 + z_2\,z_6\,z_7 +
z_3\,z_6\,z_7 + z_6^2\,z_7 - z_7^2 -
   3\,z_1\,z_7^2 - 2\,z_1^2\,z_7^2 - z_2\,z_7^2 + z_3\,z_7^2 - z_5\,z_7^2 -
z_6\,z_7^2 - z_7^3 - z_1\,z_7^3 +
   2\,z_8 - z_1\,z_8 + z_1^4\,z_8 - 3\,z_2\,z_8 - 2\,z_1\,z_2\,z_8 +
z_1^2\,z_2\,z_8 - z_2^2\,z_8 - 4\,z_3\,z_8 -
   3\,z_1^2\,z_3\,z_8 - z_2\,z_3\,z_8 + z_3^2\,z_8 - z_4\,z_8 + z_5\,z_8 +
z_1^2\,z_5\,z_8 - 3\,z_3\,z_5\,z_8 -
   5\,z_6\,z_8 - z_1\,z_6\,z_8 - z_1^2\,z_6\,z_8 - z_2\,z_6\,z_8 +
z_1\,z_2\,z_6\,z_8 + z_2^2\,z_6\,z_8 + 3\,z_3\,z_6\,z_8 -
   z_4\,z_6\,z_8 - 2\,z_5\,z_6\,z_8 + 2\,z_6^2\,z_8 - z_1\,z_6^2\,z_8 -
z_7\,z_8 - 2\,z_1\,z_7\,z_8 + z_1^3\,z_7\,z_8 -
   3\,z_2\,z_7\,z_8 + z_1\,z_2\,z_7\,z_8 - z_2^2\,z_7\,z_8 -
3\,z_3\,z_7\,z_8 - 3\,z_1\,z_3\,z_7\,z_8 - z_2\,z_3\,z_7\,z_8 +
   z_4\,z_7\,z_8 - z_5\,z_7\,z_8 + z_1\,z_5\,z_7\,z_8 - 3\,z_6\,z_7\,z_8 +
2\,z_1\,z_6\,z_7\,z_8 - z_2\,z_6\,z_7\,z_8 -
   4\,z_7^2\,z_8 - z_1\,z_7^2\,z_8 + z_1^2\,z_7^2\,z_8 + z_2\,z_7^2\,z_8 +
2\,z_8^2 + 2\,z_1^2\,z_8^2 +
   3\,z_2\,z_8^2 + 2\,z_1\,z_2\,z_8^2 + 2\,z_1^2\,z_2\,z_8^2 - z_3\,z_8^2 -
3\,z_2\,z_3\,z_8^2 + z_3^2\,z_8^2 -
   z_1\,z_4\,z_8^2 + 3\,z_5\,z_8^2 - 2\,z_1\,z_5\,z_8^2 + z_2\,z_5\,z_8^2 -
2\,z_6\,z_8^2 + 3\,z_1\,z_6\,z_8^2 -
   2\,z_1^2\,z_6\,z_8^2 - 3\,z_2\,z_6\,z_8^2 + 2\,z_3\,z_6\,z_8^2 +
z_6^2\,z_8^2 + z_7\,z_8^2 + 3\,z_1\,z_7\,z_8^2 +
   2\,z_1^2\,z_7\,z_8^2 + 4\,z_2\,z_7\,z_8^2 - z_1\,z_2\,z_7\,z_8^2 -
3\,z_3\,z_7\,z_8^2 - z_6\,z_7\,z_8^2 - z_8^3 -
   z_1\,z_8^3 - z_1^3\,z_8^3 + 3\,z_2\,z_8^3 - 2\,z_1\,z_2\,z_8^3 +
z_2^2\,z_8^3 + z_3\,z_8^3 +
   2\,z_1\,z_3\,z_8^3 - z_5\,z_8^3 + z_1\,z_6\,z_8^3 + 3\,z_7\,z_8^3 -
3\,z_1\,z_7\,z_8^3 - z_8^4 - z_1\,z_8^4 -
   2\,z_2\,z_8^4 + z_3\,z_8^4 + z_6\,z_8^4 + z_1\,z_8^5\\ $

$\bchi_{0, 2, 0, 0, 1, 0, 0, 0} =
  z_1^2 + z_1^3 + z_2 + 3\,z_1\,z_2 + 2\,z_1^2\,z_2 + z_2^3 + z_1^2\,z_3 -
z_2\,z_3 + z_1\,z_2\,z_3 - z_1\,z_4 -
   2\,z_2\,z_4 + z_5 + 2\,z_1\,z_5 + z_1^2\,z_5 + z_2\,z_5 + z_2^2\,z_5 -
z_3\,z_5 - z_4\,z_5 + z_5^2 + z_6 +
   z_1\,z_6 + 2\,z_1^2\,z_6 + z_1^3\,z_6 - 3\,z_2\,z_6 - 2\,z_1\,z_2\,z_6 -
z_3\,z_6 - z_1\,z_3\,z_6 - z_2\,z_3\,z_6 -
   z_4\,z_6 - 2\,z_6^2 - 2\,z_1\,z_6^2 - z_2\,z_6^2 - z_7 + z_1\,z_7 +
2\,z_1^2\,z_7 + z_1^3\,z_7 + 3\,z_2\,z_7 +
   3\,z_1\,z_2\,z_7 - z_1\,z_3\,z_7 + z_3^2\,z_7 - z_1\,z_4\,z_7 +
2\,z_1\,z_5\,z_7 + z_2\,z_5\,z_7 + 2\,z_6\,z_7 +
   z_1\,z_6\,z_7 + z_3\,z_6\,z_7 - z_7^2 + z_1\,z_7^2 + z_1^2\,z_7^2 +
z_2\,z_7^2 - z_3\,z_7^2 + 2\,z_1\,z_8 -
   z_1^2\,z_8 - z_1^3\,z_8 + 4\,z_2\,z_8 - z_1\,z_2\,z_8 -
2\,z_1^2\,z_2\,z_8 + z_2^2\,z_8 - z_1\,z_2^2\,z_8 -
   z_3\,z_8 - z_1^2\,z_3\,z_8 + z_2\,z_3\,z_8 + z_3^2\,z_8 - z_4\,z_8 +
2\,z_5\,z_8 - z_1^2\,z_5\,z_8 +
   2\,z_2\,z_5\,z_8 + z_6\,z_8 - 2\,z_1\,z_6\,z_8 - 2\,z_1^2\,z_6\,z_8 +
2\,z_3\,z_6\,z_8 - z_7\,z_8 + 2\,z_1\,z_7\,z_8 -
   3\,z_1^2\,z_7\,z_8 - z_1^3\,z_7\,z_8 + 4\,z_2\,z_7\,z_8 +
z_1\,z_2\,z_7\,z_8 - z_3\,z_7\,z_8 + z_1\,z_3\,z_7\,z_8 +
   z_4\,z_7\,z_8 + 3\,z_6\,z_7\,z_8 + z_1\,z_6\,z_7\,z_8 - z_7^2\,z_8 -
z_1\,z_7^2\,z_8 + z_2\,z_7^2\,z_8 + 2\,z_8^2 -
   z_1^2\,z_8^2 - z_2\,z_8^2 + z_1^2\,z_2\,z_8^2 - z_3\,z_8^2 +
z_1\,z_3\,z_8^2 - z_2\,z_3\,z_8^2 - z_5\,z_8^2 -
   3\,z_6\,z_8^2 - 2\,z_2\,z_6\,z_8^2 + z_7\,z_8^2 + z_1^2\,z_7\,z_8^2 +
z_2\,z_7\,z_8^2 - z_7^2\,z_8^2 -
   z_1\,z_8^3 + z_1^2\,z_8^3 - z_1\,z_2\,z_8^3 + z_3\,z_8^3 +
2\,z_7\,z_8^3 - z_8^4 - z_2\,z_8^4\\ $

$\bchi_{0, 2, 1, 0, 0, 0, 0, 1} =
  -1 - z_1 + z_1^2 + z_1^3 - z_1^2\,z_2 - z_1^3\,z_2 + z_2^2 + 2\,z_1\,z_2^2
+ z_1^2\,z_2^2 - z_1\,z_2^3 -
   z_1\,z_3 - z_1^2\,z_3 + z_2\,z_3 + z_1\,z_2\,z_3 - z_2^2\,z_3 + z_3^2 -
z_2\,z_3^2 + z_1\,z_4 + z_1^2\,z_4 +
   z_2\,z_4 + 2\,z_1\,z_2\,z_4 - z_2\,z_5 - 2\,z_1\,z_2\,z_5 + z_2^2\,z_5 +
z_1\,z_3\,z_5 - z_4\,z_5 + z_5^2 -
   z_1\,z_6 - 2\,z_1^2\,z_6 - z_1^3\,z_6 + 2\,z_2\,z_6 + 4\,z_1\,z_2\,z_6 +
2\,z_1^2\,z_2\,z_6 + 2\,z_3\,z_6 +
   z_1\,z_3\,z_6 - 2\,z_2\,z_3\,z_6 - 2\,z_5\,z_6 - 2\,z_1\,z_5\,z_6 +
2\,z_6^2 + 2\,z_1\,z_6^2 - z_7 + z_1^2\,z_7 -
   z_1\,z_2\,z_7 - z_1^2\,z_2\,z_7 + z_2^2\,z_7 + z_1\,z_2^2\,z_7 - z_3\,z_7
+ z_2\,z_3\,z_7 + z_4\,z_7 + z_1\,z_4\,z_7 +
   z_5\,z_7 - z_2\,z_5\,z_7 - z_6\,z_7 - z_1\,z_6\,z_7 - z_1^2\,z_6\,z_7 +
z_2\,z_6\,z_7 + z_6^2\,z_7 - z_6\,z_7^2 +
   z_1\,z_8 - z_1^3\,z_8 - 2\,z_2\,z_8 - 4\,z_1\,z_2\,z_8 + z_1^2\,z_2\,z_8
+ z_1^3\,z_2\,z_8 + z_2^2\,z_8 -
   z_1\,z_2^2\,z_8 - 2\,z_3\,z_8 - z_2\,z_3\,z_8 - 2\,z_1\,z_2\,z_3\,z_8 +
z_2^2\,z_3\,z_8 - z_4\,z_8 + z_1\,z_4\,z_8 -
   z_3\,z_4\,z_8 + 4\,z_5\,z_8 - z_1\,z_2\,z_5\,z_8 + z_5^2\,z_8 -
4\,z_6\,z_8 - z_1\,z_6\,z_8 + 3\,z_1^2\,z_6\,z_8 -
   z_2\,z_6\,z_8 - z_1\,z_2\,z_6\,z_8 - 2\,z_3\,z_6\,z_8 - z_4\,z_6\,z_8 -
z_5\,z_6\,z_8 + z_7\,z_8 - 2\,z_1\,z_7\,z_8 -
   z_1^2\,z_7\,z_8 - 4\,z_2\,z_7\,z_8 - 5\,z_1\,z_2\,z_7\,z_8 -
z_1^2\,z_2\,z_7\,z_8 - 2\,z_3\,z_7\,z_8 + z_2\,z_3\,z_7\,z_8 -
   z_4\,z_7\,z_8 + 3\,z_5\,z_7\,z_8 + z_1\,z_5\,z_7\,z_8 -
5\,z_6\,z_7\,z_8 - 2\,z_1\,z_7^2\,z_8 - 2\,z_2\,z_7^2\,z_8 +
   5\,z_8^2 - 2\,z_1\,z_8^2 - 3\,z_1^2\,z_8^2 + z_2\,z_8^2 +
2\,z_1\,z_2\,z_8^2 + 2\,z_3\,z_8^2 + z_1\,z_3\,z_8^2 -
   z_2\,z_3\,z_8^2 - z_1\,z_4\,z_8^2 + z_5\,z_8^2 - z_1\,z_5\,z_8^2 +
z_2\,z_5\,z_8^2 + 4\,z_1\,z_6\,z_8^2 -
   z_2\,z_6\,z_8^2 - z_6^2\,z_8^2 + 6\,z_7\,z_8^2 - 3\,z_1\,z_7\,z_8^2 -
z_1^2\,z_7\,z_8^2 + z_2\,z_7\,z_8^2 +
   z_1\,z_2\,z_7\,z_8^2 + 3\,z_3\,z_7\,z_8^2 + z_5\,z_7\,z_8^2 +
3\,z_6\,z_7\,z_8^2 + 2\,z_7^2\,z_8^2 +
   z_1\,z_7^2\,z_8^2 - z_8^3 - z_1\,z_8^3 + 2\,z_1^2\,z_8^3 + 2\,z_2\,z_8^3
+ z_3\,z_8^3 - 2\,z_5\,z_8^3 +
   2\,z_6\,z_8^3 - z_1\,z_6\,z_8^3 - z_7\,z_8^3 + 2\,z_1\,z_7\,z_8^3 +
z_2\,z_7\,z_8^3 - 3\,z_8^4 + 3\,z_1\,z_8^4 -
   z_2\,z_8^4 - z_3\,z_8^4 - z_6\,z_8^4 - z_7\,z_8^4 + z_8^5 - z_1\,z_8^5\\ $

$\bchi_{1, 0, 0, 1, 0, 1, 0, 0} =
  z_1 + 2\,z_1^2 + 2\,z_1^3 + z_1^4 + z_2 + 2\,z_1\,z_2 + 3\,z_1^2\,z_2 +
z_1^3\,z_2 + z_2^2 + z_1\,z_2^2 -
   z_3 - z_1\,z_3 - 2\,z_1^2\,z_3 - z_2\,z_3 - 2\,z_1\,z_2\,z_3 - z_2^2\,z_3
+ z_3^2 - z_4 - z_1\,z_2\,z_4 + z_5 +
   z_1\,z_5 + z_1^2\,z_5 + z_2\,z_5 + z_1\,z_2\,z_5 - z_3\,z_5 + z_4\,z_5 -
z_6 - z_1\,z_6 + z_1^2\,z_6 +
   z_1^3\,z_6 - z_2\,z_6 - 2\,z_1\,z_3\,z_6 - z_2\,z_3\,z_6 + z_1\,z_4\,z_6
+ z_5\,z_6 - z_2\,z_5\,z_6 - z_6^2 +
   z_2\,z_6^2 + z_6^3 + 3\,z_1\,z_7 + 4\,z_1^2\,z_7 + 3\,z_1^3\,z_7 +
z_1^4\,z_7 + 2\,z_2\,z_7 + 3\,z_1\,z_2\,z_7 +
   2\,z_1^2\,z_2\,z_7 + z_2^2\,z_7 - 2\,z_3\,z_7 - 2\,z_1\,z_3\,z_7 -
2\,z_1^2\,z_3\,z_7 - z_2\,z_3\,z_7 -
   z_1\,z_2\,z_3\,z_7 + z_3^2\,z_7 - z_4\,z_7 + z_2\,z_5\,z_7 +
z_3\,z_5\,z_7 - z_6\,z_7 - 3\,z_1\,z_6\,z_7 -
   z_1^2\,z_6\,z_7 - z_2\,z_6\,z_7 + z_1\,z_2\,z_6\,z_7 - z_5\,z_6\,z_7 -
z_6^2\,z_7 + 3\,z_1\,z_7^2 + 2\,z_1^2\,z_7^2 +
   z_1^3\,z_7^2 + z_2\,z_7^2 + z_1\,z_2\,z_7^2 - z_3\,z_7^2 -
z_1\,z_3\,z_7^2 - z_5\,z_7^2 - 2\,z_1\,z_6\,z_7^2 +
   z_1\,z_7^3 + z_8 + z_1\,z_8 - 2\,z_1^3\,z_8 + 3\,z_2\,z_8 +
2\,z_1\,z_2\,z_8 - z_1^2\,z_2\,z_8 - z_1^3\,z_2\,z_8 +
   2\,z_2^2\,z_8 - 2\,z_1\,z_2^2\,z_8 - 2\,z_3\,z_8 + z_1\,z_3\,z_8 -
z_1^2\,z_3\,z_8 - 3\,z_2\,z_3\,z_8 +
   2\,z_1\,z_2\,z_3\,z_8 + z_2^2\,z_3\,z_8 + z_3^2\,z_8 - z_4\,z_8 +
z_1^2\,z_4\,z_8 - z_3\,z_4\,z_8 + z_5\,z_8 +
   3\,z_2\,z_5\,z_8 - z_1\,z_2\,z_5\,z_8 - z_3\,z_5\,z_8 - z_6\,z_8 -
2\,z_1\,z_6\,z_8 - z_1^2\,z_6\,z_8 - 3\,z_2\,z_6\,z_8 +
   z_1\,z_2\,z_6\,z_8 - z_3\,z_6\,z_8 - z_4\,z_6\,z_8 - 3\,z_6^2\,z_8 +
z_1\,z_6^2\,z_8 + z_7\,z_8 + 2\,z_1\,z_7\,z_8 -
   3\,z_1^2\,z_7\,z_8 - 2\,z_1^3\,z_7\,z_8 + 2\,z_2\,z_7\,z_8 -
z_1^2\,z_2\,z_7\,z_8 - z_3\,z_7\,z_8 + z_1\,z_3\,z_7\,z_8 +
   z_2\,z_3\,z_7\,z_8 - 2\,z_5\,z_7\,z_8 + z_1\,z_5\,z_7\,z_8 +
z_6\,z_7\,z_8 - 4\,z_1\,z_6\,z_7\,z_8 - z_2\,z_6\,z_7\,z_8 -
   z_7^2\,z_8 + z_1\,z_7^2\,z_8 - 2\,z_1^2\,z_7^2\,z_8 - z_2\,z_7^2\,z_8 +
z_3\,z_7^2\,z_8 + z_6\,z_7^2\,z_8 -
   z_7^3\,z_8 + z_8^2 - 3\,z_1\,z_8^2 - 4\,z_1^2\,z_8^2 - z_1^3\,z_8^2 -
6\,z_1\,z_2\,z_8^2 +
   2\,z_1^2\,z_2\,z_8^2 + 2\,z_3\,z_8^2 + 2\,z_1\,z_3\,z_8^2 + z_4\,z_8^2 -
z_1\,z_4\,z_8^2 - z_1\,z_5\,z_8^2 +
   z_6\,z_8^2 - z_1\,z_6\,z_8^2 - z_2\,z_6\,z_8^2 + z_3\,z_6\,z_8^2 +
z_6^2\,z_8^2 - z_7\,z_8^2 -
   3\,z_1\,z_7\,z_8^2 - 4\,z_1^2\,z_7\,z_8^2 - z_2\,z_7\,z_8^2 +
z_1\,z_2\,z_7\,z_8^2 + 4\,z_3\,z_7\,z_8^2 +
   5\,z_6\,z_7\,z_8^2 - 2\,z_7^2\,z_8^2 + z_1\,z_7^2\,z_8^2 - z_8^3 +
z_1\,z_8^3 + 2\,z_1^2\,z_8^3 - z_2\,z_8^3 +
   z_3\,z_8^3 + 2\,z_1\,z_6\,z_8^3 + 3\,z_1\,z_7\,z_8^3 + z_1\,z_8^4 +
z_1^2\,z_8^4 - z_3\,z_8^4 - z_6\,z_8^4 +
   z_7\,z_8^4 - z_1\,z_8^5\\ $

$\bchi_{1, 0, 1, 0, 0, 1, 0, 1} =
  -(z_1\,z_2) + z_1^3\,z_2 + z_1\,z_2^2 - z_1^2\,z_2^2 + z_1^2\,z_3 -
2\,z_1\,z_2\,z_3 + z_2^2\,z_3 - z_3^2 -
   z_1\,z_4 + z_2\,z_4 - 2\,z_1\,z_5 - 2\,z_1^2\,z_5 - z_2\,z_5 +
2\,z_1\,z_2\,z_5 + 2\,z_3\,z_5 - z_1\,z_3\,z_5 +
   z_4\,z_5 - z_5^2 - z_6 - z_1\,z_6 - z_1\,z_2\,z_6 + z_2^2\,z_6 -
3\,z_3\,z_6 + 2\,z_5\,z_6 + z_1\,z_5\,z_6 -
   z_6^2 + z_1\,z_7 + z_1^2\,z_7 - z_2\,z_7 - 3\,z_1\,z_2\,z_7 +
z_1\,z_2^2\,z_7 - z_3\,z_7 - z_1\,z_3\,z_7 +
   z_2\,z_3\,z_7 - z_5\,z_7 - 3\,z_1\,z_5\,z_7 - z_2\,z_5\,z_7 -
2\,z_6\,z_7 - z_1\,z_6\,z_7 + z_2\,z_6\,z_7 + z_1\,z_7^2 -
   z_2\,z_7^2 - z_1\,z_2\,z_7^2 - z_3\,z_7^2 - z_5\,z_7^2 - z_6\,z_7^2 + z_8
+ z_1\,z_8 - z_1^2\,z_8 -
   z_1^4\,z_8 - z_2\,z_8 - 2\,z_1^2\,z_2\,z_8 + z_1^3\,z_2\,z_8 + z_2^2\,z_8
+ z_1\,z_2^2\,z_8 - z_2^3\,z_8 +
   2\,z_3\,z_8 - z_1\,z_3\,z_8 + 3\,z_1^2\,z_3\,z_8 + z_2\,z_3\,z_8 -
2\,z_1\,z_2\,z_3\,z_8 - 2\,z_3^2\,z_8 +
   z_2\,z_4\,z_8 - 2\,z_5\,z_8 + 3\,z_1\,z_5\,z_8 + z_1\,z_6\,z_8 +
5\,z_1^2\,z_6\,z_8 + 3\,z_2\,z_6\,z_8 -
   2\,z_1\,z_2\,z_6\,z_8 - 6\,z_3\,z_6\,z_8 + z_1\,z_3\,z_6\,z_8 -
z_4\,z_6\,z_8 + 2\,z_5\,z_6\,z_8 - 3\,z_6^2\,z_8 -
   z_1\,z_6^2\,z_8 + 2\,z_7\,z_8 - 3\,z_2\,z_7\,z_8 + z_1\,z_2\,z_7\,z_8 -
z_1^2\,z_2\,z_7\,z_8 + z_3\,z_7\,z_8 -
   z_5\,z_7\,z_8 + z_1\,z_5\,z_7\,z_8 + 3\,z_1\,z_6\,z_7\,z_8 +
z_2\,z_6\,z_7\,z_8 + z_7^2\,z_8 - z_1\,z_7^2\,z_8 +
   z_1^2\,z_7^2\,z_8 - z_2\,z_7^2\,z_8 - z_3\,z_7^2\,z_8 + z_8^2 -
z_1^2\,z_8^2 + 2\,z_1^3\,z_8^2 - z_2\,z_8^2 +
   z_1\,z_2\,z_8^2 - 2\,z_1^2\,z_2\,z_8^2 + z_1\,z_2^2\,z_8^2 +
2\,z_3\,z_8^2 - 3\,z_1\,z_3\,z_8^2 +
   2\,z_2\,z_3\,z_8^2 + 2\,z_5\,z_8^2 - z_2\,z_5\,z_8^2 + 4\,z_6\,z_8^2 -
4\,z_1\,z_6\,z_8^2 - z_1^2\,z_6\,z_8^2 +
   2\,z_2\,z_6\,z_8^2 + z_3\,z_6\,z_8^2 + z_7\,z_8^2 - z_1^2\,z_7\,z_8^2 -
z_2\,z_7\,z_8^2 + z_1\,z_2\,z_7\,z_8^2 +
   2\,z_3\,z_7\,z_8^2 + 2\,z_6\,z_7\,z_8^2 + z_7^2\,z_8^2 -
z_1\,z_7^2\,z_8^2 - z_8^3 + z_1\,z_8^3 +
   z_1\,z_2\,z_8^3 - z_2^2\,z_8^3 - 2\,z_3\,z_8^3 - 3\,z_6\,z_8^3 -
z_7\,z_8^3 + z_1\,z_7\,z_8^3 - z_1\,z_8^4 +
   z_2\,z_8^4\\ $

$\bchi_{1, 0, 1, 0, 1, 0, 0, 0} =
  -z_1 - 3\,z_1^2 - 2\,z_1^3 - 2\,z_1\,z_2 - z_2^2 + z_3 + 2\,z_1\,z_3 -
z_1\,z_2\,z_3 - z_3^2 + z_4 +
   2\,z_1\,z_4 - z_1\,z_5 + z_1^2\,z_5 - z_2\,z_5 + z_3\,z_5 +
z_1\,z_3\,z_5 - z_4\,z_5 + z_6 + 3\,z_1\,z_6 +
   2\,z_1^2\,z_6 + z_1^3\,z_6 - z_2\,z_6 - z_1^2\,z_2\,z_6 + z_2^2\,z_6 -
3\,z_3\,z_6 - 2\,z_1\,z_3\,z_6 - 2\,z_6^2 -
   2\,z_1\,z_6^2 + z_2\,z_6^2 - z_7 - 5\,z_1\,z_7 - 6\,z_1^2\,z_7 -
z_1^3\,z_7 - z_2\,z_7 - 2\,z_1\,z_2\,z_7 -
   z_2^2\,z_7 + z_1\,z_2^2\,z_7 + 2\,z_3\,z_7 + z_1\,z_3\,z_7 +
z_2\,z_3\,z_7 + 2\,z_4\,z_7 - z_5\,z_7 + z_1\,z_5\,z_7 -
   z_2\,z_5\,z_7 + 2\,z_6\,z_7 + z_1\,z_6\,z_7 + z_1^2\,z_6\,z_7 +
z_2\,z_6\,z_7 - z_3\,z_6\,z_7 - z_6^2\,z_7 -
   2\,z_7^2 - 5\,z_1\,z_7^2 - 2\,z_1^2\,z_7^2 - z_2\,z_7^2 + z_3\,z_7^2 -
z_7^3 - z_1\,z_7^3 - z_8 -
   4\,z_1\,z_8 + 2\,z_1^2\,z_8 + z_1^3\,z_8 - z_2\,z_8 + 3\,z_1\,z_2\,z_8 +
z_1^2\,z_2\,z_8 + z_1\,z_2^2\,z_8 +
   3\,z_3\,z_8 + z_1^2\,z_3\,z_8 + z_2\,z_3\,z_8 - z_3^2\,z_8 + z_4\,z_8 -
z_2\,z_4\,z_8 - z_5\,z_8 - z_1^2\,z_5\,z_8 -
   z_2\,z_5\,z_8 + 2\,z_3\,z_5\,z_8 + 3\,z_6\,z_8 + 4\,z_1\,z_6\,z_8 +
3\,z_2\,z_6\,z_8 - z_1\,z_2\,z_6\,z_8 -
   4\,z_3\,z_6\,z_8 + 2\,z_5\,z_6\,z_8 - 2\,z_6^2\,z_8 - 4\,z_7\,z_8 -
2\,z_1\,z_7\,z_8 + 2\,z_1^2\,z_7\,z_8 +
   3\,z_1\,z_2\,z_7\,z_8 - z_2^2\,z_7\,z_8 + 3\,z_3\,z_7\,z_8 +
z_1\,z_3\,z_7\,z_8 + 4\,z_6\,z_7\,z_8 + z_1\,z_6\,z_7\,z_8 -
   2\,z_7^2\,z_8 + 2\,z_1\,z_7^2\,z_8 + z_7^3\,z_8 - z_8^2 + 4\,z_1\,z_8^2 +
z_1^2\,z_8^2 + z_1^3\,z_8^2 +
   z_1\,z_2\,z_8^2 - 2\,z_1^2\,z_2\,z_8^2 - z_2^2\,z_8^2 + z_3\,z_8^2 -
3\,z_1\,z_3\,z_8^2 + z_2\,z_3\,z_8^2 -
   z_4\,z_8^2 - 2\,z_5\,z_8^2 + z_1\,z_5\,z_8^2 + 3\,z_6\,z_8^2 -
6\,z_1\,z_6\,z_8^2 + 2\,z_2\,z_6\,z_8^2 +
   2\,z_7\,z_8^2 + 3\,z_1\,z_7\,z_8^2 + z_1^2\,z_7\,z_8^2 -
2\,z_2\,z_7\,z_8^2 - z_6\,z_7\,z_8^2 + 2\,z_7^2\,z_8^2 +
   z_8^3 + 2\,z_1\,z_8^3 - 2\,z_1^2\,z_8^3 - 2\,z_2\,z_8^3 +
z_1\,z_2\,z_8^3 - 2\,z_3\,z_8^3 + z_5\,z_8^3 -
   2\,z_6\,z_8^3 + z_7\,z_8^3 - z_1\,z_8^4 + z_2\,z_8^4 - z_7\,z_8^4\\ $

$\bchi_{1, 1, 0, 0, 0, 1, 1, 0} =
  -z_1^2 - 2\,z_1^3 - z_1^4 - z_1\,z_2 - 2\,z_1^2\,z_2 - z_2^2 +
z_1^2\,z_2^2 + z_2^3 + 2\,z_1^2\,z_3 +
   z_1\,z_2\,z_3 - z_2^2\,z_3 - z_1\,z_4 - 2\,z_2\,z_4 + z_1\,z_5 -
z_1\,z_2\,z_5 + z_2^2\,z_5 + z_1\,z_3\,z_5 -
   z_4\,z_5 + z_5^2 - 2\,z_1\,z_6 - z_1^2\,z_6 - 3\,z_2\,z_6 -
2\,z_1\,z_2\,z_6 - z_2\,z_3\,z_6 - z_4\,z_6 -
   2\,z_5\,z_6 - z_1\,z_5\,z_6 - z_6^2 - z_1\,z_6^2 - 2\,z_1^2\,z_7 -
2\,z_1^3\,z_7 - 2\,z_1\,z_2\,z_7 -
   z_1^2\,z_2\,z_7 - z_2^2\,z_7 - z_1\,z_5\,z_7 - z_2\,z_5\,z_7 -
z_1\,z_6\,z_7 - z_1^2\,z_6\,z_7 + z_1\,z_2\,z_6\,z_7 -
   z_5\,z_6\,z_7 + z_6^2\,z_7 - z_1^2\,z_7^2 - z_1\,z_2\,z_7^2 -
z_2^2\,z_7^2 - z_1\,z_3\,z_7^2 + z_4\,z_7^2 -
   z_5\,z_7^2 + z_6\,z_7^2 + z_1\,z_8 - 3\,z_1^2\,z_8 + z_1^3\,z_8 +
z_1^4\,z_8 + 2\,z_2\,z_8 - 2\,z_1\,z_2\,z_8 -
   z_1^3\,z_2\,z_8 - z_2^2\,z_8 - 2\,z_1\,z_2^2\,z_8 + z_3\,z_8 +
3\,z_1\,z_3\,z_8 - 2\,z_1^2\,z_3\,z_8 +
   3\,z_2\,z_3\,z_8 - z_3^2\,z_8 + z_4\,z_8 + 4\,z_1\,z_4\,z_8 +
z_2\,z_4\,z_8 + z_5\,z_8 - 3\,z_1\,z_5\,z_8 +
   z_1^2\,z_5\,z_8 - z_1\,z_2\,z_5\,z_8 + z_3\,z_5\,z_8 + z_5^2\,z_8 +
2\,z_6\,z_8 + 4\,z_1\,z_6\,z_8 + z_1^2\,z_6\,z_8 +
   2\,z_2\,z_6\,z_8 + 3\,z_1\,z_2\,z_6\,z_8 - 2\,z_5\,z_6\,z_8 +
2\,z_6^2\,z_8 - z_1\,z_7\,z_8 + z_1^3\,z_7\,z_8 +
   z_2\,z_7\,z_8 - 2\,z_2^2\,z_7\,z_8 + z_3\,z_7\,z_8 +
2\,z_2\,z_3\,z_7\,z_8 + 4\,z_4\,z_7\,z_8 + z_1\,z_5\,z_7\,z_8 +
   3\,z_6\,z_7\,z_8 + 4\,z_1\,z_6\,z_7\,z_8 + z_2\,z_6\,z_7\,z_8 -
2\,z_7^2\,z_8 - z_1\,z_7^2\,z_8 + z_1^2\,z_7^2\,z_8 -
   z_2\,z_7^2\,z_8 - z_6\,z_7^2\,z_8 - 2\,z_7^3\,z_8 - z_8^2 - 6\,z_1\,z_8^2
+ 6\,z_1^2\,z_8^2 + z_1^3\,z_8^2 -
   3\,z_2\,z_8^2 + 3\,z_1\,z_2\,z_8^2 + 2\,z_1^2\,z_2\,z_8^2 - z_2^2\,z_8^2
+ z_3\,z_8^2 - 3\,z_1\,z_3\,z_8^2 +
   z_1^2\,z_3\,z_8^2 - z_3^2\,z_8^2 + 2\,z_4\,z_8^2 - z_1\,z_4\,z_8^2 -
z_1\,z_5\,z_8^2 + 4\,z_1\,z_6\,z_8^2 -
   z_1^2\,z_6\,z_8^2 - 2\,z_3\,z_6\,z_8^2 - 7\,z_7\,z_8^2 -
3\,z_1\,z_7\,z_8^2 + 3\,z_1^2\,z_7\,z_8^2 -
   z_3\,z_7\,z_8^2 + z_5\,z_7\,z_8^2 - 2\,z_6\,z_7\,z_8^2 -
4\,z_7^2\,z_8^2 - z_1\,z_7^2\,z_8^2 - 2\,z_8^3 +
   3\,z_1\,z_8^3 - 2\,z_1^2\,z_8^3 - z_1^3\,z_8^3 + z_2\,z_8^3 -
z_1\,z_2\,z_8^3 + z_2^2\,z_8^3 - z_3\,z_8^3 -
   2\,z_4\,z_8^3 - 2\,z_6\,z_8^3 - z_1\,z_6\,z_8^3 + z_7\,z_8^3 -
3\,z_1\,z_7\,z_8^3 + z_2\,z_7\,z_8^3 +
   z_7^2\,z_8^3 + 4\,z_8^4 - z_1\,z_8^4 + 4\,z_7\,z_8^4 + z_1\,z_8^5 -
z_8^6\\ $

$\bchi_{1, 1, 0, 0, 1, 0, 0, 1} =
  -z_1^3 - z_1^4 + z_1\,z_2 + z_1^2\,z_2 + z_1^3\,z_2 + z_2^2 + z_1\,z_2^2 +
2\,z_1\,z_3 + 3\,z_1^2\,z_3 +
   z_2\,z_3 - z_3^2 + z_1^2\,z_4 - z_2\,z_4 - z_1\,z_2\,z_4 - z_3\,z_4 -
z_1\,z_5 - 2\,z_1^2\,z_5 + z_1\,z_2\,z_5 +
   2\,z_3\,z_5 + z_4\,z_5 - z_5^2 + 3\,z_1\,z_6 + 3\,z_1^2\,z_6 +
2\,z_2\,z_6 - z_1\,z_2\,z_6 - z_2^2\,z_6 -
   2\,z_3\,z_6 + z_1\,z_3\,z_6 + z_2\,z_3\,z_6 - z_4\,z_6 + 2\,z_5\,z_6 -
z_6^2 + z_1\,z_7 + z_1^2\,z_7 - z_1^3\,z_7 +
   z_2\,z_7 + 2\,z_1\,z_2\,z_7 + z_1^2\,z_2\,z_7 + z_2^2\,z_7 + z_3\,z_7 +
3\,z_1\,z_3\,z_7 + z_1^2\,z_3\,z_7 -
   z_2\,z_3\,z_7 - z_3^2\,z_7 - 2\,z_4\,z_7 - 2\,z_1\,z_5\,z_7 + z_6\,z_7 +
2\,z_1\,z_6\,z_7 - z_1^2\,z_6\,z_7 -
   z_3\,z_6\,z_7 + z_7^2 + 3\,z_1\,z_7^2 + z_1^2\,z_7^2 + z_2\,z_7^2 +
z_6\,z_7^2 + z_7^3 + z_1\,z_7^3 -
   z_1\,z_8 - 2\,z_1^2\,z_8 + z_1^3\,z_8 - z_1\,z_2\,z_8 -
3\,z_1^2\,z_2\,z_8 - z_2^2\,z_8 + 3\,z_3\,z_8 -
   z_1\,z_3\,z_8 + z_2\,z_3\,z_8 - z_1\,z_2\,z_3\,z_8 - z_3^2\,z_8 +
z_4\,z_8 + z_2\,z_4\,z_8 - 3\,z_5\,z_8 -
   z_1^2\,z_5\,z_8 - z_2\,z_5\,z_8 + z_1\,z_2\,z_5\,z_8 + 2\,z_3\,z_5\,z_8 -
z_5^2\,z_8 + 3\,z_6\,z_8 - 4\,z_1\,z_6\,z_8 -
   z_2^2\,z_6\,z_8 - z_3\,z_6\,z_8 - z_1\,z_3\,z_6\,z_8 + z_4\,z_6\,z_8 +
2\,z_5\,z_6\,z_8 + z_1\,z_6^2\,z_8 + 2\,z_7\,z_8 -
   z_1\,z_7\,z_8 - 3\,z_1^2\,z_7\,z_8 - z_1^3\,z_7\,z_8 - 2\,z_2\,z_7\,z_8 -
4\,z_1\,z_2\,z_7\,z_8 + 2\,z_2^2\,z_7\,z_8 +
   2\,z_3\,z_7\,z_8 + z_2\,z_3\,z_7\,z_8 - z_1\,z_5\,z_7\,z_8 -
3\,z_6\,z_7\,z_8 + z_2\,z_6\,z_7\,z_8 + 2\,z_7^2\,z_8 -
   2\,z_1\,z_7^2\,z_8 - z_2\,z_7^2\,z_8 - z_6\,z_7^2\,z_8 - z_7^3\,z_8 -
z_8^2 + 3\,z_1\,z_8^2 + z_1^3\,z_8^2 -
   2\,z_2\,z_8^2 + 2\,z_1\,z_2\,z_8^2 + z_2^2\,z_8^2 - 2\,z_1\,z_3\,z_8^2 +
z_2\,z_3\,z_8^2 - z_4\,z_8^2 -
   z_5\,z_8^2 + 2\,z_1\,z_5\,z_8^2 - z_2\,z_5\,z_8^2 - 3\,z_6\,z_8^2 -
z_1\,z_6\,z_8^2 + z_1^2\,z_6\,z_8^2 +
   z_2\,z_6\,z_8^2 - z_3\,z_6\,z_8^2 + 2\,z_7\,z_8^2 +
2\,z_1^2\,z_7\,z_8^2 - z_3\,z_7\,z_8^2 - 3\,z_6\,z_7\,z_8^2 +
   3\,z_7^2\,z_8^2 - z_1\,z_7^2\,z_8^2 + 2\,z_8^3 - z_1\,z_8^3 +
z_2\,z_8^3 - z_2^2\,z_8^3 - z_3\,z_8^3 +
   2\,z_5\,z_8^3 + z_6\,z_8^3 + z_7\,z_8^3 + z_2\,z_7\,z_8^3 +
z_7^2\,z_8^3 - z_8^4 - 2\,z_7\,z_8^4\\ $

$\bchi_{1, 3, 0, 0, 0, 0, 0, 0} =
  z_1 + z_1^2 + z_2 + 2\,z_1\,z_2 + 3\,z_1^2\,z_2 + z_1^3\,z_2 -
z_1\,z_2^2 - z_1^2\,z_2^2 + z_1\,z_2^3 +
   2\,z_1\,z_3 + z_1^2\,z_3 + z_1^3\,z_3 - z_2\,z_3 + z_2^2\,z_3 -
2\,z_1\,z_3^2 - z_4 + z_1\,z_4 + z_1^2\,z_4 -
   2\,z_1\,z_2\,z_4 + z_5 - z_2^2\,z_5 - z_3\,z_5 + z_1\,z_3\,z_5 + z_4\,z_5
+ 2\,z_1\,z_6 + 2\,z_1^2\,z_6 +
   z_1^3\,z_6 - 2\,z_2\,z_6 - 2\,z_1\,z_2\,z_6 - 2\,z_1^2\,z_2\,z_6 +
z_2^2\,z_6 - z_3\,z_6 - 3\,z_1\,z_3\,z_6 +
   z_2\,z_3\,z_6 - z_5\,z_6 + z_1\,z_5\,z_6 - z_6^2 - 2\,z_1\,z_6^2 +
z_2\,z_6^2 + z_7 + 3\,z_1\,z_7 + z_1^2\,z_7 +
   2\,z_2\,z_7 + 4\,z_1\,z_2\,z_7 + 2\,z_1^2\,z_2\,z_7 - z_1\,z_2^2\,z_7 +
z_3\,z_7 + 2\,z_1\,z_3\,z_7 + z_1^2\,z_3\,z_7 -
   z_3^2\,z_7 - z_4\,z_7 + z_1\,z_4\,z_7 + z_6\,z_7 + z_1\,z_6\,z_7 +
z_1^2\,z_6\,z_7 - z_2\,z_6\,z_7 - 2\,z_3\,z_6\,z_7 -
   z_6^2\,z_7 + 2\,z_7^2 + 2\,z_1\,z_7^2 + z_2\,z_7^2 + z_1\,z_2\,z_7^2 +
z_3\,z_7^2 + z_6\,z_7^2 + z_7^3 +
   2\,z_8 - 2\,z_1\,z_8 - 2\,z_1^2\,z_8 - z_1^3\,z_8 + z_1^4\,z_8 + z_2\,z_8
+ z_1\,z_2\,z_8 - 2\,z_1^3\,z_2\,z_8 -
   z_2^2\,z_8 - z_1\,z_2^2\,z_8 + z_3\,z_8 + 4\,z_1\,z_3\,z_8 -
3\,z_1^2\,z_3\,z_8 + 3\,z_1\,z_2\,z_3\,z_8 - z_3^2\,z_8 +
   z_1\,z_4\,z_8 - z_2\,z_4\,z_8 + z_5\,z_8 - 3\,z_1\,z_5\,z_8 +
z_1^2\,z_5\,z_8 + z_2\,z_5\,z_8 + 4\,z_1\,z_6\,z_8 -
   2\,z_1^2\,z_6\,z_8 - z_2\,z_6\,z_8 + 2\,z_1\,z_2\,z_6\,z_8 -
z_3\,z_6\,z_8 + 2\,z_7\,z_8 - 3\,z_1\,z_7\,z_8 -
   3\,z_1^2\,z_7\,z_8 + z_1^3\,z_7\,z_8 + z_2\,z_7\,z_8 +
z_1\,z_2\,z_7\,z_8 - z_2^2\,z_7\,z_8 + 4\,z_3\,z_7\,z_8 -
   2\,z_1\,z_3\,z_7\,z_8 + 3\,z_6\,z_7\,z_8 - z_1\,z_6\,z_7\,z_8 -
4\,z_1\,z_8^2 + 2\,z_1^2\,z_8^2 - z_1^3\,z_8^2 -
   z_2\,z_8^2 - 5\,z_1\,z_2\,z_8^2 + z_1^2\,z_2\,z_8^2 + z_3\,z_8^2 -
3\,z_1\,z_3\,z_8^2 + 2\,z_2\,z_3\,z_8^2 -
   z_1\,z_5\,z_8^2 - z_6\,z_8^2 + 2\,z_2\,z_6\,z_8^2 - 2\,z_7\,z_8^2 -
z_1\,z_7\,z_8^2 - z_1^2\,z_7\,z_8^2 -
   2\,z_2\,z_7\,z_8^2 - z_3\,z_7\,z_8^2 - z_6\,z_7\,z_8^2 - z_7^2\,z_8^2 -
z_8^3 + 2\,z_1\,z_8^3 - 2\,z_2\,z_8^3 +
   2\,z_1\,z_2\,z_8^3 - z_3\,z_8^3 + z_7\,z_8^3 - z_1\,z_7\,z_8^3 +
z_1\,z_8^4 + z_2\,z_8^4\\ $

$\bchi_{2, 0, 0, 0, 0, 2, 0, 0} =
  -z_1 - 2\,z_1^2 - z_1^3 - z_2^2 - z_1\,z_2^2 - z_2^3 + z_3 + z_1\,z_3 +
z_2\,z_3 + z_4 + 2\,z_1\,z_4 +
   2\,z_2\,z_4 + z_3\,z_4 - z_5 - 2\,z_1\,z_5 - z_1^2\,z_5 - 3\,z_2\,z_5 -
2\,z_1\,z_2\,z_5 - z_2^2\,z_5 + z_4\,z_5 -
   z_5^2 + 2\,z_6 + 3\,z_1\,z_6 + 2\,z_2\,z_6 - z_1^2\,z_2\,z_6 +
2\,z_3\,z_6 + z_1\,z_3\,z_6 + z_2\,z_3\,z_6 +
   z_4\,z_6 + z_1\,z_5\,z_6 + z_6^2 + z_1^2\,z_6^2 - z_3\,z_6^2 - z_6^3 -
3\,z_1\,z_7 - 4\,z_1^2\,z_7 -
   2\,z_1^3\,z_7 + z_1\,z_2\,z_7 - z_2^2\,z_7 - z_1\,z_2^2\,z_7 +
2\,z_3\,z_7 + 3\,z_1\,z_3\,z_7 + z_2\,z_3\,z_7 +
   z_4\,z_7 + 2\,z_1\,z_4\,z_7 - z_5\,z_7 - 3\,z_1\,z_5\,z_7 -
z_1^2\,z_5\,z_7 + z_3\,z_5\,z_7 + 3\,z_6\,z_7 +
   4\,z_1\,z_6\,z_7 - z_1\,z_2\,z_6\,z_7 + z_3\,z_6\,z_7 +
2\,z_5\,z_6\,z_7 - 3\,z_1\,z_7^2 - 2\,z_1^2\,z_7^2 -
   z_1^3\,z_7^2 + z_1\,z_2\,z_7^2 + z_3\,z_7^2 + 2\,z_1\,z_3\,z_7^2 +
z_6\,z_7^2 + z_1\,z_6\,z_7^2 - z_1\,z_7^3 -
   2\,z_8 - 4\,z_1\,z_8 - 4\,z_2\,z_8 + z_1\,z_2\,z_8 + z_1^2\,z_2\,z_8 -
2\,z_2^2\,z_8 + z_1\,z_2^2\,z_8 + z_4\,z_8 -
   3\,z_5\,z_8 - 2\,z_1\,z_5\,z_8 - 2\,z_2\,z_5\,z_8 + z_1\,z_2\,z_5\,z_8 +
z_3\,z_5\,z_8 - z_5^2\,z_8 + 2\,z_6\,z_8 +
   3\,z_1\,z_6\,z_8 - z_1^2\,z_6\,z_8 + z_1^3\,z_6\,z_8 + 2\,z_2\,z_6\,z_8 +
2\,z_1\,z_2\,z_6\,z_8 + z_2^2\,z_6\,z_8 +
   3\,z_3\,z_6\,z_8 - 2\,z_1\,z_3\,z_6\,z_8 - z_4\,z_6\,z_8 +
3\,z_5\,z_6\,z_8 + 2\,z_6^2\,z_8 - 2\,z_1\,z_6^2\,z_8 -
   4\,z_7\,z_8 - 8\,z_1\,z_7\,z_8 + z_1^2\,z_7\,z_8 - 4\,z_2\,z_7\,z_8 +
z_1\,z_2\,z_7\,z_8 + z_1^2\,z_2\,z_7\,z_8 -
   z_2^2\,z_7\,z_8 - z_2\,z_3\,z_7\,z_8 - 3\,z_5\,z_7\,z_8 +
2\,z_6\,z_7\,z_8 + 3\,z_1\,z_6\,z_7\,z_8 - 2\,z_7^2\,z_8 -
   4\,z_1\,z_7^2\,z_8 + z_1^2\,z_7^2\,z_8 + z_6\,z_7^2\,z_8 - 2\,z_8^2 +
2\,z_1\,z_8^2 + 2\,z_1^2\,z_8^2 -
   2\,z_2\,z_8^2 - z_3\,z_8^2 - z_2\,z_3\,z_8^2 - z_1\,z_4\,z_8^2 -
2\,z_5\,z_8^2 + 3\,z_1\,z_5\,z_8^2 + z_6\,z_8^2 +
   z_1\,z_6\,z_8^2 - z_1^2\,z_6\,z_8^2 - z_3\,z_6\,z_8^2 - 2\,z_6^2\,z_8^2 -
3\,z_7\,z_8^2 + 3\,z_1\,z_7\,z_8^2 +
   2\,z_1^2\,z_7\,z_8^2 - 2\,z_2\,z_7\,z_8^2 - 2\,z_1\,z_2\,z_7\,z_8^2 -
2\,z_3\,z_7\,z_8^2 - z_7^2\,z_8^2 +
   z_1\,z_7^2\,z_8^2 + z_8^3 + 4\,z_1\,z_8^3 + 2\,z_2\,z_8^3 -
z_1\,z_2\,z_8^3 + z_2^2\,z_8^3 + 2\,z_5\,z_8^3 +
   z_6\,z_8^3 - 2\,z_1\,z_6\,z_8^3 + z_7\,z_8^3 + 3\,z_1\,z_7\,z_8^3 +
z_8^4 - z_1\,z_8^4 + z_2\,z_8^4 -
   z_6\,z_8^4 - z_1\,z_8^5\\ $

$\bchi_{2, 0, 0, 0, 1, 0, 1, 0} =
  2\,z_1^2 + 3\,z_1^3 + z_1^4 + z_1\,z_2 - z_1^3\,z_2 - z_3 - 3\,z_1\,z_3 -
z_1^2\,z_3 - z_2\,z_3 +
   z_1\,z_2\,z_3 + z_2^2\,z_3 - z_1\,z_4 - z_2\,z_4 + z_1\,z_2\,z_4 +
2\,z_1\,z_5 + 3\,z_1^2\,z_5 + z_1^3\,z_5 -
   z_3\,z_5 - 2\,z_1\,z_3\,z_5 - z_4\,z_5 - z_6 - 3\,z_1\,z_6 -
2\,z_1^2\,z_6 - z_2\,z_6 - z_1\,z_2\,z_6 - z_3\,z_6 -
   z_5\,z_6 - z_1\,z_5\,z_6 + z_1\,z_7 + 4\,z_1^2\,z_7 + 2\,z_1^3\,z_7 +
z_2\,z_7 + 2\,z_1\,z_2\,z_7 - z_1^2\,z_2\,z_7 +
   z_2^2\,z_7 - z_3\,z_7 - 2\,z_1\,z_3\,z_7 - z_2\,z_3\,z_7 - z_4\,z_7 -
z_1\,z_4\,z_7 + 2\,z_5\,z_7 + 4\,z_1\,z_5\,z_7 +
   z_1^2\,z_5\,z_7 + z_2\,z_5\,z_7 - z_3\,z_5\,z_7 - 3\,z_6\,z_7 -
3\,z_1\,z_6\,z_7 - z_2\,z_6\,z_7 - z_1\,z_2\,z_6\,z_7 -
   z_3\,z_6\,z_7 + z_1\,z_7^2 + z_1^2\,z_7^2 + z_2\,z_7^2 + z_2^2\,z_7^2 -
z_4\,z_7^2 + 2\,z_5\,z_7^2 -
   2\,z_6\,z_7^2 + 4\,z_1\,z_8 + 3\,z_1^2\,z_8 - z_1^3\,z_8 + 2\,z_2\,z_8 -
z_1^2\,z_2\,z_8 + z_2^2\,z_8 +
   z_1\,z_2^2\,z_8 - 2\,z_3\,z_8 - z_1\,z_3\,z_8 - z_1^2\,z_3\,z_8 +
z_2\,z_3\,z_8 + z_3^2\,z_8 - 3\,z_4\,z_8 -
   z_1\,z_4\,z_8 - z_1^2\,z_4\,z_8 - z_2\,z_4\,z_8 + z_3\,z_4\,z_8 +
3\,z_5\,z_8 + 2\,z_1\,z_5\,z_8 - z_1^2\,z_5\,z_8 +
   z_2\,z_5\,z_8 - z_3\,z_5\,z_8 - 4\,z_6\,z_8 - 5\,z_1\,z_6\,z_8 -
3\,z_1^2\,z_6\,z_8 - z_1^3\,z_6\,z_8 - z_2\,z_6\,z_8 +
   z_1\,z_2\,z_6\,z_8 + 4\,z_3\,z_6\,z_8 + 2\,z_1\,z_3\,z_6\,z_8 +
z_4\,z_6\,z_8 - z_5\,z_6\,z_8 + 3\,z_6^2\,z_8 +
   z_1\,z_6^2\,z_8 + 4\,z_7\,z_8 + 9\,z_1\,z_7\,z_8 + z_1^2\,z_7\,z_8 +
5\,z_2\,z_7\,z_8 - 2\,z_1\,z_2\,z_7\,z_8 +
   z_1^2\,z_2\,z_7\,z_8 + 2\,z_2^2\,z_7\,z_8 - 2\,z_3\,z_7\,z_8 -
z_2\,z_3\,z_7\,z_8 - 4\,z_4\,z_7\,z_8 + 6\,z_5\,z_7\,z_8 -
   2\,z_1\,z_5\,z_7\,z_8 - 6\,z_6\,z_7\,z_8 - 3\,z_1\,z_6\,z_7\,z_8 -
z_2\,z_6\,z_7\,z_8 + 7\,z_7^2\,z_8 +
   4\,z_1\,z_7^2\,z_8 + 2\,z_2\,z_7^2\,z_8 + 3\,z_7^3\,z_8 + 5\,z_8^2 +
z_1\,z_8^2 - 5\,z_1^2\,z_8^2 -
   2\,z_1^3\,z_8^2 + 2\,z_2\,z_8^2 - 2\,z_1\,z_2\,z_8^2 +
2\,z_1^2\,z_2\,z_8^2 - z_1\,z_2^2\,z_8^2 - z_3\,z_8^2 +
   4\,z_1\,z_3\,z_8^2 - z_2\,z_3\,z_8^2 - 2\,z_4\,z_8^2 + 2\,z_1\,z_4\,z_8^2
+ 2\,z_5\,z_8^2 - 4\,z_1\,z_5\,z_8^2 +
   z_2\,z_5\,z_8^2 - 4\,z_6\,z_8^2 + 4\,z_1\,z_6\,z_8^2 +
z_1^2\,z_6\,z_8^2 - 3\,z_2\,z_6\,z_8^2 + z_3\,z_6\,z_8^2 +
   13\,z_7\,z_8^2 - 2\,z_1\,z_7\,z_8^2 + 2\,z_2\,z_7\,z_8^2 -
2\,z_1\,z_2\,z_7\,z_8^2 - 2\,z_3\,z_7\,z_8^2 -
   4\,z_6\,z_7\,z_8^2 + 8\,z_7^2\,z_8^2 + 2\,z_8^3 - 9\,z_1\,z_8^3 +
2\,z_1^2\,z_8^3 - z_2\,z_8^3 +
   2\,z_3\,z_8^3 + 2\,z_4\,z_8^3 - 2\,z_5\,z_8^3 + 2\,z_6\,z_8^3 +
3\,z_1\,z_6\,z_8^3 - 6\,z_1\,z_7\,z_8^3 -
   z_7^2\,z_8^3 - 6\,z_8^4 + 3\,z_1\,z_8^4 - z_2\,z_8^4 + z_6\,z_8^4 -
6\,z_7\,z_8^4 + z_1\,z_8^5 + z_8^6\\ $

$\bchi_{2, 2, 0, 0, 0, 0, 0, 1} =
  z_1^2 + z_1^3 + z_1^4 + z_1^5 + 3\,z_1\,z_2 + 3\,z_1^2\,z_2 + z_1^3\,z_2 +
2\,z_2^2 + 3\,z_1\,z_2^2 +
   z_1^2\,z_2^2 + z_2^3 - z_1\,z_2^3 - z_1^2\,z_3 - 3\,z_1^3\,z_3 -
z_1^2\,z_2\,z_3 - z_2^2\,z_3 +
   2\,z_1\,z_3^2 + z_2\,z_3^2 - z_1\,z_4 + z_1^2\,z_4 - 2\,z_2\,z_4 +
2\,z_1\,z_2\,z_4 + z_1\,z_5 + z_1^3\,z_5 +
   z_2\,z_5 - z_1\,z_2\,z_5 + z_2^2\,z_5 - z_1\,z_3\,z_5 - z_4\,z_5 +
z_1\,z_6 - 2\,z_1^3\,z_6 + z_2\,z_6 +
   z_1\,z_2\,z_6 + z_1^2\,z_2\,z_6 - z_2^2\,z_6 + 3\,z_1\,z_3\,z_6 -
z_4\,z_6 - 2\,z_1\,z_5\,z_6 + z_1\,z_6^2 -
   z_2\,z_6^2 + z_1\,z_7 + z_1^2\,z_7 + z_1^3\,z_7 + z_1^4\,z_7 +
3\,z_2\,z_7 + 4\,z_1\,z_2\,z_7 +
   2\,z_1^2\,z_2\,z_7 + 2\,z_2^2\,z_7 + z_1\,z_2^2\,z_7 - z_1\,z_3\,z_7 -
2\,z_1^2\,z_3\,z_7 - z_2\,z_3\,z_7 +
   z_3^2\,z_7 + z_5\,z_7 + z_2\,z_5\,z_7 + z_6\,z_7 + z_1\,z_6\,z_7 -
2\,z_1^2\,z_6\,z_7 + z_2\,z_6\,z_7 + z_3\,z_6\,z_7 +
   z_6^2\,z_7 - z_7^2 + 2\,z_2\,z_7^2 + z_1\,z_2\,z_7^2 + z_6\,z_7^2 - z_7^3
+ 2\,z_1\,z_8 - z_1^2\,z_8 -
   z_1^4\,z_8 + 3\,z_2\,z_8 - 3\,z_1\,z_2\,z_8 - 2\,z_1^2\,z_2\,z_8 -
2\,z_1\,z_2^2\,z_8 + z_1^2\,z_2^2\,z_8 -
   2\,z_1\,z_3\,z_8 - z_1\,z_2\,z_3\,z_8 - z_2^2\,z_3\,z_8 + z_1\,z_4\,z_8 -
z_1^2\,z_4\,z_8 + z_2\,z_4\,z_8 +
   z_3\,z_4\,z_8 + z_5\,z_8 - z_1^2\,z_5\,z_8 - z_1\,z_2\,z_5\,z_8 +
z_3\,z_5\,z_8 + z_6\,z_8 - 2\,z_1\,z_6\,z_8 -
   z_1^3\,z_6\,z_8 + 2\,z_1\,z_3\,z_6\,z_8 + z_4\,z_6\,z_8 +
z_1\,z_6^2\,z_8 - z_7\,z_8 - z_1\,z_7\,z_8 - z_1^2\,z_7\,z_8 +
   z_2\,z_7\,z_8 - 4\,z_1\,z_2\,z_7\,z_8 - z_1^2\,z_2\,z_7\,z_8 -
z_3\,z_7\,z_8 - 2\,z_1\,z_3\,z_7\,z_8 - z_2\,z_3\,z_7\,z_8 +
   z_4\,z_7\,z_8 + z_1\,z_5\,z_7\,z_8 - z_1\,z_6\,z_7\,z_8 - 3\,z_7^2\,z_8 -
z_1\,z_7^2\,z_8 + z_1^2\,z_7^2\,z_8 -
   z_2\,z_7^2\,z_8 - z_3\,z_7^2\,z_8 - z_6\,z_7^2\,z_8 - z_7^3\,z_8 -
2\,z_1\,z_8^2 - z_1^3\,z_8^2 - z_1^4\,z_8^2 -
   3\,z_2\,z_8^2 - z_1\,z_2\,z_8^2 - z_1^2\,z_2\,z_8^2 - z_2^2\,z_8^2 -
z_3\,z_8^2 + 3\,z_1^2\,z_3\,z_8^2 +
   z_2\,z_3\,z_8^2 - z_3^2\,z_8^2 - z_5\,z_8^2 - z_2\,z_5\,z_8^2 -
2\,z_6\,z_8^2 + z_1\,z_6\,z_8^2 +
   2\,z_1^2\,z_6\,z_8^2 + z_2\,z_6\,z_8^2 - z_3\,z_6\,z_8^2 -
2\,z_7\,z_8^2 - 2\,z_2\,z_7\,z_8^2 -
   z_1\,z_2\,z_7\,z_8^2 - z_3\,z_7\,z_8^2 - z_1\,z_8^3 + z_1^3\,z_8^3 -
z_2\,z_8^3 + 2\,z_1\,z_2\,z_8^3 +
   z_3\,z_8^3 + z_6\,z_8^3 + 2\,z_7\,z_8^3 - z_1\,z_7\,z_8^3 + z_1\,z_8^4 +
z_2\,z_8^4\\ $

$\bchi_{3, 0, 0, 0, 0, 1, 0, 1} =
  -z_1^3 - z_1^4 + z_2 + z_1\,z_2 - z_1\,z_2^2 - z_1^2\,z_2^2 + 2\,z_1\,z_3
+ 3\,z_1^2\,z_3 + z_1\,z_2\,z_3 +
   z_2^2\,z_3 - z_3^2 + z_1^2\,z_4 - z_2\,z_4 - z_3\,z_4 + z_5 -
3\,z_1^2\,z_5 - z_1^3\,z_5 + 2\,z_2\,z_5 +
   2\,z_3\,z_5 + 2\,z_1\,z_3\,z_5 - z_4\,z_5 + z_5^2 + z_6 + 3\,z_1\,z_6 +
3\,z_1^2\,z_6 - z_2\,z_6 - z_1\,z_2\,z_6 +
   z_2^2\,z_6 - 3\,z_3\,z_6 - 2\,z_4\,z_6 + 3\,z_5\,z_6 + z_1\,z_5\,z_6 -
3\,z_6^2 - z_1\,z_6^2 - z_7 - z_1\,z_7 -
   z_1^3\,z_7 - z_1^4\,z_7 + z_2\,z_7 + z_1\,z_2\,z_7 - z_2^2\,z_7 +
z_3\,z_7 + 2\,z_1\,z_3\,z_7 + 3\,z_1^2\,z_3\,z_7 -
   z_3^2\,z_7 + z_4\,z_7 - z_1\,z_4\,z_7 - z_5\,z_7 - z_2\,z_5\,z_7 +
3\,z_6\,z_7 + 3\,z_1\,z_6\,z_7 + z_1^2\,z_6\,z_7 -
   2\,z_3\,z_6\,z_7 - 2\,z_7^2 - z_1\,z_7^2 + z_3\,z_7^2 - z_5\,z_7^2 +
z_6\,z_7^2 - z_7^3 - z_8 - 3\,z_1\,z_8 -
   3\,z_1^2\,z_8 + z_1^3\,z_8 + 2\,z_2\,z_8 - 2\,z_1\,z_2\,z_8 -
z_1^2\,z_2\,z_8 + z_1^3\,z_2\,z_8 - z_2^2\,z_8 +
   z_1\,z_2^2\,z_8 - z_2^3\,z_8 + 3\,z_3\,z_8 + 2\,z_2\,z_3\,z_8 -
2\,z_1\,z_2\,z_3\,z_8 - z_3^2\,z_8 + 2\,z_4\,z_8 +
   z_1\,z_4\,z_8 + 2\,z_2\,z_4\,z_8 - 2\,z_5\,z_8 - 3\,z_1\,z_5\,z_8 -
3\,z_2\,z_5\,z_8 + 2\,z_3\,z_5\,z_8 + 7\,z_6\,z_8 +
   z_1\,z_6\,z_8 + 2\,z_1^2\,z_6\,z_8 + z_1^3\,z_6\,z_8 + 2\,z_2\,z_6\,z_8 +
z_1\,z_2\,z_6\,z_8 - 2\,z_3\,z_6\,z_8 -
   2\,z_1\,z_3\,z_6\,z_8 + z_4\,z_6\,z_8 - 2\,z_6^2\,z_8 - z_1\,z_6^2\,z_8 -
7\,z_7\,z_8 - 5\,z_1\,z_7\,z_8 -
   3\,z_1^2\,z_7\,z_8 + z_1^3\,z_7\,z_8 - z_1^2\,z_2\,z_7\,z_8 -
z_2^2\,z_7\,z_8 + 4\,z_3\,z_7\,z_8 - z_1\,z_3\,z_7\,z_8 +
   z_2\,z_3\,z_7\,z_8 + 3\,z_4\,z_7\,z_8 - 5\,z_5\,z_7\,z_8 +
z_1\,z_5\,z_7\,z_8 + 9\,z_6\,z_7\,z_8 + z_1\,z_6\,z_7\,z_8 +
   z_2\,z_6\,z_7\,z_8 - 8\,z_7^2\,z_8 - 2\,z_1\,z_7^2\,z_8 - z_2\,z_7^2\,z_8
+ z_3\,z_7^2\,z_8 - 2\,z_7^3\,z_8 -
   5\,z_8^2 - z_1\,z_8^2 + 2\,z_1^2\,z_8^2 + z_1^3\,z_8^2 - 5\,z_2\,z_8^2 +
2\,z_1\,z_2\,z_8^2 -
   z_1^2\,z_2\,z_8^2 + z_1\,z_2^2\,z_8^2 - 3\,z_1\,z_3\,z_8^2 + z_4\,z_8^2 -
z_1\,z_4\,z_8^2 - 3\,z_5\,z_8^2 +
   4\,z_1\,z_5\,z_8^2 - z_2\,z_5\,z_8^2 + z_6\,z_8^2 - z_1^2\,z_6\,z_8^2 +
3\,z_2\,z_6\,z_8^2 - z_3\,z_6\,z_8^2 -
   12\,z_7\,z_8^2 + z_1\,z_7\,z_8^2 + z_1^2\,z_7\,z_8^2 -
4\,z_2\,z_7\,z_8^2 - z_3\,z_7\,z_8^2 + 2\,z_6\,z_7\,z_8^2 -
   7\,z_7^2\,z_8^2 + z_1\,z_7^2\,z_8^2 + 5\,z_1\,z_8^3 - 2\,z_3\,z_8^3 -
z_4\,z_8^3 + 2\,z_5\,z_8^3 -
   2\,z_6\,z_8^3 - 3\,z_1\,z_6\,z_8^3 + 3\,z_7\,z_8^3 + 2\,z_1\,z_7\,z_8^3 +
z_7^2\,z_8^3 + 5\,z_8^4 - z_1\,z_8^4 +
   2\,z_2\,z_8^4 - z_6\,z_8^4 + 5\,z_7\,z_8^4 - z_1\,z_8^5 - z_8^6\\ $

$\bchi_{3, 0, 0, 0, 1, 0, 0, 0} =
  -z_1 - z_1^2 + z_1^3 + z_1^4 - z_2 - 2\,z_1\,z_2 - z_2^2 + z_1\,z_2^2 -
z_1\,z_3 - 2\,z_1^2\,z_3 -
   2\,z_1\,z_2\,z_3 + z_2\,z_4 - z_5 + 3\,z_1^2\,z_5 + z_1^3\,z_5 -
2\,z_2\,z_5 + z_1\,z_2\,z_5 - z_3\,z_5 -
   2\,z_1\,z_3\,z_5 + z_4\,z_5 - z_5^2 - 2\,z_1\,z_6 - 2\,z_1^2\,z_6 -
z_1\,z_2\,z_6 - z_1^2\,z_2\,z_6 + z_3\,z_6 +
   z_2\,z_3\,z_6 - z_5\,z_6 - z_1\,z_7 - 2\,z_1^2\,z_7 + z_1^3\,z_7 -
z_2\,z_7 - 3\,z_1\,z_2\,z_7 + z_1\,z_2^2\,z_7 -
   z_1\,z_3\,z_7 - z_1\,z_4\,z_7 + z_1\,z_5\,z_7 - 4\,z_1\,z_6\,z_7 -
z_2\,z_6\,z_7 + z_3\,z_6\,z_7 - z_1^2\,z_7^2 -
   z_1\,z_2\,z_7^2 - z_8 + z_1\,z_8 + 2\,z_1^2\,z_8 - 2\,z_2\,z_8 +
2\,z_1\,z_2\,z_8 - z_1^2\,z_2\,z_8 + z_2^2\,z_8 +
   2\,z_1\,z_2^2\,z_8 - z_2^3\,z_8 - 2\,z_1\,z_3\,z_8 + z_2\,z_3\,z_8 +
z_3^2\,z_8 + 2\,z_4\,z_8 - z_1\,z_4\,z_8 +
   2\,z_2\,z_4\,z_8 - 3\,z_5\,z_8 + z_1\,z_5\,z_8 - z_1^2\,z_5\,z_8 -
2\,z_2\,z_5\,z_8 - 2\,z_3\,z_5\,z_8 + 2\,z_6\,z_8 -
   z_1\,z_6\,z_8 - 2\,z_1^2\,z_6\,z_8 + 5\,z_2\,z_6\,z_8 +
2\,z_1\,z_2\,z_6\,z_8 + 3\,z_3\,z_6\,z_8 + 2\,z_6^2\,z_8 +
   5\,z_1\,z_7\,z_8 + z_1^2\,z_7\,z_8 - z_2\,z_7\,z_8 + z_3\,z_7\,z_8 -
2\,z_5\,z_7\,z_8 + 2\,z_7^2\,z_8 +
   4\,z_1\,z_7^2\,z_8 + z_7^3\,z_8 - 2\,z_8^2 + 3\,z_1\,z_8^2 -
2\,z_1^2\,z_8^2 - z_1^3\,z_8^2 - 3\,z_2\,z_8^2 +
   z_1\,z_2\,z_8^2 - z_2^2\,z_8^2 + 2\,z_1\,z_3\,z_8^2 - z_2\,z_3\,z_8^2 -
2\,z_5\,z_8^2 - z_1\,z_5\,z_8^2 +
   z_6\,z_8^2 + 2\,z_1\,z_6\,z_8^2 + z_2\,z_6\,z_8^2 - z_7\,z_8^2 +
z_1\,z_7\,z_8^2 + z_1^2\,z_7\,z_8^2 -
   7\,z_2\,z_7\,z_8^2 - z_3\,z_7\,z_8^2 - 2\,z_6\,z_7\,z_8^2 +
z_7^2\,z_8^2 - 3\,z_1\,z_8^3 + z_3\,z_8^3 +
   z_5\,z_8^3 + 2\,z_6\,z_8^3 - 2\,z_7\,z_8^3 - 2\,z_1\,z_7\,z_8^3 +
z_1\,z_8^4 + 2\,z_2\,z_8^4 - z_7\,z_8^4 + z_8^5\\ $

$\bchi_{0, 0, 0, 0, 0, 0, 0, 6} =
  z_1^2 + z_1^3 + z_2 + 2\,z_1\,z_2 - 2\,z_1\,z_3 - z_2\,z_3 + z_4 + z_5 +
2\,z_1\,z_5 + z_6 + z_1\,z_6 +
   z_2\,z_6 + z_6^2 - z_7 - z_1\,z_7 + z_1^2\,z_7 + z_2\,z_7 - z_3\,z_7 +
2\,z_5\,z_7 - z_6\,z_7 - 2\,z_7^2 -
   z_1\,z_7^2 - z_7^3 - z_8 - z_1\,z_8 - 2\,z_1^2\,z_8 + 2\,z_4\,z_8 -
z_6\,z_8 - 2\,z_1\,z_6\,z_8 - 2\,z_7\,z_8 -
   4\,z_1\,z_7\,z_8 - 2\,z_2\,z_7\,z_8 - 6\,z_6\,z_7\,z_8 - z_7^2\,z_8 -
z_8^2 - 2\,z_2\,z_8^2 - 3\,z_5\,z_8^2 -
   2\,z_6\,z_8^2 + 3\,z_7\,z_8^2 + 3\,z_1\,z_7\,z_8^2 + 6\,z_7^2\,z_8^2 +
2\,z_8^3 + 2\,z_1\,z_8^3 + z_2\,z_8^3 +
   4\,z_6\,z_8^3 + 3\,z_7\,z_8^3 - z_8^4 - z_1\,z_8^4 - 5\,z_7\,z_8^4 -
z_8^5 + z_8^6\\ $

\section*{Acknowledgements} This paper was completed during the visit of one
of the authors
(A. M. P.) to the Max-Planck-Institut f\"ur Gravitationsphysik,
and he thanks the institute staff for the hospitality. This work was partially supported by  Spanish Government under grants MTM2006-10532 (JFN) and FIS2006-09417 (WGF and AMP).

\end{document}